\documentclass[phd,bottom,nosig]{usbthesis}  %
\usepackage{graphicx}
\usepackage{color}
\usepackage{bm}
\usepackage{amsmath}
\usepackage{amssymb}
\usepackage{setspace}
\usepackage{array}
\usepackage{tabularx}
\usepackage{multirow}
\usepackage{afterpage}
\usepackage{hhline}
\usepackage{url}
\usepackage{comment}
\usepackage{subcaption}

\usepackage{multirow}
\usepackage{tabu}                      
\usepackage{booktabs}                  
\usepackage{algorithm}
\usepackage{algorithmic}
\usepackage[hidelinks]{hyperref}
\usepackage{footmisc} 
\usepackage{xspace}
\usepackage[numbers,sort&compress]{natbib}
\usepackage{pdfpages}
\usepackage{footnote}					
\usepackage{microtype}					
\author{Bhavya Ghai}%
\title{Towards Fair and Explainable AI using a Human-Centered AI Approach}%

\month{\textbf{May}}
\year{\textbf{2023}}%
\program{Computer Science}%
\director{\textbf{Klaus~Mueller}}{Professor, Department of Computer Science}%
\chairman{\textbf{H.~Andrew~Schwartz}}{Professor, Department of Computer Science}%
\fstmember{\textbf{Susan~E.~Brennan}}{Professor, Department of Psychology}%
\outmember{\textbf{Q. Vera Liao}}{Principal Researcher}{Microsoft Research}%
\dean{Celia Marshik}%

\begin{document}

\singlespacing %
\pagenumbering{roman} %
\thispagestyle{empty}
\begin{center}
\noindent {\LARGE \textbf{Towards Fair and Explainable AI using}}

\vspace{0.3cm}

\noindent {\LARGE \textbf{a Human-Centered AI Approach}}

\vspace{4.5cm}

\noindent A Dissertation Presented

\vspace{0.6cm}

\noindent by 

\vspace{0.6cm}

\noindent {\large \textbf{Bhavya Ghai}}

\vspace{0.6cm}

\noindent to

\vspace{0.6cm}

\noindent The Graduate School

\vspace{0.6cm}

\noindent in Partial Fulfillment of the Requirements

\vspace{0.6cm}

\noindent for the degree of 

\vspace{1.6cm}

\noindent {\large \textbf{Doctor of Philosophy}}

\vspace{0.6cm}

\noindent in

\vspace{0.6cm}

\noindent {\large \textbf{Computer Science}}

\vspace{1.6cm}

\noindent Stony Brook University

\vspace{0.6cm}

\noindent {\textbf{May 2023}}


\end{center}

\begin{center}
\vspace{1.2cm}

\noindent \textbf{Stony Brook University}

\vspace{0.45cm}

\noindent The Graduate School

\vspace{1.2cm}

\noindent {\large \textbf{Bhavya Ghai}}

\vspace{1.2cm}

\noindent {We}, the dissertation committee for the above candidate for the

\vspace{0.2cm}
\noindent Doctor of Philosophy degree, hereby recommend

\vspace{0.2cm}
\noindent acceptance of this dissertation.


\vspace{1.5cm}

\noindent \textbf{Klaus Mueller} -- Dissertation Advisor

\noindent Professor, Department of Computer Science

\vspace{1cm}

\noindent \textbf{H. Andrew Schwartz} -- Dissertation Committee Chair

\noindent Associate Professor, Department of Computer Science

\vspace{1cm}

\noindent \textbf{Susan E. Brennan}

\noindent Professor, Department of Psychology

\vspace{1cm}

\noindent \textbf{Q. Vera Liao}

\noindent Principal Researcher, Microsoft Research
\end{center}

\begin{center}
\vspace{2cm}

\noindent This dissertation is accepted by the Graduate School

\vspace{1cm}

\noindent Celia Marshik \\
\noindent Dean of the Graduate School

\end{center}

\begin{abstract}

With the rise of machine learning, people are being increasingly impacted by algorithms that are getting deployed to different areas including high-stake domains like education, healthcare, law enforcement, hiring, credit, etc. This rise is accompanied by several high-profile cases that have stressed the need for fairness, accountability, explainability and trust in ML systems. The existing literature has largely focused on fully automated ML approaches that try to optimize for some performance metric. However, human-centric measures like fairness, trust, explainability, etc. are subjective in nature, context-dependent, and might not correlate with conventional performance metrics. To deal with these challenges, we explore a human-centered AI approach that empowers people by providing more transparency and human control.   

In this dissertation, we present 5 research projects that aim to enhance explainability and fairness in classification systems and word embeddings. The first project explores the utility/downsides of introducing local model explanations as interfaces for machine teachers (crowd workers). Our study found that adding explanations supports trust calibration for the resulting ML model and enables rich forms of teaching feedback. The second project presents D-BIAS, a causality-based human-in-the-loop visual tool for identifying and mitigating social biases in tabular datasets. Apart from fairness, we found that our tool also enhances trust and accountability. The third project presents WordBias, a visual interactive tool that helps audit pre-trained static word embeddings for biases against groups, such as females, or subgroups, such as Black Muslim females. The fourth project presents DramatVis Personae, a visual analytics tool that helps identify social biases in creative writing. Finally, the last project presents an empirical study aimed at understanding the cumulative impact of multiple fairness-enhancing interventions at different stages of the ML pipeline on fairness, utility and different population groups. 

We conclude this dissertation by discussing some of the future directions. This includes research problems like mitigating social biases in word embedding using a human-centered AI approach, developing an AI-powered writing assistant for gender-inclusive writing, curating fairer datasets by measuring and accounting for social biases of crowd workers, etc.
 
\clearpage
\end{abstract}
\chapter*{List of Publications}
\begin{itemize}
  \item D-BIAS: A Causality-Based Human-in-the-Loop System for Tackling Algorithmic Bias \\
  \textbf{Bhavya Ghai} and Klaus Mueller \\
  IEEE Transactions on Visualization and Computer Graphics, 2022.

  \item Cascaded Debiasing: Studying the Cumulative Effect of Multiple Fairness-Enhancing Interventions \\
\textbf{Bhavya Ghai}, Mihir Mishra, Klaus Mueller \\
ACM Conference on Information and Knowledge Management, Atlanta, CIKM 2022
 
  \item DramatVis Personae: Visual Text Analytics for Identifying Social Biases in Creative Writing \\
Md Naimul Hoque, \textbf{Bhavya Ghai}, Niklas Elmqvist \\
ACM Conference on Designing Interactive Systems, DIS 2022
 
  \item Fluent: An AI Augmented Writing Tool for People who Stutter \\
\textbf{Bhavya Ghai} and Klaus Mueller \\
ACM SIGACCESS Conference on Computers and Accessibility, 2021

\item Explainable Active Learning (XAL): Toward AI Explanations as Interfaces for Machine Teachers \\
\textbf{Bhavya Ghai}, Q. Vera Liao, Yunfeng Zhang, Rachel K. E. Bellamy, Klaus Mueller \\
ACM Conference on Computer-Supported Cooperative Work and Social Computing, CSCW 2020

\item WordBias: An Interactive Visual Tool for Exploring Intersectional Social Biases Encoded in Word Embeddings \\
\textbf{Bhavya Ghai}, Md Naimul Hoque, Klaus Mueller \\
Late Breaking Work, ACM Conference on Human Factors in Computing Systems, CHI 2021

\end{itemize}
\tableofcontents %
\listoffigures %
\listoftables %

\begin{acknowledgements}
\emph{My PhD journey would not have been possible without the support and blessings of a number of important people in my life. Firstly, I would like to thank my late grandparents, especially my grandmother Mrs. Rajrani Ghai, who have played a fundamental role in my upbringing and shaped my value system and outlook towards life. Their blessings and nurturing motivated me towards academic excellence and propelled me so far in my academic journey. I would like to thank my parents Mr. Anil Kumar Ghai and Mrs. Sunanda Ghai for their constant support and motivation throughout my PhD journey. Along with my sister Ms. Divya Ghai, they provided the much needed emotional support to deal with difficult times and bounce back.}

\emph{Next, I would like to thank my advisor Prof. Klaus Mueller and my lab mates for helping me transform from a newly minted engineer to a budding researcher. A big credit goes to my advisor who took me under his wing and gave me the independence to pursue my own research ideas without worrying about funding. I really appreciate his cool temperament, his management style and I am deeply grateful for his belief in me.    
}

\emph{Apart from my research lab, I want to thank different researchers who mentored me at some point and had a significant impact on my research life. I would like to start with Prof. Suresh Venkatasubramanian from Brown University who introduced me to the world of algorithmic fairness which became the primary theme of my dissertation. I want to thank Prof. Ellen W. Zegura from Georgia Institute of Technology, Mrs. Buvana Ramanan from Nokia Bell Labs, Dr. Q. Vera Liao from Microsoft Research and Dr. Yunfeng Zhang from Twitter for their support and mentorship during different summer internships. I want to thank Prof. Robert J. Harrison and the Institute of Advanced Computational Sciences (IACS) for supporting my research. Also, I want to thank my dissertation committee members Prof. Susan E. Brennan and Prof. H. Andrew Schwartz for their time, encouragement and support. }

\emph{Lastly, I want to thank my friends, especially Dr. Pranjal Sahu and Md. Naimul Hoque, for their personal and technical support. }
\end{acknowledgements}

\pagestyle{thesis}
\pagenumbering{arabic}
\newpage

%
%
\let\textcircled=\pgftextcircled
\chapter{Introduction}
\label{chap:introduction}
Phrases like ``Responsible AI", ``Trustworthy AI", ``Ethical AI", ``Explainable AI", etc. have emerged and gained popularity in the past few years. The rise of these paradigms can be predicated on the challenges emanating from the rapid advancements and adoption of AI systems in different important spheres of life like hiring, education, criminal justice, etc. With algorithms touching the lives of millions of people on a daily basis, there is a need to understand and regulate such algorithms better. Different technology companies and governmental organizations have come up with guidelines and laws that emphasize the need for fairness, transparency, privacy, accountability, robustness, etc. in AI systems. As indicated by the title of this dissertation, our focus will be on fairness and explainability in AI systems.    

\section{Algorithmic Fairness}
``Algorithmic bias occurs when a computer system reflects the implicit values of the humans who are involved in coding, collecting, selecting, or using data to train the algorithm." -- Wikipedia
\\
Cathy O' Neil, author of Weapons of Math Destruction, describes algorithms as opinions expressed in code \cite{o2016weapons}. 
Intrinsically, algorithms are just a set of mathematical formulations designed to solve a problem. The issue arises when humans intentionally or unintentionally encode on their own social biases into algorithms. Unlike humans, algorithms are capable of propagating such biases at lightning speed on a massive scale. Furthermore, there is lack of regulations/laws to audit/question such algorithms because algorithms are generally perceived as unbiased. With such suitable conditions and rise of artificial intelligence, algorithmic bias is spreading rapidly. 
In the past few years, some of its manifestations came to public light which caused huge public outrage. There has also been an emergence of several high profile studies that have called into question the fairness, accountability, and transparency of ML systems. We have highlighted some of the examples in Sec. 1.1.3.

In the next few sections, we will talk about the sources of bias, types of algorithmic biases and how its impacting our society. 

\subsection{Sources of Algorithmic Bias}
Social biases based on race, gender, etc. can get encoded in the ML pipeline via two primary sources: training dataset and the people developing/interacting with it. Like humans, algorithms also learn from textbooks and teachers. For algorithms, training datasets act as textbooks and code developers act as teachers. If any of them is biased, then algorithms also become biased.
On their own, algorithms are just mathematical formulations designed to work under a set of assumptions. They have no knowledge about how the world works. They are devoid of morals, ethics, emotions, etc. They just try to learn patterns from data and sometimes learn social biases in the process. Bias in training data might due to historical discrimination or lack of diversity in training data. Similarly, if the team of developers writing code have similar backgrounds in terms of race, academic disciplines, gender, etc., they will have common blindspots which might cause bias.

\subsection{Types of Algorithmic Bias}
Algorithmic bias has roughly influenced all disciplines that are impacted by machine learning. If we classify algorithmic bias with respect to technical domains, we can observe bias in natural language processing \cite{100years,200sentiment,ghai2021wordbias}, computer vision \cite{genderShades}, clustering, speech \cite{accentBias}, ranking, recommender systems and classification ML systems \cite{ghai2022cascaded}. In natural language processing, we can observe bias in word embeddings. Word embeddings like Glove are shown to have gender and racial bias. Male pronouns are closer to computer programmers, doctors, etc. and female pronouns are closer to nurse, teacher, etc. In the case of computer vision, there is racial bias in identifying faces. For example, face recognition API by Face++ is more likely to identify white male over black females by over 33\% ~\cite{buolamwini2018gender}. Similarly, there are reports where Google Photos misidentified the face of an African American female as gorilla. In the world of speech processing, it seems like not all accents are created equal \cite{accentBias}. Talking in a different regional accent like southern, west coast or an international accent like spanish, chinese, etc. can seriously impact how well the voice assistant understands you. Recommender systems like Google Ads are found to be racially biased \cite{sweeney2013discrimination}.            

If we analyze bias with respect to its impact on society, it can be categorized into \textit{allocation harm} and \textit{representation harm} \cite{kateNIPS}. Kate Crawford, Principal Researcher at Microsoft Research, talked about allocative harm and representation harm in her keynote talk at NIPS 2017 conference. Allocative harm occurs when a system allocates or withholds a resource/opportunity from a section of society. It includes biases in decisions like who gets a job, college admission, mortgage, loan, etc. On the other hand, representation harm occurs when a system reinforces subordination of certain sections of society. The impact of representation bias is not immediately felt. It slowly percolates into the society and primarily impacts minorities.     

\subsection{Impact on Society}
\label{sec:impact}
Assessing the impact of algorithmic bias on society is a research problem in itself that needs further investigation. Most ML models do not have a customer facing end so it is very difficult to track biases in those systems. However, in this section, I have tried to give various real world examples on how algorithmic bias is impacting our society.  

\textbf{Recidivism Score}
\label{subsec:propublica}
A proprietary software system called COMPAS is used to predict recidivism or the likelihood of a person committing future crime. This system assigns a number on a 1-10 scale that determines how risky a person is to commit future crimes. 1 being least risky and 10 being highly risky. These numbers are called risk scores and are used by judges while deciding if a person can be set free at every stage of the criminal justice system. For 9 states in the US, namely Arizona, Colorado, Delaware, Kentucky, Louisiana, Oklahoma, Virginia, Washington and Wisconsin, these risk scores are even used for criminal sentencing. ProPublica, a non-profit based in New York city, investigated these risk scores. They found that COMPAS risk scores are racially biased \cite{angwin2016machine}. In other words, for similar criminal history a black person might be assigned a higher risk score than a while one.

For example, an 18-year old black girl Brisha borden was assigned a high-risk score of 8 after she was caught taking a kid’s bike and scooter. She has also been involved with juvenile misdemeanors when she was juvenile. On the other hand, a 41-year old white man Vernon Prater was assigned a low-risk score of 3 when he was caught shoplifting from a Home depot store. Vernon is a seasoned criminal who has been convicted of 2 armed robberies and 1 attempted armed robbery. He has served 5 years in prison for the same. After few years, it was observed that the high risk black girl did not commit any future crimes. On the other hand, the low risk white guy broke in a warehouse and stole electronic goods worth thousands of dollars. 
   Such risk scores directly influence the judgements as they are data driven and perceptually unbiased. Even if the COMPAS algorithm has been created with best of intentions, racial bias can creep in through the biased dataset it was trained on. The story of Brisha and Vernon is just one example. ProPublica studied risk scores for 7,000 people in Broward county, Florida. Their investigation showed that the risk scores are inaccurate and biased. Only 20 percent of people who were predicted to commit crimes actually went on to do so. The error rate for committing crime was about the same for different races but the nature of the error was biased. The system was twice as likely to flag black defendants as future criminals than white criminals. White defendants are more likely to be mislabeled low risk than black defendants. Such algorithms are poisoning our criminal system and trying to propagate racial bias at a massive scale. 

\textbf{Youtube Recommender System}
Youtube is a for-profit company whose revenue is directly dependent on the time a user spends on its platform. Youtube recommender system uses machine learning techniques to keep a user engaged. In the quest of keeping users engaged, youtube's recommender system has inferred that people are more attracted towards more extreme/incendiary content \cite{youtube}. This recommender system guides the autoplay feature and the set of recommended videos. Zeynep Tufekci, Professor at UNC, observed that watching Donald Trump rallies videos lead to white extremist videos. Thereafter, she watched Hillary Clinton and Bernie Sanders videos which directed her towards leftish conspiratorial videos. This pattern generalizes to other genres well. For example, watching vegetarianism videos will lead to veganism. Jogging videos can lead to videos about running marathons. End users might feel that they are never good enough for youtube standards as it always keep pushing to more extreme content. ``YouTube leads viewers down a rabbit hole of extremism, while Google racks up the ad sales" \cite{youtube}.

\textbf{Google Translate}
Google Translate is a free online tool that translates text from one language to another. It supports over 100 languages and serves over 500 million users daily \cite{translate}. This service is based on machine learning algorithms trained on millions of documents. It's sometimes criticized for inaccuracy but recently it has been found to encourage gender stereotypes. 

\begin{figure}[tb]
 \centering 
 \includegraphics[width=1.0\columnwidth]{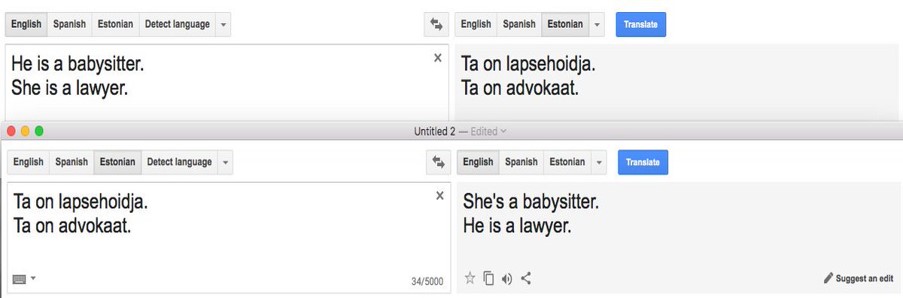}
  \setlength{\belowcaptionskip}{-8pt}
\setlength{\abovecaptionskip}{-4pt}
 \caption{Google translate tries to enforce that 'He' should be a lawyer and 'She' should be associated to babysitter}
 \label{fig:translate}
\end{figure}
 
On converting two sentences, i.e., ``He is a babysitter" and ``She is a lawyer" from english to estonian and back, we observe that the pronoun is reversed as shown in figure \ref{fig:translate}. It is a manifestation of the inherent gender bias in the training dataset. Such biases are not just limited to professions. When a turkish poem containing gender-neutral words along with adjectives was translated to english, it was found that words like hard working, pessimistic, strong were associated to males while beautiful, lazy, optimistic were associated to females \cite{translate2}.

\textbf{Amazon Prime}
In 2016, Amazon extended its free same day delivery to 27 metropolitan cities \cite{amazonPrime}. It tried to target regions that constitute high return customers. Bloomberg conducted a study where it analyzed different regions covered under this program. They found that the zipcodes not covered were mostly black neighborhoods \cite{amazonBloomberg}. This pattern was observed in New York city, Chicago, Atlanta, Washington D.C., Boston and Dallas. 
``The Amazon delivery pattern replicates long-standing demographic realities in the United States that in many ways come out of the the National Housing Act of 1934" - USA Today \cite{amazonPrime}. After this study became public, Amazon extended this service to black regions as well. 

\textbf{Facebook News Feed}
Facebook, one of the social media giants, recently started to serve news as well. Given its popularity, Facebook news became one of the primary news sources for millions of people around the world especially millennials. Similar to other popular online services like youtube, netflix, etc, Facebook also want its users to spend more time on the platform as part of its business model. To achieve this objective, it personalizes news feed and other posts. This is where machine learning comes in the picture and so does bias. Facebook is biased towards user's past behavior, i.e., likes, comments, shares, posts, etc. and their friends/relatives. It's possible that a user no longer affirm to their past views but facebook will rank posts pertaining to their previous behavior as more important. This will create a tunnel vision for the user who is less likely to see diverse posts. Secondly, Facebook rank posts shared/liked by friends/relatives as more important. For example, if a user's friend group have different political inclinations, then a given user might hesitate to surface their own opinions. This will create an echo chamber where one point of view is reinforced recursively. Such echo chambers will reduce diversity of opinions and urge users to join the popular view \cite{facebookNews}.

\textbf{Voice Assistants}
Voice Assistants like Amazon Alexa, Google Home, etc. are a revolution in the domain of human-computer interaction. They promise a new era where users can talk with computers to get things done. Such voice assistants are built of different components like wake word engine, automatic speech recognition, text-to-speech, intent processing, etc. All of these components heavily rely on machine learning to function well. It is no surprise that such machine learning systems can be biased as well.
Washington Post along with two other research groups tested Google home \& Amazon echo with thousands of voice commands dictated by more than 100 people across nearly 20 cities \cite{accentBias}. They found that Google home and Amazon echo performed better than average for US accents. On the other hand, their performance is below average for someone with Indian, Chinese or Spanish accents. People with different accents can find these virtual assistants unresponsive, inaccurate and isolating. 

\textbf{Google Images}
Google Images is a free service provided by Google where a user can search for relevant images corresponding to a specified text. Every day, billions of people of people interact with Google Images and other similar interfaces. The results shown on such websites impacts user's perception about the world which ultimately impacts their behavior. 

According a study conducted at University of Washington \cite{kay2015unequal}, Google images propagate gender bias and stereotypes with respect to profession. For example, on searching for CEO's on google images, the percentage of women CEO's in top 100 results is just 11\%. The actual figure is 27\%. Similarly, on searching for telemarketers, percentage of women telemarketers in top 100 results is 64\% (the actual figure is 50\%). It seems Google images is trying to exaggerate gender stereotypes w.r.t. profession. Professions like doctor, CEO are more associated with males while teachers, telemarketers are associated with females. 

\textbf{Google Ads}
Latanya Sweeney, Professor at Harvard University, conducted a study \cite{sweeney2013discrimination} which showed that Google's Ad recommender system was biased against African Americans. Searching for a first name associated with African Americans on Google yielded more ads related to criminal background checks than white first names. Furthermore, those results were more likely to suggest searches for arrest records.
For example, she showed that ads related to `Latanya Sweeney' were about fetching public records, background checks, etc. Such instances clearly portray one community in poor light. Sustained exposure to such discriminatory practices may affect people's perceptions and behavior.

\textbf{Face Recognition}
Many tech companies like Microsoft, Google, Amazon, etc. have launched their face recogntion APIs. These APIs can be used by the police for criminal identification. MIT Media lab conducted a study to evaluate these APIs for gender/racial bias \cite{buolamwini2018gender}. They found that such APIs are more likely to identify whites over blacks and males over females. Black females are the worst hit. White males are more likely to be identified than black females by as much as 34\%. 
Another incident happened when Google Photos mislabeled a black girl as gorilla. After reporting this issue, Google removed gorillas from the set of labels to prevent any further cases of mislabeling \cite{buolamwini2018gender}.   
Recently, Amazon face recognition software confused 28 congressman with publicly available mugshots. Nearly 40\% of matches were people of color although they make up only 20\% of congress \cite{lawmakers}.

\section{Explainable AI}
The field of explainable AI (XAI))~\cite{gunning2017explainable,guidotti2019survey}, often referred to interchangeably as interpretable Machine Learning~\cite{carvalho2019machine,doshi2017towards},  started as a sub-field of AI that aims to produce methods and techniques that make AI's decisions understandable by people. The field has surged in recent years as complex and opaque AI technologies such as deep neural networks are now widely used. Explanations of AI are sought for various reasons, such as by regulators to assess model compliance, or by end users to support their decision-making~\cite{zhang2020effect,liao2020questioning,tomsett2018interpretable}. Most relevant to our work, explanations allow model developers to detect a model's faulty behaviors and evaluate its capability, fairness, and safety~\cite{doshi2017towards,dodge2019explaining}. Explanations are therefore increasingly incorporated in ML development tools supporting debugging tasks such as performance analysis~\cite{ren2016squares}, interactive debugging~\cite{kulesza2015principles}, feature engineering~\cite{krause2014infuse}, instance inspection and model comparison~\cite{hohman2019gamut,zhang2018manifold}.
Recent policy regulations like GDPR (European Union General Data Protection Regulation) requires that people impacted by algorithmic systems be provided with some logic/reasoning for the decision they receive. Such regulations make XAI even more important in today's context.

\section{Human-Centered AI approach}
Historically, all decisions were made by humans. Humans are good at decision making because they possess domain expertise accumulated over the years. They can interpret their decisions, and understand emotions, morals and ethics. The downside is that they are slow, expensive to hire and might be biased. To circumvent these issues, algorithms were introduced. Algorithms are fast, cheap and might be perceived as unbiased. On these lines, different automation tools were developed to replace humans with algorithms. It is important to note that the current AI systems are far from perfect and are incapable of understanding complex sociological concepts like race, culture, religion, etc. They lack accountability and are not suitable for many important applications.

A better alternative might be the Human-Centered AI (HCAI) approach \cite{shneiderman2022human}. HCAI is an emerging research area that aims to augment or amplify human abilities rather than replacing them. Two of the primary tenets of this approach includes: AI systems should understand humans and AI systems should help humans understand itself.
From a ML perspective, HCAI can be imagined as a semi-automated approach that leverages the best of human and machine intelligence to solve a problem. Traditionally, this approach has been used to expedite the learning process for machine learning systems via techniques like active learning. In this work, we have sought to adopt this methodology for tackling algorithmic bias and enhancing explainability. If the training dataset is not large or diverse enough to be representative of the population, the model might learn the wrong associations. Similar results can be expected if the data is skewed due to historical biases. For example, the model might learn that high-achieving software developers are men due to the skewed gender distribution in the tech industry. Humans can leverage their social and domain knowledge to debunk such biased associations. In the following, we provide more detailed thoughts on the added fairness, accountability, explainability, and trust this approach can invoke.

\textbf{Fairness.} In the literature, multiple definitions \cite{arvindTalk} of fairness are proposed that can be mutually incompatible and entail unavoidable tradeoffs \cite{kleinberg2016inherent}. There is no consensus on a single most appropriate definition of fairness \cite{gajane2017formalizing}. For a given application in a given context, algorithms can not be expected to determine the most appropriate definition of fairness and decide a desirable tradeoff between different metrics that is acceptable to all stakeholders. On the other hand, a human trusted by the majority of stakeholders can make an informed decision when presented with the required information. Hence, introducing a human in the loop might improve perceived fairness.

\textbf{Accountability.} Another major concern with algorithmic bias is accountability. Who should be held accountable if an algorithm does something wrong? Should software developers be held accountable or the people who curated/selected the dataset? Given that software developers themselves are not fully aware of the behavior of the black box models under different conditions, it would not be entirely justified to blame them. 
Having a human in the loop who can understand/monitor the underlying state of the AI system and who is empowered to guide the ML process as they see fit might enhance accountability.  


\textbf{Explainability} Similar to fairness, explainability is also context-dependent and there is no one size fits all solution \cite{liao2021human}. The need for explanations can be diverse based on the task and the target audience. For example, models developers might have different requirements than decision makers or impacted groups. Based on the unique requirements, one can choose from the set of possible explanation techniques. A human-centered approach might enhance explainability compared to a fully-automated approach as a human can better understand the nuanced needs of a particular situation.   


\textbf{Trust.} People are more likely to trust a system if they can tinker with it. This was shown in \cite{dietvorst2016overcoming}, where it was found that people were more likely to use an algorithm and accept its errors if they were given the opportunity to modify it themselves, even if it meant making it perform imperfectly. Our human-centric AI approach allows human(s) to supervise/participate in the entire process and make changes as they feel fit. Hence, our approach also instills more trust.

In the following chapters, this dissertation presents five research projects that we have undertaken during the course of my PhD \cite{ghai2022d, ghai2020explainable, ghai2021wordbias, ghai2022cascaded, hoque2022dramatvis}. Also, we discuss some of the venues for future work.  
\let\textcircled=\pgftextcircled
\chapter{Data Curation using AI Explanations as Interfaces for Crowd Workers}
\label{chap:xal}

The wide adoption of Machine Learning (ML) technologies has created a growing demand for people who can train ML models. Some advocated the term ``machine teacher'' to refer to the role of people who inject domain knowledge into ML models. This ``teaching'' perspective emphasizes supporting the productivity and mental wellbeing of machine teachers through efficient learning algorithms and thoughtful design of human-AI interfaces. One promising learning paradigm is Active Learning (AL) (a form of semi-supervised learning), by which the model intelligently selects instances to query a machine teacher for labels, so that the labeling workload could be largely reduced. However, in current AL settings, the human-AI interface remains minimal and opaque. A dearth of empirical studies further hinders us from developing teacher-friendly interfaces for AL algorithms. In this work, we begin considering AI explanations as a core element of the human-AI interface for teaching machines. When a human student learns, it is a common pattern to present one's own reasoning and solicit feedback from the teacher. When a ML model learns and still makes mistakes, the teacher ought to be able to understand the reasoning underlying its mistakes. When the model matures, the teacher should be able to recognize its progress in order to trust and feel confident about their teaching outcome. Toward this vision, we propose a novel paradigm of explainable active learning (XAL), by introducing techniques from the surging field of explainable AI (XAI) into an AL setting. We conducted an empirical study comparing the model learning outcomes, feedback content and experience with XAL, to that of traditional AL and coactive learning (providing the model's prediction without explanation). Our study shows benefits of AI explanation as interfaces for machine teaching--supporting trust calibration and enabling rich forms of teaching feedback, and potential drawbacks--anchoring effect with the model judgment and additional cognitive workload. Our study also reveals important individual factors that mediate a machine teacher's reception to AI explanations, including task knowledge, AI experience and Need for Cognition. By reflecting on the results, we suggest future directions and design implications for XAL, and more broadly, machine teaching through AI explanations. 

\section{Introduction}
While Machine Learning technologies are increasingly used in a wide variety of domains ranging from critical systems to everyday consumer products, currently only a small group of people with formal training possess the skills to develop these technologies. Supervised ML, the most common type of ML technology, is typically trained with knowledge input in the form of labeled instances, often produced by subject matter experts (SMEs). Current ML development process presents at least two problems. First, the work to produce thousands of instance labels is tedious and time-consuming, and can impose high development costs. Second, the acquisition of human knowledge input is isolated from other parts of ML development, and often has to go through asynchronous iterations with data scientists as the mediator. For example, seeing suboptimal model performance, a data scientist has to spend extensive time obtaining additional labeled data from the SMEs, or gathering other feedback which helps in feature engineering or other steps in the ML development process~\cite{amershi2014power,brooks2015featureinsight}. 

The research community and technology industry are working toward making ML more accessible through the recent movement of ``democratizing data science''~\cite{chou2014democratizing}. Among other efforts, interactive machine learning (iML) is a research field at the intersection of HCI and ML. iML work has produced a variety of tools and design guidelines~\cite{amershi2014power} that enable SMEs or end users to interactively drive the model towards desired behaviors so that the need for data scientists to mediate can be relieved. More recently, a new field of ``machine teaching" was called for to make the process of developing ML models as intuitive as teaching a student, with its emphasis on supporting ``the teacher and the teacher's interaction with
data''~\cite{simard2017machine}.

The technical ML community has worked on improving the efficiency of labeling work, for which Active Learning (AL) came to become a vivid research area. AL could reduce the labeling workload by having the model select instances to query a human annotator for labels. However, the interfaces to query human input are minimal in current AL settings, and there is surprisingly little work that studied how people interact with AL algorithms. Algorithmic work of AL assumes the human annotator to be an oracle that provides error-free labels~\cite{settles2009active}, while in reality annotation errors are commonplace and can be systematically biased by a particular AL setting. Without understanding and accommodating these patterns, AL algorithms can break down in practice. Moreover, this algorithm-centric view gives little attention to the needs of the annotators, especially their needs for transparency~\cite{amershi2014power}. For example, "stopping criteria", knowing when to complete the training with confidence remains a challenge in AL, since the annotator is unable to monitor the model's learning progress. Even if performance metrics calculated on test data are available, it is difficult to judge whether the model will generalize in the real-world context or is bias-free.



Meanwhile, the notion of model transparency has moved beyond the scope of descriptive characteristics of the model studied in prior iML work (e.g., output, performance, features used~\cite{kulesza2015principles,rosenthal2010towards, fails2003interactive,fogarty2008cueflik}). Recent work in the field of explainable AI (XAI)~\cite{gunning2017explainable} focuses on making the \textit{reasoning} of model decisions understandable by people of different roles, including those without formal ML training. In particular, \textit{local explanations} (e.g.~\cite{lundberg2017unified,ribeiro2016should}) is a cluster of XAI techniques that explain how the model arrived at a particular decision. Although researchers have only begun to examine how people actually interact with AI explanations, we believe explanations should be a core component of the interfaces to teach learning models. 

Explanations play a critical role in human teaching and learning~\cite{wellman2004theory,meyer1997consensually}. Prompting students to generate explanations for a given answer or phenomenon is a common teaching strategy to deepen students' understanding. The explanations also enable the teacher to gauge the students' grasp of new concepts, reinforce successful learning, correct misunderstanding, repair gaps, as well as adjust the teaching strategies~\cite{lombrozo2012explanation}. Intuitively, the same mechanism could enable machine teachers to assess the model logic, oversee the machine learner's progress, and establish trust and confidence in the final model. Well-designed explanations could also allow people without ML training to access the inner working of the model and identify its shortcomings, thus potentially reducing the barriers to provide knowledge input and enriching teaching strategies, for example by giving direct feedback for the model's explanations.  

Toward this vision of ``machine teaching through model explanations'', we propose a novel paradigm of \textit{explainable active learning} (XAL), by providing local explanations of the model's predictions of selected instances as the interface to query an annotator's knowledge input. We conduct an empirical study to investigate how local explanations impact the annotation quality and annotator experience. It also serves as an elicitation study to explore how people naturally want to teach a learning model with its explanations. The contributions of this work are threefold:
\begin{itemize}
    \item We provide insights into the opportunities for explainable AI (XAI) techniques as an interface for machine teaching, specifically feature importance based local explanation. We illustrate both the benefits of XAI for machine teaching, including supporting trust calibration and enabling rich teaching feedback, and challenges that future XAI work should tackle, such as anchoring judgment and cognitive workload. We also identify important individual factors mediating one's reception to model explanations in the machine teaching context, including task knowledge, AI experience and Need for Cognition.
    
    \item We conduct an in-depth empirical study of interaction with an active learning algorithm. Our results highlight several problems faced by annotators in an AL setting, such as increasing challenge to provide correct labels as the model matures and selects more uncertain instances, difficulty to know when to stop with confidence, and desire to provide knowledge input beyond labels. We claim that some of these problems can be mitigated by explanations.
    
    \item We propose a new paradigm to teach ML models, \textit{explainable active learning (XAL)}, that has the model selectively query the machine teacher, and meanwhile allows the teacher to understand the model's reasoning and adjust their input. The user study provides a systematic understanding on the feasibility of this new model training paradigm. Based on our findings, we discuss future directions of technical advancement and design opportunities for XAL.
\end{itemize}{}

In the following, we first review related literature, then introduce the proposal for XAL, research questions and hypotheses for the experimental study. Then we discuss the XAL setup, methodology and results. Finally, we reflect on the results and discuss possible future directions.

\section{Related work}
\label{literature}
Our work is motivated by prior work on AL, interactive machine learning and explainable AI. 

\subsection{Active learning}
 The core idea of AL is that if a learning algorithm intelligently selects instances to be labeled, it could perform well with much less training data~\cite{settles2009active}. This idea resonates with the critical challenge in modern ML, that labeled data are time-consuming and expensive to obtain~\cite{zhu2005semi}.  AL can be used in different scenarios like stream based~\cite{cohn1994improving} (from a stream of incoming data),  pool based~\cite{lewis1994sequential} (from a large set of unlabeled instances), etc.~\cite{settles2009active}. To select the next instance for labeling, multiple query sampling strategies have been proposed in the literature \cite{qbc, qbc2, unc, dasgupta2008hierarchical, quire, entropy, confidence}. Most commonly used is \textit{Uncertainty sampling} \cite{unc, entropy, confidence, margin}, which selects instances the model is most uncertain about.  Different AL algorithms exploit different notions of uncertainty, e.g. entropy \cite{entropy}, confidence \cite{confidence}, margin \cite{margin}, etc. 
 

While the original definition of AL is concerned with instance labels, it has been broadened to query other types of knowledge input. Several works explored querying feedback for features, such as asking whether the presence of a feature is an indicator for the target concept~\cite{raghavan2006active,druck2009active,settles2011closing}. For example, DUALIST~\cite{settles2011closing} is an active learning tool that queries annotators for labels of both instances (e.g., whether a text document is about ``baseball'' or ``hockey'') and features (which keywords, if appeared in a document, are likely indicators that the document is about ``baseball''). Other AL paradigms include \textit{active class selection}~\cite{lomasky2007active} and \textit{active feature acquisition}~\cite{zheng2002active}, which query the annotator for additional training examples and missing features, respectively.  

Although AL by definition is an interactive annotation paradigm, the technical ML community tends to simply assume the human annotators to be mechanically queried oracles. The above-mentioned AL algorithms were mostly experimented with simulated human input providing error-free labels. But labeling errors are inevitable, even for simple perceptual judgment tasks~\cite{cheng2015measuring}. Moreover, in reality, the targeted use cases for AL are often ones where high-quality labels are costly to obtain either because of knowledge barriers or effort to label. For example, AL can be used to solicit users' labels for their own records to train an email spam classifier or context-aware sensors ~\cite{kapoor2010interactive,rosenthal2010towards}, but a regular user may lack the knowledge or contextual information to make all judgments correctly. Many have criticized the unrealistic assumptions that AL algorithms make. For example, by solving a multi-instance, multi-oracle optimization problem, \textit{proactive learning}~\cite{donmez2008proactive} relaxes the assumptions that the annotator is infallible, indefatigable (always answers with the same level of quality), individual (only one oracle), and insensitive to costs.

Despite the criticism, we have a very limited understanding on how people actually interact with AL algorithms, hindering our ability to develop AL systems that perform in practice and provide a good annotator experience. Little attention has been given to the annotation interfaces, which in current AL works are undesirably minimal and opaque. To our knowledge, there has been little HCI work on this topic. One exception is in the field of human-robot interaction (HRI), where AL algorithms were used to develop robots that continuously learn by asking humans questions~\cite{cakmak2010designing,cakmak2012designing,chao2010transparent,gonzalez2014asking,saponaro2011generation}. In this context, the robot and its natural-language queries \textit{is} the interface for AL. For example,  Cakmak et al. explored robots that ask three types of AL queries~\cite{cakmak2010designing,cakmak2012designing}: instance queries, feature queries and demonstration queries. The studies found that people were more receptive to feature queries and perceived robots asking about features to be more intelligent. The study also pointed out that a constant stream of queries led to a decline in annotators' situational awareness~\cite{cakmak2010designing}. This kind of empirical results challenged the assumptions made by AL algorithms, and inspired follow-up work proposing mixed-initiative AL: the robot only queries when certain conditions were met, e.g., following an uninformative label. Another relevant study by Rosenthal and Dey ~\cite{rosenthal2010towards} looked at information design for an intelligent agent that queries labels to improve its classification. They found that contextual information, such as keywords in a text document or key features in sensor input, and providing the system's prediction (so people only need to confirm or reject labels) improved labeling accuracy. Although this work cited the motivation for AL, the study was conducted with an offline questionnaire without interacting with an actual AL algorithm.

We argue that it is necessary to study annotation interactions with a real-time AL algorithm
because temporal changes are key characteristics of AL settings. With an interactive learning algorithm, every annotation impacts the subsequent model behaviors, and the model should become better aligned with the annotator's knowledge over time. Moreover, systematic changes could happen in the process in both the type of queried instances, depending on the sampling strategy, and the annotator behaviors, for example fatigue~\cite{settles2011closing}. These complex patterns could only be understood by holistically studying the annotation and and the evolving model in real time. 

Lastly, it is a nontrivial issue to understand how annotator characteristics impact their reception to AL system features. For example, it would be instrumental to understand what system features could narrow the performance gaps of people with different levels of domain expertise or AI experience, thus reducing the knowledge barriers to teach ML models.

\subsection{Interactive machine learning}

Active learning is sometimes considered a technique for iML. iML work is primarily motivated by enabling non-ML-experts to train a ML model
through ``rapid, focused, and incremental model updates''~\cite{amershi2014power}. However, conventional AL systems, with a minimum interface asking for labels, lack the fundamental element in iML--a tight interaction loop that transparently presents how every human input impacts the model, so that the non-ML-experts could adapt their input to drive the model into desired directions~\cite{amershi2014power,fails2003interactive}.  Our work aims to move AL in that direction.

Broadly, iML encompasses all kinds of ML tasks including supervised ML, unsupervised ML (e.g., clustering ~\cite{choo2013utopian,smith2018closing}) and reinforcement learning~\cite{cakmak2010designing}. To enable interactivity, iML work has to consider two coupled aspects: \textit{what information} the model presents to people, and \textit{what input} people give to the model. Most iML systems present users with \textit{performance} information as impacted by their input, either performance metrics~\cite{kapoor2010interactive,amershi2015modeltracker}, or model output, for example by visualizing the output for a batch of instances~\cite{fogarty2008cueflik} or allowing users to select instances to inspect. An important lesson from the bulk of iML work is that users value \textit{transparency} beyond performance~\cite{rosenthal2010towards,kulesza2013too}, such as descriptive information about how the algorithm works or what features are used~\cite{kulesza2015principles,rosenthal2010towards}. Transparency not only improves users' mental model of the learning model but also satisfaction in their interaction outcomes~\cite{kulesza2013too}. Also, it helps provide more effective input. 

iML research has studied a variety of user input into the model such as providing labels and training examples~\cite{fails2003interactive}, as well as specifying model and algorithm choice~\cite{talbot2009ensemblematrix}, parameters, error preferences~\cite{kapoor2010interactive}, etc. A promising direction for iML to out-perform traditional approaches to training ML models is to enable feature-level human input. Intuitively,  direct manipulation of model features represents a much more efficient way to inject domain knowledge into a model~\cite{simard2017machine} than providing labeled instances. For example, FeatureInsight~\cite{brooks2015featureinsight} supports ``feature ideation'' for users to create dictionary features (semantically related groups of words) for text classification. EluciDebug~\cite{kulesza2015principles} allows users to add, remove and adjust the learned weights of keywords for text classifiers. Several interactive topic modeling systems allow users to select keywords or adjust keyword weights for a topic~\cite{choo2013utopian,smith2018closing}. Although the empirical results on whether feature-level input from end users improves performance per se have been mixed~\cite{kulesza2015principles,ahn2007open,wu2019local,stumpf2009interacting}, the consensus is that it is more efficient (i.e., fewer user actions) to achieve comparable results to instance labeling, and that it could produce models better aligned with an individual's needs or knowledge about a domain.

It is worth pointing out that all of the above-mentioned iML and AL systems supporting feature-level input are for text-based models~\cite{settles2011closing,raghavan2006active,stumpf2007toward,smithno,kulesza2015principles}. We suspect that, besides algorithmic interest, the reason is that it is much easier for lay people to consider keywords as top features for text classifiers compared to other types of data. For example, one may come up with keywords that are likely indicators for the topic of ``baseball'', but it is challenging to rank the importance of attributes in a tabular database of job candidates.  One possible solution is to allow people to access the model's own reasoning with features and then make incremental adjustments. This idea underlies recent research into visual analytical tools that support debugging or feature engineering work~\cite{krause2016interacting,hohman2019gamut,wexler2019if}. However, their targeted users are data scientists who would then go back to the model development mode. For non-ML-experts, they would need more accessible information to understand the inner working of the model and provide direct input that does not require heavy work of programming or modeling. Therefore, we propose to leverage recent development in the field of explainable AI as interfaces for non-ML experts to understand and teach learning models.

\subsection{Explainable AI} 
There have been many recent efforts to categorize the ever-growing collection of explanation techniques~\cite{guidotti2019survey,mohseni2018multidisciplinary,anisi03,lim2019these,wang2019designing,lipton2018mythos,arya2019one}. We focus on those explaining ML classifiers (as opposed to other types of AI system such as planning~\cite{chakraborti2020emerging} or multi-agent systems~\cite{rosenfeld2019explainability}). Guidotti et al. summarized the many forms of explanations as solving three categories of problems: \textit{model explanation} (on the whole logic of the classifier), \textit{outcome explanation} (on the reasons of a decision on a given instance) and \textit{model inspection} (on how the model behaves if changing the input). The first two categories, model and outcome explanations, are also referred as \textit{global} and \textit{local}  explanations~\cite{lipton2018mythos,mohseni2018multidisciplinary,arya2019one}. The HCI community have defined explanation taxonomies based on different types of user needs, often referred as intelligibility types~\cite{lim2009and,lim2019these,liao2020questioning} . Based on Lim and Dey's foundational work~\cite{lim2009and,lim2010toolkit}, intelligibility types can be represented by prototypical user questions to understand the AI, including inputs, outputs, certainty, why, why not, how to, what if and when. A recent work by Liao et al.~\cite{liao2020questioning} attempted to bridge the two streams of work by mapping the user-centered intelligibility types to existing XAI techniques. For example, global explanations answer the question ``\textit{how} does the system make predictions'', local explanations respond to ``\textit{why} is this instance given this prediction'', and model inspection techniques typically addresses \textit{why not}, \textit{what if} and \textit{how to}.

Our work leverages local explanations to accompany AL algorithms' instance queries. Compared to other approaches including example based and rule based explanations~\cite{guidotti2019survey}, \textit{Feature importance}~\cite{ribeiro2016should,guidotti2019survey} is the most popular form of local explanations. It justifies the model's decision for an instance by the instance's important features indicative of the decision (e.g., ``because the patient shows symptoms of sneezing, the model diagnosed him having a cold''). Local feature importance can be generated by different XAI algorithms depending on the underlying model and data. Some algorithms are model-agnostic~\cite{ribeiro2016should,lundberg2017unified}, making them highly desirable and popular techniques. Local importance can be presented to users in different formats~\cite{lipton2018mythos}, such as described in texts~\cite{dodge2019explaining}, or by visualizing the feature importance values~\cite{poursabzi2018manipulating,cheng2019explaining}. 

While recent studies of XAI often found explanations to improve users' understanding of AI systems~\cite{cheng2019explaining,kocielnik2019will,buccinca2020proxy}, empirical results regarding its impact on users' subjective experience such as trust~\cite{cheng2019explaining,poursabzi2018manipulating,zhang2020effect} and acceptance~\cite{kocielnik2019will} have been mixed. One issue, as some argued~\cite{zhang2020effect}, is that explanation is not meant to enhance trust or satisfaction, but rather to appropriately \textit{calibrate} users' perceptions to the model quality. If the model is under-performing, explanations should work towards exposing the algorithmic limitations; if a model is on par with the expected capability, explanation should help foster confidence and trust. Calibrating trust is especially important for AL settings: if explanations could help the annotator appropriately increase their trust and confidence as the model learns, it could help improve their satisfaction with the teaching outcome and confidently apply stopping criteria (knowing when to stop). Meanwhile, how people react to flawed explanations generated by early-stage, naive models, and changing explanations as the model learns, remain open questions~\cite{smithno}. We will empirically answer these questions by comparing annotation experiences in two snapshots of an AL process: an \textit{early stage} annotation task with the initial model, and a \textit{late stage} when the model is close to the stopping criteria.

On the flip side, explanations present additional information and the risk of overloading users~\cite{narayanan2018humans}, although some showed that their benefit justifies the additional effort~\cite{kulesza2015principles}. Explanations were also found to incur over-reliance~\cite{stumpf2016explanations,poursabzi2018manipulating} which makes people less inclined or able to scrutinize AI system's errors. It is possible that explanations could bias, or \textit{anchor} annotators' judgment to the model's. While anchoring judgment is not necessarily counter-productive if the model predictions are competent,  we recognize that the most popular sampling strategy of AL--uncertainty sampling--focuses on instances the model is most uncertain of.  To test this, it is necessary to decouple the potential anchoring effect of the model's predictions~\cite{rosenthal2010towards}, and the model's explanations, as an XAL setting entails both. Therefore, we compare the model training results with XAL to two baseline conditions: traditional AL and \textit{coactive learning} (CL)~\cite{shivaswamy2015coactive}. CL is a sub-paradigm of AL, in which the model presents its predictions and the annotator is only required to make corrections if necessary. CL is favored for reducing annotator workload, especially when their availability is limited. 

Last but not least, recent XAI work emphasizes that there is no ``one-fits-all'' solution and different user groups may react to AI explanations differently~\cite{arya2019one,liao2020questioning,dodge2019explaining}. Identifying individual factors that mediate the effect of AI explanation could help develop more robust insights to guide the design of explanations.
Our study provides an opportunity to identify key individual factors that mediate the preferences for model explanations in the machine teaching context. Specifically, we study the effect of \textit{Task (domain) Knowledge} and \textit{AI Experience} to test the possibilities of XAL for reducing knowledge barriers to train ML models. We also explore the effect of \textit{Need for cognition}~\cite{cacioppo1982need}, defined as an individual's tendency to engage in thinking or complex cognitive activities. Need for cognition has been extensively researched in social and cognitive psychology as a mediating factor for how one responds to cognitively demanding tasks (e.g.~\cite{cacioppo1983effects,haugtvedt1992personality}). Given that explanations present additional information, we hypothesize that individuals with different levels of Need for Cognition could have different responses.

\section{Explainable Active Learning and Research Questions}

We propose \textit{explainable active learning (XAL)} by combining active learning and \textit{local explanations}, which fit naturally with the AL workflow without requiring additional user input:  instead of opaquely requesting instance labels, the model presents its own decision accompanied by its explanation for the decision, answering the question ``\textit{why} am I giving this instance this prediction". It then requests the annotator to confirm or reject. For the user study, we make the design choice of explaining AL with \textit{local feature importance} instead of other forms of local explanations (e.g., example or rule-based explanations~\cite{guidotti2019survey}), given the former approach's popularity and intuitiveness--it reflects how the model weighs different features and gives people direct access to the inner working of the model. We also make the design choice of presenting local feature importance with a visualization (Figure~\ref{fig:interface_2}) instead of in texts, in the hope of reading efficiency.

Our idea differentiates from prior work on feature-querying AL and iML in two aspects. First, we present the model's own reasoning for a particular instance to query user feedback instead of requesting global feature weights from people~\cite{settles2011closing,raghavan2006active,kulesza2015principles,brooks2015featureinsight}. Recent work demonstrated that, while ML experts may be able to reason with model features globally, lay people prefer local explanations grounded in specific cases~\cite{arya2019one,kulesza2013too,hohman2019gamut,kulesza2011oriented}. Second, we look beyond text-based models as in existing work as discussed above, and consider a generalizable form of explanation--visualizing local feature importance. While we study XAL in a setting of tabular data, this explanation format can be applied to any type of data with model-agnostic explanation techniques (e.g.~\cite{ribeiro2016should}).

At a high level, we posit that this paradigm of presenting explanations and requesting feedback better mimics how humans teach and learn, allowing transparency for the annotation experience. Explanations can also potentially improve the teaching quality in two ways. First, it is possible that explanations make it easier for one to reject a faulty model decision and thus provide better labels, especially for challenging situations where the annotator lacks contextual information or complete domain knowledge~\cite{rosenthal2010towards}. Second, explanations could enable new forms of teaching feedback based on the explanation. These benefits were discussed in a very recent paper by Teso and Kersting~\cite{teso2018should}, which explored soliciting corrections for the model's explanation, specifically feedback that a mentioned feature should be considered irrelevant instead. This correction feedback is then used to generate counter examples as additional training data, which are identical to the instance except for the mentioned feature. While this work is closest to our idea, empirical studies were absent to understand how adding explanations impacts AL interactions. 

We believe a user study is necessary for two reasons. First, accumulating evidence, as reviewed in the previous section, suggests that explanations have both benefits and drawbacks relevant to an AL setting. They merit a user study to test its feasibility. Second, a design principle of iML recommends that algorithmic advancement should be driven by people's natural tendency to interact with models~\cite{amershi2014power,cakmak2012designing,stumpf2009interacting}. Instead of fixing on a type of input as in Teso and Kersting~\cite{teso2018should}, an \textit{interaction elicitation study} could map out desired interactions for people to teach models based on its explanations and then inform algorithms that are able to take advantage of these interactions. A notable work by Stumpf et al.~\cite{stumpf2009interacting} conducted an elicitation study for interactively improving text-based models, and developed new training algorithms for Naïve Bayes models. Our study explores how people naturally want to teach a model with a local-feature-importance visualization, a popular and generalizable form of explanation. Based on the above discussions, this work sets out to answer the following research questions and test the following hypotheses:

 

\begin{itemize}
    \item \textbf{RQ1}: How do local explanations impact the annotation and training outcomes of AL?
    \item \textbf{RQ2}: How do local explanations impact annotator experiences?
    
    \begin{itemize}
        \item \textbf{H1}: Explanations support \textit{trust calibration}, i.e. there is an interactive effect between the presence of explanations and the model learning stage (early v.s. late stage model) on annotator's trust in deploying the model. 
        \item \textbf{H2}: Explanations improve \textit{annotator satisfaction}.
        \item \textbf{H3}: Explanations increase perceived \textit{cognitive workload}. 
        
    \end{itemize}{}
 \item \textbf{RQ3}: How do individual factors, specifically \textit{task knowledge}, \textit{AI experience}, and \textit{Need for Cognition}, impact annotation and annotator experiences with XAL?
 
 \begin{itemize}
        \item \textbf{H4}: Annotators with lower task knowledge benefit more from XAL, i.e., there is an interactive effect between the presence of explanations and annotators' task knowledge on some of the annotation outcome and experience measures  (trust, satisfaction or cognitive workload). 
        
        \item \textbf{H5}:  Annotators inexperienced with AI benefit more from XAL, i.e., there is an interactive effect between the presence of explanations and annotators' experience with AI on some of the annotation outcome and experience measures (trust, satisfaction or cognitive workload).
        
        \item \textbf{H6}: Annotators with lower Need for Cognition have a less positive experience with XAL, i.e., there is an interactive effect between the presence of explanations and annotators' Need for Cognition on some of the annotation outcome and experience measures (trust, satisfaction or cognitive workload),

    \end{itemize}{}
 \item \textbf{RQ4}: What kind of feedback do annotators naturally want to provide upon seeing local explanations?
\end{itemize}{}

\section{XAL Setup}

\subsection{Prediction task}
We aimed to design a prediction task that would not require deep domain expertise, where common-sense knowledge could be effective for teaching the model. The task should also involve decisions by weighing different features so explanations could potentially make a difference (i.e., not simple perception based judgment). Lastly, the instances should be easy to comprehend with a reasonable number of features. With these criteria, we chose the Adult Income dataset \cite{adultIncome} for the task of predicting whether the annual income of an individual is more or less than \$80,000~\footnote{After adjusting for inflation (1994-2019)~\cite{inflation}, while the original dataset reported on the income level of \$50,000}. The dataset is based on a Census survey database and consists of 48,842 rows and 14 columns. Each row in the dataset characterizes a person with a mix of numerical and categorical variables like age, gender, education, occupation, etc., and a binary annual income variable, which was used as our ground truth.

\begin{figure*}[ht]
    \centering
\begin{subfigure}{.47\textwidth}
  \centering
  \includegraphics[width=1.1\linewidth]{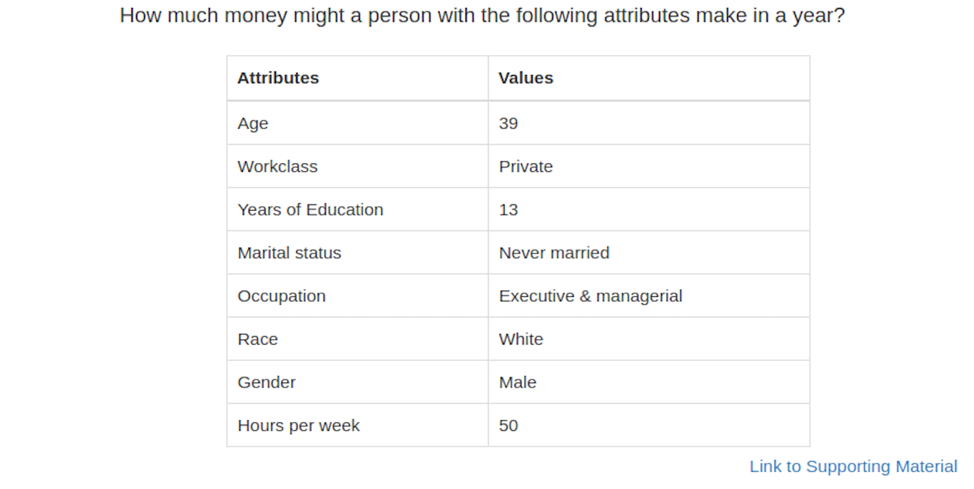}
  \caption{Customer profile presented in all conditions for annotation}
  \label{fig:interface_1}
  \label{fig:sub1}
\end{subfigure} \hspace{5mm}
\begin{subfigure}{.47\textwidth}
  \centering
  \includegraphics[width=.8\linewidth]{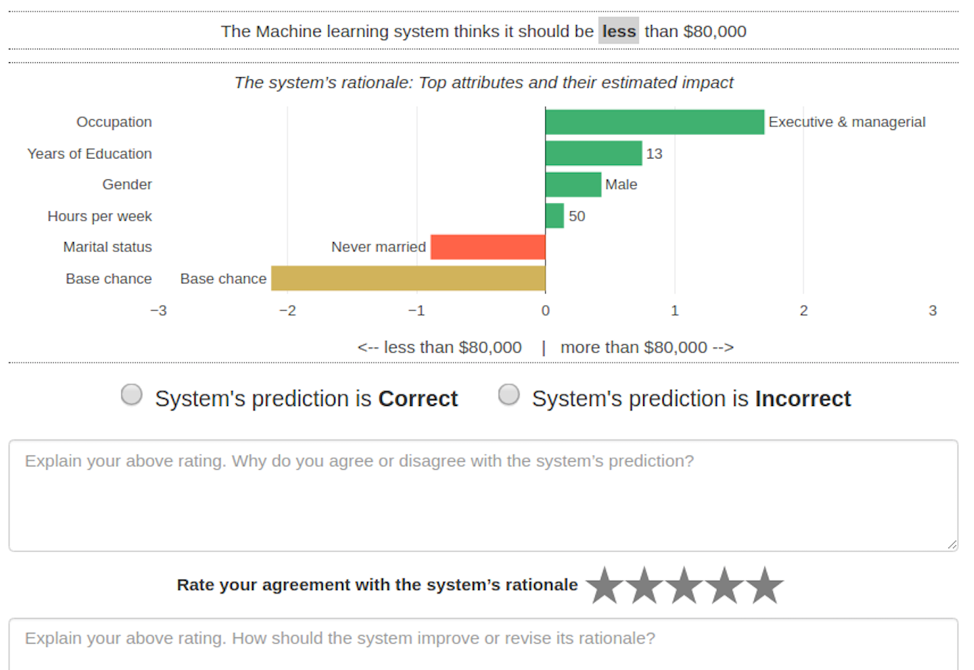}
  \caption{Explanation and questions presented in the XAL condition}
  \label{fig:interface_2}
  \label{fig:sub2}
\end{subfigure}
    \caption{Experiment interface }
    \label{fig:interface}
\end{figure*}

In the experiment, we presented participants with a scenario of building an ML classification system for a customer database. Based on a customer's background information, the system predicts the customer's income level for a targeted service. The task for the participants was to judge the income level of instances that the system selected to learn from, as presented in Figure~\ref{fig:interface_1}. This is a realistic AL task where annotators might not provide error-free labels, and explanations could potentially help reveal faulty model beliefs. To improve participants' knowledge about the domain, we provided a practice task before the trials, which will be discussed in  Section~\ref{domain}.

\subsection{Active learning setup}

AL requires the model to be retrained after new labels are fetched, so the model and explanations used for the experiment should be computationally inexpensive to avoid latency. Therefore we chose logistic regression (with L2 regularization), which was used extensively in the AL literature \cite{settles2009active, yang2018benchmark}. Logistic regression is considered directly interpretable, i.e., its local feature importance could be directly generated, as to be described in Section~\ref{explanation}. We note that this form of explanation could be generated by post-hoc techniques for any kind of ML model~\cite{ribeiro2016should}.

\begin{figure}[h!]
    \centering
    \includegraphics[scale=0.5]{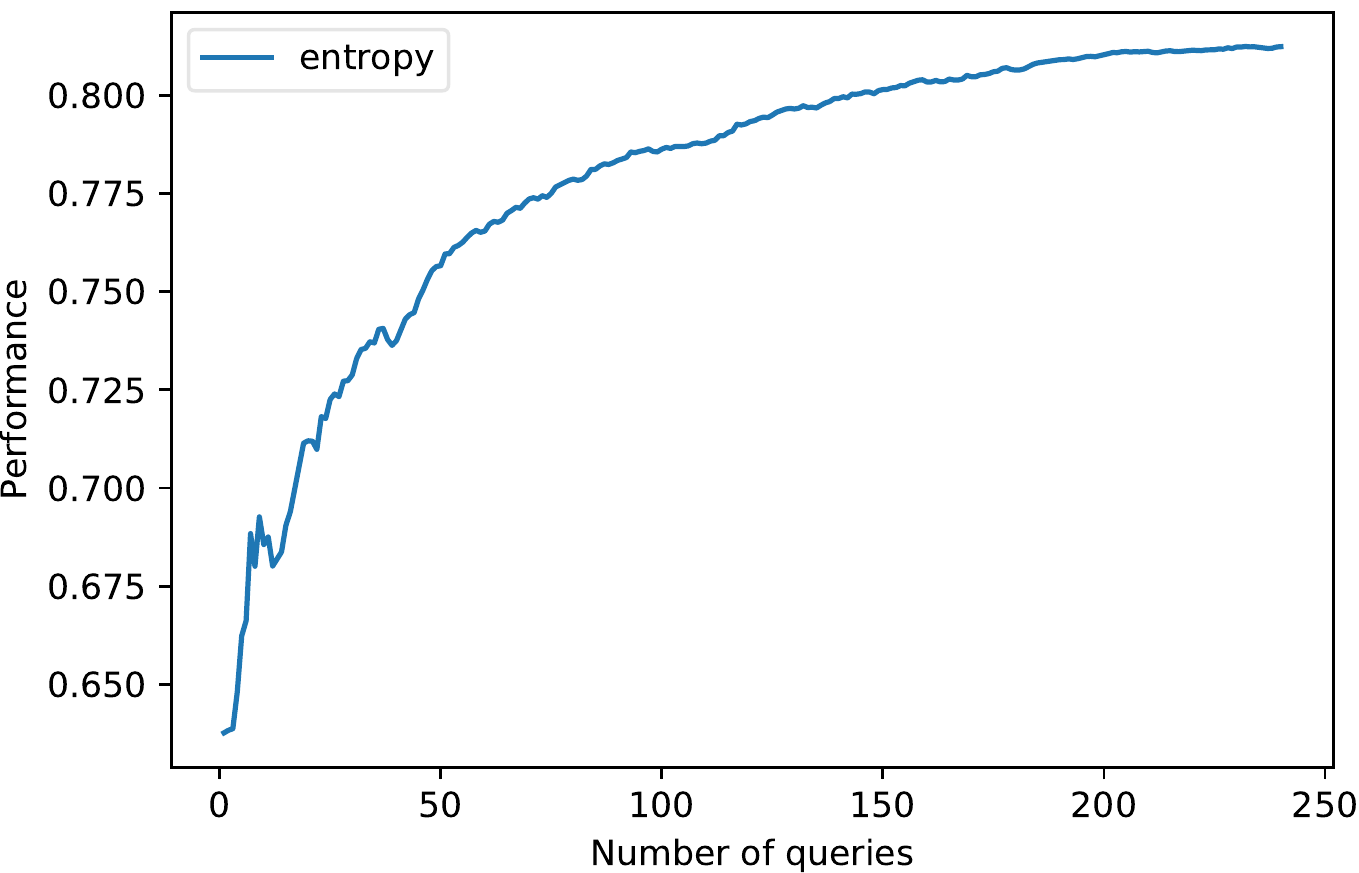}
    \caption{Accuracy as a function of number of queries in the simulation experiment}
    \label{fig:accuracy_ch2}
\end{figure}

Building an AL pipeline involves the design choices of sampling strategy, batch size, the number of initial labeled instances and test data. For this study, we used entropy-based uncertainty sampling to select the next instance to query, as it is the most commonly used sampling strategy \cite{yang2018benchmark} and also computationally inexpensive. We used a batch size of 1~\cite{batchSize}, meaning the model was retrained after each new queried label. We initialized the AL pipeline with two labeled instances. To avoid tying the experiment results to a particular sequence of data, we allocated different sets of initial instances to different participants, by randomly drawing from a pool of more than 100 pairs of labeled instances. The pool was created by randomly picking two instances with ground-truth labels, and being kept in the pool only if they produced a model with initial accuracy between 50\%-55\%. This was to ensure that the initial model would perform worse than humans and did not vary significantly across participants. 25\% of all data were reserved as test data for evaluating the model learning outcomes.

As discussed, we are interested in the effect of explanations at different stages of AL. We took two snapshots of an AL process--an early-stage model just started with the initial labeled instances, and a late-stage model that is close to the stopping criteria. We define the stopping criteria as plateau of accuracy improvement on the test data with more labeled data. To determine where to take the late-stage snapshot, we ran a simulation where AL queried instances were given the labels in the ground truth. The simulation was run with 10 sets of initial labels and the mean accuracy is shown in Figure \ref{fig:accuracy_ch2}. Based on the pattern, we chose the late stage model to be where 200 queries were executed. To create the late-stage experience without having participants answer 200 queries, we took a participant's allocated initial labeled instances and simulated an AL process with 200 queries answered by the ground-truth labels. The model was then used in the late-stage task for the same participant. This also ensured that the two tasks a participant experienced were independent of each other (i.e., a participant's performance in the early-stage task did not influence the late-stage task). In each task, participants were queried for 20 instances. Based on the simulation result in Figure~\ref{fig:accuracy_ch2}, we expected an improvement of 10\%-20\% accuracy with 20 queries in the early stage, and a much smaller increase in the late stage.

\subsubsection{Explanation method}\label{explanation}
Figure~\ref{fig:interface_2} shows a screenshot of the local explanation presented in the XAL condition, for the instance shown in Figure~\ref{fig:sub1}. The explanation was generated based on the coefficients of the logistic regression, which determine the impact of each feature on the model's prediction. To obtain the \textit{feature importance} for a given instance, we computed the product of each of the instance's feature values with the corresponding coefficients in the model. The higher the magnitude of a feature's importance, the more impact it had on the model's prediction for this instance. A negative value implied that the feature value was tilting the model's prediction towards less than \$80,000 and vice versa. We sorted all features by their absolute importance and picked the top 5 features responsible for the model's prediction. 

The selected features were shown to the participants in the form of a horizontal bar chart as in Figure~\ref{fig:interface_2}. The importance of a feature was encoded by the length of the bar where a longer bar meant greater impact and vice versa. The sign of the feature importance was encoded with color (green-positive, red-negative), and sorted to have the positive features at the top of the chart. Apart from the top contributing features, we also displayed the intercept of the logistic regression model as an orange bar at the bottom. Because it was a relatively skewed classification task (the majority of the population has an annual income of less than \$80,000), the negative base chance (intercept) needed to be understood for the model's decision logic. For example, in Figure \ref{fig:interface}, Occupation is the most important feature. Martial status and base chance are pointing towards less than \$80,000. While most features are tilting positively, the model prediction for this instance is still less than \$80,000 because of the large negative value of base chance.   



\section{Experimental design}

We adopted a 3 $\times$ 2 experimental design, with the learning condition (AL, CL, XAL) as a between-subject treatment, and the learning stage (early v.s. late) as a within-subject treatment. That is, participants were randomly assigned to one of the conditions to complete two tasks, with queries from an early and a late stage AL model, respectively. The order of the early and late stage tasks was randomized and balanced for each participant to avoid order effect and biases from knowing which was the "improved" model.

We posted the experiment as a human intelligence task (HIT) on Amazon Mechanical Turk. We set the requirement to have at least 98\% prior approval rate and each worker could participate
only once. Upon accepting the HIT, a participant was assigned to one of the three conditions. The annotation task was given with a scenario of building a classification system for a customer database to provide targeted service for high- versus low-income customers, with a ML model that queries and learns in real time. Given that the order of the learning stage was randomized, we instructed the participants that they would be teaching two configurations of the system with different initial performance and learning capabilities. 

With each configuration, a participant was queried for 20 instances, in the format shown in Figure~\ref{fig:interface_1}. A minimum of 10 seconds was enforced before they could proceed to the next query. In the AL condition, participants were presented with a customer's profile and asked to judge whether his or her annual income was above 80K. In the CL condition, participants were presented with the profile and the model's prediction. In the XAL condition, the model's prediction was accompanied by an explanation revealing the model's "rationale for making the prediction" (the top part of Figure~\ref{fig:interface_2}). In both the CL and XAL conditions, participants were asked to judge whether the model prediction was correct and optionally answer an open-form question to explain that judgement (the middle part of Figure~\ref{fig:interface_2}). In the XAL condition, participants were further asked to also give a rating to the model explanation and optionally explain their ratings with an open-form question (the bottom part of Figure~\ref{fig:interface_2}).  After a participant submitted a query, the model was retrained, and performance metrics of accuracy and F1 score (on the 25\% reserved test data) were calculated and recorded, together with the participant's input and the time stamp.
 
After every 10 trials, the participants were told the percentage of their answers matching similar cases in the Census survey data, as a measure to help engaging the participants. An attention-check question was prompted in each learning stage task, showing the customer's profile in the prior query with two other randomly selected profiles as distractors. The participants were asked to select the one they just saw. Only one participant failed both attention-check questions, and was excluded from the analysis. 

After completing 20 queries for each learning stage task, the participants were asked to fill out a survey regarding their subjective perception of the ML model they just finished teaching and the annotation task. The details of the survey will be discussed in Section ~\ref{survey}. At the end of the HIT we also collected participants' demographic information and factors of individual differences, to be discussed in Section~\ref{individual}.

\subsubsection{Domain knowledge training} \label{domain}
We acknowledge that MTurk workers may not be experts of an income prediction task, even though it is a common topic. Our study is close to \textit{human-grounded evaluation} proposed in ~\cite{doshi2017towards} as an evaluation approach for explainability, in which lay people are used as proxy to test general notions or patterns of the target application (i.e., by comparing outcomes between the baseline and the target treatment). 

To improve the external validity, we took two measures to help participants gain domain knowledge. First, throughout the study, we provided a link to a supporting document with statistics of personal income based on the Census survey. Specifically, chance numbers--the chance of people with a feature-value to have income above 80K--were given for all feature-values the model used (by quantile if numerical features). Second, participants were given 20 practice trials of income prediction tasks and encouraged to utilize the supporting material. The ground truth--income level reported in the Census survey--was revealed after they completed each practice trial. Participants were told that the model would be evaluated based on data in the Census survey, so they should strive to bring the knowledge from the supporting material and the practice trials into the annotation task. They were also incentivized with a \$2 bonus if the consistency between their predictions and similar cases reported in the Census survey were among the top 10\% of all participants.

After the practice trials, the agreement of the  participants' predictions with the ground-truth in the Census survey for the early-stage trials reached a mean of 0.65 (SE=0.08). We note the queried instances in AL using uncertainty-based sampling are challenging by nature. The agreement with ground truth by one of the authors, who is highly familiar with the data and the task, was 0.75.

\subsubsection{Survey measuring subjective experience}\label{survey}
To understand how explanation impacts annotators' subjective experiences (\textbf{RQ2}), we designed a survey for the participants to fill after completing each learning stage task. We asked the participants to self report the following (all based on a 5-point Likert Scale):

\textit{Trust} in deploying the model: We asked participants to assess how much they could trust the model they just finished teaching to be deployed for the target task (customer classification). Trust in technologies is frequently measured based on McKnight’s framework on Trust~\cite{mcknight1998initial,mcknight2002developing}, which considers the dimensions of \textit{capability}, \textit{benevolence}, \textit{integrity} for trust belief, and multiple action-based items (e.g., "I will be able to rely on the system for the target task") for trust intention. We also consulted a recent paper on trust scale for automation~\cite{korber2018theoretical} and added the dimension of \textit{predictability} for trust belief. We picked and adapted one item in each of the four trust belief dimensions (e.g., for benevolence, "Using predictions made by the system will harm customers’ interest") , and four items for trust intention, and arrived at an 8-item scale to measure trust (3 were reversed scale).  The Cronbach's alpha is 0.89.

\textit{Satisfaction} of the annotation experience, by five items adapted from After-Scenario Questionnaire~\cite{lewis1995computer} and User Engagement Scale~\cite{o2018practical} (e.g. "I am satisfied with the ease of completing the task", "It was an engaging experience working on the task"). The Cronbach's alpha is 0.91

\textit{Cognitive workload} of the annotation experience, by selecting two applicable items from the NASA-TLX task load index (e.g., "How mentally demanding was the task: 1=very low; 5=very high"). The Cronbach's alpha is 0.86.


\subsubsection{Individual differences}\label{individual}
\textbf{RQ3} asks about the mediating effect of individual differences, specifically the following:

\textit{Task knowledge} to perform the income prediction judgement correctly. We used one's performance in the practice trails as a proxy, calculated by the percentage of trials judged correctly based on the ground truth of income level in the Census database.

\textit{AI experience}, for which we asked participants to self-report ``How much do you know about artificial Intelligence or machine learning algorithms.'' The original questions had four levels of experience. With few answered higher level of experience, we decided to combine the answers into a binary variable--without AI experience v.s. with AI experience.

\textit{Need for Cognition} measures individual differences in the tendency to engage in thinking and cognitively complex activities. To keep the survey short, we selected two items from the classic Need for Cognition scale developed by Cacioppo and Petty~\cite{cacioppo1982need}. The Cronbach's alpha is 0.88.

\subsubsection{Participants}
37 participants completed the study. One participant did not pass both attention-check tests and was excluded. The analysis was conducted with 12 participants in each condition. Among them, 27.8\% were female; 19.4\% under the age 30, and 13.9\% above the age 50; 30.6\% reported having no knowledge of AI, 52.8\% with little knowledge ("know basic concepts in AI"), and the rest to have some knowledge ("know or used AI algorithms").  In total, participants spent about 20-40 min on the study and were compensated for \$4 with a 10\% chance for additional \$2 bonus, as discussed in Section~\ref{domain}

\section{Results}
 For all analyses, we ran mixed-effects regression models to test the hypotheses and answer the research questions, with participants as random effects,
 learning \textit{Stage}, \textit{Condition}, and individual factors (\textit{Task Knowledge}, \textit{AI Experience}, and \textit{Need for Cognition}) as fixed effects. RQ2 and RQ3 are concerned with interactive effects of Stage or Individual factors with learning Conditions. Therefore for every dependant variable we are interested in, we started with including all two-way interactions with Condition in the model, then removed insignificant interactive terms in reducing order. A VIF test was run to confirm there was no multicollinearity issue with any of the variables (all lower than 2). In each sub-section, we report statistics based on the final model and summarize the findings at the end.

\subsection{Annotation and learning outcomes (RQ1, RQ3)}

First, we examined the model learning outcomes in different conditions. In Table~\ref{tab:performance} (the third to sixth columns), we report the statistics of performance metrics--\textit{Accuracy} and \textit{F1} scores-- after the 20 queries in each condition and learning stage. We also report the performance improvement, as compared to the initial model performance before the 20 queries.

For each of the performance and improvement metrics, we ran a mixed-effect regression model as described earlier. In all the models, we found only significant main effect of Stage for all performance and improvement metrics ($p<0.001$). The results indicate that participants were able to improve the early-stage model significantly more than the later-stage model, but the improvement did not differ across learning conditions. 


\begin{table}
    \centering
  \caption{Results of model performance and labels }\label{tab:performance}
   
  \begin{tabular}{p{1cm}p{1.2cm}p{1.4cm}p{1.4cm}p{1.4cm}p{1.4cm}p{1.4cm}p{1.4cm}}
    \toprule
    Stage&Condition&Acc.&Acc. improve&F1&F1 improve&\%Agree&Human Acc.\\
    \midrule
    &AL & 67.0\% & 13.7\% & 0.490 & 0.104 & 55.0\% & 66.7\%\\
    Early &CL & 64.2\% & 11.7\% & 0.484 & 0.105 & 58.3\% & 62.1\%\\
    &XAL & 64.0\% & 11.8\% & 0.475 & 0.093 & 62.9\% & 63.3\%\\
    \midrule
    &AL & 80.4\% & 0.1\% & 0.589 & 0.005 & 47.9\% & 54.2\%\\
    Late &CL & 80.8\% & 0.2\% & 0.587 & 0.007 & 55.8\% & 58.8\%\\
    &XAL & 80.3\% & -0.2\% & 0.585 & -0.001 & 60.0\% & 55.0\%\\
  \bottomrule
\end{tabular}

\end{table}

\begin{figure}
    \centering
    \includegraphics[width=0.7\linewidth]{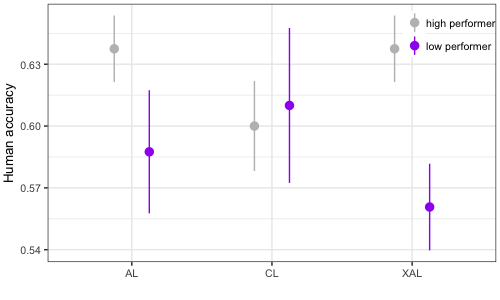}
    \caption{Human accuracy across conditions and task knowledge levels. All error bars represent +/- one standard error.}
    \label{fig:human_acc}
\end{figure}

In addition to the performance metrics, we looked at the \textit{Human accuracy}, defined as the percentage of labels given by a participant that were consistent with the ground truth. Interestingly, we found a significant interaction effect between Condition and participants' Task Knowledge (calculated as one's accuracy score in the training trials): taking CL condition as a reference level, XAL had a positive interaction effect with Task Knowledge ($\beta=0.67,SE=0.29, p=0.03$). In Figure~\ref{fig:human_acc}, we plot the pattern of the interaction effect by first performing a median split on Task Knowledge scores to categorize participants into \textit{high performers} and \textit{low performers}. The figure shows that, compared to the CL condition, adding explanations had a reverse effect for those with high or low task knowledge. While explanations helped those with high task knowledge to provide better labels, it impaired the judgment of those with low task knowledge. There was also a main negative effect of late Stage ($SE=0.21,t=3.87,p<0.001$), confirming that queried instances in the later stage were more challenging for participants to judge correctly. 


\begin{figure}
    \centering
    \includegraphics[width=0.7\linewidth]{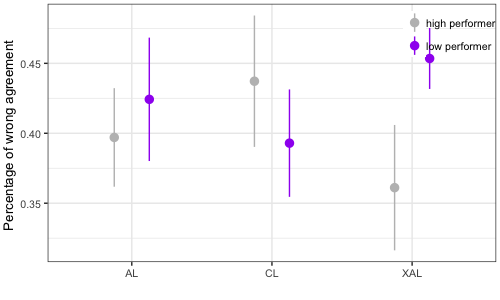}
    \caption{Percentage incorrect among all judgments that matched model predictions across Conditions and Task Knowledge levels. All error bars represent +/- one standard error.}
    \label{fig:wrong_agree}
\end{figure}

We conducted the same analysis on the \textit{Agreement} between each participant's labels and the model predictions and found a similar trend: using the CL condition as the reference level, there was a marginally significant interactive effect between XAL and Task Knowledge ($\beta=-0.75, SE=0.45,p=0.10$) \footnote{We consider $p<0.05$ as significant, and $0.05 \leq p<0.10$ as marginally significant, following statistical convention~\cite{cramer2004sage}}. The result suggests that explanations might have an "anchoring effect" on those with low task knowledge, making them more inclined to accept the model's predictions. Indeed, we zoomed in on trials where participants agreed with the model predictions, and looked at the percentage of \textit{wrong agreement} where the judgment was inconsistent with the ground truth. We found a significant interaction between XAL and Task Knowledge, using CL as a reference level ($\beta=-0.89, SE=0.45,p=0.05$). We plot this interactive effect in Figure~\ref{fig:wrong_agree}: adding explanations had a reverse effect for those with high or low task knowledge, making the latter more inclined to mistakenly agree with the model's predictions.  We did not find such an effect for \textit{incorrect disagreement} looking at trials where participants disagreed with the model's predictions. 

Taken together, to our surprise, we found the opposite results of \textbf{H4}: local explanations further polarized the annotation outcomes of those with high or low task knowledge, compared to only showing model predictions without explanations. While explanations may help those with high task knowledge to make better judgment, they have a \textbf{negative anchoring effect for those with low task knowledge by making them more inclined to agree with the model even if it is erroneous}. This could be a potential problem for XAL, even though we did not find this anchoring effect to have statistically significant negative impact on the model's learning outcome. We also showed that with uncertainty sampling of AL, \textbf{as the model matured, it became more challenging for annotators to make correct judgment and improve the model performance}.


\subsection{Annotator experience (RQ2, RQ3)}
We then investigated how participants' self-reported experience differed across conditions by analyzing the following survey scales (measurements discussed in Section ~\ref{survey}): trust in deploying the model, interaction satisfaction, and perceived cognitive workload. Table~\ref{tab:survey} reports the mean ratings in different conditions and learning stage tasks. For each self-reported scale, we ran a mixed-effects regression model as discussed in the beginning of this section.

\begin{table}
    \centering
  \caption{Survey results }\label{tab:survey}

  \begin{tabular}{p{0.8cm}p{1.2cm}p{1.5cm}p{1.5cm}p{1.5cm}}
    \toprule
        Stage&Condition&Trust&Satisfaction&Workload\\
    \midrule
    &AL &3.14 &4.23  &2.08\\
    Early&CL &3.83 &3.69 &2.71 \\
    &XAL &2.42 &3.31 &3.00 \\
    \midrule
    &AL &3 & 4.18&2.25\\
    Late&CL &2.71 &3.63 &2.67\\
    &XAL &2.99 &3.35&3.14  \\

  \bottomrule
\end{tabular}

\end{table}

\begin{figure}
    \centering
    \includegraphics[width=0.7\linewidth]{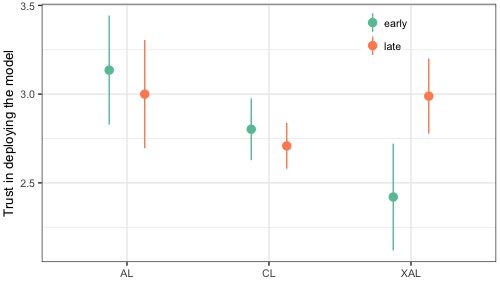}
    \caption{Trust in deploying the model across Conditions and Stages. All error bars represent +/- one standard error.}

    \label{fig:trust_stage}
\end{figure}

First, for trust in deploying the model, using AL as the reference level, we found a significant positive interaction between XAL Condition and Stage ($\beta=0.70, SE=0.31,p=0.03$). As shown in Table~\ref{tab:survey} and Figure~\ref{fig:trust_stage}, compared to the other two conditions, participants in the XAL Condition had significantly lower trust in deploying the early stage model, but enhanced their trust in the later stage model. The results confirmed \textbf{H1 }that \textbf{explanations  help calibrate annotators' trust} in the model at different stages of the training process, while showing model predictions alone (CL) was not able to have that effect.

\begin{figure}[t!]
    \centering
    \includegraphics[width=0.7\linewidth]{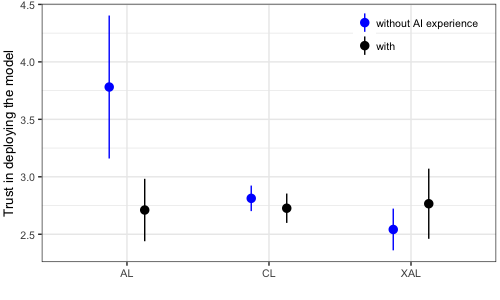}
    \caption{Trust in deploying the model across conditions and experience with AI. All error bars represent +/- one standard error.}
    \label{fig:trust_AI}
\end{figure}

We also found a two-way interaction between XAL Condition and participants' AI Experience (with/without experience) on trust in deploying the model ($\beta=1.43, SE=0.72,p=0.05$) (AL as the reference level). Figure ~\ref{fig:trust_AI} plots the effect: people without AI experience had exceptionally high ``blind'' trust and high variance of the trust (error bar) in deploying the model in the AL condition. With XAL they were able to an appropriate level of trust. The result highlight the \textbf{challenge for annotators to assess the trustworthiness of the model to be deployed, especially for those inexperienced with AI. Providing explanations could effectively appropriate their trust}, supporting \textbf{H5}.

\begin{figure}[t!]
    \centering
    \includegraphics[width=0.7\linewidth]{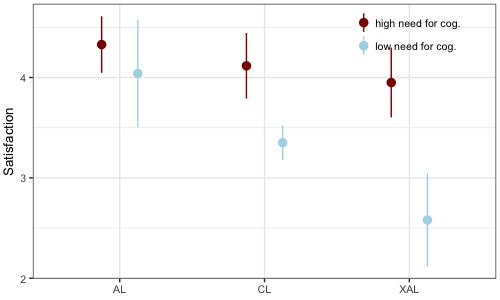}
    \caption{Satisfaction across conditions and experience with AI. All error bars represent +/- one standard error.}
    \label{fig:satisfaction_nc}
\end{figure}

For interaction satisfaction, the descriptive results in Table~\ref{tab:survey} suggests a decreasing trend of satisfaction in XAL condition compared to baseline AL. By running the regression model we found a significant two-way interaction between XAL Condition and Need for Cognition ($\beta=0.54, SE=0.26, p=0.05$) (AL as reference level). Figure ~\ref{fig:satisfaction_nc} plots the interactive effect, with median split on Need for Cognition scores. It demonstrates that \textbf{explanations negatively impacted satisfaction, but only for those with low Need for Cognition}, supporting \textbf{H6} and rejecting \textbf{H2}. We also found a positive main effect of Task Knowledge ($SE=1.31,t=2.76,p=0.01$), indicating that people who were good at the annotation task reported higher satisfaction.


\begin{figure}[t]
    \centering
    \includegraphics[width=0.7\linewidth]{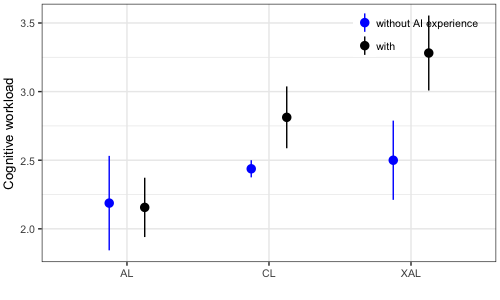}
    \caption{Cognitive workload across conditions and experience with AI. All error bars represent +/- one standard error.}
    \label{fig:workload_AI}
\end{figure}

For self-reported cognitive workload, the descriptive results in Table~\ref{tab:survey} suggests an increasing trend in XAL condition compared to baseline AL. Regression model found an interactive effect between the condition XAL and AI experience ($\beta=1.30, SE=0.59,p=0.04$). As plotted in Figure~\ref{fig:workload_AI}, the \textbf{XAL condition presented higher cognitive workload compared to baseline AL, but only for those with AI experience}. This partially supports \textbf{H3}, and potentially suggests that those with AI experience were able to more carefully examine the explanations. 

We also found an interactive effect between CL condition and Need for Cognition on cognitive workload ($\beta=0.53, SE=0.19,p=0.01$), and the remaining negative main effect of Need for Cognition ($\beta=-0.41, SE=0.14,p=0.01$). Pair-wise comparison suggests that participants with low Need for Cognition reported higher cognitive workload than those with high Need for Cognition, except in the CL condition, where they only had to accept or reject the model's predictions. Together with the results on satisfaction, \textbf{CL may be a preferred choice for those with low Need for Cognition}.


In summary, to answer \textbf{RQ2}, participants' self-reported experience confirmed the benefit of explanations for calibrating trust and judging the maturity of the model. Hence XAL could potentially help annotators form stopping criteria with more confidence. Evidence was found that explanations increased cognitive workload, but only for those experienced with AI. We also identified an unexpected effect of explanations in reducing annotator satisfaction, but only for those self-identified to have low Need for Cognition, suggesting that the additional information and workload of explanation may avert annotators who have little interest or capacity to deliberate on the explanations.

 The quantitative results with regard to \textbf{RQ3} confirmed the mediating effect of individual differences in Task Knowledge, AI Experience and Need for Cognition on one's reception to explanations in an AL setting. Specifically, people with better Task Knowledge and thus more capable of detecting AI's faulty reasoning, people inexperienced with AI who might be otherwise clueless about the model training task, and people with high Need for Cognition, may benefit more from XAL compared to traditional AL.

   



\subsection{Feedback for explanation (RQ4)}
In the XAL condition, participants were asked to rate the system's rationale based on the explanations and respond to an optional question to explain their ratings. Analyzing answers to these questions allowed us to understand what kind of feedback participants naturally wanted to give the explanations (\textbf{RQ4}).

\begin{figure}[t]
    \centering
    \includegraphics[scale=0.9]{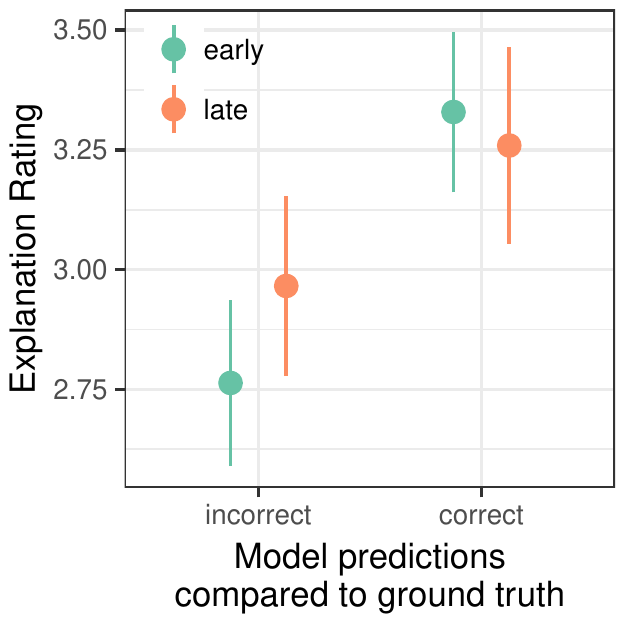}
    \caption{Explanation ratings for correct and incorrect model predictions}
    \label{fig:model_pred}
\end{figure}

First, we inspected whether participants' explanation ratings could provide useful information for the model to learn from. Specifically, if the ratings could distinguish between correct and incorrect model predictions, then they could provide additional signals. Focusing on the XAL condition, we calculated, for each participant, in each learning stage task, the \textit{average explanation ratings} given to instances where the model made correct and incorrect predictions (compared to ground truth). The results are shown in Figure~\ref{fig:model_pred}. By running an ANOVA on the \textit{average explanation ratings}, with \textit{Stage} and \textit{Model Correctness} as within-subject variables, we found the main effect of \textit{Model Correctness} to be significant, $F(1, 11)=14.38$, $p<0.01$. This result indicates that participants were able to distinguish the rationales of correct and incorrect model predictions, in both the early and late stages, confirming the utility of annotators' rating on the explanations.

\begin{figure}[t]
    \centering
    \includegraphics[scale=0.9]{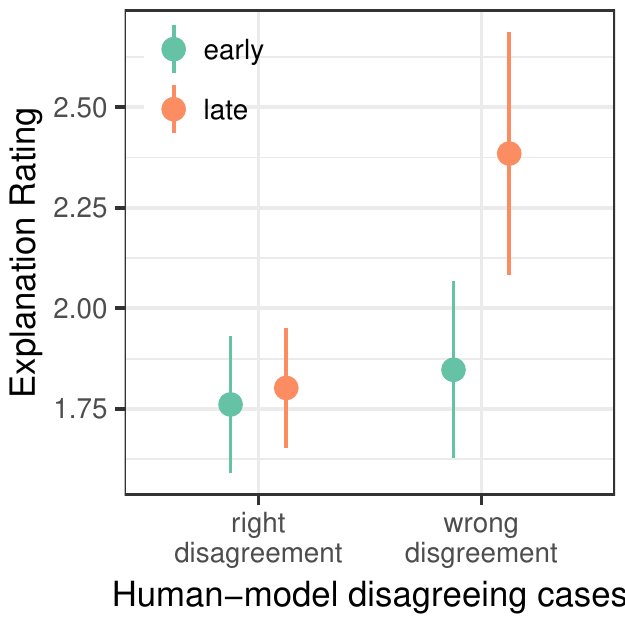}
    \caption{Explanation ratings for disagreeing instances}
    \label{fig:disagree}
\end{figure}

One may further ask whether explanation ratings provided additional information beyond the judgement expressed in the labels. For example, among cases where the participants disagreed (agreed) with the model predictions, some of them could be correct (incorrect) predictions, as compared to the ground truth. If explanation ratings could distinguish right and wrong disagreement (agreement), they could serve as additional signals that supplement instance labels. Indeed, as shown in Figure~\ref{fig:disagree}, we found that among the \textit{disagreeing instances}, participants' average explanation rating given to \textit{wrong disagreement} (the model was making the correct prediction and should not have been rejected) was higher than those to the \textit{right disagreement} ($F(1, 11)=3.12$, $p=0.10$), especially in the late stage (interactive effect between \textit{Stage} and \textit{Disagreement Correctness} $F(1, 11)=4.04$, $p=0.07$). We did not find this differentiating effect of explanation for agreeing instances. 

The above results are interesting as Teso and Kersting proposed to leverage feedback of ``weak acceptance'' to train AL ("right decision for the wrong reason"~\cite{teso2018should}), in which people agree with the system's prediction but found the explanation to be problematic. Empirically, we found that the tendency for people to give weak acceptance may be less than weak rejection. Future work could explore utilizing weak rejection to improve model learning, for example, with AL algorithms that can consider probabilistic annotations~\cite{song2018active}. 

\subsubsection{Open form feedback} 
\label{feedback}
We conducted content analysis on participants' open form answers to provide feedback, especially by comparing the ones in the CL and XAL conditions. In the XAL condition, participants had two fields as shown in Figure~\ref{fig:interface_2} to provide their feedback for the model decision and explanation. We combined them for the content analysis as some participants filled everything in one text field. In the CL condition, only the first field on model decision was shown. Two authors performed iterative coding on the types of feedback until a set of codes emerged and were agreed upon. In total, we gathered 258 entries of feedback on explanations in the XAL conditions (out of 480 trials). 44.96\% of them did not provide enough information to be considered valid feedback (e.g. simply expressing agreement or disagreement with the model). 

The most evident pattern contrasting the CL and XAL conditions is a shift from commenting on the top features to determine an income prediction to more diverse types of comments based on the explanation. For example, in the CL condition, the majority of comments were concerned with the job category to determine one's income level, such as ``\textit{Craft repair likely doesn't pay more than 80000}.'' However, for the model, job category is not necessarily the most important feature for individual decisions, suggesting that people's direct feature-level input may not be ideal for the learning model to consume. In contrast, feedback based on model explanations is not only more diverse in their types, but also covers a wider range of features. Below we discuss the types of feedback, ranked by the occurring frequency.


\begin{itemize}

    \item \textit{Tuning weights} ($N=81$): The majority of feedback focused on the weights bars in the explanation visualization, expressing disagreement and adjustment one wanted to make. E.g.,"\textit{marital status should be weighted somewhat less}". It is noteworthy that while participants commented on between one to four features, the median number of features was only one. Unlike in the CL condition where participants overly focused on the feature of job category, participants in the XAL condition often caught features that did not align with their expectation, e.g. ``\textit{Too much weight put into being married}'', or ``\textit{Age should be more negatively ranked}''. Some participants kept commenting on a feature in consecutive queries to keep tuning its weights, showing that they had a desired range in mind. 
    
    \item \textit{Removing, changing direction of, or adding features} ($N=28$): Some comments suggested, qualitatively, to remove, change the impact direction of, or add certain features. This kind of feedback often expressed surprise, especially on sensitive features such as race and gender, e.g."\textit{not sure why females would be rated negatively}", or "\textit{how is divorce a positive thing}". Only one participant mentioned \textit{adding} a feature not shown, e.g., "\textit{should take age into account}". These patterns echoed observations from prior work that local explanation heightens people's attention towards unexpected, especially sensitive features~\cite{dodge2019explaining}. We note that ``removing a feature to be irrelevant" is the feedback Teso and Kersting's AL algorithm incorporates~\cite{teso2018should}.

    \item \textit{Ranking or comparing multiple feature weights} ($N=12$) : A small number of comments explicitly addressed the ranking or comparison of multiple features, such as "\textit{occupation should be ranked more positively than marital status}".

    \item \textit{Reasoning about combination and relations of features} ($N=10$): Consistent with observation in Stumpf et al.'s study~\cite{stumpf2007toward}, some comments suggested the model to consider combined or relational effect of features--e.g., "\textit{years of education over a certain age is negligible}", or ``\textit{hours per week not so important in farming, fishing}''. This kind of feedback is rarely considered by current AL or iML systems.

    \item \textit{Logic to make decisions based on feature importance} ($N=6$): The feature importance based explanation associates the model's prediction with the combined weights of all features. Some comments ($N=6$) expressed confusion, e.g. "\textit{literally all of the information points to earning more than 80,000}" (while the base chance was negative). Such comments highlight the need for a more comprehensible design of explanations, and also indicate people's natural tendency to provide feedback on the model's overall logic.
    
    \item \textit{Changes of explanation} ($N=5$): Interacting with an online AL algorithm, some participants paid attention to the changes of explanations. For example, one participant in the condition seeing the late-stage model first noticed the declining quality of the system's rationale. Another participant commented that the weights in the model explanation ``\textit{jumps back and fourth, for the same job}''. Change of explanation is a unique property of the AL setting. Future work could explore interfaces that explicitly present changes or progress in the explanation and utilize the feedback. 
 
\end{itemize}{}

To summarize, we identified opportunities to use local explanations to elicit knowledge input beyond instance labels. By simply soliciting a rating for the explanation, additional signals for the instance could be obtained for the learning model. Through qualitative analysis of the open-form feedback, we identified several categories of input that people naturally wanted to give by reacting to the local explanations. Future work could explore algorithms and systems that utilize annotators' input based on local explanations for the model's features, weights, feature ranks, relations, and changes during the learning process. 

\section{Discussions and Future Directions}
 Our work is motivated by the vision of creating natural experiences to teach learning models by seeing and providing feedback for the model's explanations of selected instances. While the results show promises and illuminate key considerations of user preferences, it is only a starting point. To realize the vision, supporting the needs of machine teachers and fully harnessing their feedback for model explanations, requires both algorithmic advancement and refining the ways to explain and interact. Below we provide recommendations for future work of XAL as informed by the study.

\subsection{Explanations for machine teaching}
Common goals of AI explanations, as reflected in much of the XAI literature, are to support a complete and sound understanding of the model~\cite{kulesza2015principles,carvalho2019machine}, and to foster trust in the AI~\cite{poursabzi2018manipulating,cheng2019explaining}. These goals may have to be revised in the context of machine teaching. First, explanations should aim to \textit{calibrate} trust, and in general the perception of model capability, by accurately and efficiently communicating the model's current limitations. 

Second, while prior work often expects explanations to enhance adherence or persuasiveness~\cite{poursabzi2018manipulating}, we highlight the opposite problem in machine teaching, as an ``anchoring'' effect to a naive model's judgment could be counter-productive and impair the quality of human feedback. Future work should seek alternative designs to mitigate the anchoring effect. For example, it would be interesting to use a partial explanation that does not reveal the model's judgment (e.g., only a subset of top features~\cite{lai2019human}), or have people first make their own judgment before seeing the explanation.  

Third, the premise of XAL is to make the teaching task accessible by focusing on individual instances and eliciting incremental feedback. It may be unnecessary to target a complete understanding of the model, especially as the model is constantly being updated. Since people have to review and judge many instances in a row, \textit{low cognitive workload} without sacrificing the quality of feedback should be a primary design goal of explanations for XAL. One potential solution is \textit{progressive disclosure} by starting from simplified explanations and progressively provide more details~\cite{springer2019progressive}. Since the early-stage model is likely to have obvious flaws, using simpler explanations could suffice and demand less cognitive resource. Another approach is to design explanations that are sensitive to the targeted feedback, for example by only presenting features that the model is uncertain about or people are likely to critique, assuming some notion of uncertainty or likelihood information could be inferred.

While we used a local feature importance visualization to explain the model, we could speculate on the effect of alternative designs based on the results. We chose a visualization design to show the importance values of multiple features at a glance. While it is possible to describe the feature importance with texts as in ~\cite{dodge2019explaining}, it is likely to be even more cognitively demanding to read and comprehend. We do not recommend further limiting the number of features presented, since people are more inclined to critique features they see rather than recalling ones not presented. Other design choices for local explanations include presenting similar examples with the same known outcome~\cite{bien2011prototype,gurumoorthy2017protodash}, and rules that the model believes to guarantee the prediction~\cite{ribeiro2018anchors} (e.g., ``someone with an executive job above the age of 40 is highly likely to earn more than 80K``). We suspect that the example based explanation might not present much new information for feedback. The rule-based explanation, on the other hand, could be an interesting design for future work to explore, as annotators may be able to approve or disapprove the rules, or judge between multiple candidate rules~\cite{hanafi2017seer}. This kind of feedback could be leveraged by the learning model. Lastly, we fixed on local explanations for the model to self-address the \textit{why} question (intelligibility type). We believe it fits naturally with the workflow of AL querying selected instances. A potential drawback is that it requires annotators to carefully reason with the explanation for every new queried instance. It would be interesting to explore using a global explanation so that annotators would only need to attend to changes of overall logic as the model learns. But it is unknown whether a global explanation is as easy for non-AI-experts to make sense of and provide feedback on. 

 There are also opportunities to develop new explanation techniques by leveraging the temporal nature of AL. One is to \textit{explain model progress}, for example by explicitly showing changes in the model logic compared to prior versions. This could potentially help the annotators better assess the model progress and identify remaining flaws. Second is to utilize \textit{explanation and feedback history} to both improve explanation presentation (e.g., avoiding repetitive explanations) and infer user preferences (e.g., how many features is ideal to present).

Lastly, our study highlights the needs to tailor explanations based on the characteristics of the teacher. People from whom the model seeks feedback may not be experienced with ML algorithms, and not necessarily possess the complete domain knowledge or contextual information. Depending on their cognitive style or the context to teach, they may have limited cognitive resources to deliberate on the explanations. These individual characteristics may impact their preferences for the level of details, visual presentation, and whether explanation should be presented at all.

\subsection{Learning from explanation based feedback}
Our experiment intends to be an elicitation study to gather the types of feedback people naturally want to provide for model explanations. An immediate next step for future work is to develop new AL algorithms that could incorporate the types of feedback presented in Section~\ref{feedback}. Prior work, as reviewed in Section~\ref{literature}, proposed methods to incorporate feedback on top features or boosting the importance of features~\cite{raghavan2006active,druck2009active,settles2011closing,stumpf2007toward}, and removing features~\cite{teso2018should,kulesza2015principles}. However most of them are for text classifiers. Since feature-based feedback for text data is usually binary (a keyword should be considered a feature or not), prior work often did not consider the more quantitative feedback shown in our study, such as tuning the weights of features, comparatively ranking features, or reasoning about the logic or relations of multiple features. While much technical work is to be done, it is beyond the scope of this work. Here we highlight a few key observations from people's natural tendency to provide feedback for explanations, which should be reflected in the assumptions that future algorithmic work makes.

First, people's natural feedback for explanations is \textit{incremental} and \textit{incomplete}. It tends to focus on a small number of features that are most evidently unaligned with one's expectation, instead of the full set of features. Second, people's natural feedback is  \textit{imprecise}. For example, feature weights were suggested to be qualitatively increased, decreased, added, removed, or changing direction. It may be challenging for a lay person to accurately specify a quantitative correction for a model explanation, but a tight feedback loop should allow one to quickly view how an imprecise correction impacts the model and make follow-up adjustment. Lastly, people's feedback is \textit{heterogeneous}. Across individuals there are vast differences on the types of feedback, the number of features to critique, and the tendency to focus on specific features, such as whether a demographic feature should be considered fair to use~\cite{dodge2019explaining}. 

Taken together, compared to providing instance labels, feedback for model explanations can be noisy and frail. Incorporating the feedback ``as it is'' to update the learned features may not be desirable. For example, some have warned against ``local decision pitfalls''~\cite{wu2019local} of human feedback in iML that overly focuses on modifying a subset of model features, commonly resulting in an overfitted model that fails to generalize. Moreover, not all ML models are feasible to update the learned features directly. While prior iML work often builds on directly modifiable models such as regression or naïve Bayes classifiers, our approach is motivated by the possibility to utilize popular \textit{post-hoc} techniques to generate local explanations~\cite{ribeiro2016should,lundberg2017unified} for any kind of ML models, even those not directly interpretable such as neural networks. It means that an explanation could give information about how the model weighs different features but it is not directly connected to its inner working. How to incorporate human feedback for post-hoc explanations to update the original model remains an open challenge. It may be interesting to explore approaches that take human feedback as weighted signals, constraints, a part of a co-training model or ensemble~\cite{stumpf2009interacting}, or impacting the data~\cite{teso2018should} or the sampling strategy. 

A coupled aspect to make human feedback more robust and consumable for a learning algorithm is to design interfaces that scaffold the elicitation of high-quality, targeted type of feedback. This is indeed the focus of the bulk of iML literature. For example, allowing people to drag-and-drop to change the ranks of features, or providing sliders to change the feature weights, may encourage people to provide more precise and complete feedback. It would also be interesting to leverage the explanation and feedback history to extract more reliable signals from multiple entries of feedback, or purposely prompt people for confirmation of prior feedback. Given the heterogeneous nature of people's feedback, future work could also explore methods to elicit and cross-check input from multiple people to obtain more robust teaching signals.

\subsection{Explanation- and explainee-aware sampling}
Sampling strategy is the most important component of an AL algorithm to determine its learning efficiency. But existing AL work often ignores the impact of sampling strategy on annotators' experience. For example, our study showed that uncertainty sampling (selecting instance the model is most uncertain about to query) led to an increasing challenge for annotators to provide correct labels as the model matures. 

For XAL algorithms to efficiently gather feedback and support a good teaching experience, sampling strategy should move beyond the current focus on decision uncertainty to considering the explanation for the next instance and what feedback to gain from that explanation. For the machine teacher, desired properties of explanations may include easiness to judge, non-repetitiveness, tailored to their preferences and tendency to provide feedback, etc.~\cite{sokol2020explainability}. For the learning model, it may gain value from explaining and soliciting feedback for features that it is uncertain about, have not been examined by people, or have high impact on the model performance. Future work should explore sampling strategies that optimize for these criteria of explanations and explainees.

\section{Limitations}
We acknowledge several limitations of the study. First, the participants were recruited on Mechanical Turk and not held accountable for consequences of the model, so their behaviors may not generalize to all SMEs. However, we attempted to improve the ecological validity by carefully designing the domain knowledge training task and reward mechanism (participants received bonus if among 10\% performer). Second, this is a relatively small-scale lab study. While the quantitative results showed significance with a small sample size, results from the qualitative data, specifically the types of feedback may not be considered an exhaustive list. Third, the dataset has a small number of features and the model is relatively simple. For more complex models, the current design of explanation with feature importance visualization could be more challenging to judge and provide meaningful feedback for.  \\

 While active learning has gained popularity for its learning efficiency, it has not been widely considered as an HCI problem despite its interactive nature. We propose explainable active learning (XAL), by utilizing a popular local explanation method as the interface for an AL algorithm. Instead of opaquely requesting labels for selected instances, the model presents its own prediction and explanation for its prediction, then requests feedback from the human. We posit that this new paradigm not only addresses annotators' needs for model transparency, but also opens up opportunities for new learning algorithms that learn from human feedback for the model explanations. Broadly, XAL allows training ML models to more closely resemble a ``teaching'' experience, and places explanations as a central element of machine teaching. We conducted an experiment to both test the feasibility of XAL and serve as an elicitation study to identify the types of feedback people naturally want to provide. The experiment demonstrated that explanations could help people monitor the model learning progress and calibrate their trust in the teaching outcome. But our results cautioned against the adverse effect of explanations in anchoring people's judgment to the naive model's, if the annotator lacks adequate knowledge to detect the model's faulty reasoning, and the additional workload that could avert people with low Need for Cognition. Besides providing a systematic understanding of user interaction with AL algorithms, our results have three broad implications for using model explanations as the interface for machine teaching. First, we highlight the design goals of explanations applied to the context of teaching a learning model, as distinct from common goals in XAI literature, including calibrating trust, mitigating anchoring effect and minimizing cognitive workload. Second, we identify important individual factors that mediate people's preferences and reception to model explanations, including task knowledge, AI experience and Need for Cognition. Lastly, we enumerate on the types of feedback people naturally want to provide for model explanations, to inspire future algorithmic work to incorporate such feedback.

\let\textcircled=\pgftextcircled
\chapter{Tackling Social Biases in Tabular Datasets}
\label{chap:allocative}

With the rise of AI, algorithms have become better at learning underlying patterns from the training data including ingrained social biases based on gender, race, etc.
Deployment of such algorithms to domains such as hiring, healthcare, law enforcement, etc. has raised serious concerns about fairness, accountability, trust and interpretability in machine learning algorithms.
To alleviate this problem, we propose D-BIAS, a visual interactive tool that embodies human-in-the-loop AI approach for auditing and mitigating social biases from tabular datasets.
It uses a graphical causal model to represent causal relationships among different features in the dataset and as a medium to inject domain knowledge.
A user can 
detect the presence of bias against a group, say females, or a subgroup, say black females, by identifying unfair causal relationships in the causal network and using an array of fairness metrics.
Thereafter, the user can mitigate bias by refining the causal model and acting on the unfair causal edges. For each interaction, say weakening/deleting a biased causal edge, the system uses a novel method to simulate a new (debiased) dataset based on the current causal model while ensuring a minimal change from the original dataset.  
Users can visually assess the impact of their interactions on different fairness metrics, utility metrics, data distortion, and the underlying data distribution. Once satisfied, 
they can download the debiased dataset and use it for any downstream application for fairer predictions.
We evaluate D-BIAS by conducting experiments on 3 datasets and also a formal user study. We found that D-BIAS helps reduce bias significantly compared to the baseline debiasing approach across different fairness metrics while incurring little data distortion and a small loss in utility. 
Moreover, our human-in-the-loop based approach significantly outperforms an automated approach on trust, interpretability and accountability.

\section{Introduction} 
When computer systems discriminate based upon an individual's inherent characteristic such as gender, or acquired traits such as nationality, which are both irrelevant to the decision making process, it constitutes algorithmic bias. A simple way to deal with this problem can be to remove the sensitive attribute such as race before training the machine learning (ML) model. However, algorithmic bias can still persist via proxy variables such as zipcode that are correlated with the sensitive attribute. Recent years have seen a huge surge in research papers that deal with this problem. These papers have largely focused on pure algorithmic means to detect and remove bias at different stages of the ML pipeline.
However, fairness is contextual and thus cannot be achieved using fully automated methods \cite{wachter2021fairness}. Moreover, existing techniques are largely black boxes, offering only limited insight on the proxy variables and how bias is mitigated. Finally, they are also limited in providing capabilities that allow users to actively steer and control the debiasing process. Given this limited transparency and human control, accountability and trust become a major concern.

To address these needs, we hypothesize that a human-in-the-loop (HITL) approach is the way forward. A human expert can determine what fairness means in a given context. Such domain knowledge can be incorporated effectively via the HITL approach and hence improve perceived fairness. Introducing a human into the loop will only be effective when a person can understand the underlying state of the system and provide useful feedback. Hence, this approach is naturally inclined towards interpretability. 
On the trust aspect, people are more likely to trust a system if they can tinker with it, even if it means making it perform imperfectly \cite{dietvorst2016overcoming}.
Human interaction is a core part of the HITL approach, so it might instill more trust.
Lastly, this approach should also foster accountability as the human has a much bigger role to play, which can significantly impact the results.

\begin{figure}[tb]
 \centering 
 \includegraphics[width=0.8\columnwidth]{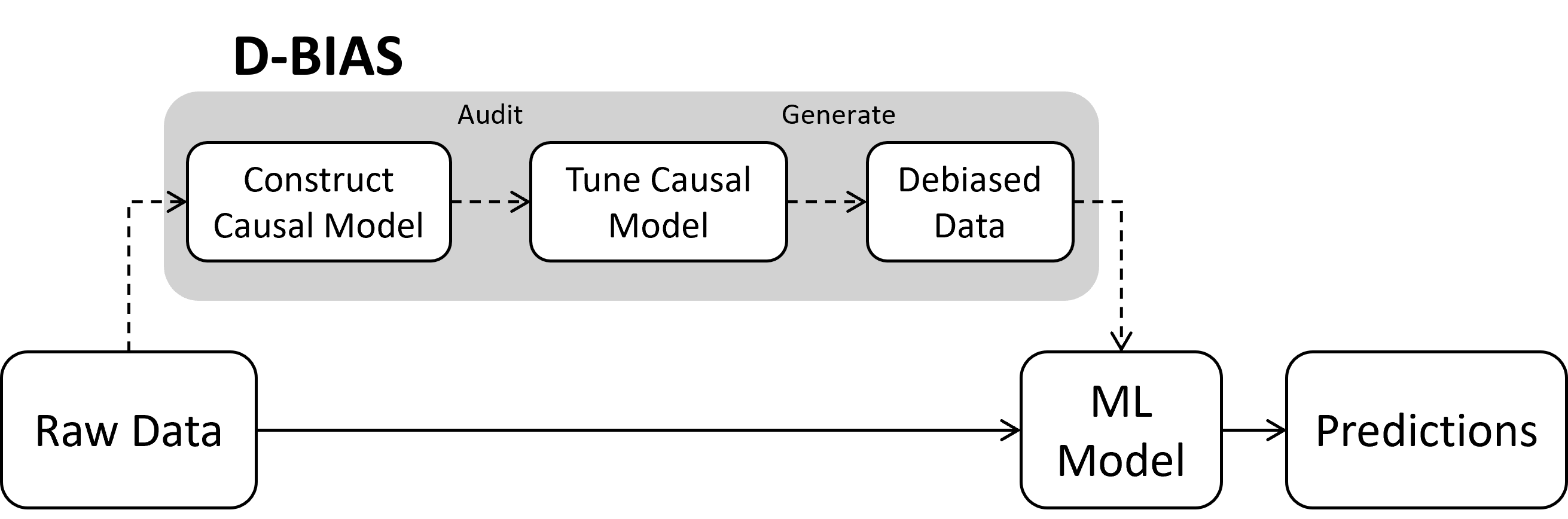}
 \caption{This figure shows how D-BIAS fits into the typical ML pipeline. D-BIAS allows users to act on the raw data and returns its debiased version which can be fed into a ML model for fairer predictions.}
 \label{fig:stages}
\end{figure}

We present D-BIAS, a visual interactive tool
that embodies a HITL approach for bias identification and mitigation. Given a tabular dataset as input, D-BIAS assists users in auditing the data for different biases and then helps generate its debiased version (see Fig. \ref{fig:stages}). It uses a graphical causal model as a medium for users to visualize the causal structures inherent in the data and  to inject their domain knowledge. We have made use of causal models since discrimination is inherently causal, and causal models also have a natural inclination towards interpretability \cite{zhang2017anti, silva,hoque2021outcome}. Apart from causal model, D-BIAS also includes multiple statistical fairness metrics to help identify bias. Users can choose to compare between two groups based on a single variable like gender (Male, Female), or a combination of attributes like race and gender (Black Females, White Males). Thereafter, they can inject their domain knowledge by acting on the edges of the causal network, for instance by deleting or weakening biased causal edges. Since the causal model encodes the data-generating mechanism, any user intervention modifies that process. Following each change, the system generates a new dataset based on the current causal model while keeping track and visualizing the impact of the user interventions on utility, data distortion and various fairness metrics. 
Users can interact with the system until they are satisfied with the outcome and then download the debiased dataset for use in any downstream ML application to achieve fairer predictions.
The major contributions of our work are:

\begin{itemize}
\setlength{\itemsep}{-3pt}
  \item A novel human-in-the-loop method to debias tabular datasets.
  \item An end to end interactive visual tool for algorithmic bias identification and mitigation.
  \item A demonstration of the effectiveness of our tool in reducing bias using three datasets.
  \item A user study to evaluate our tool on human-centric measures like usability, trust, interpretability, accountability, etc. 
\end{itemize}



\section{Background and Related Work}
\subsection{Causal Discovery}
\label{sec:causal_dis}
Randomized controlled trials (RCTs) have been used across disciplines to establish causal relations \cite{glymour2019review}. However, conducting such experiments can cost a lot of money, time or might simply be impractical \cite{glymour2019review}. Moreover, in this work, we are dealing with observational data which renders RCTs ineffective. Hence, we have relied on a popular data-driven alternative, namely Causal Discovery Algorithms (CDAs). It is important to note that there are different frameworks for causality like the Neyman-Rubin model, also known as the potential outcomes framework \cite{zeng2022survey}. CDAs fall under Judea Pearl's causality framework  and all further discussion will focus on this framework \cite{Pearl2000}.

Given observational data, CDAs find causal relationships among different attributes in the dataset \cite{glymour2019review}.
Causal relations returned by CDAs can be graphically represented via a directed acyclic graph (DAG), also known as causal graph, where each node represents a data attribute and each edge represents a causal relation. For example, a directed edge from X to Y signifies that X is the cause of Y, i.e., a change in X will cause a change in Y. The set of causal discovery algorithms can be broadly classified into constraint-based methods such as PC \cite{Spirtes2000}, Fast Causal Inference (FCI), etc., and score-based algorithms such as Greedy Equivalence Search (GES) \cite{chickering2002optimal}, Fast GES \cite{fges}, etc. Among the set of possible causal DAGs, constraint-based CDAs rely on a clever schedule of conditional independence tests whereas score-based CDAs rely on some fitness metric like BIC score to filter to the final causal graph. 

In this work, we have used PC algorithm~\cite{Spirtes2000} 
as it can handle mixed data types and provide asymptotically correct results. PC algorithm, named after its inventors \underline{P}eter Spirtes and \underline{C}lark Glymour, is a constraint-based CDA. PC algorithm starts with a complete undirected graph where each node is an attribute in the tabular dataset. Thereafter, it starts filtering edges based on conditional independence tests. For example, if nodes X and Y are unconditionally independent, then the edge X $\to$ Y  is removed. Here, the tabular dataset can consist of numeric and categorical attributes so we have used the \textit{Symmetric Conditional Independence Test} which can deal with mixed data types \cite{tsagris2018constraint}. After filtering, it uses a set of orientation rules to direct edges. Each directed edge represents a causal relationship from cause to effect. PC algorithm returns a partially directed acyclic graph, i.e., a DAG with some undirected edges.

Despite recent advances, the realm of automated causal discovery is an active research area. Causal discovery algorithms operate under a set of assumptions and limitations to yield the true causal model. However, in a real-world setting, one or more of the prerequisite conditions might be violated. For example, PC algorithm assumes that no confounders (direct common cause of two variables) are missing from the dataset. In a practical setting with a limited number of attributes, this can lead to a noisy causal model. Other possible factors that might add noise include sampling bias, measurement error, non-stationary data (data generated from different or evolving data-generating processes), etc. \cite{glymour2019review}. To effectively leverage the potential of causal models and interpret the outcome of CDAs, it is important for the end user to understand the limitations and assumptions of CDAs. A good resource is the practical guide by Malinsky and Danks \cite{malinsky2018causal}, and the paper by Glymour et al. \cite{glymour2019review}.
 One way to address the shortcomings of CDAs is to inject domain knowledge \cite{challenges}. In line with existing work \cite{Wang2016, wang2017visual, hoque2021outcome}, we have employed a visual analytics approach to augment CDAs with human knowledge to get to the true causal graph.

\subsection{Bias Identification}
\label{sec:identification}
The existing literature on bias identification mostly revolves around different fairness metrics. Numerous fairness metrics have been proposed that capture different facets of fairness, such as group fairness, individual fairness, counterfactual fairness, etc. 
\cite{dwork2018group,aif360, bechavod2017penalizing, kusner2017counterfactual,arvindTalk}.
Another way to classify fairness metrics can be on the level they operate on. For example, dataset based metrics are solely computed using the dataset and are independent of any ML model, such as statistical parity difference. On the other hand, classifier based metrics are computed over the predictions of a trained ML model, such as false negative rate difference. 
So far, there is not a single best fairness metric. Moreover, some of the fairness metrics can be mutually incompatible, i.e., it is impossible to optimize different metrics simultaneously\cite{kleinberg2016inherent}. 
In line with existing visual tools \cite{silva, aif360}, our tool also uses a diverse set of fairness metrics to present a more comprehensive picture.

Many fairness metrics solely focus on the aggregate relationship between the sensitive attribute and the output variable.  
This can lead to misleading conclusions as the aggregate trend might disappear or reverse when accounting for other relevant factors. A prime example of this phenomenon, also known as Simpson's paradox \cite{Pearl2009}, is the Berkeley Graduate Admission dataset \cite{bickel1975sex}. There it appeared as if the admission process was biased against women since the overall admit rate for men (44\%) was significantly higher than for women (30\%) \cite{barocas2017fairness}. However, this correlation/association did not account for the fact that women typically applied for more competitive departments than men. After correcting for this factor, it was found that the admission process had a small but statistically significant bias \textit{in favor of} women \cite{bickel1975sex}.
Causal models can be an effective tool for dealing with such a situation as they can decipher the different intermediate factors (indirect effects) along with their respective contributions behind an aggregate trend. 
Hence, our tool also employs causal model for bias identification.

\begin{figure*}
 \centering 
 \includegraphics[width=\columnwidth]{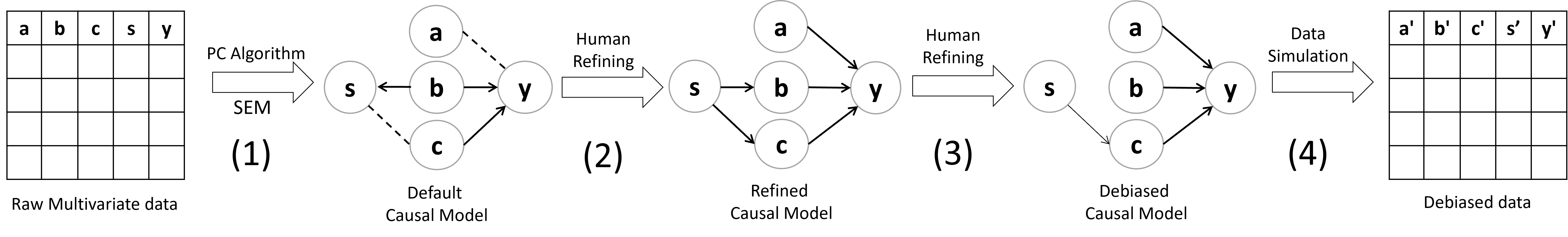}
  \setlength{\belowcaptionskip}{-8pt}
\setlength{\abovecaptionskip}{-4pt}
 \caption{The workflow of D-BIAS has 4 parts: (1) The default causal model is inferred using the PC algorithm and SEM. Here, dotted edges represent undirected edges (2) Refine stage - the default causal model is refined by directing undirected edges, reversing edges, etc. (3) Debiasing stage - the causal model is debiased by deleting/weakening biased edges (4) The debiased causal model is used to simulate the debiased dataset.}  
 \label{fig:workflow}
\end{figure*}

\subsection{Bias Mitigation}
\label{sec:mitigation}
The existing literature on algorithmic bias mitigation can be broadly categorized into the three different stages in which they operate within the ML pipeline, namely pre-processing, in-processing and post-processing. In the pre-processing stage, the dataset is modified such that the bias with respect to an attribute or set of attributes is reduced or eliminated \cite{zemel2013learning, kamiran2009classifying, hajian2013dataModification, nipsIBM, rajabi2021tabfairgan}. This can be achieved by either modifying the output label \cite{kamiran2012, kamiran2009classifying} or by modifying the input attributes \cite{nipsIBM, zemel2013learning}. 
In the in-processing stage, the algorithm is designed to take in biased data but still generate unbiased predictions. This can be achieved by tweaking the learning objectives such that accuracy is optimized while discrimination is minimized 
\cite{inprocess}. Finally, in the post-processing stage, predictions from ML algorithms are modified to counter bias 
\cite{lohia2018bias}. 
Our work relates closely with the pre-processing stage where we 
make changes to the input attributes and the output label.  

There is also a growing set of work at the intersection of bias mitigation and causality \cite{zhang2017causal, wu2018discrimination, chiappa2019path}. The general idea is to generate the causal network, modify it, and then simulate debiased data. These approaches fully rely on automated techniques to yield the true causal network, and assume a priori knowledge about fair/unfair causal relationships. 
Our work draws inspiration from this line of work and presents a more general solution where human domain knowledge is leveraged to refine the causal network and generate the debiased dataset. 



\subsection{Visual Tools}
Recent years have seen visual tools like \textit{Silva}\cite{silva}, \textit{FairVis}\cite{fairvis}, \textit{FairRankVis}\cite{xie2021fairrankvis}, \textit{DiscriLens}\cite{DiscriLens}, \textit{FairSight}\cite{fairsight}, \textit{WordBias}\cite{ghai2021wordbias}, etc. which are all aimed at tackling algorithmic bias. Although most of these tools are focused on bias identification, a few of them, such as FairSight, also permit debiasing. However, the debiasing strategy used in such tools is fairly basic, like eliminating proxy variable(s). Simple measures like this can lead to high data distortion and can have a high negative impact on data utility. Our work relates more closely with \textit{Silva} which also features a graphical causal model in its interface. Using an empirical study, it showed that users can interpret causal networks and found them helpful in identifying algorithmic bias. However like most other visual tools, Silva is limited to bias identification. Our work advances the state of the art by presenting a tool that supports both bias identification and mitigation, with our debiasing strategy being more nuanced and sensitive toward data distortion. 
\section{Methodology}
Given raw tabular data as input, D-BIAS outputs its debiased version. The entire process can be understood in four steps as shown in \autoref{fig:workflow}. In step 1, the raw multivariate data is read in and is used to compute a causal model using automated techniques. 
In step 2, the user interacts with the edges of the initial causal model like directing undirected edges to reach to a reliable causal model. In step 3, the user identifies and modifies biased causal relationships. Finally, in step 4, our tool computes a new dataset based on user interactions in the previous step. The user can thereafter click on `Evaluate Metrics' to compute and plot different evaluation metrics. 
If the user is satisfied with the evaluation metrics, they can download the debiased dataset and safely use it for any downstream application. Otherwise, the user may return to the causal network and make further changes. This cycle continues until the user is satisfied. A detailed discussion follows next.

\subsection{Generating the Causal Model}
\label{sec:causal}
A causal model generally consists of two parts: causal graph (skeleton) and statistical model. Given a tabular dataset as input, we first infer an initial causal graph (directed acyclic graph) using the PC algorithm  
\cite{Spirtes2000,Colombo2014Order-independentLearning} (see \autoref{sec:causal_dis}).
Each node in the causal graph represents a data attribute and each edge represents a causal relation. 
The PC algorithm provides qualitative information about the presence of causal relationship between variables. To quantitatively estimate the strength of each causal relationship, 
we use linear structural equation models (SEM) where a node is modeled as a linear function of its parent nodes. 

\begin{equation}
y = \sum_{i}^{parents(y)}{\beta_{i} x_i} \;+ \;\varepsilon 
\end{equation}

In the above equation, variable y is modeled as a linear combination of its parents $x_i$, their regression coefficients ($\beta_i$) and the intercept term ($\varepsilon$). 
If y is a numerical variable, we use linear regression else we use the multinomial logit model to estimate the values of $\beta_i$ and $\varepsilon$. Here, $\beta_i$ represents the strength of causal relation between $x_i$ and y. 
We repeat this modeling process for each node with non-zero parent nodes. Such nodes that have at least one edge leading into them are termed as endogenous variables.  
Other nodes correspond to exogenous variables or independent variables that have no parent nodes. In \autoref{fig:workflow} (Refined Causal Model), s and a are exogenous variables while b, c and y are endogenous variables. Here, y will be modeled as a function of its parent nodes, i.e., a, b and c. Similarly, b and c will be modeled as function of s.
 After this process, we arrive at a causal graph whose different edges are parameterized using SEM. This constitutes a Structural Causal Model (SCM) or simply causal model (see Fig. \ref{fig:workflow} (1)).  

This causal model, generated using automated algorithm, might have some undirected/erroneous causal edges due to different factors like sampling bias, missing attributes, etc.   
To achieve a reliable causal model, we have taken a similar approach as Wang and Mueller \cite{wang2017visual} where we leverage user knowledge to refine the causal model via operations like adding, deleting, directing, and reversing causal edges (see Fig. \ref{fig:workflow} (2)). For every operation, the system computes a score (Bayesian Information Criterion (BIC)) of how well the modified causal model fits the underlying dataset. Similar to \cite{wang2017visual}, our system assists the user in refining the causal model by providing the difference in BIC score before and after the change. A negative BIC score suggests a better fit.  
After achieving a reliable causal model, we enter into the debiasing stage where any changes made to the causal model will reflect on the debiased dataset (see Fig. \ref{fig:workflow} (3)).

\subsection{Auditing and Mitigating Social Biases}
From a causal perspective, discrimination can be defined as an unfair causal effect of the sensitive attribute (say, race) on the outcome variable (say, getting a loan)\cite{Pearl2000}. A direct causal path from a sensitive variable to the output variable constitutes disparate treatment while an unfair indirect path via some proxy attribute (say, zipcode) constitutes disparate impact \cite{chiappa2018causal}. 
A direct path is certainly unfair but an indirect path may be fair or unfair (as in the case of Berkeley Admission dataset \cite{Pearl2009}). Our system computes all paths from a sensitive attribute to the outcome variable. Thereafter, it is the job of the user to decide if a causal path is fair or unfair. If a causal path is unfair, the user should identify which constituting causal relationship(s) are unfair and act on them. The user can do this by deleting or weakening such biased causal relationships to reduce/mitigate the impact of the sensitive attribute on the outcome variable. For example, in Fig. \ref{fig:workflow} (Refined Causal Model), s is the sensitive attribute and y is the outcome variable. Here, the user deletes the edge $s \to b$ and weakens the edges $s \to c$ (shown as a thin edge). Once the user has dealt with the biased causal relationships, we achieve what we call the \textit{Debiased Causal Model}.  


\begin{algorithm}
\caption{Generate Debiased Dataset}
\label{algo:1}
\begin{algorithmic}[1]
\STATE $ D \leftarrow $ Original Dataset
\STATE $G(V, E) \leftarrow $ refined causal model   
\STATE $E_a \leftarrow$ set of edges added during the debiasing stage
\STATE $E_m \leftarrow $ subset of E that were deleted/strengthened/weakened during the debiasing stage
\STATE
\FOR{\textbf{each} e in $E_a$}
\STATE $n \leftarrow $node pointed by head of e
\STATE retrain linear model for n as a function of its parents 
\ENDFOR
\STATE
\STATE $V_{sim} \leftarrow \emptyset$ \hfill // Attributes that need to be simulated
\FOR{\textbf{each} e in $(E_m \cup E_a)$}
\STATE $n \leftarrow $node pointed by head of e
\STATE $V_{sim} \leftarrow V_{sim} + n + all\_descendent\_nodes(n)$
\ENDFOR
\STATE $V_{sim} \leftarrow remove\_duplicates(V_{sim})$
\STATE
\STATE $D_{deb} \leftarrow \emptyset $ \hfill // Debiased Dataset
\FOR{\textbf{each} v in topological\_sort(V)}
\IF{v present in $V_{sim}$}
\STATE $D_{deb}[v] \leftarrow $ generate values based on \autoref{eq:gen}
\ELSE
\STATE $D_{deb}[v] \leftarrow D[v]$
\ENDIF
\ENDFOR
\STATE
\STATE // Rescale values for the simulated attributes
\FOR{\textbf{each} v in $V_{sim}$} 
\IF{v is a categorical variable}
\STATE $D_{deb}[v] \leftarrow $ rescale values based on Algorithm 2
\ELSE 
\STATE // v is a numerical variable 
\STATE $\mu, \sigma^2 \leftarrow mean(D[v]), variance(D[v])$
\STATE $\mu_{deb}, \sigma_{deb}^2 \leftarrow mean(D_{deb}[v]), variance(D_{deb}[v])$
\STATE $D_{deb}[v] \leftarrow \mu + (D_{deb}[v] - \mu_{deb})/\sigma_{deb}*\sigma$ 
\ENDIF
\ENDFOR
\STATE \textbf{Result} $D_{deb}$
\end{algorithmic}
\end{algorithm}

\subsection{Generating the Debiased Dataset}
We simulate the debiased dataset based on the debiased causal model (see Fig. \ref{fig:workflow} (4)). 
The idea is that if the user weakens/removes biased edges from the causal network, then the simulated dataset might also contain less bias. The standard way to simulate a dataset based on a causal model is to generate random numbers for the exogenous variables \cite{sofrygin2017simcausal}. Thereafter, each endogenous variable is simulated as a function of its parent nodes (variables) in the causal network. 
In this work, we have adapted this procedure to suit our needs, i.e., simulating a fair dataset while having minimum distortion from the original dataset. 



Our approach to generate the debiased dataset, as illustrated in Algorithm \ref{algo:1}, can be broken down into 4 steps.
At step 1 (lines 6--9), we focus on the set of edges added during the debiasing stage ($E_{a}$). We retrain regression models corresponding to each of the target nodes of $E_{a}$. This will update the weights (regression coefficients ($\beta_i$)) for all edges leading into any target node of $E_{a}$. At step 2 (lines 11--16), we identify the set of nodes (attributes) that need to be simulated ($V_{sim}$). Unlike the standard procedure, we only simulate selective nodes that are directly/indirectly impacted by the user's interaction to minimize distortion. This set includes the target nodes of all edges that the user has interacted with along with their descendent nodes. For example, in \autoref{fig:workflow}(3), the user deletes the edge $s \to b$ and weakens the edge $s \to c$. So, we will only simulate variables $b$, $c$ and $y$. At step 3 (lines 18--25), we actually simulate all nodes that are a part of $V_{sim}$ using \autoref{eq:gen}. All other nodes are left untouched and their values are simply copied from the original dataset into the debiased dataset. 
It should be noted that all parent nodes must be simulated before their child node as the values for a node are computed using their parent nodes. To ensure that we simulate all endogenous variables in a topological order.
For example, in \autoref{fig:workflow}(4), variable \textit{b} and \textit{c} will always be simulated before variable \textit{y}.  

\vspace{-7pt}
\begin{equation}
\label{eq:gen}
y = \sum_{i}^{parents(y)}{\alpha_{i}\beta_{i} x_{i}} \;+\;\; \varepsilon  \;\;+\; \sum_{i}^{parents(y)}{(1-\alpha_{i})\beta_{i}r}
\end{equation}
\vspace{-7pt}

In the above equation, node y is simulated as a sum of 3 terms. The first term is the weighted linear combination of parent nodes. Here, $\alpha_i$ is the scaling factor that has a default value of 1 and can range between [0, 2] as determined by user interaction. 
Strengthening an edge, sets $\alpha_i>$1; weakening an edge, sets $\alpha_i<$1. For example, if the user weakens the edge between node $x_i$ and y by -35\%, then $\alpha_i$=0.65, strengthening it by +35\% will set $\alpha_i$=1.35, and deleting it will set $\alpha_i$=0. The second term is the intercept that was computed when the regression model for y was last trained. The third term adds randomness in proportion to the degree to which the user has altered an incoming edge. Here, r is a normal random variable that has a similar distribution as y ($r \sim \mathcal{N}({\mu_{y}, \sigma_{y}^{2}})$). This term adds fairness as it is random and alleviates distortion for y as it follows the same distribution. 

\begin{algorithm}[t]
\caption{Rescale Categorical Variable v}
\begin{algorithmic}[1]
\STATE D[v] $\leftarrow$ categorical variable v in the original dataset
\STATE prob\_mat $\leftarrow$ probability matrix for v post-debiasing 
\STATE
\STATE lr $\leftarrow $ 0.1  \hfill // learning rate
\STATE iterations $\leftarrow $ 0
\LOOP
    \STATE // DPD: Discrete Probability Distribution 
    \STATE $dist_{ori} \leftarrow$ DPD( D[v] )
    \STATE $dist_{deb} \leftarrow$ DPD( argmax(prob\_mat) )
    \STATE diff $\leftarrow \sum{\left\|(dist_{ori} - dist_{deb})/dist_{deb}\right\|}$
    \STATE scale\_factor $\leftarrow 1 + lr*\sum{(dist_{ori} - dist_{deb})/dist_{deb})}$
    \STATE prob\_mat $\leftarrow$ scale\_factor * prob\_mat
    \STATE $dist_{deb} \leftarrow$ DPD( argmax(prob\_mat) )
    \STATE new\_diff $\leftarrow \sum{\left\|(dist_{ori} - dist_{deb})/dist_{deb}\right\|}$
    \IF{new\_diff $>$ diff or iterations $>$ 50}
    \STATE break
    \ENDIF
    \STATE iterations $\leftarrow$ iterations + 1
\ENDLOOP
\STATE $D_{deb}[v] \leftarrow$ argmax(prob\_mat)
\STATE \textbf{Result} $D_{deb}[v]$

\end{algorithmic}
\label{algo:2}
\end{algorithm}

\begin{figure}[t]
 \centering 
 \includegraphics[width=0.65\columnwidth]{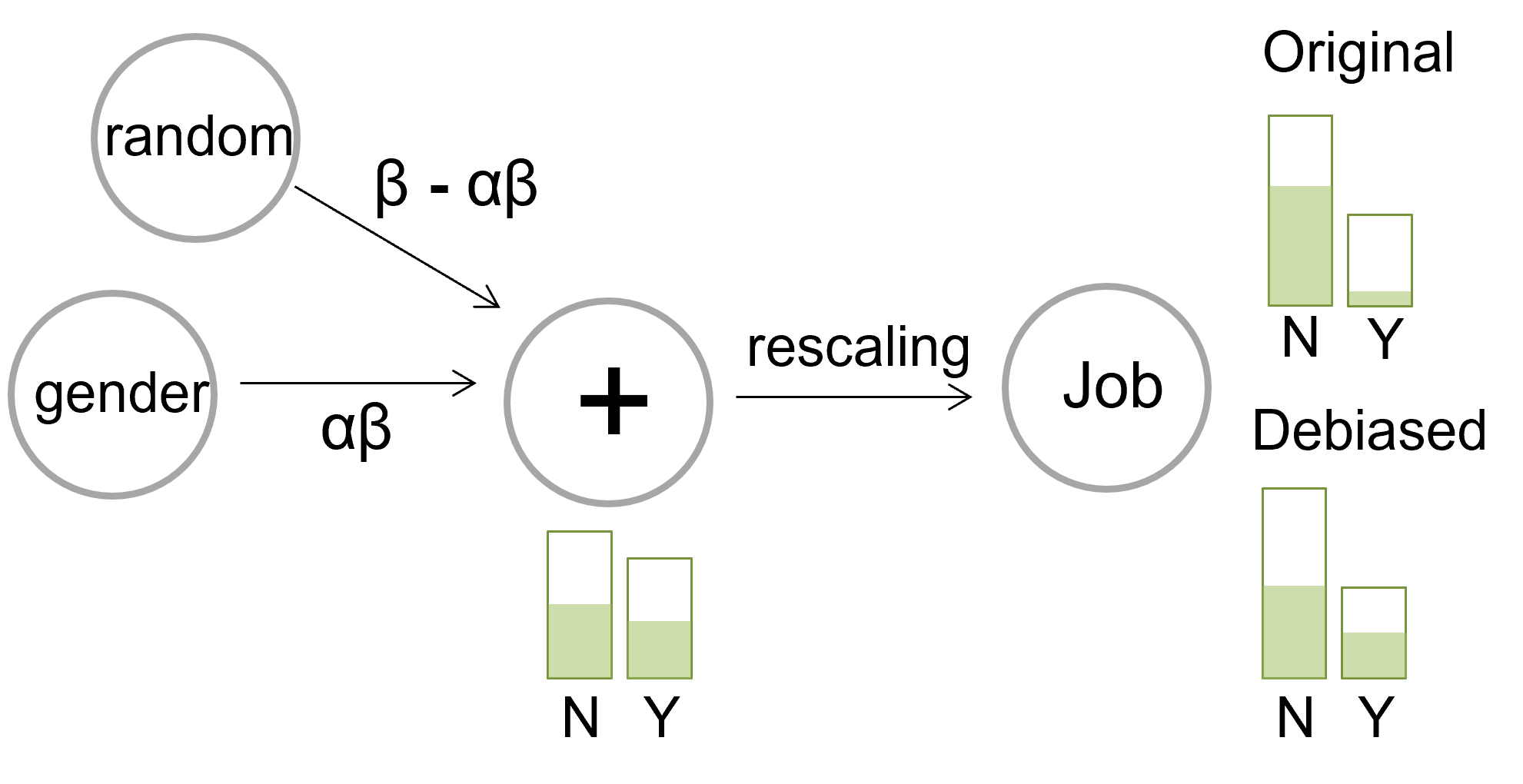}
  \setlength{\belowcaptionskip}{-4pt}
  \setlength{\abovecaptionskip}{-2pt}
 \caption{Illustration of the data debiasing process using a toy example where the node Job is caused by a single variable, namely gender. Green color marks the proportion of females who got the job (Y) or not (N).}  
 \label{fig:deb_eqn}
\end{figure}

Fig. \ref{fig:deb_eqn} illustrates the case where the node \textit{Job} has a single parent node (\textit{Gender}). Let's say that the user chooses to delete this edge ($\alpha$=0). Going the conventional route (without the third term), the attribute \textit{Job} will be reduced to a constant value (the intercept ($\varepsilon$)) which is undesirable.
Adding the (third) random term generates the \textit{Job} distribution below the `+' node which is far more balanced (fair) in terms of \textit{Gender} than the `Original' distribution on the top right. However,
the number of people getting the job (or not) is distorted compared to the `Original' distribution. 

This is corrected in Step 4 (lines 27--37) in Algorithm \ref{algo:1}, where we rescale each simulated variable so that its distribution is close to its original distribution. For numerical variable(s), we simply standardize values to their original mean and standard deviation. For categorical variables, we modulate their predicted probability matrix to match the original distribution. As described in Algorithm \ref{algo:2}, we iteratively scale the probability matrix so that it inches toward the original distribution. We continue this process until a fixed number of iterations or when the difference between original and debiased distribution starts increasing. In Fig. \ref{fig:deb_eqn}, we observe that the resulting `Debiased' distribution matches the `Original' distribution in terms of \textit{Job} allocation, while maintaining gender parity. 
It should be noted that simulating each attribute adds a corresponding modeling error to the process. This modeling error is typically small but it can potentially overpower the impact of the user's intervention, especially when a user makes a small change, say weakening an edge by 5\%. 
In such a case, the results may not be in strict accordance with the user’s expectations.



\subsection{Evaluation Metrics}
\label{sec:eval_metrics}
Once the debiased dataset is generated, it can be evaluated using different metrics that operate at the dataset level and the classifier level. For the second case, the debiased dataset is used to train a ML model chosen by the user, and a set of metrics are computed over the model's predictions. Here, the idea is to evaluate the downstream effects of debiasing.
All the evaluation metrics can be grouped into three broad categories, namely utility, fairness and distortion. It should be noted that there might be a trade-off among the three categories. For example, reducing bias might cause high data distortion or lower utility. For comparison, we have used a baseline debiasing strategy which just removes the sensitive attribute(s) from the dataset.

\textbf{Fairness.} Our tool presents a diverse set of 5 popular fairness metrics, namely statistical parity difference (Parity diff), individual fairness (Ind. bias), accuracy difference (Accuracy diff), false negative rate difference (FNR diff) and false positive rate difference (FPR diff) \cite{aif360}. Two of these metrics operate at the dataset level (Parity diff, Individual bias) and the rest operate on the classifier’s predictions.
Here, Ind. bias is defined as the mean percentage of a data point's k-nearest neighbors that have a different output label. A lower value for Ind. bias means is desirable as it means that similar data points have similar output labels. For the four other fairness metrics, we compute some statistic for the two groups (e.g. males and females) and then report their absolute difference. This statistic can be ML model dependent, such as accuracy,
or model independent, such as the likelihood of getting a positive output label.
Lower values for such metrics suggest more equality between groups. For computing model-based metrics, we omit the sensitive attributes(s) and perform 3-fold cross validation using the user-specified ML model with 50:50 train test split ration, and then report the mean absolute difference between groups across the three folds.
\begin{figure}[t]
 \centering 
 \includegraphics[width=\linewidth]{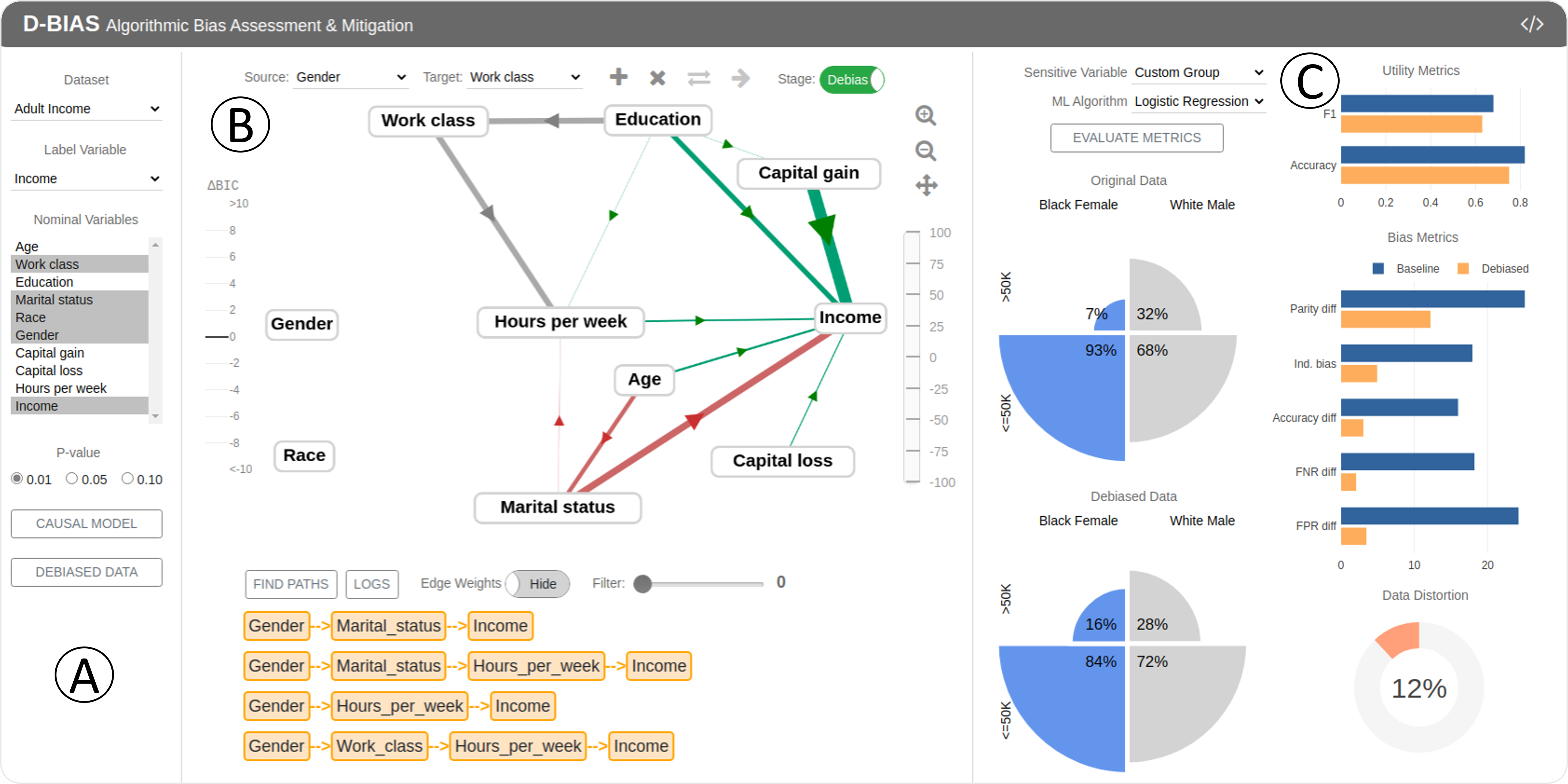}
   \caption{The visual interface of our D-BIAS tool. (A) The Generator panel: used to create the causal network and download the debiased dataset (B) The Causal Network view: shows the causal relations between the attributes of the data, allows user to inject their prior in the system (C) The Evaluation panel: used to choose the sensitive variable, the ML model and displays different evaluation metrics. }  
 \label{fig:teaser}
\end{figure}

\textbf{Utility.} The utility of ML models is typically measured using metrics like accuracy, f1 score, etc. 
In our context, we are interested in measuring the utility of a ML model when it is trained over the debiased dataset instead of the original dataset.
To compute the utility for the original dataset, we perform 3-fold cross validation using the user-specified ML model and report the mean accuracy and f1 score. For the debiased dataset, we follow a similar procedure where we train the user-specified ML model using the debiased dataset. However, we use the output variable from the original dataset as the ground truth for validation. Sensitive attribute(s) are removed from both datasets before training.
Ideally, we would like the utility metrics for the debiased dataset to be close to the corresponding metrics for the original dataset.   

\textbf{Data Distortion.} Data distortion is the magnitude of deviation of the debiased dataset from the original dataset. 
Since the dataset can have a mix of continuous and categorical variables, we have used the \textit{Gower distance}\cite{gower} metric. 
We compute the distance between corresponding rows of the original and debiased dataset, and then report the mean distance as data distortion. This metric is easy to interpret as it has a fixed lower and upper bound ([0,1]). It is 0 if the debiased dataset is the same as the original while higher values signify higher distortion. Lower values for data distortion are desirable.



\begin{figure}
 \centering 
 \includegraphics[scale=0.3]{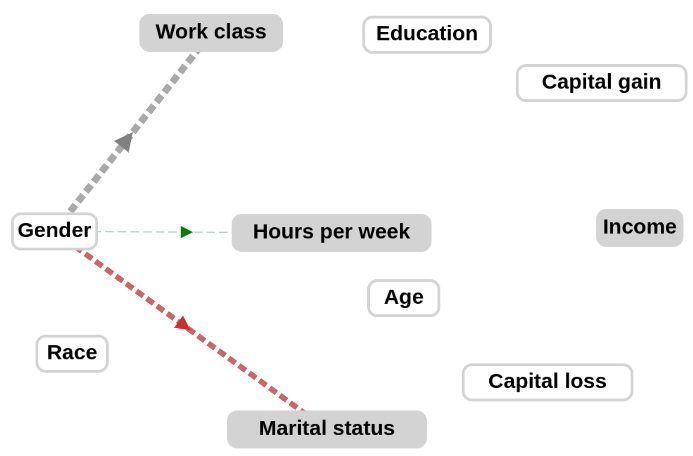}

 \caption{Logs view highlighting the changes made to the causal network for the Adult Income dataset. Dotted lines represent deleted edges and nodes in grey represent the impacted nodes.}
 \label{fig:logs}
\vspace{-10pt}
\end{figure}

\section{The D-BIAS Tool}

\subsection{Generator Panel}
This is the first component of the tool the user interacts with (see Fig. \ref{fig:teaser} (A)). The user starts off by choosing a dataset from the dropdown menu. This triggers an event that updates the set of possible options for the label and nominal variables to match the selected dataset. The user then selects the label variable which should be a binary categorical variable as we are considering a classification setting. Next, the user selects all nominal variables which is required for fitting the SEM model. Lastly, the user chooses a p-value and clicks the `Causal Model' button to generate the causal network. 
Here, the p-value is used by the PC algorithm to conduct independence tests. We set $p=0.01$ for all our demonstrations. It can be changed to 0.05 or 0.10 for smaller datasets.
The `Debiased Data' button downloads the debiased dataset. 

\subsection{Causal Network View}
This is the most critical piece of the interface where the user will spend most of the time (see Fig. \ref{fig:teaser}(B)). The center of this view contains the actual causal network which is surrounded by four panels on all sides. 

\textbf{Causal Network.} All features in the dataset are represented as nodes in the network and each edge represents a causal relation. The width of an edge encodes the magnitude of the corresponding standardized beta coefficient. It signifies the feature importance of the source node in predicting the target node. The color of an edge encodes the sign of the corresponding standardized beta coefficient. Green (red) represents positive (negative) influence of the source node on the target node. If an edge is undirected, it doesn't have a beta coefficient and is colored orange. Finally, gray color encode edges 
that represent relationships which can't be represented by a single beta coefficient. This occurs when the target node is a categorical variable with more than two levels. 

The causal network supports many interactions to enhance the user's overall experience and productivity. It supports operations like zooming and panning. The user can move nodes around if they are not satisfied with the default layout. On clicking a node, all directly connected edges and nodes are highlighted. Similarly, on clicking an edge, its source and target nodes are highlighted. Moreover, clicking a node or an edge also visualizes their distribution in the \textit{Comparison View} (see \autoref{subsec:eval_panel}). 

\begin{figure}[h!]
 \centering 
 \includegraphics[width=0.95\columnwidth]{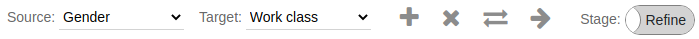}
 \label{fig:top_chart}
\end{figure}

\textbf{Top panel.} The panel right above the causal network (as shown above) allows the selection of an edge by choosing the source and target nodes. Next to the dropdown menus are a series of buttons which allow a user to inject their prior into the system. Going from left to right, they represent operations like adding an edge, deleting an edge, reversing the edge direction
and directing an undirected edge. The toggle at the end represents the current stage (Refine/Debias) and helps transitioning from one to the other.

\textbf{Left panel.} The bar to the left of the causal network shows the change in BIC score. This bar is updated each time the user performs operations like directing an undirected edge, adding/deleting an edge, etc. A negative value means that the change made to the causal network is in sync with the underlying dataset; positive values mean the opposite. Negative (positive) values are represented in green (red).   

\textbf{Right panel.} The panel to the right of the causal network offers four functionalities. Going from top to bottom, they represent zooming in, zooming out, reset layout and changing weight of an edge. D-BIAS supports zooming in and out via buttons apart from mouse scrolling to support uniform zooming across different hardware. The slider at the bottom gets activated when the user clicks on an edge during the debiasing stage. It allows the user to weaken/strengthen an edge depending on the selected value between -100\% to 100\%. Moving the slider changes $\alpha_i$ and also impacts the effective beta coefficient for the selected edge ($\alpha_i\beta_i$). This change manifests visually in the form of proportional change in the corresponding edge width. Moving the slider to -100\% will result in deletion of the selected edge.

\begin{figure}[h!]
 \centering 
 \includegraphics[scale=0.3]{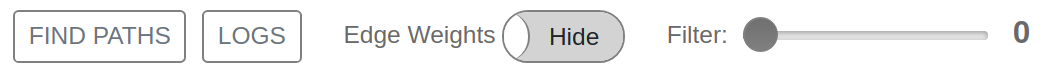}
 \label{fig:bottom_chart}
\end{figure}

\textbf{Bottom panel.} As shown above, the bottom panel offers 4 functionalities. The ``Find Paths" button triggers depth first traversal of the causal graph to compute all directed paths between the source and target node as selected in the top panel. This will be especially helpful 
when the graph is big and complex. All the computed paths are then displayed below the bottom panel. A user can click on a displayed path to highlight it and see an animated view of the path going from the source to the target node. The ``Logs" button highlights changes made to the causal network during the debiasing stage (see Fig. \ref{fig:logs}). All edges are hidden except for the newly added edges, deleted edges and edges that were weakened/strengthened. Nodes impacted by such operations ($V_{sim}$) are highlighted in grey. 

If a user is interested in knowing the exact beta coefficients for all edges, the edge weights toggle will help in doing just that. By default, it is set to `hide' to enhance readability. If turned to `show', the beta coefficients will be displayed on each edge. The filter slider helps user focus on important edges by hiding edges with absolute beta coefficients less than the chosen value. 

\subsection{Evaluation Panel}
\label{subsec:eval_panel}
\begin{figure}[t]
 \centering 
 \includegraphics[width=0.6\columnwidth]{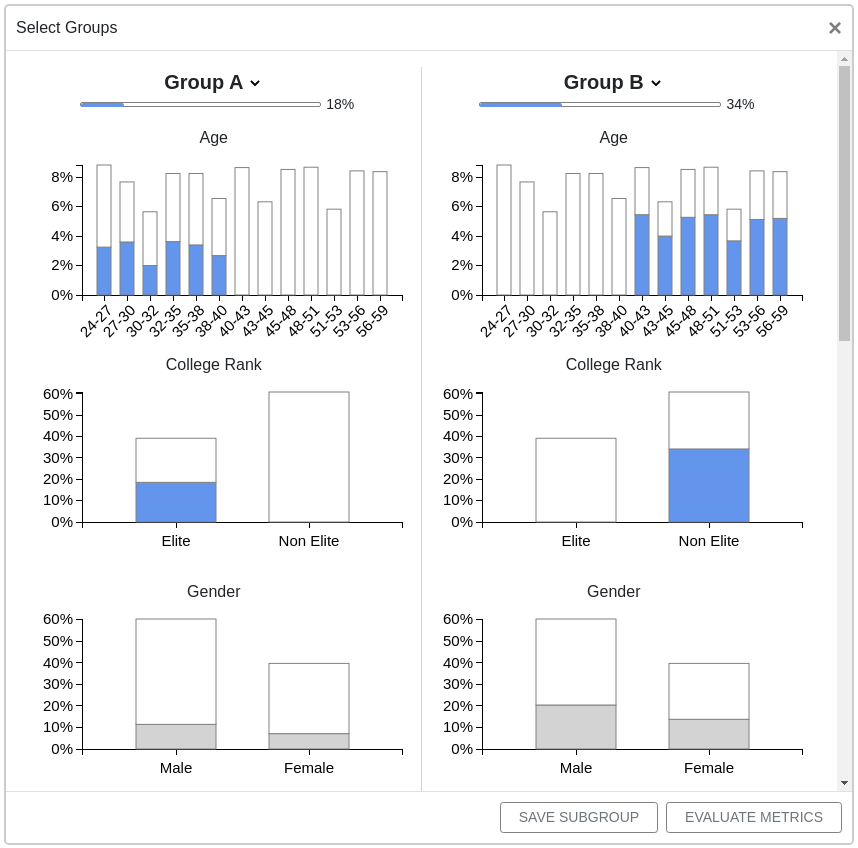}
 \caption{The visual interface for selecting subgroups (Group A and B). Each column consists of a list of bar charts/histograms representing all attributes in the dataset. By default, all bars are colored gray. The user can click on multiple bars to select a subgroup. Selected bars are colored blue. Each bar is filled in proportion of their representation in the selected subgroup as a ratio of the entire dataset. In this picture, Group A consists of individuals who went to elite universities and whose age lies between 24-40. It represents 18\% of the dataset.}  
 \label{fig:custom_group}
\vspace{-10pt}
\end{figure}

This panel, at the right (see Fig. \ref{fig:teaser}(C)), provides different options and visual plots for comparing and evaluating the changes made to the original dataset. From the top left, users can select the sensitive variable and the ML algorithm from their respective dropdown menus. This selection will be used for computing different fairness and utility metrics.  
For the sensitive variable, the dropdown menu consists of all binary categorical variables in the dataset along with a \textit{Custom Group} option. Selecting the \textit{Custom Group} option opens a new window (see Fig. \ref{fig:custom_group}) which allows the user to select groups composed of multiple attributes. This interface facilitates comparison between intersectional groups say black females and white males. 

Clicking the ``Evaluate Metrics" button triggers the computation of evaluation metrics that are displayed on the right half of this panel (Performance View). It  also visualizes the relationship between the sensitive attribute or selected groups and the output variable using a 4-fold display \cite{friendly1994fourfold} on the left half of this panel (Comparison View). 

\textbf{Comparison View.} This view comprises two plots aligned vertically where the top plot represents the original dataset and the bottom plot represents the debiased dataset. 
This view has two functions. It first aids the user in the initial exploration of different features and relationships. When a node or edge is clicked in the causal network, the summary statistics of the corresponding attributes is visualized (see Fig. \ref{fig:adult}(b) for an example). For binary relationships, we use either a scatter plot, a grouped bar chart or an error bar chart depending on the data type of the attributes. The second function of the Comparison View is to visualize the differences between the original and the debiased dataset. Initially, the original data is the same as the debiased data and so identical plots are displayed. However, when the user injects their prior into the system, the plots for the original and debiased datasets start to differ. We added this view to provide more transparency and interpretability to the debiasing process and also help detect sampling bias in the data. 

When the user clicks the ``Evaluate Metrics" button, the Comparison View visualizes the binary relation between the sensitive attribute or selected groups and the output variable via the 4-fold plot as shown in Fig. \ref{fig:teaser}(C), left panel). 
We chose a 4-fold display over a more standard  brightness-coded confusion matrix since the spatial encoding aids in visual quantitative assessments. The left/right half of this display represents two groups based on the chosen sensitive variable (say males and females) or as defined in the Custom Group interface, while the top/bottom half represents different values of the output variable say getting accepted/rejected for a job. Here, symmetry along the y-axis means better group fairness.


\textbf{Performance View.}
This view houses all the evaluation metrics as specified in Sec. \ref{sec:eval_metrics}. It uses a horizontal grouped bar chart to visualize two utility and five fairness metrics. Lower values for the fairness metrics mean better fairness. Higher values for utility metrics means better utility. Data distortion is visualized using a donut chart. On hovering over any of these charts, a tooltip shows the exact values.   

\section{Case Study - Hiring Dataset}
\label{sec:syn_data_case_study}
\textbf{Dataset} We generated a synthetic hiring dataset fraught with gender and racial bias to test our tool.
Unlike real world datasets, we exactly know the underlying data generation process for a synthetic dataset. This helps better gauge the overall performance of our tool.

\begin{figure}[tb]
 \centering 
 \includegraphics[width=0.7\columnwidth]{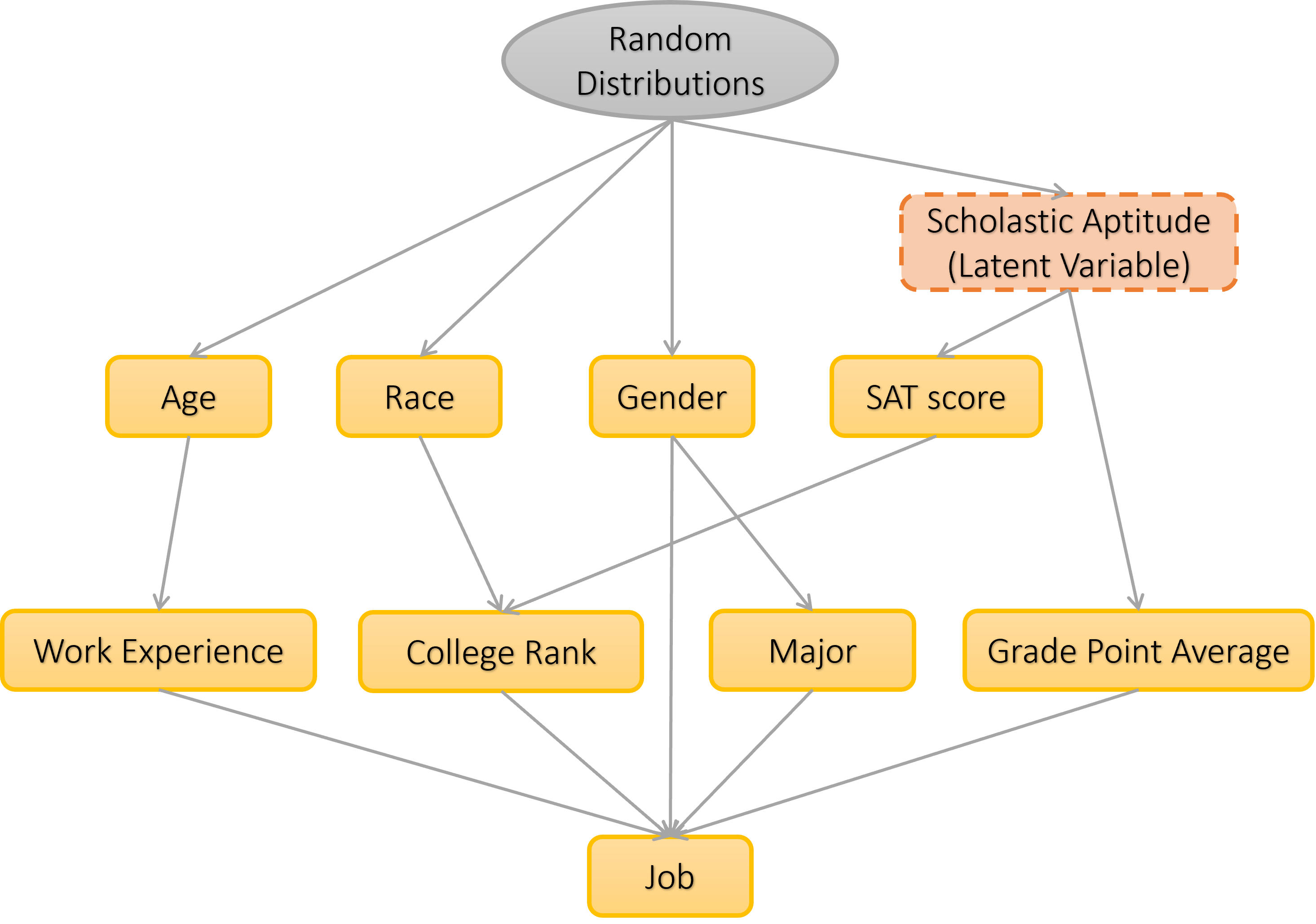}
 \caption{Data generating mechanism for the synthetic hiring dataset. Each node, colored in gold, represents a feature in the dataset; each arrow represents a causal relation between features.}  
 \label{fig:syn_data_gen}
\end{figure}

This dataset consists of 4,000 rows and 9 columns (3 numeric and 6 categorical). Each row represents a job candidate with features like gender, work experience, age, etc. The output variable `Job' represents whether a candidate got the job or not. 
We generated the exogenous variables such as race, gender, age, etc. by drawing random samples from different distributions. Here, we used binomial distribution for categorical variables and uniform distribution for numeric variables. Thereafter, each endogenous variable is generated as a linear function of its parent nodes.  
The inter-dependencies between all features is represented in \autoref{fig:syn_data_gen}. In an effort to mimic the real world challenges, we introduced sampling bias for the gender variable such that there are more males than females. Moreover, we introduced a latent confounding variable (scholastic aptitude), i.e., a relevant variable that is missing from the dataset. 

Jane works as a Data Scientist for a big software company. Her company receives an abundance  of job applications each day. It is virtually impossible to manually go through each of those applications, so she's tasked with building an AI tool that can filter out weak applications. Learning from Amazon's example \cite{amazonHiringSexist}, she's aware that such recruiting tools can be discriminatory towards minorities and pose serious challenges for her organization. Hence, she decides to use D-BIAS to check if there exists some bias in the training data, and if so, mitigate its effects by debiasing the data before moving forward with the tool.

\begin{figure*}[t]
 \centering 
 \includegraphics[width=\columnwidth]{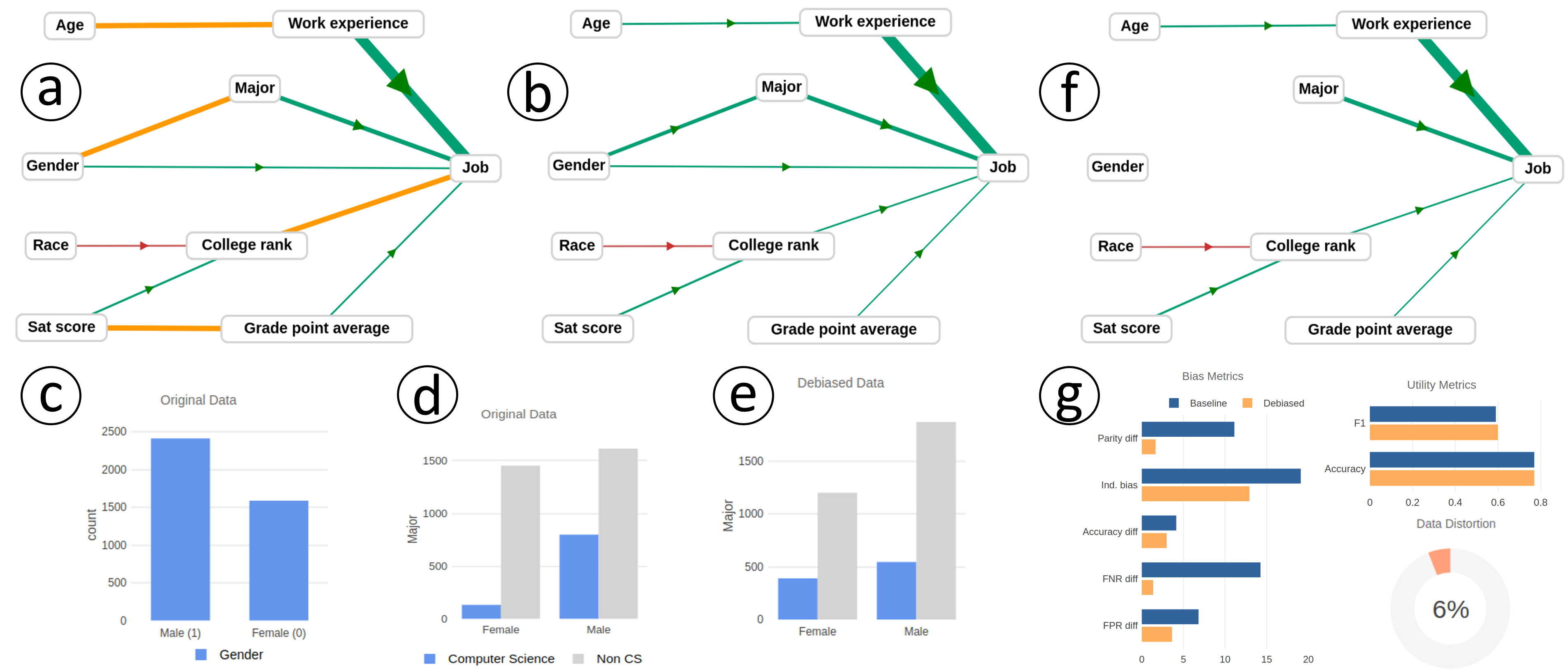}

 \caption{Different steps for the case study of the Synthetic Hiring dataset (a) Causal model generated using automated techniques (b) Refined causal model (c) Clicking the Gender node visualizes its distribution as a bar chart (d) Bivariate relation between Major and Gender for the original dataset (e) Bivariate relation between Major and Gender for the debiased dataset (f) Debiased causal model (g) Evaluation metrics to compare our results against the baseline debiasing approach.} 
 \label{fig:synthetic}
\end{figure*}

\textbf{Generating the causal network.} Jane uploads the synthetic hiring dataset which contains records of past applicants profiles and whether they got the job. She selects \textit{Job} as the label variable and chooses \textit{Gender}, \textit{Race}, \textit{Major}, \textit{Grade Point Average}, \textit{College Rank}, \textit{Job} as the nominal attributes. Next, she clicks on the ``Causal Model" button to generate the causal network with p=0.01 (see \autoref{fig:synthetic}(a)). As we are dealing with a synthetic dataset, we can compare the auto-generated causal network using PC algorithm with the ground truth (see \autoref{fig:syn_data_gen}). On comparison, we observe that the automated causal discovery algorithm was able to correctly identify many causal relations such as \textit{Work experience} $\to$ \textit{Job} and \textit{SAT score} $\to$ \textit{College rank}; there are other causal relations which were correctly identified but whose direction could not be determined like \textit{Age} $\to$ \textit{Work experience}. Moreover, there is a wrongly identified causal relation, between \textit{SAT score} and \textit{Grade point average}, that might due to the missing attribute (Scholastic Aptitude). Jane assesses different causal edges and directs undirected edges based on her domain knowledge.
She directs edges like \textit{Gender} $\to$ \textit{Major}, \textit{College Rank} $\to$ \textit{Job}, etc. to get to the refined causal model (see \autoref{fig:synthetic} (b)).

\begin{figure*}[h!]
 \centering 
 \includegraphics[width=\columnwidth]{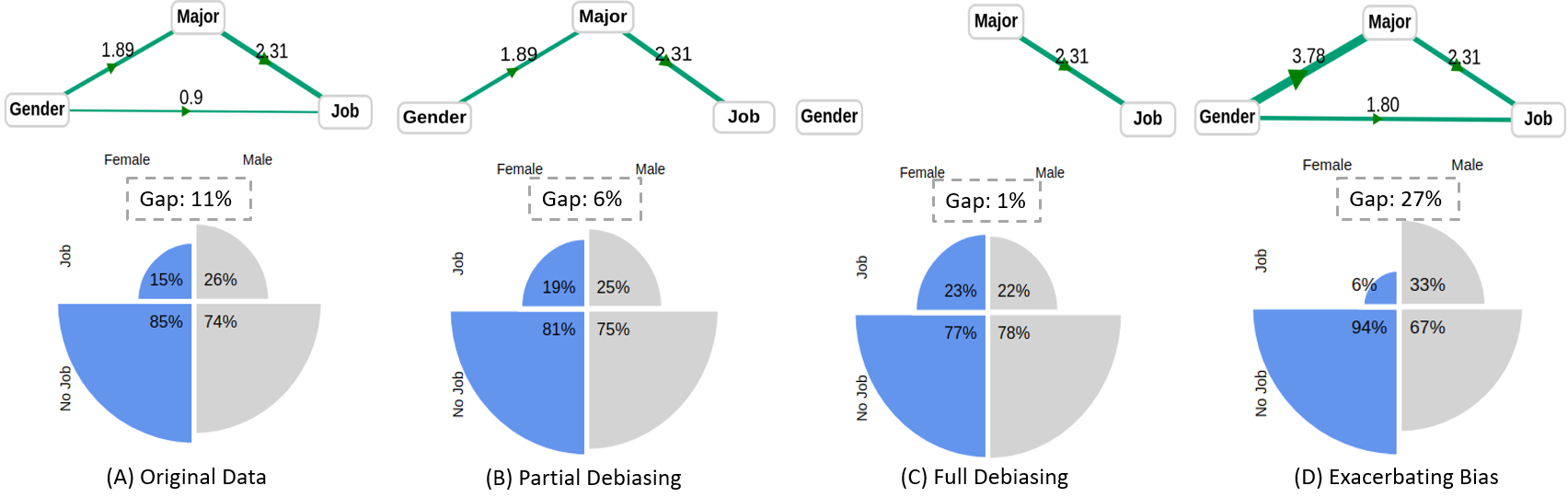}

 \caption{Fourfold displays corresponding to different user interactions for the synthetic hiring dataset. Here, we have we have shown a subset of the causal network which connects the Gender node with the Job node. The left half of each fourfold display represents females and the right half represents males. The top left quadrant represents the percentage of females who got the job, the top right quadrant represents the percentage of males who got the job, and so on.} 
 \label{fig:syn_fourfold}
\end{figure*}

\textbf{Auditing for social biases.}
Post-refining, she starts exploring the dataset by clicking on different nodes and edges in the causal network to see their underlying distributions. On clicking the \textit{Gender} node, she finds that there are far more males than females (2412 vs 1588) in the dataset (see \autoref{fig:synthetic}(c)). Such differences in representation can also lead to algorithmic bias. 
On further exploration, she finds two causal paths connecting the gender node with the job node. She is surprised to find that one of those paths is a direct link between the two nodes. This shows that her company's hiring decisions in the past have been directly influenced by gender. The other path connects the two nodes via the major node (see \autoref{fig:synthetic}(b)). Overall, these two causal paths confirms the presence gender bias and gender based disparities inherent in the dataset.  
To determine the extent of gender bias, she clicks the ``Evaluate Metrics" button from the evaluation panel. As shown in \autoref{fig:syn_fourfold}(A), she observes a 11\% gap in the likelihood of getting a job between the two genders; 26\% of males in the dataset got the job versus 15\% for females. To ensure that the AI tool does not learn such historical biases, she decides to debias the dataset first before training the ML model. 

\textbf{Mitigating bias} 
She starts the debiasing process by flipping the stage toggle on the top panel from \textit{Refine} to
\textit{Debias}. To mitigate the impact of direct gender discrimination, she deletes the edge \textit{Gender} $\to$ \textit{Job}. On evaluation, this results in a significant reduction of bias across all fairness metrics, a small loss in accuracy (1\%), and a slight increase in data distortion (2\%). As shown in \autoref{fig:syn_fourfold}(B), the percentage of females with job increased from 15\% to 19\% while the percentage of males with job decreased from 26\% to 25\%. In totality, the disparity between genders reduced from 11\% to 6\%. Given the minimal loss in utility and the scope of mitigating bias further, Jane decides to continue with the debiasing process. She focuses on the indirect path from \textit{Gender} to \textit{Job} via \textit{Major} node. She selects the edge \textit{Gender} $\to$ \textit{Major} to see their bivariate distribution. As shown in see \autoref{fig:synthetic}(d), she finds that the female representation in computer science is much lower than that of males which is indirectly contributing to bias against women in hiring. This is not something that her organization is directly responsible for. However, she does not want the ML model to learn such a socially undesirable pattern that might eventually lead to gender-based disparate impact. So, she deletes this edge to reach to the debiased causal model (see \autoref{fig:synthetic}(f)). Post-deletion, she focuses on the comparison view to understand the impact of her last intervention. \autoref{fig:synthetic}(d) and \autoref{fig:synthetic}(e) shows the bivariate distribution between gender and major in the original and debiased dataset, respectively. She finds that the number of females in computer science has increased from 135 to 389 while the number of males in computer science has decreased from 799 to 545. Overall, this intervention reduced the disparity between females and males opting for computer science.     

Finally, she re-evaluates the net impact of her interventions as captured by different evaluation metrics. She finds that the disparity between gender has reduced to just 1\% (as shown in \autoref{fig:syn_fourfold} (C)); data distortion has increased to 6\%; utility metrics are virtually the same, and fairness increases across the board as captured by all the 5 fairness metrics (see \autoref{fig:synthetic}(g)). Now, she can simply download the debiased data and use it to train the AI tool for fairer predictions. 
In this case study, we have limited ourselves to mitigating gender bias. However, similar process can be followed for mitigating racial bias.  

\textbf{Incorporating institutional goals.} D-BIAS facilitates incorporation of human prior in the system. Human prior is not just limited to social biases. It can be used to implement policy decisions or other institutional goals as well. Let's say Jane's organization changes its hiring policies and now looks for candidates with much higher work experience than before. One way to accomplish this task is by building a custom ML algorithm which incorporates this objective. Another way this can be accomplished is by modifying the dataset such that candidates with higher experience get the Job. Jane can accomplish this objective by simply selecting the edge \textit{Work Experience} $\to$ \textit{Job} and strengthening that edge by some percentage points depending on the policy. This will ensure that the debiased data inherits this policy. Now, Jane can train any vanilla ML algorithm over the debiased dataset to accomplish this goal.

It is important to note that the flexibility offered by the D-BIAS tool can also be leveraged by malicious users to exacerbate bias. For example, one might strengthen the edges \textit{Gender} $\to$ \textit{Job} and \textit{Gender} $\to$ \textit{Major} by a 100\% (as shown in the \autoref{fig:syn_fourfold}(D)). This will lead to a surge in disparity between males and females from 11\% to 27\%.

\section{Case Study - Adult Income Dataset}
\label{sec:case_study}
We demonstrate the utility of our tool for bias identification and mitigation using the Adult Income dataset. Each data point in the dataset represents a person described by 14 attributes recorded from the US 1994 census. Here, the prediction task to classify if a person's income will be greater or less than \$50k based on attributes like age, sex, education, marital status, etc. We chose this dataset as it is widely used in the algorithmic fairness literature \cite{inprocess,nipsIBM,ghai2022cascaded}. Here, we have chosen a random sample of 3000 points from this dataset for faster computation. 

\begin{figure*}[tb]
 \centering 
 \includegraphics[width=\columnwidth]{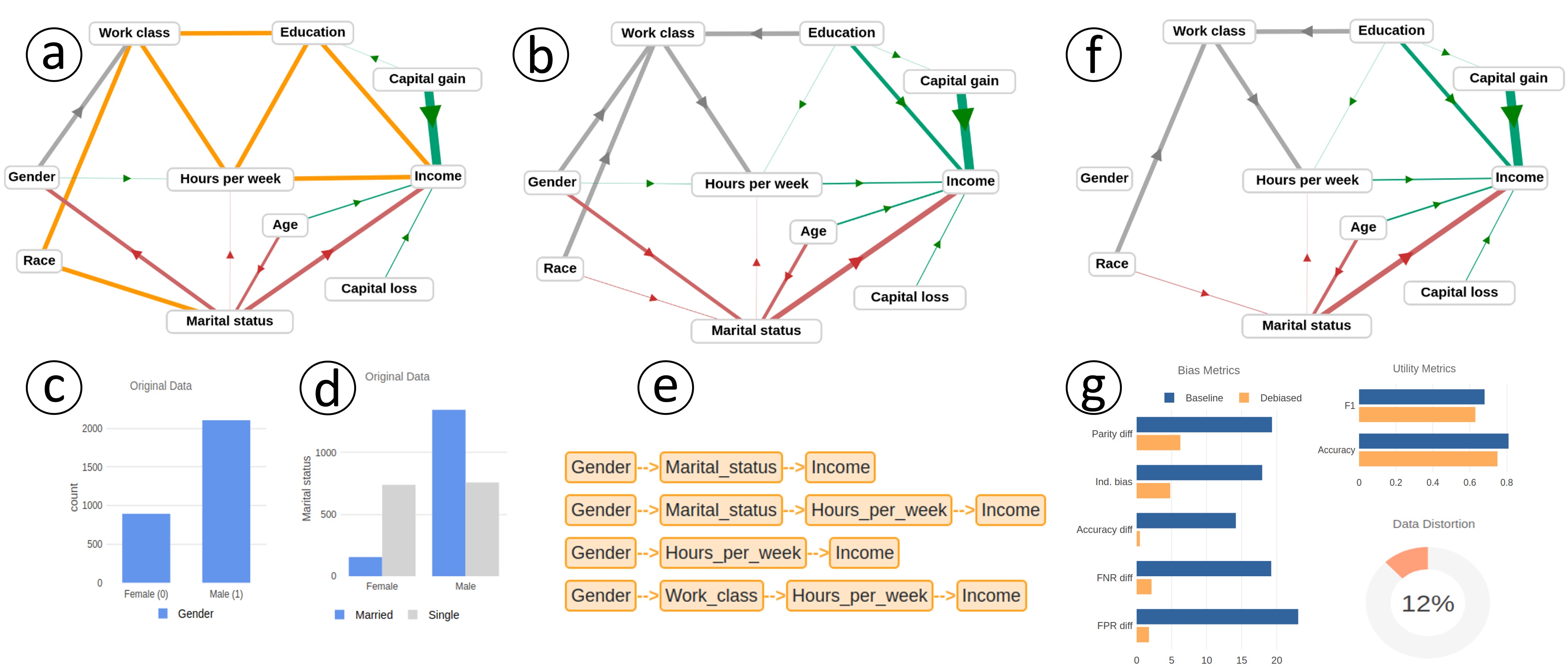}
  \setlength{\belowcaptionskip}{-2pt}
  \setlength{\abovecaptionskip}{-1pt}
 \caption{Case study: Adult Income dataset (a) Causal model generated using automated techniques (b) Refined causal model (c) Clicking the Gender node visualizes its distribution as a bar chart (d) Bivariate distribution between gender and marital status (e) All paths from Gender to Income in the refined causal model (f) Debiased causal model (g) Evaluation metrics to compare our results against the baseline debiasing approach. }  
 \label{fig:adult}
\end{figure*}

\textbf{Generating the causal network.}
We start off by selecting the Adult Income dataset from the respective dropdown menu in the Generator panel. 
We select \textit{Income} as the label variable and \textit{Work class}, \textit{Marital Status}, \textit{Race}, \textit{Gender}, \textit{Income} as the nominal variables. Next, we click on the \textit{Causal Model} button which generates the default causal model (see Fig. \ref{fig:adult} (a)). 
Here, we examine different edges of the causal model and act on them as needed to reach to a reliable causal model. We start with the 7 undirected edges encoded in orange. We direct edges based on our domain knowledge like \textit{Hours per Week} $\to$ \textit{Income},
\textit{Education} $\to$ \textit{Income}, \textit{Education} $\to$ \textit{Hours per week}, etc. After each of these operations, we observe a green bar in the left panel of the Causal Network View. This indicates that the resulting causal model is a better fit over the underlying dataset.  
Next, we examine other directed edges. Many of them align with our domain knowledge like \textit{Capital Gain} $\to$ \textit{Income}, \textit{Age} $\to$ \textit{Income}, etc. However, we found a couple of them to be counter-intuitive, namely \textit{Capital Gain} $\to$ \textit{Education} and \textit{Marital Status} $\to$ \textit{Gender}. In a causal relation, cause always precedes effect. Hence, immutable personal characteristics like sex, race, etc., which are assigned at birth, can not be the effect of a later life event like marriage or work class. So, in this case, we chose to reverse both these edges to get to the refined causal model (see Fig. \ref{fig:adult}(b)). 

\begin{figure*}
 \centering 
 \includegraphics[width=\columnwidth]{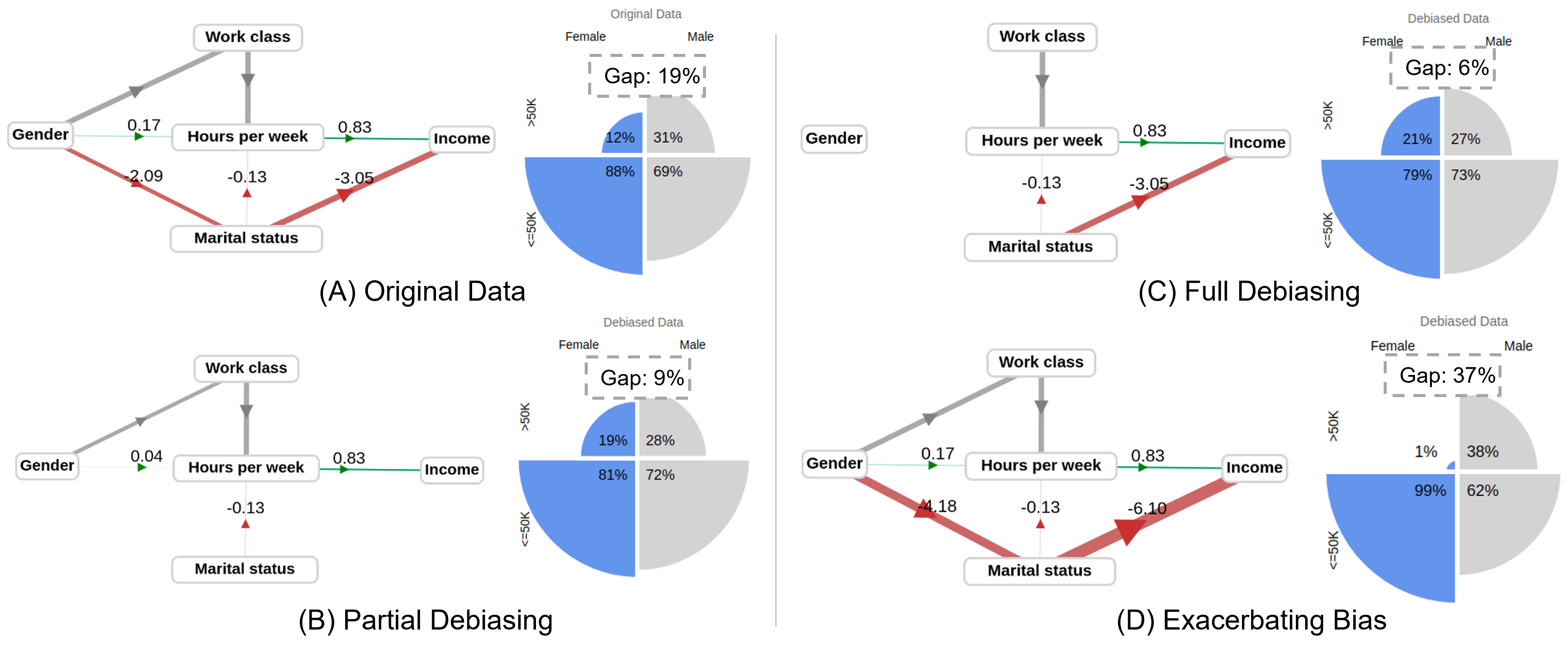}
  \setlength{\belowcaptionskip}{-8pt}
  \setlength{\abovecaptionskip}{-4pt}

 \caption{The above picture shows the impact of 3 types of user interaction as captured by the 4-fold display. Due to space constraints, we have only shown a subset of the causal network which connects the Gender node with the Income node. For details, please refer to the description in Section \ref{sec:case_study}.} 
 \label{fig:adult_fourfold}
\end{figure*}

\textbf{Auditing for social biases.}
Once we get to a reliable causal model, we start auditing for different kinds of biases. We click on different nodes and edges to explore their distributions. For example, clicking the \textit{Gender} node visualizes its distribution in the \textit{Comparison View}. We observe that females are underrepresented in the dataset (895 females vs 2105 males) (see Fig. \ref{fig:adult} (c)). 
Given this representation bias and the fact that gender pay gap is a well-known issue, we decided to investigate further.  
We found an indirect path from \textit{Gender} to \textit{Income} via \textit{Hours per week}. This indicates a possible disparity in income based on gender. To probe further, we click the ``Evaluate Metrics" button with \textit{Gender} as the sensitive variable to compute different fairness metrics. We observe significant gender bias as captured by metrics like Accuracy diff (14\%), FPR diff (22\%), FNR diff (17\%), etc.
The 4-fold display for the original data in Fig. \ref{fig:adult_fourfold} (A) reveals that only 12\% of the females earn more than \$50k compared to 31\% for males. Thus there is a significant income disparity based on gender. To have a more comprehensive understanding of the issue, we search for all possible paths by selecting \textit{Gender} and \textit{Income} as the source and target from the Top panel and then clicking the ``Find Paths" button from the bottom panel. This populates 4 different causal paths below the bottom panel (see Fig. \ref{fig:adult} (e)). We will now focus on the different causal relationships in these paths and try to make minimal changes to achieve more fairness.      

\textbf{Bias mitigation.} To mitigate gender bias, we first enter the debiasing stage by flipping the Stage toggle from Refine to Debias. From here on, any changes to the causal network will simulate a new debiased dataset. 
Among the 4 paths we previously discovered, the top 2 paths have a common causal edge, i.e., \textit{Gender} $\to$ \textit{Marital status}. On clicking this edge, we find that most males in the dataset are married while most females are single (see Fig. \ref{fig:adult}(d)). This pattern indicates sampling bias.
Ideally, we would like no relation between these attributes so we delete this causal edge. 
Next, we assess the remaining two causal paths. Based on our domain knowledge, we find the causal edge \textit{Hours per week} $\to$ \textit{Income} to be socially desirable, and the edges \textit{Gender} $\to$ \textit{Work class} and \textit{Gender} $\to$ \textit{Hours per week} to be socially undesirable. We delete the two biased edges to get to the debiased causal model (see Fig. \ref{fig:adult}(f)). To verify all changes made so far, we click on the \textit{Logs} button. As shown in Fig. \ref{fig:logs}, it shows 3 dotted lines for the removed edges and highlight the impacted attributes. 

Lastly, we click on the ``Evaluate Metrics" to see the effect of our interventions. The 4-fold display for the debiased data in Fig. \ref{fig:adult_fourfold}(C) shows that the disparity between the two genders has now decreased from 19\% to 6\% (compare Fig. \ref{fig:adult_fourfold}(A)). The percentage of females who make more than \$50k has undergone a massive growth of 75\% (from 12\% to 21\%). 
Moreover, as shown in Fig. \ref{fig:adult}(g), we find that all fairness metrics have vastly improved, with only a slight decrease in the utility metrics and an elevated distortion (12\%). 
These results clearly indicate the efficacy of our debiasing approach.



\textbf{Partial debiasing.} Considering the tradeoff between different metrics, one might choose to debias data partially based on their context, i.e., weaken biased edges instead of deleting them or keeping certain unfair causal paths from the sensitive variable to the label variable intact.
For example, one might choose to delete the edge \textit{Gender} $\to$ \textit{Marital status} and weaken the edges \textit{Gender} $\to$ \textit{Work class} and \textit{Gender} $\to$ \textit{Hours per week} by 25\% and 75\%, respectively. On evaluation, we find this setup to sit somewhere between the original and the full debiased dataset (see Fig. \ref{fig:adult_fourfold} (B)). It performs better on fairness than the original dataset (gap: 9\% vs 19\%) but worse than the full debiased version (gap: 9\% vs 6\%). Similarly, it incurs more distortion than the original dataset but less than the full debiased version (11\% vs 12\%).  

\begin{table}
  \caption{Evaluation metrics to compare the debiased dataset generated using our tool against the baseline debiasing approach for different datasets.}
  \label{tab:table}
  \scriptsize%
	\centering%
	\resizebox{\columnwidth}{!}{%
  \begin{tabu}{%
	r%
	*{10}{c}%
	*{2}{c}%
	}
  \toprule
   \multirow{2}{*}{Dataset} & Sensitive & \multirow{2}{*}{version} & \multirow{2}{*}{ML model} & \multirow{2}{*}{Accuracy} & \multirow{2}{*}{F1} & Parity & Individual & Accuracy &  FNR &  FPR & Data \\ 
   & attribute & & & & & difference & Bias & difference &  difference &  difference & Distortion \\ 
  \midrule \\
  \multirow{2}{*}{Synthetic Hiring} & \multirow{2}{*}{Gender} & baseline & \multirow{2}{*}{SVM} & 77\% & 0.59 & 11.12 & 19.09 & 4.14 & 14.26 & 6.82 & 0\% \\ 
  & & debiased & & 77\% & 0.60 & 1.66 & 12.93 & 2.99 & 1.37 & 3.63 & 6\% \\ \\
  
  \multirow{2}{*}{Adult Income} & \multirow{2}{*}{Gender} & baseline & Logistic & 82\% & 0.69 & 19.32 & 17.92 & 14.35 & 17.98 & 22.53 & 0\%\\ 
  & & debiased & Regression & 75\% & 0.63 & 6.24 & 4.8 & 0.88 & 2.33 & 1.9 & 12\% \\ \\
  
  \multirow{2}{*}{COMPAS} & \multirow{2}{*}{Race} & baseline & Random & 67\% & 0.64 & 12.07 & 33.9 & 0.17 & 23.07 & 16.89 & 0\%\\
  & & debiased & Forest & 63\% & 0.59 & 11.12 & 2.19 & 0.44 & 0.68 & 1.55 & 13\% \\ \\
  
  \bottomrule
  \end{tabu}%
  }
\vspace{-5pt}
\end{table}

\textbf{Intersectional groups.} D-BIAS facilitates auditing for biases against intersectional groups using the ``Custom Group" option from the sensitive variable dropdown.
Here, we choose \textit{Black Females} and \textit{White Males} as the two groups. At the outset, there is a great disparity between the groups as reflected in the 4-fold display and the fairness metrics (see Fig. \ref{fig:teaser}). 
As these subgroups are defined by \textit{Gender} and \textit{Race}, we focus on the unfair causal paths from these nodes to the label variable (\textit{Income}). 
For debiasing, we first perform the same operations we did for gender debiasing. Thereafter, we reduce the impact of race by deleting the edges \textit{Race} $\to$ \textit{Work class} and \textit{Race} $\to$ \textit{Marital status} which we deem as socially undesirable. On evaluation (see Fig. \ref{fig:teaser}), we find a significant decrease in bias across all fairness metrics for the debiased dataset compared to the conventional debiasing practice (blue bars) which just trains the ML model with the sensitive attributes (here \textit{Gender} and \textit{Race}) simply removed.
Finally, the two 4-fold displays reveal that the participation of the disadvantaged group more than doubled, while the privileged group experienced only a modest loss.

\textbf{Exacerbating bias.} The flexibility offered by D-BIAS to refine the causal model can be misused to increase bias as well. 
Bias can be exacerbated by strengthening/adding biased causal edges and weakening/deleting other relevant causal edges.
For example, 
one can exacerbate gender bias by strengthening the edges \textit{Gender} $\to$ \textit{Marital status} and \textit{Marital status} $\to$ \textit{Income} by a 100\%. On evaluation, we find that the proportion of females making $>$\$50k has shrunk to just 1\% while the proportion of males has surged to 38\%. In effect, this has broadened the gap between males and females making more than \$50k by about 2x from 19\% to 37\% (see Fig. \ref{fig:adult_fourfold} (D)). 

\textbf{Results.} Apart from the Adult Income dataset, we also tested our tool using the synthetic hiring dataset and the COMPAS recidivism dataset (see appendix D for details). 
The evaluation metrics for all 3 datasets after full debiasing are reported in Table \ref{tab:table}. As we can observe, our tool is able to reduce bias significantly compared to the baseline across the 3 datasets for a small loss in utility and data distortion. These results validate the potential of HITL approach in mitigating bias. It is interesting to observe that the F1 score for the synthetic hiring dataset is slightly higher than the baseline. However, this is line with the existing literature \cite{ghai2022cascaded} where similar instances have been recorded.

\section{Case Study - COMPAS dataset}
\label{sec:compas_case_study}
COMPAS is a popular dataset often used in the fairness literature \cite{aif360}. It pertains to the criminal defendants from Broward County, Florida. The task is to predict recidivism within the next 2 years. In other words, we are to classify whether a convicted criminal will reoffend in the next two years. After preprocessing, we ended up with 6,150 rows and 7 columns. Here, each row corresponds to an individual described by attributes such as number of juvenile offenses, charge degree (felony, misdemeanor), race (Caucasian, African-American), age, gender, etc.  The sensitive attribute is race. In the following, we will demonstrate how D-BIAS can help identify and alleviate racial bias from this dataset. 

\begin{figure*}[t]
 \centering 
 \includegraphics[width=\columnwidth]{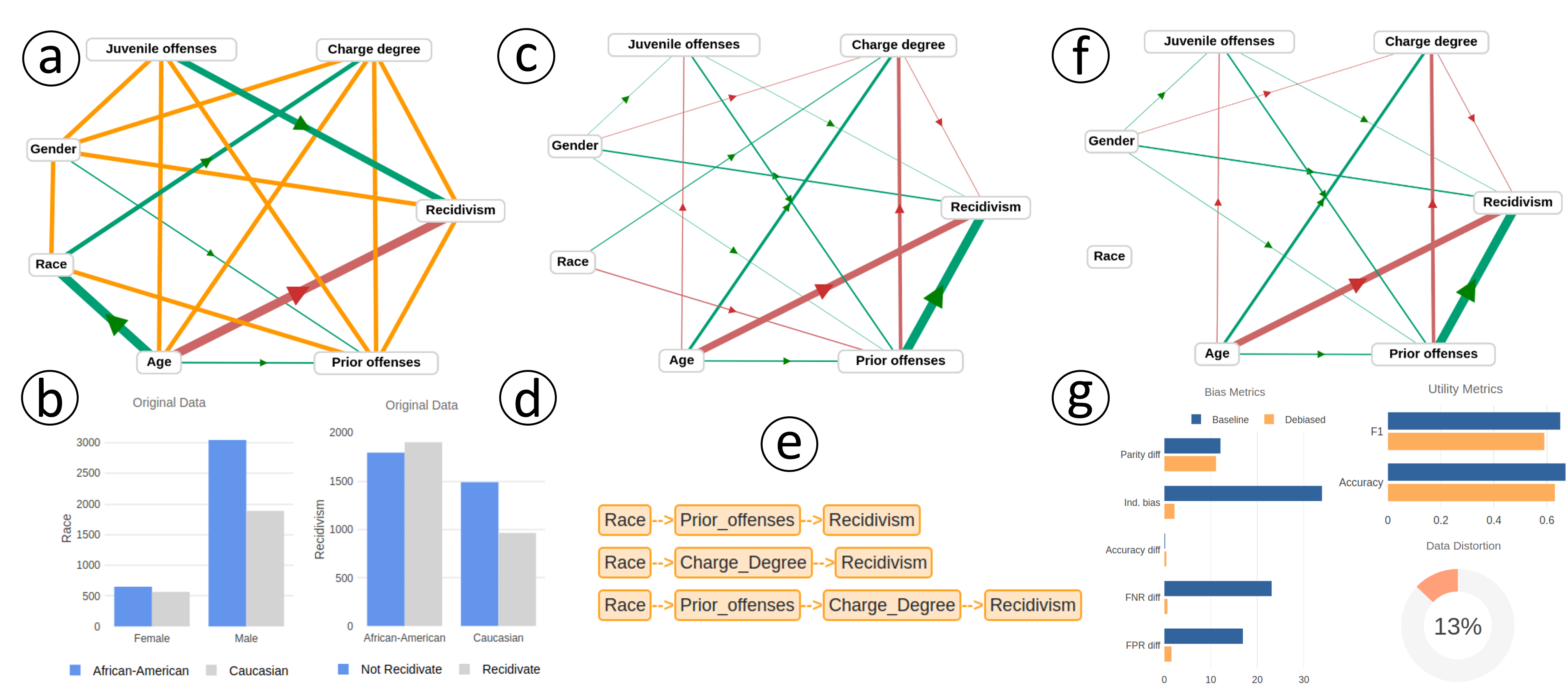}
 \caption{Use case: COMPAS dataset (a) Initial causal network returned by the PC Algorithm (b) Bivariate relationship between Race and Gender (c) Refined causal network (d) Grouped bar chart between Race and Recidivism showing the inherent racial bias (e) All paths between Race and Recidivism (f) Debiased causal network (g) Evaluation metrics to compare our results against the baseline debiasing approach. }  
 \label{fig:compas}
\end{figure*}

   \textbf{Generating the causal network} We start by selecting COMPAS as the dataset, $Recidivism$ as the label variable, and \textit{Gender}, \textit{Race}, \textit{Recidivism}, \textit{Charge degree} as the nominal variables from the generator panel. Thereafter, we click ``Causal Model" to generate the causal network as shown in \autoref{fig:compas}(a). We observe that there are a lot of undirected edges (in orange). This can be due to missing attributes or due to the limitation of the PC algorithm in determining their direction. We go through all these edges and deal with them based on our domain knowledge. For example, we direct edges such as \textit{Prior Offences} $\to$ \textit{Recidivism}, \textit{Charge degree} $\to$ \textit{Recidivism}, etc. We direct these edges in this way since such a relationship is more plausible than the contrary. Interestingly, we observe an undirected edge between \textit{Gender} and \textit{Race}, and a directed edge between \textit{Age} and \textit{Race}. From our domain knowledge, we know that these relationships do not exist in the real world. To investigate further, we click on these edges to see their corresponding distributions. We observed a sampling bias for both the cases that might have led to the false detection of these edges. As shown in \autoref{fig:compas}(b), we observe that there are a lot more males than females in the dataset. Also, there are more african-americans than caucasians for each gender. This observation hints the need to have a more diverse dataset with respect to gender and race. However, it may not always be possible to gather additional data points. Considering such a scenario, we decided to remove these edges in the refining stage to prevent unnecessary data distortion. The refined causal model can be seen in \autoref{fig:compas}(c).     

   \textbf{Auditing and mitigating bias} As previously observed, African Americans are over-represented in the dataset (60\%) relative to their their population as a whole in Florida (17\%). To further probe the presence of racial bias, we selected \textit{Race} and \textit{Recidivism} from the top panel to see their bivariate distribution in the comparison view (as shown in \autoref{fig:compas}(d)). From the grouped bar chart, we can clearly observe a disparity in the likelihood to recidivate (recommit a crime) between African Americans and Caucasians. As seen in the case of Berkeley's admission dataset, disparity between two groups does not necessarily mean systematic discrimination, so we tried to understand the underpinnings of this disparity via the causal network. 

   We used the find paths functionality to find all paths from \textit{Race} to \textit{Recidivism}. We obtain 3 paths as shown in \autoref{fig:compas}(e). Among the causal paths, we found causal edges such as \textit{Race} $\to$ \textit{Prior offences} and \textit{Race} $\to$ \textit{Charge degree}. From our domain knowledge, we find these causal relationships, inherent in the dataset, to be biased and socially undesirable. Training a ML model over such a dataset can potentially result in replicating, and even amplifying such racial biases. To prevent that, we deleted these causal edges to obtain our debiased causal network (see \autoref{fig:compas}(f)). 
   Since both these operations were executed during the debiasing stage, the system generates a new (debiased) dataset which accounts for these changes.   
   Lastly, we evaluated the debiased data in terms of different evaluation metrics. We found that the debiased data performs well for 4 out of the 5 fairness metrics (\autoref {fig:compas}(g)). More specifically, there has been a vast improvement in fairness as captured by Ind. bias, FNR diff and FPR diff metrics. On the flip side, this process has incurred a small loss in accuracy (4\%) and f1 score (5\%), and the data distortion rose to $13\%$. The debiased dataset can be downloaded from the generator panel and can be used in place of the original data for fairer predictions. 

\section{User Study}
\label{sec:user_study}
We conducted a user study to evaluate two primary goals: (1) usability of our tool, i.e., if participants can comprehend and interact with our tool effectively to identify and mitigate bias, (2) compare our tool with the state of the art in terms of human-centric metrics like accountability, interpretability, trust, usability, etc.

\textbf{Participants.} We recruited 10 participants aged 24-36; gender: 7 Male and 3 Female; profession: 8 graduate students and 2 software engineers.
The majority of the participants are computer science majors with no background in data visualization or algorithmic fairness. 80\% of the participants trust AI and ML technologies in general.
The participation was voluntary with no compensation. 

\begin{figure*}[t]
 \centering 
 \includegraphics[width=\columnwidth]{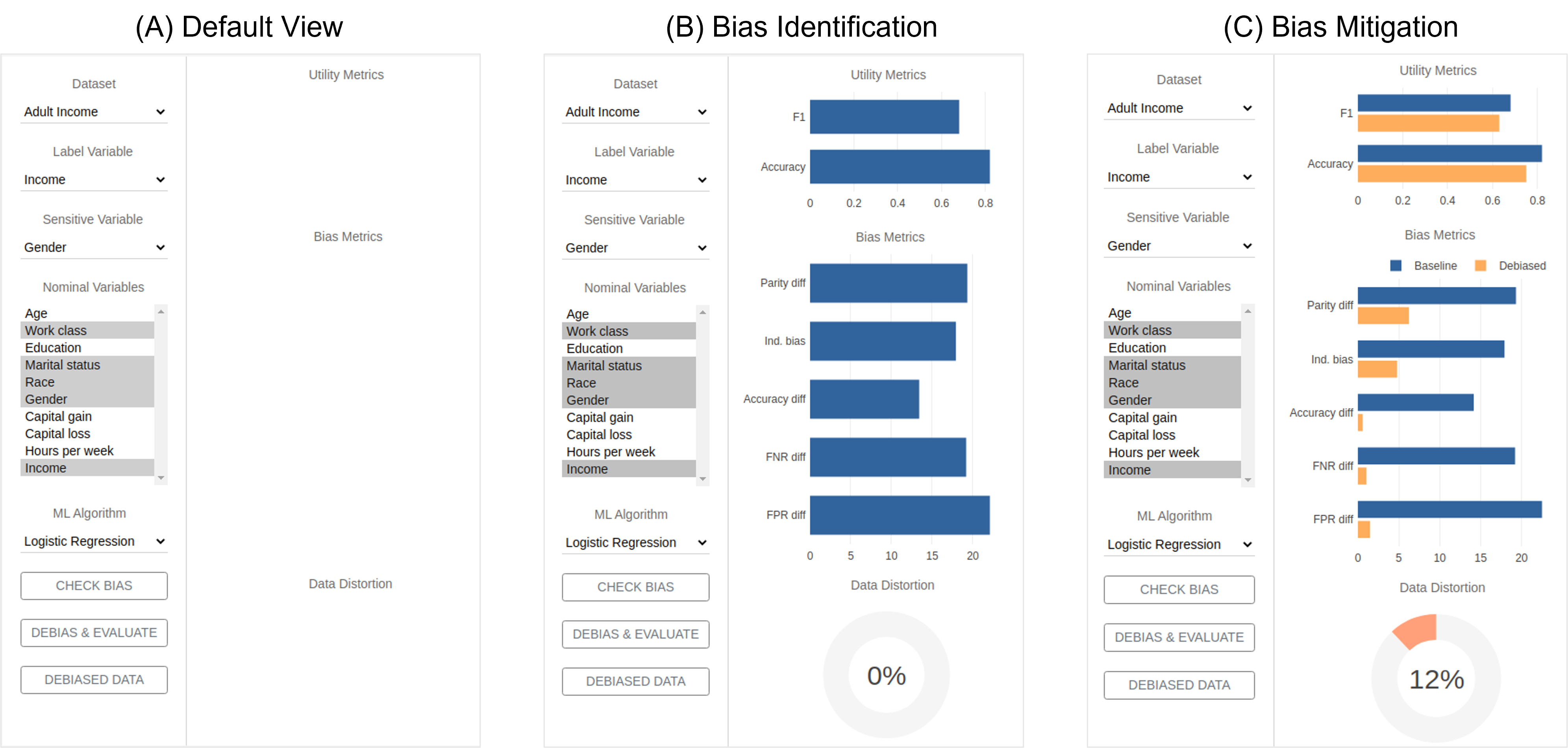}
 \caption{Visual interface of the baseline tool (A) Default view (B) Bias Identification (C) Bias mitigation }  
 \label{fig:baseline_stages}
\end{figure*}

\textbf{Baseline Tool.} To have an even comparison, we looked for existing tools with a visual interface that support bias identification and mitigation. This led us to IBM’s AI Fairness 360 \cite{aif360} toolkit whose visual interface
can be publicly accessed online\footnote{https://aif360.mybluemix.net/data}. However, we didn't go further with this toolkit as the baseline because it has a significantly different look and feel which is difficult to control for. Instead, we took inspiration from this toolkit and built a baseline visual tool (not to be confused with the baseline debiasing strategy) which mimics its workflow but matches the design of our D-BIAS tool (see Fig. \ref{fig:baseline_stages}). The visual interface of the baseline tool is roughly equivalent to the D-BIAS tool except for the causal network view.

IBM’s AI Fairness toolkit allows the user to choose from a set of fairness enhancing interventions with varying impact on the evaluation metrics. However, this study is focused on important human-centric measures such as trust, accountability, etc. So, in order to have a tightly controlled experiment, we imagine a hypothetical automated debiasing algorithm whose performance exactly matches the peak performance of our tool for all evaluation metrics. Here, peak performance refers to the state where all unfair causal edges are deleted. 

Using the baseline tool is quite simple. 
The user first encounters the default view and selects the dataset, label variable, nominal variables and the sensitive variable (see \autoref{fig:baseline_stages} (A)). For bias identification, the user simply clicks the ``Check Bias" button to compute and visualize a set of bias and utility metrics as shown in \autoref{fig:baseline_stages}(B). If the bias scores are within an acceptable range, then no further action is needed as there is no significant bias with respect to the sensitive variable. Otherwise, the user can click on the ``Debias \& Evaluate" button to debias the dataset and compute a new set of evaluation metrics. As shown in \autoref{fig:baseline_stages} (C), the user can compare how the debiased dataset performs on different evaluation metrics relative to the baseline (original dataset without the sensitive attribute). 
To have a tightly controlled experiment, we ensured that the evaluation metrics (utility metrics, fairness metrics and data distortion) for the baseline tool exactly matches with the peak performance of our D-BIAS tool. In other words, our study compares both tools (representing two different approaches) while controlling for their performance on different evaluation metrics. This helps better evaluate the design of our tool and helps understand if human interaction can enhance trust, accountability, etc. in the context of bias mitigation.   

\textbf{Study design.}  We conducted a within subject study where each participant was asked to use the baseline tool and D-BIAS in random order. The study was conducted remotely, i.e., each participant could access and interact with the tools via their own machine. For each tool, a small tutorial was given using the Synthetic Hiring dataset (see \autoref{sec:syn_data_case_study})
to demonstrate the workflow and features of the tool. Each participant was then given some time to explore and interact with the system. Next, the participants were asked to identify and mitigate bias for the Adult Income dataset. 
For the D-BIAS tool, participants were asked to complete a set of 5 tasks to evaluate usability. The task set was carefully designed to cover our testing goals and had a verifiable correct solution.
Tasks included: generate a causal network, direct undirected edges, identify if bias exists with respect to an attribute, identify proxy variables and finally debias the dataset. After using each tool, the participants were asked to answer a set of survey questions. Lastly, we collected subjective feedback from each participant regarding their overall experience with both the tools. 
Throughout the study, participants were in constant touch with the moderator for any assistance.
Participants were encouraged to think aloud during the user study. 

\textbf{Survey Design.} Each participant was asked to answer a set of 13 survey questions to quantitatively measure usability, interpretability, workload, accountability and trust. All of these questions can be answered on a 5-point Likert Scale. To capture cognitive workload, we selected two applicable questions from the NASA-LTX task load index \cite{nasa_tlx}, i.e., ``How mentally demanding was the task? and ``How hard did you have to work to accomplish your level of performance?". Participants could choose between 1 = very low to 5 = very high. For capturing usability, we picked 3 questions from the System Usability Scale (SUS) \cite{sus}. For example, ``I thought the system was easy to use", ``I think that I would need the support of a technical person to be able to use this system". Participants could choose between 1 = Strongly disagree to 5 = Strongly agree. To capture accountability, we asked two questions based on previous studies \cite{XAL, cai2019effects}. For example, ``The credit/blame for mitigating bias effectively is totally due to" (1 = System's capability, 5 = My input to the system). To capture interpretability, we consulted Madsen Gregor scale\cite{madsen2000measuring} and adopted 3 questions for our application. For example, ``I am satisfied with the insights and results obtained from the tool?", ``I understand how the data was debiased?" Answers could lie between 1 = Strongly disagree to 5 = Strongly agree. For measuring trust, we referred to McKnight's framework on Trust\cite{trust1, trust2} and other studies \cite{XAL, drozdal2020trust} to come up with 3 questions for our specific case. For example, ``I will be able to rely on this system for identifying and debiasing data" (1 = Strongly disagree, 5 = Strongly agree). 
%
\begin{figure} [t]
 \centering 
 \includegraphics[width=0.9\columnwidth]{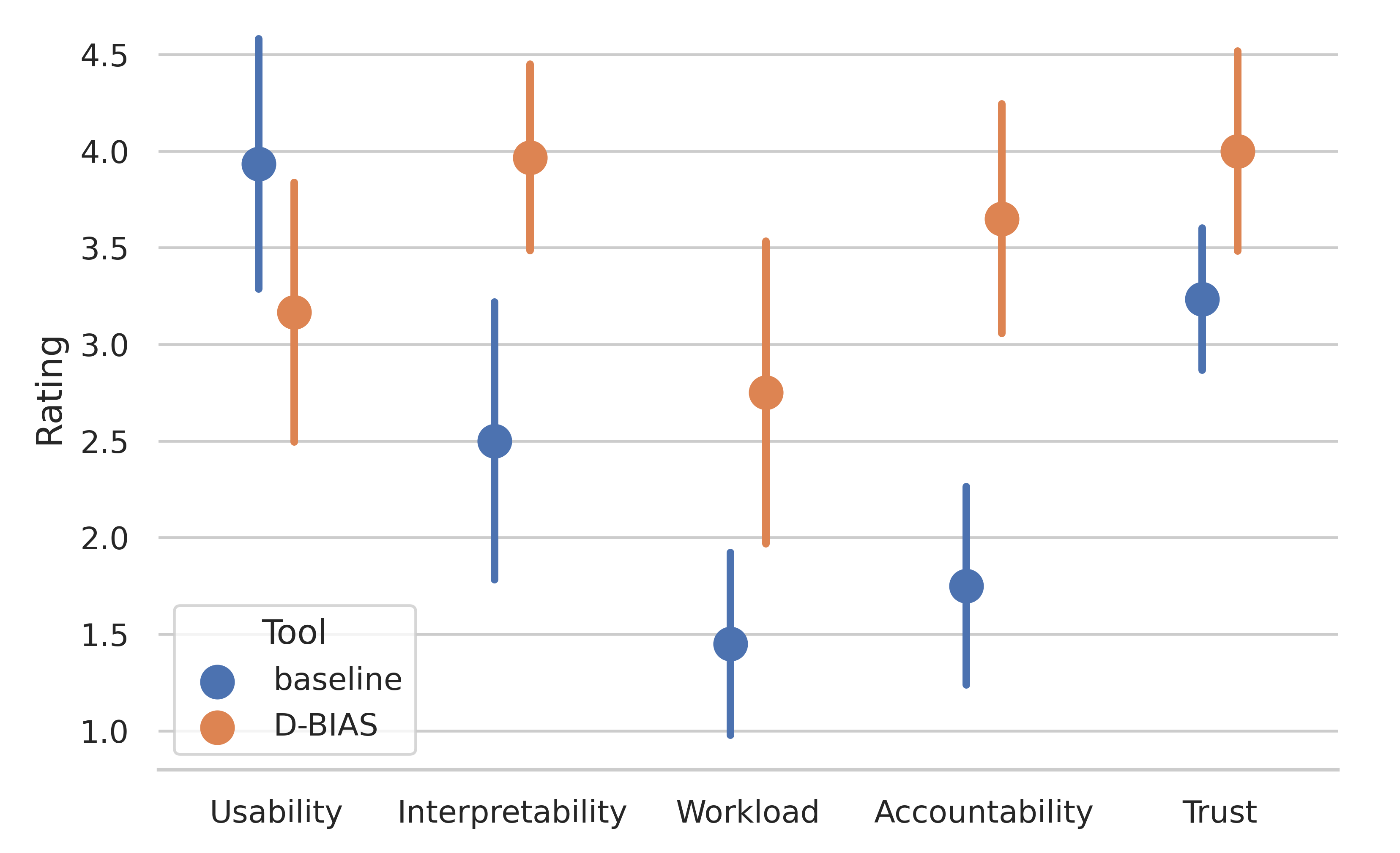}
 \caption{Mean user ratings from the survey data along with their standard deviation for different measures. 
 } 
 \label{fig:user_study}
\end{figure}

\textbf{Results.} Despite not having a background in algorithmic fairness or data visualization, all participants were able to complete all 5 tasks using the D-BIAS tool. This indicates that our tool is easy to use. 

  The survey data was analyzed to calculate usability, interpretability, workload, accountability and trust ratings for each tool by each participant. The mean ratings along with their standard deviations are plotted in Fig. \ref{fig:user_study}. 
  Using t-test, we found statistically significant differences for all measures with p$<$0.05. We found that D-BIAS outperforms the baseline tool in terms of trust, accountability and interpretability. However, it lags in usability and cognitive workload.
  So, if someone is looking for a quick fix or relies on ML algorithms more than humans, automated debiasing is the way to go. Conversely, if trust, accountability or interpretability is important, D-BIAS should be the preferred option. Looking at these results in conjunction with the results reported in Sec. 5, \textit{we find that our tool enhances fairness while fostering accountability, trust and interpretability.}     

\textbf{Subjective feedback.}
After the study, we gathered feedback from each participant about what they liked or disliked about D-BIAS. Most participants liked the overall design, especially the causal network. We got comments like ``\textit{Interface is user friendly}", ``\textit{Causal network gives control and flexibility}", ``\textit{Causal network is very intuitive and easy to understand}", ``\textit{Causal network is a great way to understand relationships between features}". Most participants agreed that after a tutorial session, it should be fairly easy for even non-experts to play with the system. Another important aspect which received a lot of attention was our human-in-the-loop approach. Participants felt that they had a lot more control over the system and that they could change things around. One of the participants commented ``\textit{It feels like I have a say}". Some of the participants said they felt more accountable because the system offered much flexibility and that they had a choice to make.

Many of the participants strongly advocated for D-BIAS over the baseline tool. For example, ``\textit{D-BIAS better than automated debiasing any day}", ``\textit{D-BIAS hand's down!}". Few of the participants had a more nuanced view. They were of the opinion that the baseline tool might be the preferred option if someone is looking for a quick fix. 
We also received concerns and suggestions for future improvement. Two of the participants raised concern about the tool's possible misuse if the user is biased. Another participant raised concern about scalability for larger datasets. Most participants felt that adding tooltips for different UI components especially the fairness metrics will be a great addon. Two participants wished they could see the exact changes in CSV file in the visual interface itself. 

\section{Discussion, Limitations \& Future Work}

\textbf{Efficacy.} The efficacy of our tool depends on how accurately the causal model captures the underlying data-generating process and the ensuing refining/debiasing process. As our tool is based on causal discovery algorithms, it inherits all its assumptions such as the causal markov condition and other limitations like missing attributes, sampling biases, etc.\cite{challenges}.
For example, Caucasians had a higher mean age than African Americans in the COMPAS dataset. So, the PC algorithm falsely detected a causal edge between \textit{Age} and \textit{Race} (see \autoref{sec:compas_case_study}). From our domain knowledge, we know that such a relation does not exist. One possible explanation for this error might be sampling bias. The user can use this insight to gather additional data points for the under-sampled group. However, it may not always be possible. In such cases, our tool can be leveraged to remove such patterns in the debiasing stage. We dealt with a similar case (\textit{Gender} $\to$ \textit{Marital status}) for the Adult Income data (see Sec. \ref{sec:case_study}). 

It is also worth noting that our tool is able to reduce the disparity between males and females from 19\% to 6\% for the adult income dataset but was unable to close the gap entirely. This can be due to missing proxy variables whose influence is unaccounted for or maybe because linear models are too simple to capture the relationship between a node and its parents. Future work might use non-linear SEMs and causal discovery algorithms like Fast Causal Inference (FCI)
that can better deal with missing attributes.


\textbf{Scalability.} As the size of the dataset increases in terms of features and rows, scalability can become an issue. 
With dataset size, the time to generate a causal network, debiasing data, finding paths between nodes and computing evaluation metrics will all increase proportionally. The next step will be to employ GPU-based parallel implementation of PC algorithm like cuPC \cite{zarebavani2019cupc} or using inherently faster causal discovery algorithms like F-GES \cite{xie2020visual}.
On the front end, the causal network will become big and complex as the number of features increase. With limited screen space, the user might find it difficult to comprehend the causal network. To alleviate this issue, we have implemented different visual analytics techniques like zooming, panning, filtering weak edges, finding paths, etc.  
Future work might optimize graph layout algorithm and explore other visual analytics techniques like node aggregation to help navigate larger graphs better \cite{xie2020visual}.


\textbf{Applications.} In this work, we have emphasized how our tool can help identify and remove social biases. However, our approach and tool is not limited to social biases. Our tool can incorporate human feedback to realize policy and institutional goals as well. For example, one might strengthen the edge between the nodes \textit{Education} and \textit{Income} to implant a policy intervention where people with higher education are incentivized.
A ML model trained over the resulting dataset is likely to reflect such policy intervention in its predictions.   
Next, we plan to extend our HITL methodology to tackle biases in other domains such as word embeddings.

\textbf{Human factors.} Involving a human in the loop for identifying and debiasing data is a double edged sword. On one hand, it is a key strength of our tool as it provides real world domain knowledge and fosters accountability and trust. On the other hand, it can also be its main weakness if the human operating this tool intentionally/unconsciously injects social biases. A user can misuse the system in two ways. Firstly, the user can choose to ignore the social biases inherent in the dataset by not acting on the unfair causal edges. Such behaviour renders the system ineffective. Secondly, a biased user can explicitly introduce their own biases in the system by adding/strengthening unfair causal edges. 
Since this is a human aided tool, the biases that are inherent to the human user cannot be avoided. Hence, we recommend choosing the the user responsibly. A human who is well versed with the domain of the data, sensitivities of the society and who is trusted by the majority of stakeholders might be a good fit. If a user tries to misuse the system, we can always check the system logs and hold the person responsible for their action/inaction. 
  
\let\textcircled=\pgftextcircled
\chapter{Auditing Social Biases in Word Embeddings}
\label{chap:wordbias}

\textit{Intersectional bias} 
is a bias caused by an overlap of multiple social factors like gender, sexuality, race, disability, religion, etc.
A recent study has shown that word embedding models can be laden with biases against intersectional groups like African American females, etc. The first step towards tackling such intersectional biases is to identify them. 
However, discovering biases against different intersectional groups remains a challenging task.   
In this work, we present \textit{WordBias}, an interactive visual tool designed to explore biases against intersectional groups encoded in static word embeddings. Given a pretrained static word embedding, WordBias computes the association of each word along different groups based on race, age, etc. and then visualizes them using a novel interactive interface.
Using a case study, we demonstrate how WordBias can help uncover biases against intersectional groups like Black Muslim males, Poor females, etc. encoded in word embedding. 
In addition, we also evaluate our tool using qualitative feedback from expert interviews.

\section{Introduction}
Word embedding models such as Glove \cite{glove} and Word2vec \cite{mikolov2013distributed} can be understood as a mapping between a word and its corresponding vector representation. They serve as the foundational unit for many NLP applications such as sentiment analysis, machine translation, etc. and could possibly be used to bootstrap any NLP task~\cite{surveyEval}. It has been shown that word embedding can learn and exhibit social biases based on race, gender, ethnicity, etc. that are encoded in the training dataset \cite{bolukbasi2016man, 100years, narayan2017semantics}. Social biases in word embeddings are manifested as stereotypes or undesirable associations between words \cite{100years}. For example, word embedding models 
might disproportionately associate male names with career and math, while female names might be associated with family and arts \cite{100years}. Existing literature has mostly focused on measuring and mitigating the \textit{individual} social biases based on race, gender, etc. encoded in word embeddings \cite{bolukbasi2016man, 100years, narayan2017semantics, vargas2020exploring, kumar2020nurse, guo2020detecting, dev2019attenuating}.

Recent studies have shown the presence of \textit{Intersectional Bias} 
in AI systems \cite{kim2020intersectional, genderShades, guo2020detecting} i.e. a bias towards a population defined by multiple sensitive attributes like `black muslim females' \cite{crenshaw2017intersectionality, intersectionality_book}. For example, facial recognition software applications have been shown to perform worse for the intersectional group `darker females' than for either darker individuals or females \cite{genderShades}. Similarly, word embedding models have also been shown to contain biases against intersectional groups like Mexican American females \cite{guo2020detecting}. When such biased word embeddings are used for any downstream application, their inherent social biases are propagated further, which can cause discrimination \cite{german,zhao2018gender}. Hence, it becomes critical to investigate the presence of different intersectional biases before using it for some application.

Stereotypes associated with an intersectional group say `Black males' are composed of stereotypes pertaining to constituting subgroups (Blacks and males) along with some unique elements \cite{ghavami2013intersectional}. The proportion of stereotypes that overlap with either of the constituting subgroups can vary based on the intersectional group. For example, a study on 627 undergraduate students found that the percentage of overlap for intersectional groups like White men is 81\%, White women is 88\%, Black women is 44\%, Black men is 73\%, Middle Eastern American men is 91\%, etc. \cite{ghavami2013intersectional}. This work focuses on this overlapping aspect of intersectionality. Given that word embedding models can consist of thousands of unique words and the number of intersectional groups can increase drastically with the number of sensitive attributes considered, it becomes challenging to explore the massive space of possible associations. Writing custom code to test the different associations can be tedious and ineffective. 
In this work, we present the first interactive visual tool, \textit{WordBias}, for exploring biases against different intersectional groups encoded in word embeddings. Given a pretrained word embedding, our tool computes the association (bias score) of each word along different social categorizations (bias types) like gender, religion, etc. and then visualizes them using a novel interactive interface (see \autoref{fig:teaser_wordbias}). Here, each categorization (bias type) e.g. race consists of two subgroups, say Blacks and Whites. Using bias metrics, WordBias computes the degree to which a word aligns with one subgroup over the other. 
The visual interface then allows the user to \textit{investigate} how a specific word associates with different individual subgroups and also \textit{discover} words that are associated with an intersectional group. Considering the overlapping aspect of intersectionality, WordBias considers a word to be associated with an intersectional group say `Christian males' if it associates strongly with each of its constituting subgroups (Christians and males).


Users can interact with our tool to explore the space of word associations and then use their real-world knowledge to determine if a given association is socially desirable. For example, the association between the word `queen' and female is desirable whereas the association between `teacher' and female is not. 
Using a case study, we demonstrate how WordBias can help discover biases against different intersectional groups like `Young Poor Blacks', `Black Muslim males', etc. in Word2Vec embedding. Identifying such biases can serve as the first step toward deterring its spread and help develop counter-strategies. 
Lastly, we evaluate the usability and utility of our tool using qualitative feedback from domain experts. 
We have made the source code for our tool along with a live demo publicly available for easy reproducibility and accessibility (\href{https://github.com/bhavyaghai/WordBias}{github.com/bhavyaghai/WordBias}).
\section{Related Work}
\subsection{Bias in Word Embeddings}
The existing literature on bias in word embeddings can be broadly classified into bias identification and mitigation. For bias identification, a number of bias metrics are proposed like \textit{Subspace Projection}~\cite{bolukbasi2016man}, \textit{Relative Norm Difference}~\cite{100years}, \textit{Word Embedding Association Test (WEAT)}~\cite{narayan2017semantics}, etc., but there is no single agreed-upon method~\cite{group_words}. 
Our tool builds upon such bias identification metrics to explore the space of word associations and help detect biased associations. More specifically, our tool uses the \textit{Relative Norm Difference} metric as it is simple to interpret and can be easily extended for different kinds of biases. Previously, this metric has been used to capture biases against individual sensitive groups like females. In this work, we have used this metric to capture biases against intersectional groups as well. Once bias has been detected, there are a host of debiasing techniques that can be used for bias mitigation \cite{bolukbasi2016man, zhao2018gender, wang2020double}. However, we will not go into these details as our work is limited to bias discovery. 
Our work relates more closely with the work of Swinger et al.~\cite{swinger2019biases} that find biases in word embeddings using purely algorithmic means compared to our visual analytics approach. Our dynamic visual interface makes the entire process more interactive and accessible to non-programmers. It also provides more flexibility by allowing the user to drive the bias discovery process as they see fit.

\subsection{Visual Tools}
Recent years have seen a spike in visual tools aimed at tackling algorithmic fairness like Silva~\cite{silva}, FairVis~\cite{fairvis}, FairSight~\cite{fairsight}, What-If~\cite{what_if}, etc. All of these tools help detect algorithmic bias but they are mostly limited to tabular datasets. Moreover, many of these tools are designed to deal with individual biases and not intersectional biases. 
Our tool, WordBias, helps fill in this gap by helping discover intersectional biases encoded in word embeddings. Our tool relates closely to Google's Embedding Projector (GEP) ~\cite{smilkov2016embedding} which supports a custom projection adopted from \cite{bolukbasi2016man} to visualize bias. As a general-purpose tool primarily aimed at visualizing high dimensional data in 2D or 3D space, GEP has several limitations when it comes to exploring biases in word embeddings: (1) it does not support any bias quantification algorithm, (2) it is limited to visualizing only two types of bias simultaneously, and (3) its custom projection only allows one word to characterize a subgroup, say 'he' for males. In this work, we have tried to overcome all these limitations by carefully designing an interactive visual platform geared towards exploring social biases. 
\begin{figure*}
  \centering
  \includegraphics[width=0.90\textwidth]{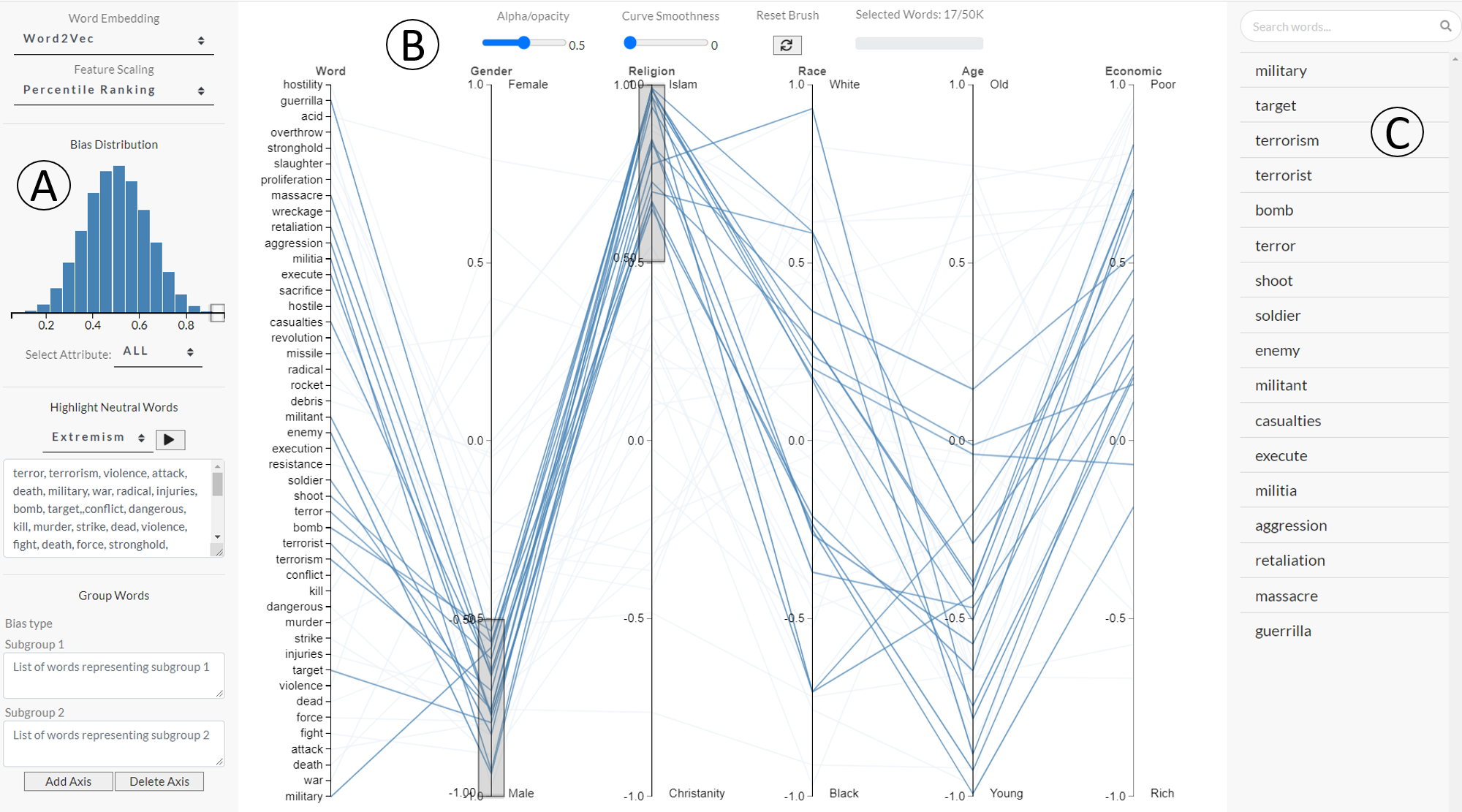}
  \caption{Visual interface of \textit{WordBias} using Word2Vec embedding. (A) The Control Panel provides options to select words to be projected on the parallel coordinates plot
  (B) The Main View shows the bias scores of selected words (polylines) along different bias types (axes)
  (C) The Search Panel enables users to search for a word and display the search/brushing results. 
  In the above figure, the user has brushed over 'Male' and 'Islam' subgroups. Words with strong association to both these subgroups are listed below the search box.}
 \label{fig:teaser_wordbias}
\end{figure*}

\section{\textsc{WordBias}}

\subsection{Design Goals}
Based on the current literature and the problem at hand, we have identified the following four design goals:

\begin{itemize}
    
    \item [\textbf{G1.}] \textbf{Bias Scores}: Our tool should compute bias scores and accurately visualize them such that the user can quickly identify the different subgroups a word is associated to along with their degree of association. 

    \item [\textbf{G2.}] \textbf{Bias Exploration}: Our tool should support quick and intuitive exploration of words associated with a single subgroup say males or an intersectional group say \textit{Rich White females}.

     \item [\textbf{G3.}] \textbf{Bias Types}: The existing literature on biases in word embeddings is heavily skewed towards gender bias (93\%) followed by racial bias (54\%) \cite{rozado2020wide}. Our tool should support the exploration of these well known biases but also under-reported biases based on physical appearance, political leanings, etc. or any user-defined bias type.

    
    \item [\textbf{G4.}] \textbf{Data Volume}: Word embedding models can consist of millions of unique words. 
    Our tool should be designed to deal with a large volume of data at both the back and the front end to ensure a smooth user experience.
    
\end{itemize}

\subsection{Bias Quantification}
\label{sec:Bias_Quantification}
We have used the \textit{Relative Norm Difference} \cite{100years} to quantify the association (bias) of a word along different bias types. Like most bias metrics, it assumes that a given bias type, say gender, consists of two subgroups, say males and females. Each of such subgroups is defined using a set of words called \textit{group words}. For example, group words for males might include he, him, etc. while for females, it might include she, her, etc. Mathematically, a subgroup is expressed as the average of word embeddings for the words which define that subgroup. For a given bias type, let $\vec{g1}$, $\vec{g2}$ represent either subgroups. 
We then define the bias score for a word w with embedding $\vec{w}$ as follows:
\begin{equation}
\label{eq:1_wordbias}
    Bias\_score(w) = cosine\_distance(\vec{w}, \vec{g1}) - cosine\_distance(\vec{w}, \vec{g2}) 
\end{equation}
A bias score can be understood as the association of a word toward a subgroup with respect to the other. The magnitude of the bias score represents the strength of the association and the sign indicates which subgroup it is associated to. We compute bias scores for each word across bias types using \autoref{eq:1_wordbias} and then repeat this process for all words.  

\subsection{Feature Scaling}
Visualizing raw bias scores might be difficult to interpret and compare because the distribution of bias scores varies across bias types. For example, a 0.3 bias score for gender bias might mean a much stronger/weaker degree of association compared to the same score for race bias. To cope, WordBias supports two kinds of feature scaling methods namely, \textit{Min-Max Normalization} and \textit{Percentile Ranking}. Min-Max Normalization ensures that bias scores across bias types share the same range by simply stretching raw bias scores over the range [-1,1]. However, it is still difficult to compare bias scores because of the different standard deviation across bias types. 
To overcome this limitation, we use \textit{Percentile Ranking}. Each word is assigned a percentile score \cite{percentile_score} based on its ranking within its subgroup. For e.g., a 0.8 percentile score means that 80\% of all words associated with the same subgroup have a bias score less than or equal to the given word. This makes it easier to interpret and compare bias scores across different bias types (\textbf{G1, G2}). It should be noted that percentile scores can sometimes be misleading as they are not equally spaced. Lets say that raw bias scores for most words along a bias type is close to 0. However, we can still obtain high percentile scores for words which otherwise have negligible raw bias scores.   
Hence, we recommend trying both feature scaling methods to get a comprehensive picture.  

\subsection{Design Rationale}
The problem of visualizing biases against intersectional groups boils down to visualizing a large multivariate dataset where each word corresponds to a row and each column corresponds to a bias type. A straightforward solution for visualizing such high-dimensional data is to use standard \textit{dimensionality reduction} techniques like MDS, TSNE, biplot, etc. and then use popular visualization techniques like scatter plot. However, algorithmic bias is a sensitive domain; we must make sure that we \textit{accurately} depict the biases of each word (\textbf{G1}). Hence, \textit{dimensionality reduction} and related techniques like the Data Context Map\cite{dataContextMap} are not an option because they almost always involve some information loss. Using such techniques might inflate/deflate real bias scores which might mislead the user. 

Next, we enumerated other possible ways to visualize multivariate dataset, like scatterplot matrix, radar chart, etc. and then started filtering these options based on the design challenges G1-G4. The scatter plot is a popular choice which is also used in Google's Embedding Projector \cite{smilkov2016embedding}, but it is limited to three dimensions. A few more dimensions can be added by encoding the radius and color of each dot yielding a plot that can visualize five dimensions; but such a plot will be virtually indecipherable. The scatterplot matrix can also be an option but it is more geared toward visualizing binary relationships than the feature value of each point. Moreover, it becomes more space inefficient as the number of dimensions grow. Another alternative can be the biplot but it can be difficult to read and involves information loss. The radar plot provides for a succinct representation to visualize multivariate data but it can only handle a few points before polygons overlap and it becomes unreadable (defeating G4). 
We ended up with the parallel coordinate (PC) plot \cite{inselberg1990parallel} based on our design goals \textbf{G1-G4}. PC can visualize a significant number of points with multiple dimensions without any information loss (G1, G4). It also facilitates bias exploration and adding new bias types. To support plotting large numbers of points, we chose canvas over SVG and also used progressive rendering \cite{progressive_rendering} (\textbf{G4}). 

\begin{figure*}
\centering
\begin{minipage}{.30\textwidth}
  \centering
    \includegraphics[scale=0.50]{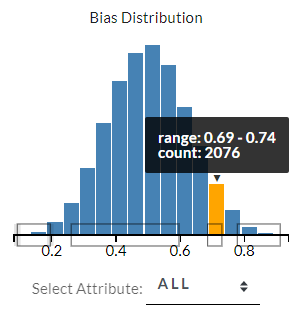}
  \caption{Select words based on their bias scores by brushing on the x-axis of the histogram.}
  \label{fig:histogram}
\end{minipage}%
\begin{minipage}{0.02\textwidth}
  \includegraphics[scale=0.05]{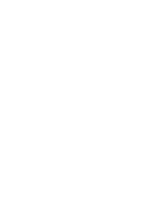}
\end{minipage}
\begin{minipage}{.58\textwidth}
  \centering
  \includegraphics[scale=0.23]{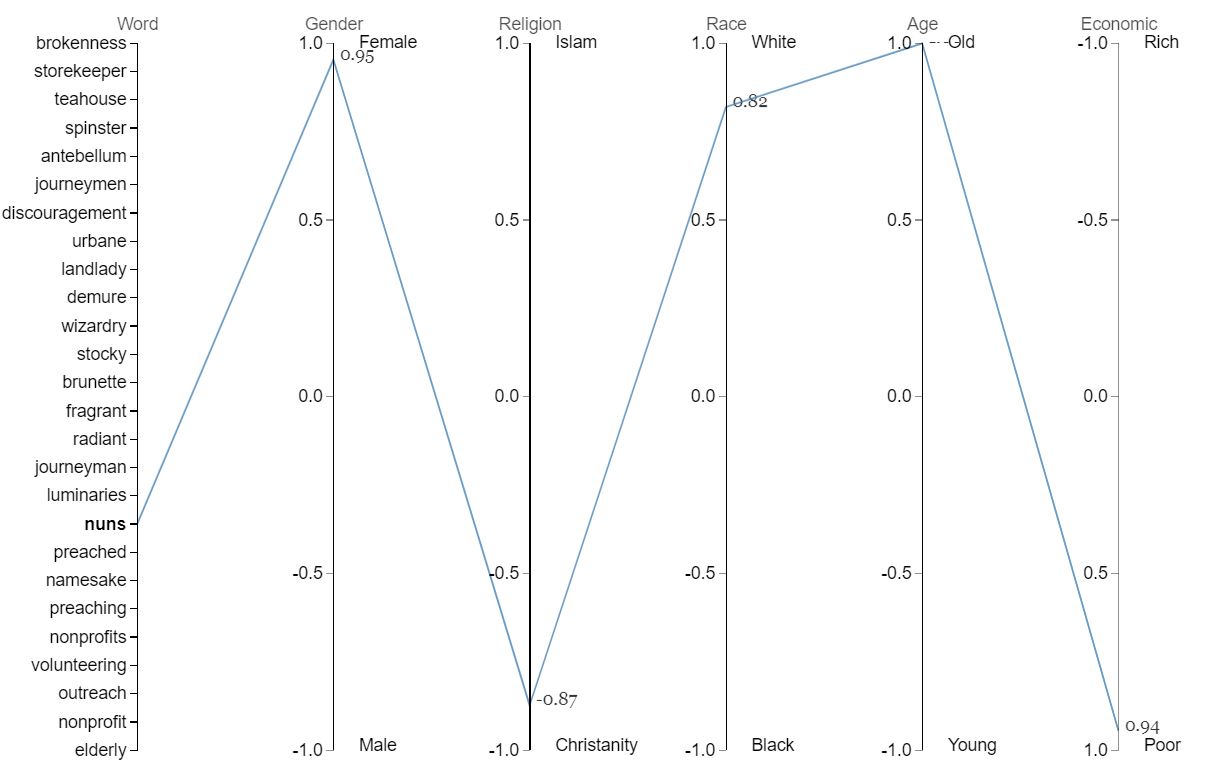}
  \caption{On Hovering over the word 'nuns', 
  we can observe its association with 'Female', 'Christianity', 'White', 'Old' and 'Poor' subgroups.
  }
  \label{fig:hover}
\end{minipage}
\end{figure*}

\subsection{Visual Interface}
\label{sec:vis_interface}
The visual interface can be classified into 3 components (see \autoref{fig:teaser_wordbias}). Next, we will discuss each component in detail.  

\subsubsection{Main View}
At the very center is the Main View which consists of a parallel coordinate plot \cite{inselberg1990parallel} (see \autoref{fig:teaser_wordbias} (B)). 
Each axis of the PC plot represents a type of bias based on gender, race, etc. and each piecewise linear curve, called \textit{polyline}, encodes a word. Either end of each axis represents a subgroup. For example, the gender axis encodes males and females on either extremes. The higher the magnitude of a word's bias score, the higher is the inclination of the corresponding polyline towards either group. We also have an additional axis, \textit{word}, which lists all the words currently displayed. On hovering over any word on the \textit{word} axis, its corresponding polyline gets highlighted (see \autoref{fig:hover}). This visualizes all different associations for the specific word (\textbf{G1}). On clicking over any word on the \textit{word} axis, the word and its synonyms get highlighted. Synonyms for a word are fetched via Thesaurus.com (using an API call) and from the nearest neighbors in the word embedding space.

To identify words with strong association toward a subgroup say females, the user can simply brush on the corresponding end of a given axis. Similarly, brushing either ends on multiple axes will help discover words associated to that intersectional group (\textbf{G2}). As shown in \autoref{fig:teaser_wordbias}, the user may brush on \textit{Male} and \textit{Islam} ends on gender and religion axes to obtain words related to the intersectional group \textit{Muslim males}. Words selected via brushing are displayed under the search box (see \autoref{fig:teaser_wordbias} (C)). 
At the top of the Main panel, there are sliders to customize \textit{Alpha/Opacity} and \textit{Curve smoothness} of the polylines. They are useful to see the underlying pattern between lots of polylines and to deal with the `crossing problem'\cite{graham2003using} respectively. \\


\subsubsection{Control Panel}
The left panel (see \autoref{fig:teaser_wordbias} (A)), the Control Panel, allows the user to control what is displayed on the parallel coordinates plot. The user can choose the word embedding and the feature scaling method from the respective dropdown menus. 
It also contains a histogram accompanied by a dropdown menu. 
The dropdown menu contains a list of all bias types currently displayed in the parallel coordinates 
along with an \textit{ALL} option. 
The histogram serves two purposes. First, it helps users understand the underlying distribution of bias scores for the selected bias type across the word embedding.  
Second, it helps users deal with the problem of over-plotting by acting as a filtering mechanism (\textbf{G4}). 
The user can select single/multiple ranges of variable length on the x-axis of the histogram (see \autoref{fig:histogram}). The words whose bias score falls in the selected range(s) are displayed on the parallel coordinates. 
The \textit{ALL} option (default) paints an aggregate picture as it corresponds to the mean absolute bias score across all bias types. 


We precomputed commonly known biases based on gender, race, religion, age, etc. to jump-start the bias discovery process when the tool first loads (\textbf{G4}). We used group words from the existing literature for each bias type \cite{100years,narayan2017semantics,kozlowski2019geometry} (see Appendix). The user is free to investigate a new bias type or drop an existing one by using the Add/Delete axis button (\textbf{G3}). To add a new bias type say \textit{political orientation}, the user needs to fill in details like axis name, subgroup names and \textit{group words} under the `Group Words' section and click `Add Axis'. Here, the \textit{Group words} should be chosen carefully as they play a critical role in computing the bias scores. 

Lastly, we included a set of neutral words corresponding to categories like professions, personality traits, etc. which should ideally have no association with bias types like gender, race, etc. These words have been derived from existing literature \cite{100years, narayan2017semantics} (see Appendix). On clicking the `play' button, the currently selected set of neutral words will be highlighted in the PC plot. This provides a quick way to audit an embedding for potential biases.

\subsubsection{Search Panel}
The right panel (see \autoref{fig:teaser_wordbias} (C)) enables a user to search for a specific word and see the respective search/brushing results. A user can simply lookup how a specific word associates with different groups by searching for it in the search box (\textbf{G1}). This will highlight the specific word and its synonyms in the parallel coordinates. The area under the search box is used to populate the list of synonyms and brushing results. 


\section{Implementation}
WordBias is implemented as a web application built over python based web framework \textit{Flask}. On the back end, we used \textit{gensim} package to deal with word embeddings and \textit{PyThesaurus}\footnote[1]{\url{pypi.org/project/py-thesaurus}} to fetch synonyms from Thesaurus.com. For the front end, we used D3 based library \textit{Parallel Coordinates}\footnote[2]{github.com/syntagmatic/parallel-coordinates} and used \textit{D3.js}, \textit{Bootstrap}, \textit{noUiSlider}, etc. for rendering different visual components. 


\begin{table*}
  \caption{Words with strong association with each intersectional group (within top 25 percentile of each constituting subgroup) in Word2vec embedding trained over Google News corpus. 
  }
  \label{tab:table_04}
  \centering%
  \begin{tabular}{p{0.25\linewidth}p{0.7\linewidth}}
    \toprule
   Intersectional Group & Associated Words \\
  \midrule
  Poor - Young - Black & disaster, struggle, tackle, chaos, woes, hunger, uprising, desperation, insecurity, rampage, roadblocks, scarcity, calamity, homophobia, shoddy, falter, jailbreak, mineworkers, marginalization, evictions \\
  
  Rich - Old - White & formal, attractive, appealing, desirable, castle, desserts, seaside, golfing, cordial, bungalow, fanciful, warmly, salty, nutty, gentler, aristocratic, snug, prim, urbane \\
  
  Black - Muslim - Male & gun, assassination, bullets, bribes, thugs, looted, dictators, electrocuted, cowards, agitating, storekeeper, looter, bleeping, lynch, strongman, disbelievers, hoodlums \\
  
  Young - Christian - Male & career, dominant, brilliant, lone, terrific, heroes, superb, epic, monster, prowess, heavyweights, excelled, superstars, supremacy, fearless, inexperience, mastery, crafty, ply, conquering, rampaging \\

Poor - Female & ostracism, brokenness, mortgages, eviction, brothels, witchcraft, traumatized, discrimination, layoffs, uninsured, sterilizations, abortion, powerlessness, sufferer, neediest, prostitution, microloans, distressed, homelessness, miscarry\\

White - Christian - Female & romantic, nuns, virgin, republicans, peachy, platonic, convent, radiant, unspoiled, unpersuasive, soppy, honeymooning, drippy, soapy\\

  \bottomrule
  \end{tabular}
\end{table*}


\section{Case Study}
Let us assume a user, Divya (she/her), who works as a data scientist for a big tech firm. 
Her team is tasked with building an automatic language translation tool. Made aware by the infamous \textit{Google Translate} example \cite{prates2019assessing,stanovsky2019evaluating}, she knows that such translation tools can be discriminatory toward minorities and can pose serious challenges for her organization. One of the ways in which bias can creep in is via word embeddings \cite{bios, bolukbasi2016man}. So, she needs to audit the word embedding for different social biases before using it. One way to explore/detect biases can be via purely algorithmic means, i.e., writing custom program to test the different associations. Given that exploration is a dynamic process, one might need to tweak and re-run the code repeatedly which can be tedious and cause delays. Moreover, analysing raw numbers for thousands of words across multiple bias types can be overwhelming and ineffective. Interactive visualization techniques excel at exploratory data analysis as they provide a faster, efficient and user friendly way to interact with massive datasets effectively \cite{keim2008visual}. So, Divya decides to use a visual analytics based tool, \textit{WordBias}, to audit her word embedding. Note that while we have used an embedding generated by word2vec \cite{mikolov2013distributed} trained over the Google News corpus, 
in a real world scenario this may be a word embedding trained over the company's private data. 

On first loading the tool, 
Divya observes that a small fraction of words are visualized that have a strong association with multiple groups. 
These words correspond to the right tail of the histogram, i.e. they are words with high mean bias score.  
She hovers over some words like storekeeper, landlady, luminaries, nuns, etc. on the word axis to see their corresponding associations. Some of the associations are accurate and align well with her real world knowledge, like `landlady' and `nuns' have a strong association with females. In contrast, other associations, like `storekeeper' and `luminaries' have a strong male orientation which she views as problematic. It indicates that this word embedding might encode gender bias.  
To make sure that it is not a one-off case, 
she searches for the word `corrupt' in the Search panel. 
Just by looking at the parallel coordinates display, she can make out that the word `corrupt' and most of its synonyms like corruption, corrupted, crooked, unscrupulous, etc. have a strong association with males and Blacks. 
This reaffirms the presence of gender bias and also indicates racial bias and bias against Black males.

She carries on her investigation using different sets of words under the `Neutral Words' section in the control panel. 
Each time she finds a strong association of `ideally neutral' words with at least one kind of subgroup. When visualizing a set of \textit{Professions}, she finds words like teacher, nurse, dancer, etc. on brushing over the female subgroup and words like farmer, mechanic, physicist, laborer, etc. on brushing over the male subgroup. 
\autoref{fig:teaser_wordbias} represents the case when she chooses to visualize words characterizing \textit{Extremism}. On brushing over the male and Islam subgroup, she observes words like terrorist, bomb, aggression, etc. in the search panel. After this exercise, she is confirmed that this embedding encodes strong social biases against different groups as well as intersectional groups like Black males, Muslim males, etc. Her team might have to use different debiasing techniques before actually using this word embedding. 

The first step towards debiasing a word embedding is to identify the different impacted groups \cite{bolukbasi2016man, manzini2019black}. 
So, she explores different intersectional biases by selecting all the words using the histogram and then brushing over different subgroups. She finds lots of positive and negative stereotypes (biases) against multiple intersectional groups. Some of the more striking associations are described in \autoref{tab:table_04}. Overall, our tool helped Divya and her team to prevent a possible disaster by making them aware about the different social biases encoded in the word embedding. From here on, they can take multiple paths like trying to mitigate these biases, using a different word embedding, etc. They also need to be cautious about other possible sources of bias \cite{mehrabi2021survey} like training dataset to make sure that bias does not creep in.       

\section{Expert Evaluation}
We conducted a set of individual 45-60 min long semi-structured interviews with five domain experts. 
All experts E1-E5 are faculty members affiliated with departments like Computer Science (E1, E3), Sociology (E4, E5) and Business School (E2) at reputed R1 Universities. They have taught course(s) and/or published research paper(s) dealing with Algorithmic Fairness/Intersectionality.
Each expert was briefed about the problem statement and existing solutions. 
Thereafter, we demonstrated the different features, interactions and the workflow of our system using a case study.
Lastly, we solicited their comments on usability, utility, and scope for future improvements which are summarized as follows.

All experts found the interface to be \textit{intuitive} and \textit{easy to use}. Some experts found the interface to be a bit `overwhelming' at first glance. They were unsure of where to start interacting with the tool. 
However, a brief tutorial neutralized these concerns.
E4 commented, \textit{"Once you understand the tool, its very useful and you know what you are seeing"}. E3 commented that \textit{the UI looks clean} and \textit{actions required to accomplish tasks are simple and straightforward}. E2 commented, \textit{"Given a brief tutorial, most people should be able to get along quickly"}. 

On the utility front, E2 and E3 found this tool \textit{"Definitely useful"} for the NLP community while E1 stressed its utility for the Socio-Linguists and as an educational tool. E3 emphasized its \textit{broad} utility for developers, researchers and consumers, and also expressed interest in using this tool for teaching about bias in their NLP class. E4 emphasized the tool's utility for researchers and showed interest in loading their own custom word embedding into the tool. 
E4 added, \textit{"Anytime we want to ask a question from the data, we need to rerun the jupyter notebook which might take some time. This tool can cut down that Long feedback loop while providing rich information".}
E2 and E4 particularly liked that with WordBias users can dynamically add a new bias type on the go. This would make WordBias capable of supporting \textit{sentiment analysis} by encoding positive and negative sentiments on either extremes of an axis. 
Another important aspect of Wordbias which received appreciation is its \textit{accessibility} i.e., our tool can be hosted on a web server and then be easily accessed via a web browser without needing to install any software or dealing with github.   

For the future, most experts suggested extending support for Contextualized word embeddings like BERT \cite{bert}, ELMo \cite{elmo}, etc. 
They pointed out that WordBias' current setup assumes a binary view of the real world since it only supports two subgroups per bias type. 
However, the real world is multi-polar.
They suggested accommodating multiple subgroups like Whites, Blacks, Hispanics, Asians, etc. under a single bias type, say race.  
E1 highlighted that some of the bias variables like race and economic status might be correlated.  
Future work should account for such correlations while computing the bias scores. 
E5 suggested to encode multiple word embeddings representing different time periods on each axes. This will help in analysing how different biases evolve over time.   
E4 suggested to add a 'Download' button which can help store all words currently displayed in the tool along with their bias scores in CSV format.  

\section{Discussion, Limitations \& Future Work}

\paragraph{\textbf{Scalability}} We will discuss scalability on two aspects, namely front-end rendering and back-end computation. On the front end, we have used the parallel coordinate plot which can get cluttered as the number of points increases beyond a threshold. 
 We have used a number of visual analytics based techniques to ameliorate this issue, such as histogram based selection, changing opacity of lines, brushing, highlighting words on hover, etc. We have also used canvas based progressive rendering instead of SVG to render large data effectively (G4). Finally, there are also natural limitations on the number of bias types (axes) that can be differentiated in terms of their word associations.    

 Our current back-end can deal with words on a scale of $10^4$ while still maintaining a smooth user experience. As the number of words increases, the time for loading the word embedding and the time to calculate bias scores for a new bias type increases proportionally. Future work might use databases to store and query word embeddings to reduce load time. Furthermore, leveraging multiple compute cores will enable faster computation for any new bias type on the fly. 


\paragraph{\textbf{Quantifying Bias}} 
Measuring bias in word embeddings is an active research area and there is no consensus on a single best metric. In our case, we have used the \textit{Relative Norm Difference} metric. So, 
  the bias scores reported by our tool 
  are susceptible to the possible limitations of this metric and the group words used.
  The feature scaling methods, especially percentile ranking, can impact the perceived strength of an association. We recommend switching between different feature scaling methods (including raw bias scores) to get an accurate picture. 
Moreover, WordBias assumes a binary view of an inherently multi-polar world. This can impact the bias scores of words that do not fit into either categories. For example, our tool reports white (race) orientation for the word `Asian' even though it is a different race altogether. One must interpret the bias scores responsibly in light of these limitations.
Future work might support multiple bias metrics 
to paint a more comprehensive picture and also include metrics that can better capture the multi-polar world.  

It is important to understand that the term \textit{Intersectionality} has a broader meaning beyond the multiplicity of identities \cite{crenshaw1989demarginalizing, crenshaw1990mapping, fleming2018less}. 
Quantifying such a complex sociological concept accurately needs more research. 
Our tool considers a narrow definition of Intersectionality where a word is linked to an intersectional group only if it relates strongly with each of the constituting subgroups. In reality, there can be cases like 'Hair Weaves' where a word is associated with an intersectional group (Black females) even though it does not relate strongly with either constituting subgroups (Blacks or females) \cite{ghavami2013intersectional}. Future work might incorporate bias metrics like EIBD \cite{guo2020detecting} which can capture such cases as well. 

\paragraph{\textbf{Utility}} Using a case study, we demonstrated how WordBias can be used as an \textit{auditing tool} by data scientists to probe for different kinds of social biases. 
Furthermore, the comments from the domain experts pointed at its possible utility for students and researchers. Given that WordBias does not require any programming expertise and can be easily accessed via a Web Browser, it can serve as an \textit{educational tool} for students and non-experts to learn how AI (word embedding model) might be plagued with multiple kinds of social biases. For researchers, our tool can expedite the bias discovery process by acting as a quick alternative to writing code. Future work might involve students, researchers and data scientists to further refine and evaluate the usability and utility of our tool for different target audiences.

\paragraph{\textbf{Group Words}} They play a critical role in computation of bias scores~\cite{group_words}. In our case, we have used group words that have been proposed in existing literature (see Appendix) to kick off the bias exploration process. The user is advised to examine the default set of group words and update them via the visual interface as required \cite{antoniak2021bad}. If the user chooses to add a new bias type (axis), they should choose the words carefully to get an accurate picture. So far, there is no objective way to choose group words. However, our tool can assist in selecting the most relevant group words by facilitating comparison against a set of alternatives (as recommended in \cite{antoniak2021bad}). Let us say the user wants to add a new axis for `political orientation' and they have multiple sets of group words to choose from. In such a case, the user can add multiple axes corresponding to each set of group words. Thereafter, the user can explore and compare bias scores for different words across these axes. Group words corresponding to the axis which best aligns with the user's domain knowledge can be chosen.

\paragraph{\textbf{Word Embedding}} We have focused on static word embeddings trained on an English language corpus (word2vec). Similar social biases based on gender, etc. have been found in embeddings trained on other languages like French, Spanish, Hindi, German, Arabic, Dutch, etc. \cite{fr_es, hindi, german, arabic, dutch}. Furthermore, contextualized word embeddings like BERT \cite{bert}, Elmo \cite{elmo}, etc. have also been found to contain social biases based on gender, etc. \cite{wang2019gender, intersectional}. Future work will involve extending support for contextualized word embeddings and embeddings trained on other languages.  \\

In this work, we designed, implemented and evaluated a novel visual interactive tool to discover intersectional biases in word embeddings. We demonstrated how our tool helped uncover biases against multiple intersectional groups encoded in Word2Vec embedding. The source of such biases can be training data, word embedding model, or they might be false positives due to limitations of the bias metric or sub-optimal group words. 
Future research might investigate the exact cause of such biases and develop effective counter strategies.   

\newcommand{\sysname}{\textsc{DramatVis Personae}}
\let\textcircled=\pgftextcircled
\chapter{Tackling Social Biases in Creative Writing}

Implicit biases and stereotypes are often pervasive in different forms of creative writing such as novels, screenplays, and children's books. To understand the kind of biases writers are concerned about and how they mitigate those in their writing, we conducted formative interviews with nine writers. The interviews suggested that despite a writer's best interest, tracking and managing implicit biases such as a lack of agency, supporting or submissive roles, or harmful language for characters representing marginalized groups is challenging as the story becomes longer and complicated. Based on the interviews, we developed \sysname{} (DVP), a visual analytics tool that allows writers to assign social identities to characters, and evaluate how characters and different intersectional social identities are represented in the story. To evaluate DVP, we first conducted think-aloud sessions with three writers and found that DVP is easy-to-use, naturally integrates into the writing process, and could potentially help writers in several critical bias identification tasks. We then conducted a follow-up user study with 11 writers and found that participants could answer questions related to bias detection more efficiently using DVP in comparison to a simple text editor.

\section{Introduction}
Gandalf.
Elizabeth Bennet. 
Hermione Granger and Ron Weasley.
Atticus Finch.
Anomander Rake, Lord of Moon's Spawn and Son of Darkness.
Holly Golightly, Lisbeth Salander, and Hannibal Lecter: literature lives and dies by its characters.
Heroes and anti-heroes, villains and bad guys, innocent bystanders or willing accomplices---fiction is arguably about conjuring more or less complete humans out of whole cloth and then providing audiences with the emotional release of \textit{catharsis}, often by having these characters go through hell and high water.
Therein lies also the secret of great literature: creating believable, nuanced, and multidimensional characters that spring out of the written page and come to life in the reader's mind.
Achieving this is no mean feat, particularly when considering that truly great fiction often requires a diverse, inclusive, and just treatment of its cast of characters; its \textit{dramatis personae}.

Creative writing or storytelling can be seen as a reflection of our societal beliefs, while at the same time societal beliefs
can be influenced by stories~\cite{correll2007influence}.
Thus, it is imperative for written stories to not promote biased representation of minority and marginalized groups.
However, current literature is filled with tired, unoriginal, and sometimes harmful stereotypes related to race, gender, sexuality, ethnicity, and age, such as the trope of the angry African-American woman, the studious Asian person, or the helpless damsel in distress~\cite{beckett2010away, layne2015zebra, hoyle-etal-2019-unsupervised, fast2016shirtless, gdblack2019, gdfilms2008}.

Change is coming, with various organizations, institutes, writers, and the publishing community working continuously to raise awareness against biases in creative writing.
For example, the Geena Davis Institute regularly publishes reports of gender and racial representation in Hollywood and foreign creative materials~\cite{gdfilms2008, gdblack2019}.
Twitter hashtag ``OwnVoices'' promotes writers from marginalized groups who write about their community.
Professional writers now seek feedback from ``Sensibility Readers'', a group of readers who especially look for harmful stereotypes before the material is published.
On the computational front, Natural Language Processing (NLP) has been used to measure stereotypes in creative writing.
Many of these studies have helped us understand how stereotypes operate in culture by analyzing corpus containing millions of books, a scale much larger than any previous analysis~\cite{hoyle-etal-2019-unsupervised, norberg2016naughty, fast2016shirtless}.

\textcolor{black}{As a result of these efforts, writers are increasingly becoming aware of harmful biases and stereotypes and we see reports of better representation in recent creative materials~\cite{parity_children, parity_films, uclareport2020}.
However, little is known about writers' current practices for addressing biases in their writing. Given that biases often taken nuanced, complicated, and intersectional forms that can be hard to detect, we speculate that computational support in this regard can help writers detect biases and write more inclusive and representative materials.
Motivated by that, this work seeks to understand creative writers' current practices for tackling biases and how computational tools can potentially help them in this regard.}
To inform our research, we conducted formative interviews with nine creative writers with published stories in their portfolios.
The interviews revealed that writers are mostly concerned about two types of biases: (1) \emph{Lack of agency} for minority characters (e.g., a female character introduced only to forward the plot for a male protagonist); (2) \emph{Stereotypes} encoded in how characters are \emph{described} and the \emph{actions} they take in the story (e.g., a female character described as beautiful and homely).
Writers mentioned that they actively look for such biases in their stories.
However, these biases are often implicit and unconscious and difficult to wheedle out even for the best and most self-reflective of authors.
The process is even more challenging for longer and complicated stories where many characters take intersectional identities. 

Based on the findings of the formative interviews, we designed \sysname{} (DVP), a web-based visual analytics system to help writers identify stereotypes in creative writing.
DVP is designed to integrate smoothly with the writer's own creative process, allowing them to analyze other writers' work for research, upload their written content as it becomes available
or even write in the tool itself, and then having its text analytics and visualizations update in real time.
Using entity recognition, co-reference resolution, dependency parsing, and other sophisticated Natural Language Processing (NLP) methods, DVP automatically detects characters in the text and collects data about them as the story progresses, including their aliases, mentions, and actions.
The author can then furnish demographic information for each character, such as their age, ethnicity, gender, etc.
The DVP dashboard uses this continually growing dataset to visualize the presence of characters and social identities over time. 

After our initial design and implementation, we approached writers from the formative interviews and conducted think-aloud sessions using the tool.
During a hands-on evaluation session conducted over videoconferencing, one writer was asked to use the tool to write a short story given a specific writing prompt.
Other participants used the tool to evaluate their own existing stories.
We observed their performance and then interviewed them with regard to their experience.
All participants expressed positive sentiment about the DVP tool, claiming that it helped them get a better grip of their characters and their story arcs throughout the process.
In particular, all participants appreciated that the tool managed and visualized character demographics, suggesting that the tool is helpful in writing a more nuanced and equitable story.
We further conducted a user study with $11$ participants to evaluate the effectiveness of DVP in detecting biases.
The study revealed that participants could answer questions related to bias detection more efficiently using DVP in comparison to a simple text editor. 

In sum, we claim the following contributions with this work: 
(1) findings on how to support the creative writing process by mitigating implicit bias, via an interactive interview session with nine fiction writers; 
(2) a visual analytics tool called \sysname{} (DVP) for supporting both online creative writing as well as offline analysis of fiction;
(3) results from three separate think-aloud sessions of deploying DVP with creative writers in both story generation as well as story analysis settings;
and (4) results from a summative user study, outlining the effectiveness of DVP in detecting biases and stereotypes.

\section{Background and Related Work}

In this work, we focus on ``creative writing,'' or the production of the written artifacts capturing the narrative, such as the book manuscript, fiction, or short story.
Creative writing falls under the umbrella of ``creative storytelling'' since authors are essentially telling stories through writing.
The rest of this section is designed to discuss research around creative writing, bias in creative writing, NLP, and text and literary visualization.

\subsection{Bias in Creative Writing}
\label{sec:story-bias}

Stereotypes in the form of art often mirror the problems, issues, thinking, and perception of different social groups in society~\cite{ross2019media, tsao2008gender}.
They can further reinforce biases and stereotypes against minority and marginalized groups in society~\cite{correll2007influence}.
The presence of biases and stereotypes, especially gender and racial bias, has been reported ubiquitously in different forms of creative writing.
We provide a brief overview of research in this area below. 

Many researchers have shown the prevalence of gender stereotypes in children's books, dating back to the early 1970s~\cite{flerx1976sex}.
Since then, several studies have reported that males are often portrayed as active and dominating, while females are instead described as passive and soft~\cite{kolbe1981sex, narahara1998gender, paynter2011gender}.
Other studies have found the presence of racial bias~\cite{layne2015zebra}, stereotypes against disability~\cite{beckett2010away}, and occupation~\cite{hamilton2006gender} in children's books.
Researchers have argued that the presence of such stereotypes in children's books is severely problematic as children are susceptible to inheriting stereotypes at an early age~\cite{kolbe1981sex, tsao2008gender}.
While the situation is improving (i.e., females are portrayed with more active roles in recent children's books) due to increased social awareness, the improvement is not significant~\cite{peterson1990gender}, and there are still reports of the prevalence of different stereotypes in children's books~\cite{adukia2021we, hamilton2006gender, CLPE}. 

Another form of creative writing medium that has been heavily criticized for promoting stereotypes is movie scripts.
The \textit{Geena Davis Institute} regularly publishes reports of gender and racial representation in Hollywood movies and is a valuable resource for current representational problems.
The institute has found underrepresentation and misrepresentation of females~\cite{gdfilms2008} and Black or African American females in Hollywood~\cite{gdblack2019}.
Beyond Hollywood, researchers have found similar sorts of biases in television shows and movies in other countries.
Emons et al.~\cite{emons2010he} found stereotypes in gender roles of males and females in U.S.-produced Dutch TV shows, misrepresenting females in Dutch society.
Madaan et al.~\cite{madaan2018analyze} has shown the existence of gender biases in Bollywood movie scripts. Similar to movie scripts, many studies have shown how biases and stereotypes operate implicitly in news articles and how they adversely affect the audience~\cite{correll2007influence, entman1992blacks, ramasubramanian2007activating, valentino1999crime}.

Finally, newer forms of writing such as blogs, online writeups, and social media posts are rife with harmful stereotypes.
Fast et al.~\cite{fast2016shirtless} analyzed fiction written by novice writers in the online community Wattpad and found it to be rampant with common gender stereotypes.
Joseph et al.~\cite{joseph2017girls} analyzed forty-five thousand Twitter users who actively tweeted about the Michael Brown and Eric Garner tragedies.
Their method can quantify semantic relations between social identities.
Other work discussed the impact of stereotypes in Reddit~\cite{ferrer2020discovering}, Facebook~\cite{matamoros2017platformed}, and U.S.\ history books~\cite{lucy2020content}.

All the above-mentioned research has been instrumental in raising awareness among writers, directors, and the general audience, a critical step towards equality.
As a result, there are reports of better representation and inclusivity in recent years~\cite{parity_children, parity_films, uclareport2020}.
Originating on Twitter in 2015, ``OwnVoices'' has become  a  campaign for promoting writers from diverse backgrounds writing about their experiences and cultures.
We strongly believe that empowering writers from historically excluded groups is tremendously important for our society.
At the same time, we believe writers from all backgrounds need to be cautious when writing creative materials that represent a culture or social identity and invest efforts to learn about relevant communities for correct portrayals.
In fact, we believe our tool will be most useful for writers who want to ensure that they represent other communities correctly and do not propagate stereotypes.
We intentionally avoided identifying a specific user group for our tool as we acknowledge that identities are intersectional, meaning a person who is considered privileged in one dimension of their identity, may well be a minority in another dimension. 
Rather, we strove to include writers from diverse backgrounds in the design of this tool.
We believe our tool can assist in the production of more inclusive and representative writings that will have a positive impact on society.

\subsection{Computational Support for Creative Writing}

\label{background:process}


Beyond the everyday-use text processors such as Microsoft Word, there are several professional and open-source software available to writers for helping them in guiding character development.
Scrivener~\cite{scrivener} is a paid service for organizing stories.
It allows flexible page breaking, adding synopsis and notes to each section, and easy merging or swapping between sections.
It also has a distraction-free writing mode where everything else on the computer is tuned out.
Granthika~\cite{granthika} is a similar sort of paid service that helps writers in tracking characters and events in a story.
It lets users integrate knowledge into the system as they write, and then use that knowledge for tracking in a timeline.
It also allows writers to apply causal constraints to the events and people in a story (e.g, ``The inquest must happen after the murder''). Grammarly is used for checking grammatical errors.

Over the years, NLP research has developed and refined different techniques such as coreference resolution, named entity recognition (NER), dependency parsing, sentiment analysis, etc, which can be instrumental in analyzing stories.
For example, NER helps extract named entities such as person names, organizations, locations, etc from the text.
In our context, this can help identify different characters of a story~\cite{dekker2019evaluating}.
Dependency parsing finds relationships between words.
This can help identify different adjectives/verbs linked to a character in a story~\cite{mitri2020story}. 
Similarly, coreference resolution can help track the representation of different characters and their demographic groups across the storyline~\cite{kraicer2019social}.     
Other studies in the NLP literature have specifically focused on analyzing stories.
This includes segmenting stories by predicting chapter boundaries~\cite{pethe2020chapter}, recognizing flow of time in a story~\cite{kim2020time}, analyzing emotional arc of a story~\cite{reagan2016emotional}, extracting character networks from novels~\cite{labatut2019extraction}, etc.
Using such methods, researchers have conducted large-scale analysis of books, and stories~\cite{norberg2016naughty, pearce2008investigating, hoyle-etal-2019-unsupervised, labatut2019extraction, fast2016shirtless, joseph2017girls}.

Finally, the invent of powerful generative language models such as GPT-3~\cite{brown2020language} have fueled research on human-AI collaboration for creative writing.
Many tools in this area generate stories iteratively: a writer provides an initial prompt for the story; AI generates a section of the story automatically; the writer edits the generated story and provides further prompts and so on~\cite{clark2018creative, coenen2021wordcraft, lee2022coauthor}.
Recently, Chung et al.~\cite{talebrush} proposed TaleBrush, an interactive tool where AI can generate a narrative arc based on a sketch drawn by a writer that outlines the expected changes in the fortunes of characters.
Other works include IntroAssist~\cite{hui2018introassist}, a web-based tool that helps novice entrepreneurs to write introductory help-requests to potential clients, investors, and stakeholders; INJECT~\cite{maiden2018making}, a tool for supporting journalists explore new creative angles for their stories under development; and an interactive tool proposed by Sterman et al.~\cite{sterman2020interacting} that allows writers to interact with literary style of an article. 



While these works provide necessary background for the technical design of our tool, none of these works have features to add social identities to characters and investigate potential biases.
Additionally, prior works have primarily used textual descriptions for summarizing and communication.
While that is helpful, we utilize data visualization, a visual communication medium, for gathering insights from the information extracted by the tool quickly and efficiently.
.



\subsection{Visualization for Text and Literature}

Harking back to some of the original approaches to visualizing ``non-visual'' text documents~\cite{DBLP:conf/infovis/WiseTPLPSC95}, data visualization has long been proposed as an alternative to reading through large document corpora~\cite{DBLP:journals/widm/AlencarOP12, Qihong2014, DBLP:conf/kes/SilicB10}. 
For example, the investigative analytics tool Jigsaw~\cite{DBLP:conf/ieeevast/StaskoGLS07} is often styled as a ``visual index'' into a document collection; while it is not a replacement for reading, it provides a linked collection of entities and documents for easy overview and navigation.
This is generally also true for document and text visualization as a whole; the goal is to be able to ``see beyond'' the raw text into content, structure, and semantics~\cite{DBLP:journals/widm/AlencarOP12}.

Beyond simplistic text visualization techniques such as word clouds, data visualization can become particularly powerful when applied to entire documents~\cite{Qihong2014}.
These ideas can also be used for literary analysis of fiction and poetry~\cite{DBLP:journals/cgf/CorrellWG11}.
For example, Rohrer et al.~\cite{DBLP:conf/infovis/RohrerSE98} used implicit 3D surfaces to show similarities between documents, such as the work of William Shakespeare.
Similarly, Keim and Oelke propose a visual fingerprinting method for performing comparative literature analysis~\cite{DBLP:conf/ieeevast/KeimO07}.
McCurdy~\cite{DBLP:journals/tvcg/McCurdyLCM16} present a organic linked visualization approach to scaffolding close reading of poetry.
The literary tool Myopia~\cite{DBLP:conf/dihu/ChaturvediGMAH12} uses color-coded entities to show the literary attributes of a poem for readers.
Abdul-Rahman et al.~\cite{DBLP:journals/cgf/Abdul-RahmanLCMMWJTC13} also apply data visualization to poetry.
Our work here is inspired by, if not the design and implementation, then at least the motivation of these tools; however, in comparison, our goal is to support the creative processing while focusing on identifying and mitigating implicit social bias.

Finally, visualization can also be applied to the stories themselves rather than the actual text.
XKCD \#657,\footnote{\url{https://xkcd.com/657/}} titled \textit{Movie Narrative Charts}, shows temporal representations of plotlines in five movies, including the original \textit{Star Wars} trilogy (1977--1983), \textit{Jurassic Park} (1993), and the complete \textit{Lord of the Rings} movie trilogy (2001--2003).
Liu et al.~\cite{DBLP:journals/tvcg/LiuWWLL13} propose an automated approach to generating such storyline visualizations called StoryFlow.
Tanahashi and Ma discuss design considerations for best utilizing storylines~\cite{DBLP:journals/tvcg/TanahashiM12}.
Some effort has also been directed towards minimizing crossings in storyline visualizations~\cite{DBLP:conf/gd/GronemannJLM16}.
VizStory~\cite{DBLP:conf/taai/HuangLS13} takes a very literal approach to visualize stories by identifying segments and themes and then searching the web for appropriate representative images.
TextFlow~\cite{Cui2011} and ThemeDelta~\cite{DBLP:journals/tvcg/GadJGEEHR15} and related topic modeling visualizations can be used to automatically extract and visualize evolving themes in a document (or document collection) over time. 
StoryPrint~\cite{DBLP:conf/iui/WatsonSSGMK19} shows polar representations of movie scripts and screenplays in a fashion similar to Keim and Oelke's fingerprints; the authors note that this approach could also be used to support the creative writing process. StoryCurves~\cite{DBLP:journals/tvcg/KimBISGP18} shows non-linear narratives in movies on a timeline representation. Finally, Story Analyzer~\cite{mitri2020story} shows several visualizations representing summary statistics of a story.
Common for all of these storylines and plot visualization tools is that they are designed mostly for retrospective analysis and not for online creative writing.
None of these tools support bias identification in an interactive environment.
Nevertheless, we draw on all of these tools in our design of DVP.

\section{Formative Study}


To understand creative writers' current practices and challenges for addressing biases, we conducted semi-structured interviews with 9 creative writers.
The study was approved by our university's Institutional Review Board (IRB).

\begin{table*}[t!]
    \centering
    \caption{Participant demographics.}
    \label{tab:participant}
    \resizebox{\columnwidth}{!}{%
    \begin{tabular}{llp{3.4cm}cp{8.5cm}c}
    \toprule
     \textbf{Id} & \textbf{Gender} & \textbf{Race} & \textbf{Age} & \textbf{Expertise} & \textbf{Yrs Exp} \\ 
    \midrule
    W1 & Male & Asian & 30 & Short stories, poems, and blogs & 10  \\
    W2 & Male & Asian & 27 & Short stories, poems, and critiques & 12 \\
    W3 & Female & White & 26 & Novels (fiction/non-fiction), short stories, screenplays, poems, blogs, critiques, and fanfiction & 15 \\
    W4 & Female & White & 49 & Novels (fiction/non-fiction), and roleplaying games & 25 \\
    W5 & Female & Black or African-American & 44 & Picture books and books for beginning readers & 6 \\
    W6 & Non-binary & Prefer not to respond & 36 & Novels (fiction/non-fiction), short stories, and poems & 10\\
    W7 & Female & Asian & 25 & Short stories and poems & 15\\
    W8 & Male & Asian & 25 & Screenplays, blogs, and critiques & 12\\
    W9 & Female & White & 34 & Screenplays and poems & 15\\ \bottomrule
    \end{tabular}}
\end{table*}

\subsection{Participants}

We posted recruitment advertisements to social media such as Twitter and Facebook, local mailing lists, and university mailing lists.
After initial pre-screening, we selected 12 responses out of 54.
Among the 12 responses, 9 participants agreed to participate in the final interviews.  
Our inclusion criteria included prior experiences in creative writing such as novels, short stories, screenplays, etc, and familiarity with writing in text editors.
Table~\ref{tab:participant} presents the participant (writer) information.
All participants had published materials in their portfolios.
Each participant received an Amazon gift card worth \$15 for their participation.


\subsection{Procedure}

We conducted semi-structured interviews over Zoom.
Each interview lasted around 1 hour and was divided into three parts.
First, after gathering informed consent, we asked the writers to share their perspectives on bias in creative writing. In the second part, we asked the writers about the challenges they faced in addressing biases in their writing, and their current approach for overcoming biases. Finally, writers brainstormed with the study administrator for outlining the potentials and requirements for digital tools that might help them in managing biases and stereotypes. 

\subsection{Analysis}

We created anonymized transcript for each interview from the recorded audio.
Two authors of this work open-coded the transcripts independently.
A code was generated by summarizing relevant phrases or sentences from the transcripts with a short descriptive text.
Both coders then conducted a thematic analysis~\cite{braun2006using} to group related codes into themes.
Throughout this process, the coders refined the themes and codes over multiple meetings by discussing disagreements and adjusting boundaries, scopes, and descriptions of the codes and themes.
The open codes and themes were also regularly discussed with the full research group.
We present the findings from the interviews next.





\subsection{Findings}

Our findings relate to the topics of implicit bias in creative writing, mitigating bias, and the potential for computational support for such issues.

\subsubsection{Current State of Biases in Creative Writing}

All participants believed that writers and publishers are increasingly becoming aware of biases in creative writing.
For example, W3 mentioned that publishers now encourage writers to seek feedback from ``Sensitivity Readers'' for potential misrepresentation.
W4 lauded the recent Twitter hashtag ``\#ownvoices'' that started the discussion about the importance of writers from marginalized groups writing about their community. 

Despite these efforts, all participants mentioned that bias in creative writing is still a present and complicated problem. Participants said that there is a need for increasing diversity among writers and characters. 
W5 said: \textit{``One recent study\footnote{https://www.theguardian.com/books/2020/nov/11/childrens-books-eight-times-as-likely-to-feature-animal-main-characters-than-bame-people} found that there were more characters that feature animals than there were of all of the minorities.
    I mean, it is horrendous, to be honest with you, in 2021... and it is really sad.''} W7 emphasized the importance of research (e.g., reading literature, understanding culture) for writing about characters that represent marginalized groups.
To summarize, the publishing and writing community need diverse writers (W1-W9) while writers from all communities need to be conscious about how their writing represents different social groups (W1-W9).

\subsubsection{Bias 1: Lack of agency for minority characters}
\label{bias1}
One critical challenge for writers is to ensure that characters representing minority groups have impact in the story and are not sidelined. W3 said: \textit{``I believe it is Gail Simone.
    She is a comic book writer and she started this website called ``Women in Refrigerators'' because there is a very famous storyline from a Green Lantern comic (vol. 3, \#54) where the Green Lantern comes home to find that his girlfriend has been murdered and shoved in his refrigerator.
    And the only reason that the woman existed was to be killed.
    So it has become a name for this type of killing of female characters to advance the man's plot, which is unfortunately very common.''} W3 also referred to the \textit{sexy lamp test}, which tests if a female character of a story is replaced with a sexy lamp whether the story still makes sense.
W4 mentioned the \textit{Bechdel} test, which asks whether a story features at least two women who talk to each other about something other than a man. 



To address this type of bias, writers employ several self-evaluation techniques. They mentally track the presence of characters as well as their social identities in the story~(\textbf{N1|W3-7, W9}).
This frequently requires them to go back and forth between different parts of the story and read through them (\textbf{N2|W3-7, W9}). While reading, writers investigate interactions between characters (\textbf{N3|W3-7}), and how actions of the characters reflect on the social identities they represent (\textbf{N4|W3-5, W7}).  However, this process becomes challenging and tiresome as the stories become longer and complicated (W3-7, W9). Moreover, the identities often take intersectional form (e.g., a Female African-American character) which makes it even more mentally demanding to track them. It is worth noting here that writers also apply these methods to analyze existing literature as part of their research; although often that process is not as thorough as correcting their own work and depends mostly on high-level subjective understanding (W2-4, W7-9). Such research helps them understand representation of different social identities as portrayed in existing literature (W2-4, W7-9).

\subsubsection{Bias 2: Stereotypes in describing characters}
\label{bias2}
Another form of bias is how characters are described in the story. For example, W1 mentioned that female characters are often described as homely, beautiful, and sacrificial.  W7 said: \textit{``One example would be Truman Capote's Breakfast at Tiffany's. I read the book as a teenager, and it's very beautiful and well-written overall. But one example of a stereotyped character is Holly's neighbor Mr.\ Yunioshi. He is a side character but is portrayed simply as the irritable neighbor who always tells Holly off for forgetting her keys. This depiction of East Asian characters as grumpy old people is very stereotypical and common in Western literature.''}

To address this type of bias, writers remain careful while writing and examine the actions (\textbf{N4 from above| W1-4, W9}) and descriptors (\textbf{N5| W1-3, W7-9})  they use for the characters and social identities. This again requires them to go back and forth between different parts to read the story (\textbf{N2 from above| W1-4, W7-9}), and constantly check the actions and descriptors. These methods are also helpful to writers for research (analyzing existing literature).

\subsubsection{Potential and recommendations for an interactive tool}
\label{sec:tool_potential}

All participants were enthusiastic for an analytic tool for identifying potential stereotypes.
Critique groups and sensitivity readers provide important feedback to writers; however, they are usually available at the advanced stage of a formal publication. During writing, a tool supporting their self-reflective process would be helpful. The tool will also be helpful to a writer to analyze existing literature as a part of their research (W2-4, W7-9).

Writers also provided a few recommendation for such a tool. First, writers suggested that the tool could support story writing since the various bias mitigation strategies discussed in Section~\ref{bias1} and~\ref{bias2} are closely integrated into the writing cycle (\textbf{N6| W2, W4, W8}). 
\textcolor{black}{Second, the tool could have a ``distraction-free'' mode, similar to Scrivener, where writers could concentrate only on writing if they want to  (W2-5, W8).} Finally, writers suggested that the tool should \textbf{avoid suggesting interpretations of its own}; rather the analytic tool should support writers in \textbf{exploring their work} so that they can make informed decisions. Writers provided several reasons behind this suggestion. W1 said:\textit{``Language around different social groups is fluid and continuously evolving. What is acceptable today, may not be acceptable a few months from now.''} W6 said: \textit{``Creative writing can be subjective to a writer's own experience, or imagination, and may elicit intended stereotypes for the plot. I want to write about discrimination against my community which if flagged by a tool would be disappointing''}

\begin{figure*}
    \centering
    \includegraphics[width=\textwidth]{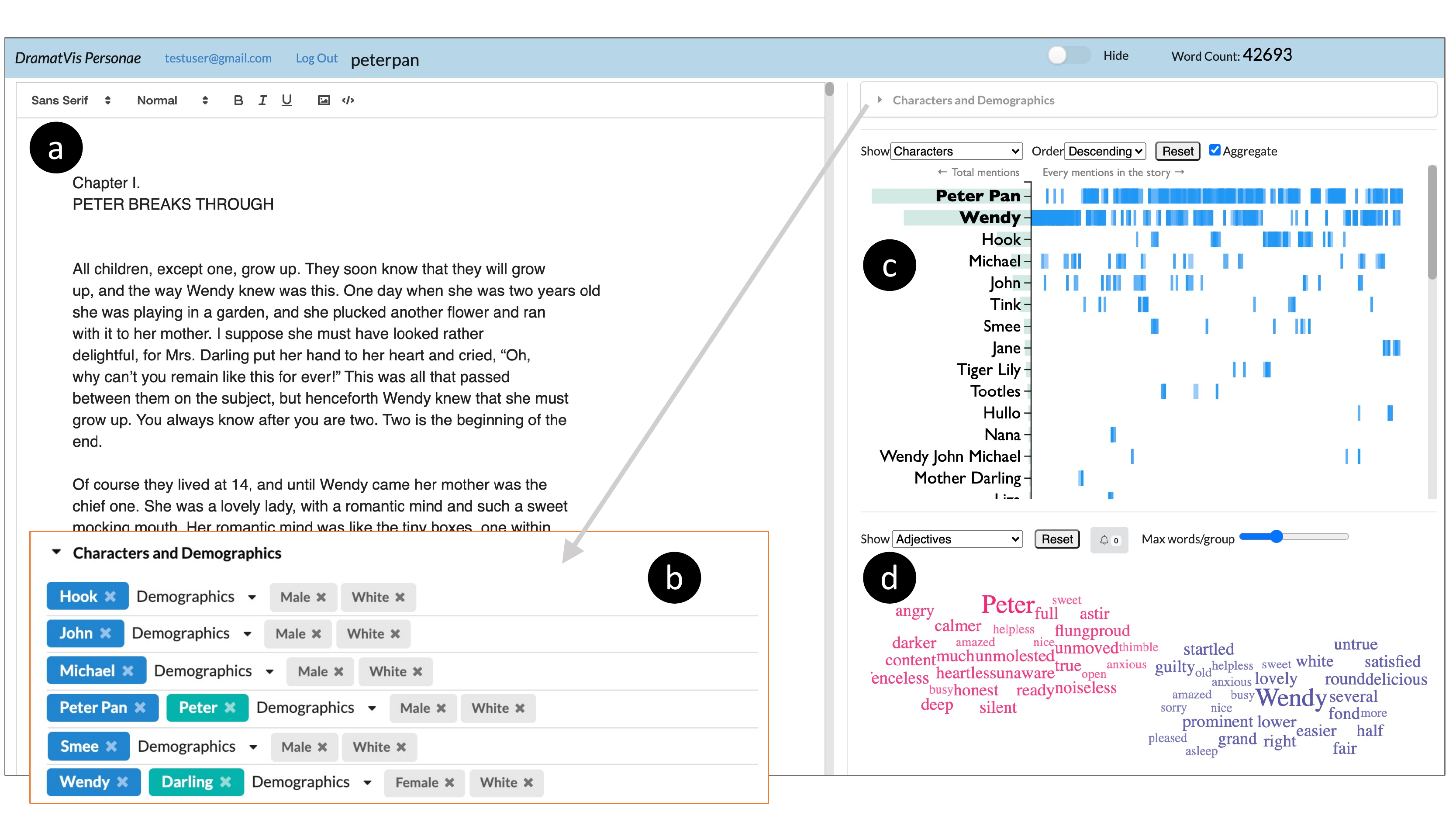}
    \caption{Overview of \sysname{}.
    (a) Rich text editor.
    (b) Characters and Demographics panel for listing characters identified by the system, merging aliases, and assigning social identities to characters.
    (c) Timeline representation of the story (\textit{Peter Pan} by J.\ M.\ Barrie (1911)) showing every mentions of characters as well as the total number of mentions of characters.
    (d) Word zone~\cite{hearst2019evaluation} showing sample adjectives used for the selected characters (Peter Pan and Wendy).}
    \label{fig:teaser_05}
\end{figure*}

\section{Design Guidelines}
\label{sec:design}

Here we outline and discuss the design guidelines and decisions that originated from the literature and formative study.

\textbf{DG1. Nature of the Support: Exploratory.}
The first design decision we made is the nature of the support.
The formative study suggested several challenges that writers face in identifying biases and stereotypes.
They apply several evaluation techniques (Table~\ref{tab:needs}) that can be mentally demanding.
All participants were enthusiastic about a tool that will support them in this process.
However, based on the findings from Section~\ref{sec:tool_potential}, it was clear that the tool should avoid flagging any writing as inappropriate and suggesting how to reduce biases.
Suggesting interventions for reducing biases without a proper understanding of a writer's goal could seriously lower user trust in the system and be counter-productive.
Another potential concern is writers using such suggestions to write about cultures and identities outside their own identity without really investing efforts to learn about them. 
Considering all these, we decided that the primary goal of the tool would be to \textbf{help writers perform \textbf{N1} to \textbf{N6} easily, without any explicit recommendation}.
We decided to use interactive visualization for this purpose which is an effective way to summarize any form of abstract data and enable exploration in an interactive environment.

\begin{table}
    \centering
    \caption{Design needs identified from the formative study, based on the evaluation techniques performed by writers to identify biases.}
    \begin{tabular}{p{3.8cm} p{1.6cm} c}
       \toprule
       \textbf{Design Need}  &  \textbf{Purpose} & \textbf{Participant} \\
       \midrule
       \textbf{N1.} Evaluate presence of characters and social identities.   & Bias 1  & W3-7, W9 \\ \hline
       \textbf{N2.} Move between different parts of the story for reading.  & General  & W1-9 \\ \hline
       \textbf{N3.} Evaluate interactions between characters-characters and social identities-social identities.  & Bias 1  &  W3-7\\ \hline
       \textbf{N4.} Evaluate and compare actions of characters and social identities.  & Bias 1, Bias 2  & W1-7, W9 \\ \hline
       \textbf{N5.} Evaluate and compare descriptions of characters and social identities.  & Bias 2  &  W1-3, W7-9\\ \hline
       \textbf{N6.} Support writing in the tool.  & General  & W2, W4, W8 \\
       \bottomrule
    \end{tabular}
    \label{tab:needs}
\end{table}

\textbf{DG2: Help writers evaluate agency for characters and social identities.}
The formative interviews revealed that writers are concerned about the lack of agency for characters that represent minority groups.
To ensure agency for characters, writers practice three evaluation techniques (\textbf{N1, N3, N4} from Table~\ref{tab:needs}).
Prior research has also used presence, and interactions between characters to quantify character agency~\cite{hoque2020toward, parity_films}.
Thus, our tool should support writers in these tasks.

\textbf{DG3: Help writers evaluate stereotypes in describing characters.}
Another important concern raised during the formative interviews was the stereotypes used for characters. 
Writers mentioned that they search for possible stereotypes in the action characters take in the story (verbs, \textbf{N4}), and the words that describe the characters (adjectives, \textbf{N5}).
Previous research has also used verbs and adjectives to quantify biases and stereotypes~\cite{hoyle-etal-2019-unsupervised,  emons2010he, fast2016shirtless, 100years}.
Thus, our tool should help writers in these tasks.


\textbf{DG4: Support writing and critical reading.}
\textcolor{black}{\textbf{N2} and \textbf{N6} are not directly related to bias identification.
However, the formative study suggests support for these features is necessary to facilitate bias identification.
They will also enable bias identification in different stages of writing a story. For example, during the pre-writing (e.g., analyzing existing literature) or post-writing stage, a writer can use the support for \textbf{N2} and other bias identification methods for critical reading. During the writing stage, writers require support for both writing (\textbf{N6}) and reading (\textbf{N2)}.}

\textcolor{black}{We also decided to include a ``distraction-free'' mode similar to Scrivener~\cite{scrivener} that will allow writers to write or read without any distraction and use our bias identification support only when they want to.
This will ensure that our tool does not obstruct the creative process for writers.}

\textbf{DG5: Inclusive design.}
Prior technological efforts in bias identification and mitigation often focused on a specific type of bias, or a binary view on identity (e.g., male, female as gender category)~\cite{100years, hoque2020toward, hoyle-etal-2019-unsupervised}.
However, the formative interviews suggest that writers may use diverse and intersectional identities for characters that may not conform to any pre-defined categories.
We anticipated that the tool would be demoralizing if a social identity that an author is writing about is not supported in the tool; especially since authors may themselves share that identity.
Thus, our design should be inclusive, and support social identities of any kind.


\textbf{DG6: Easy-to-understand visualization and scalable computation }
To ensure accessibility for writers who are not familiar with complex data visualization paradigms, our tool should use easy-to-understand visualizations.
Finally, the formative study suggests the tool would be most useful as the story becomes larger and complicated.
Thus, our tool should support stories of large size and provide feedback to the writers in a short response time.
Similarly, the visual components should be able to show information extracted from large textual data.



%
















\section{\sysname{} (DVP)}

DVP is a visual analytics system to support off-line and online creative writing and analysis, particularly in identifying biases.
We present the visual components and analysis pipeline for the tool below.

\subsection{Visual Interface}

In this section, we demonstrate four visual components (Figure~\ref{fig:teaser_05}) of the interface. We link the design guidelines from Section~\ref{sec:design} wherever applicable.
We also discuss design rationales and design alternatives considered during the development of the tool. The interface supports analysis for three different types of entity: (a) Characters, (b) Social Identities (e.g., Male, Female), and (c) Intersectional Social Identities (e.g., Muslim Males).
We use the term ``entity'' or ``entities'' to refer to the three entities together whenever applicable for brevity.
For demonstration purpose, we use several well-known western stories.

\paragraph{Text Editor}

The central component of DVP is a text editor (\textbf{DG4}). 
The text editor is equipped with traditional formatting features such as selecting fonts, font sizes, font-weights, etc.
We use QuillJS~\cite{QuillJS} as a rich text editor which has been widely used on the web including social media apps such as Slack, and Linkedin.


The visualizations are contained in a sidebar; a user can hide or show the sidebar by clicking the \textit{``Hide''} toggle button (\textbf{DG4}). This ensures that writers can see the visualizations whenever they want, but also can concentrate on writing by hiding the sidebar whenever they want.

\begin{figure*}
    \centering
    \includegraphics[width = 0.95\textwidth]{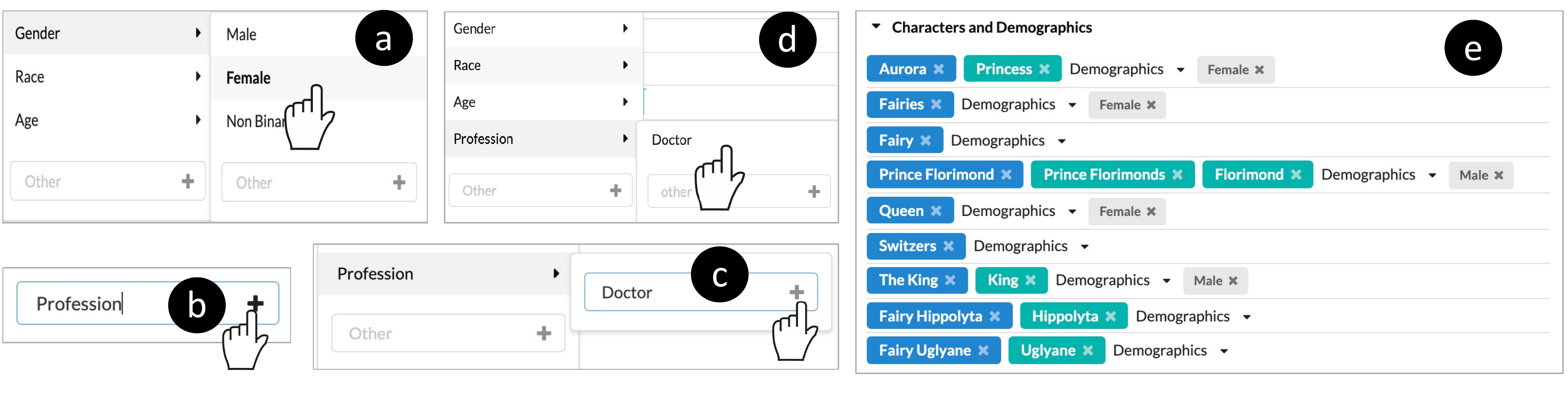}
    \caption{Characters and Demographics panel.
    (a) A dropdown menu with different social identities is available for each character in the Characters and Demographics Panel.
    (b) The dropdown can be extended dynamically.
    User can add a new identity (e.g., Profession).
    (c) User can then add categories under the newly added identity (e.g., Doctor).
    (d) The new Demographics dropdown with Doctor as a new profession.
    (e) An example demographics panel for the children's book, \textit{Sleeping Beauty}, retold by Arthur Quiller-Couch and Charles Perrault, freely available under Project Gutenberg~\cite{sleeping_beauty}.}
    \label{fig:demographics}
\end{figure*}

\paragraph{Characters and Demographics Panel}

The Characters and Demographics Panel lists all the named entities identified by our NLP pipeline. The panel supports several validation functionalities. First, a user can delete a character from the list in the case they do not wish to track that character, or if the character was wrongly identified by the NLP pipeline (e.g., an institute identified as a person). Second, a user can merge different names of the same character together (e.g., merging Peter Parker, and Spiderman as one character which might be identified as separate characters by the NLP pipeline). A user can use drag and drop for merging characters. Note that these validation functions are not necessarily a part of a writer's work process, but are needed to use the tool reliably given the nature of current NLP toolkits.


Each character in this panel has a dropdown named \textit{Demographics}. 
Using this dropdown, a user can add multiple social identities to each character (Figure~\ref{fig:demographics}a).
We populate the dropdowns with commonly used identities.
However, to support \textbf{DG5}, we made the dropdowns dynamically extendable.
A user can add any number of new identities in these dropdowns.
For example, Figure~\ref{fig:demographics}b and \ref{fig:demographics}c show how a user can add Profession as a new identity and Doctor as a profession in the dropdowns.

\paragraph{Timeline}

To support \textbf{DG2}, we designed a timeline representation of the story (see Fig.\ref{fig:teaser_05} (C)). The timeline is divided into two parts.
On the left side of the y-axis, we encode the total number of mentions for entities using bars encompassing the axis labels.
The y-axis can be sorted either in descending or ascending order.
A user can choose the sort order from the \textit{Order} dropdown.
The dropdown defaults to descending order for showing the prominent entities at the top.
On the right side of the y-axis, we show individual mentions for each character.
The x-axis represents a linear scale with a range $(1, S)$ where $S$ is the total number of sentences.
For a mention of an entity in sentence $s$, we draw a tile (rectangle) with width $(pos(s) + 0.5) - (pos(s) - 0.5) $ where $pos(s)$ represents the position of $s$ in the x-axis.
We used linear scale instead of ordinal scale to make the adjacent tiles connected and smoother.

\begin{figure*}
    \centering
    \includegraphics[width = 0.95\textwidth]{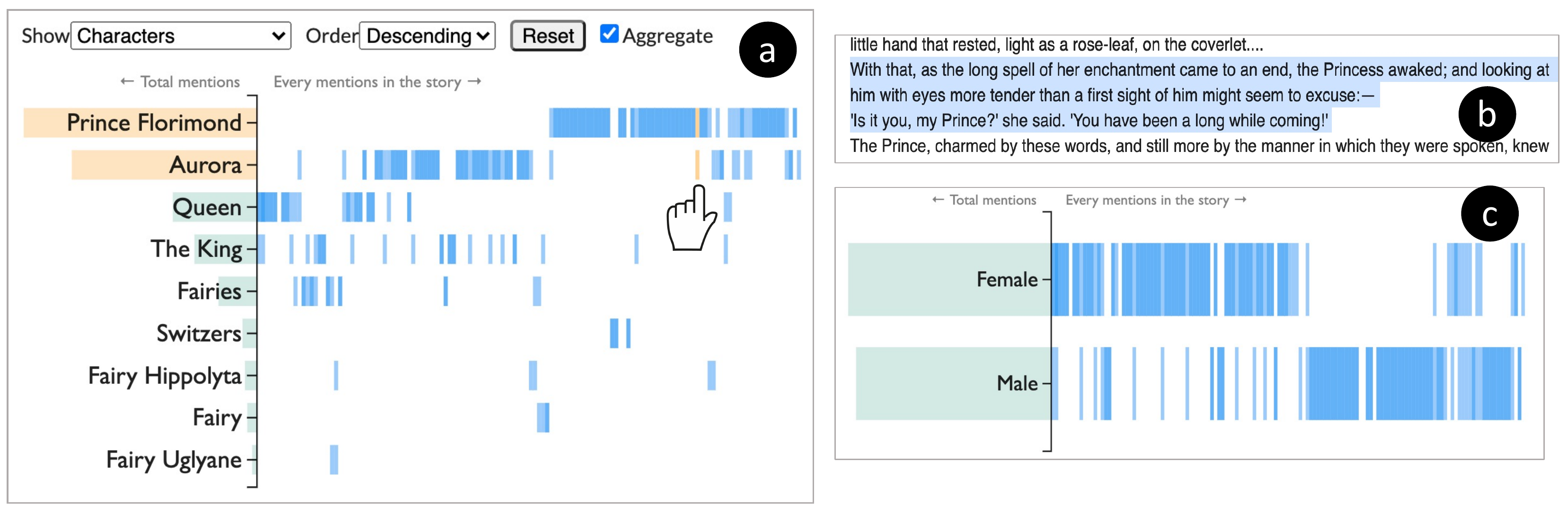}
    \caption{Timeline interface.
    (a) An example timeline based on the characters of \textit{Sleeping Beauty}.
    Note that even though the story is about Aurora, the sleeping beauty, it was Prince Florimond who had the most number of mentions.
    Note also the lack of overlaps between the timelines of two characters.
   (b) Upon hovering over a tile from the timeline, a user can see the relevant passage in the editor.
   The relevant passage in this case is the first scene after the Princess wakes up.
   (c) A user can also visualize the presence of different social identities in the timeline.
   We only used Male and Female for this example.
   Note that when aggregated, the female presence in the story is slightly larger than the male presence.}
   \label{fig:timeline}
\end{figure*}

For larger documents, due to space constraints, the tiles can become extremely small, making it difficult to interact with them.
To scale the visualization to larger documents (\textbf{DG6}), we added an \textit{Aggregate} option in the toolbar of the timeline (Figure~\ref{fig:timeline}a).
In the case of a document with more than 500 sentences $(S > 500)$, the option is automatically triggered, and the document is binned together to restrict the x-axis to range $(1,500)$.
At that point, the x-axis represents passages, instead of sentences.
The user can disable the aggregate option and see the timeline in terms of a single sentence.
As an alternative, we considered making the x-axis scrollable.
The tile size would have remained the same for any documents in that design.
However, we noticed that this design does not provide a full overview of the timeline together and it becomes difficult to compare different parts of the timeline as the user scrolls left and right in the timeline.

Figure~\ref{fig:timeline}a shows an aggregated version of the timeline for the children's book, \textit{Sleeping Beauty}, retold by Arthur Quiller-Couch and Charles Perrault, freely available under Project Gutenberg~\cite{sleeping_beauty}. The opacity of a tile represents the number of times an entity was mentioned in a particular passage. A user can hover over any tile and see the relevant passage, highlighted in the text editor (Figure~\ref{fig:timeline}b). This facilitates easily going back and forth between different parts of the story (\textbf{DG4}).

\begin{figure*}
    \centering
    \includegraphics[width = 0.9\textwidth]{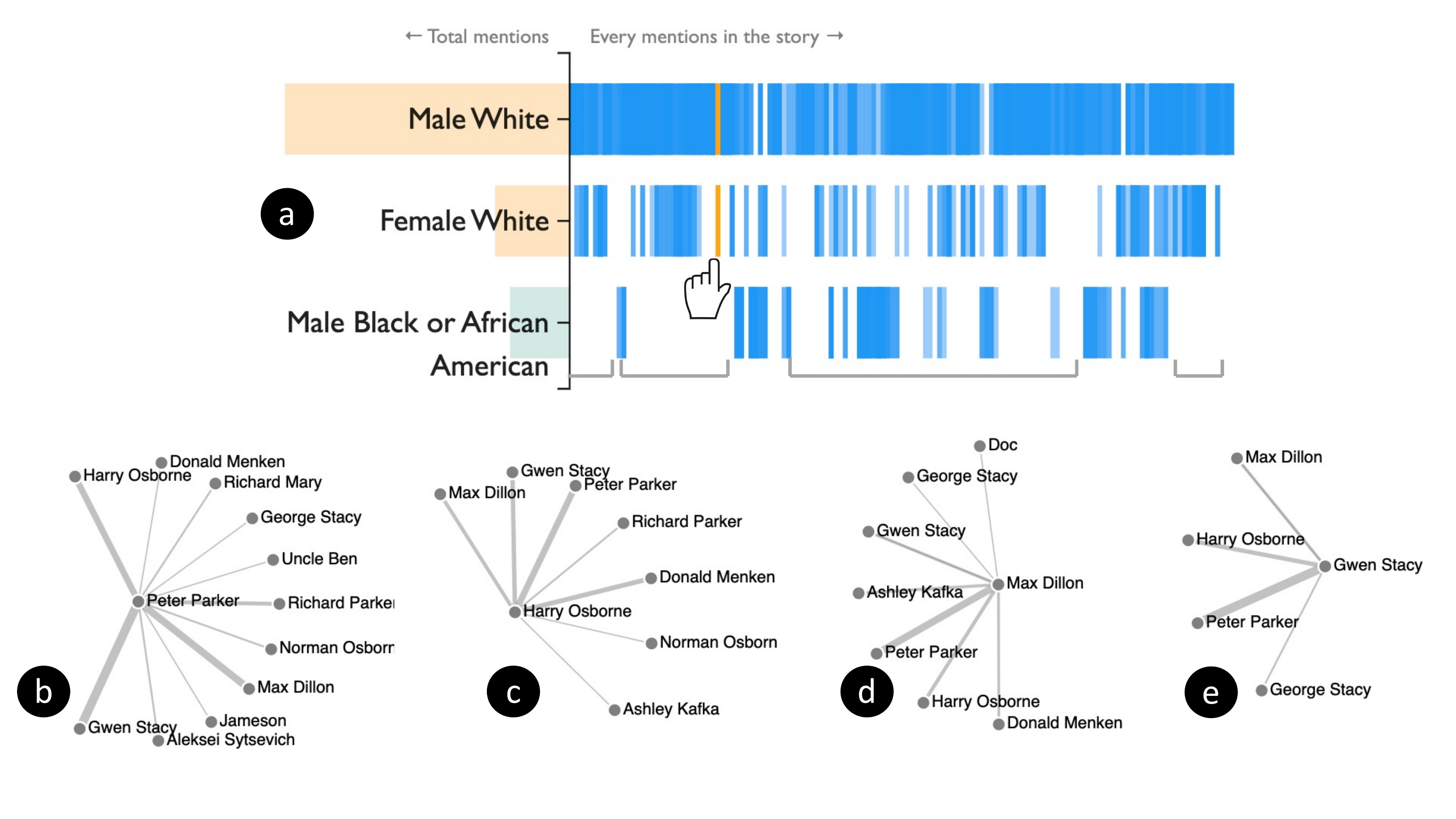}
    \caption{Timeline interface and impact graphs.
    (a) An example timeline showing the presence of three intersectional identities (\textit{Male White, Male Black or African American}, and \textit{Female White}) in the Movie \textit{The Amazing Spiderman 2} (2014).
    Male White characters are present throughout the storyline.
    Further, there appears to be a lack of interactions (identified by the grey encompassing x-axis lines) between Male Black or African-American and Female White characters except for a few aberrant ones.
    However, both groups have interactions with Male White characters.
    The orange bars show one such interaction between Male White and Female White characters.
    (b-e) Example impact graphs for the characters Peter Parker, Harry Osborne, Max Dillon, and Gwen Stacy.}
    \label{fig:timeline2}
\end{figure*}

Based on the identities defined by a user in the Characters and Demographics Panel, the timeline can also show the presence of different identities in the timeline (Figure~\ref{fig:timeline}c). Using the \textit{Show} dropdown in the timeline~(Figure~\ref{fig:timeline}a), a user can choose to show the timeline for characters or demographics.
A user can also choose to show the presence of intersectional identities in the timeline.
For example, Figure~\ref{fig:timeline2}a presents a timeline showing the presence of intersectional groups in the Movie \textit{The Amazing Spiderman-2}. 
Note that for the sake of brevity, we do not show the Characters and Demographics Panel for this movie.
We also note that the race assigned for the characters of this movie is based on the perception of the authors and may be different in reality.

In summary, the goal of the timeline visualization is to facilitate writers investigating agency for characters and social identities (\textbf{DG2}).
Since the task is subjective to a story and a writer's perspective, we intended to expose potential gaps in the story and let a writer explore and navigate the story easily (\textbf{DG1}).
We figured visualizing the mentions of the characters in a timeline can be a starting point for this task and will allow users to see gaps easily.
The user can then use the timeline to further investigate any part of the story.
To aid this process, other visual components are also connected to the timeline, as described below. 

\paragraph{Impact Graph}

The timeline visualization primarily shows the presence of entities in a story.
While a user can identify the interactions between entities by observing overlaps in their presence in the x-axis, the timeline does not give a definitive answer to how an entity interacted with the other entities.
Additionally, multiple entities can be mentioned in a passage of a story; however, that does not necessarily mean they interacted with each other.

To overcome this shortcoming, we introduced the \textit{Impact Graph}, a force-directed network visualization that shows interactions between entities (\textbf{DG2}).
We consider an interaction between two entities if they are mentioned together in a sentence.
The impact graph for an entity is available whenever a user clicks on a y-axis label in the timeline. 

Figure~\ref{fig:timeline2}b-d show impact graphs for Peter Parker, Harry Osborne, Max Dillon, and Gwen Stacy from the movie \textit{The Amazing Spiderman 2}.
The selected entity is placed at the center of the network.
The edge widths represent the strength of the interactions.
For reducing clutter, only edges with at least five interactions are shown.
We observe that the impact graph for Gwen Stacy is much smaller than the other three male characters. 
Further, Gwen Stacy has significant interactions with Peter Parker only.

\paragraph{Word Zones}

Word cloud is a popular visual representation for showing a collection of words.
They have aesthetic value to lay users and are fun, and engaging~\cite{hearst2019evaluation, viegas2009participatory}.
In contrast, researchers have shown that they are not well-suited for analytic tasks such as finding a word and comparing the frequencies of words~\cite{hearst2019evaluation}.
To balance the utility and aesthetic value of word clouds, Hearst et al.~\cite{hearst2019evaluation} proposed Word Zones, a variation of word clouds, where words are grouped based on predefined labels/categories.
Since our users will most likely be non-experts in terms of visualization expertise, we decided to use Word Zones, thus opting for a visualization that is expected to be well-known to the writers as well as has better representation for analytic purposes (\textbf{DG6}).

A writer can add an entity for seeing words used with the entity in the word zone visualization whenever a user clicks on a y-axis label in the timeline (\textbf{DG3}). A writer can add as many entities as they want in the word zone. A user can control the number of words to show for each entity in the word zone using a slider. Users also have a dropdown to see relevant adjectives, verbs, or both. The verbs and adjectives for characters are extracted using dependency parsing. They are then aggregated for the social identities. Overall, this mechanism helps in answering questions such as: \textit{How the female characters were described in the story?} \textit{What were the actions of female characters in the story?}~(\textbf{DG3}). 

 We considered each entity (a character or identity) as a document and the full story as a corpus of documents (entities). Based on that, the weight of a word ($w$) for an entity ($e$) is calculated as:

\begin{equation}
\label{eq:1}
    weight(w, e) = tf(w,e) * (1/df(w)) 
\end{equation}

where $tf(w,e)$ is the frequency of $w$ in $e$
and $df(w)$ is the frequency of $w$ in the whole story. It is essentially a normalized version of \textit{tf-idf}, popularly used for filtering out common and stop words, finding words of interest. However, \textit{tf-idf} is often applied on large corpora. In our case, the number of entities is limited to at most a few hundred. We also consider adjectives and verbs only. Thus, we opted for a simple normalization. 

\begin{figure}
    \centering
    \includegraphics[width=0.8\columnwidth]{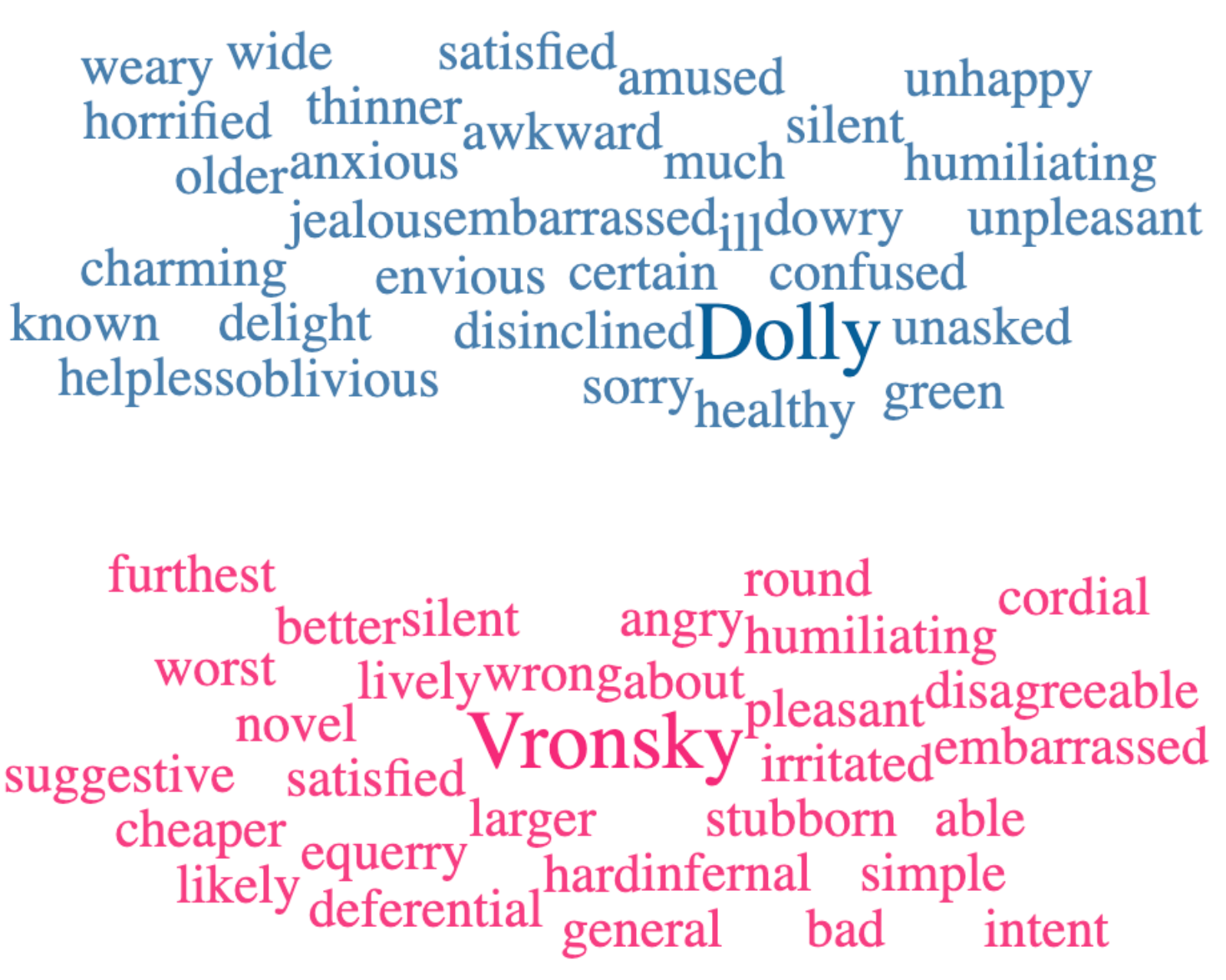}
    \caption{Word Zone representation for Dolly and Vronsky from \textit{Anna Karenina} (1877) by Leo Tolstoy, publicly available under Project Gutenberg~\cite{anna_kerreina}.}
    \label{fig:word_zone}
\end{figure}

Figure~\ref{fig:word_zone} shows an example word zone showing adjective used for Dolly (a female character), and Vronsky (a male character) from \textit{Anna Karenina} (1877) by Leo Tolstoy. Note the words such as ``charming'', ``envious'', ``jealous'', ``helpless'', ``oblivious'' and others in Dolly's word zone.








\subsection{Entity Extraction and Coreference Resolution}

We used NeuralCoref~\cite{neuralcoref} along with Spacy~\cite{spacy} for extracting named entities, and their mentions in a document (Coreference Resolution~\cite{jurafskyspeech}).
Both packages are considered state-of-the-art and widely used in different applications.
However, coreference resolution for long documents such as a book can take a lot of time and system memory \cite{toshniwal2020learning}.
This could hinder the usability of the tool as our tool is targeted as a writing tool that provides instant feedback to writers as they write (\textbf{DG6}).

\begin{figure*}
    \centering
    \includegraphics[width = 0.9\textwidth]{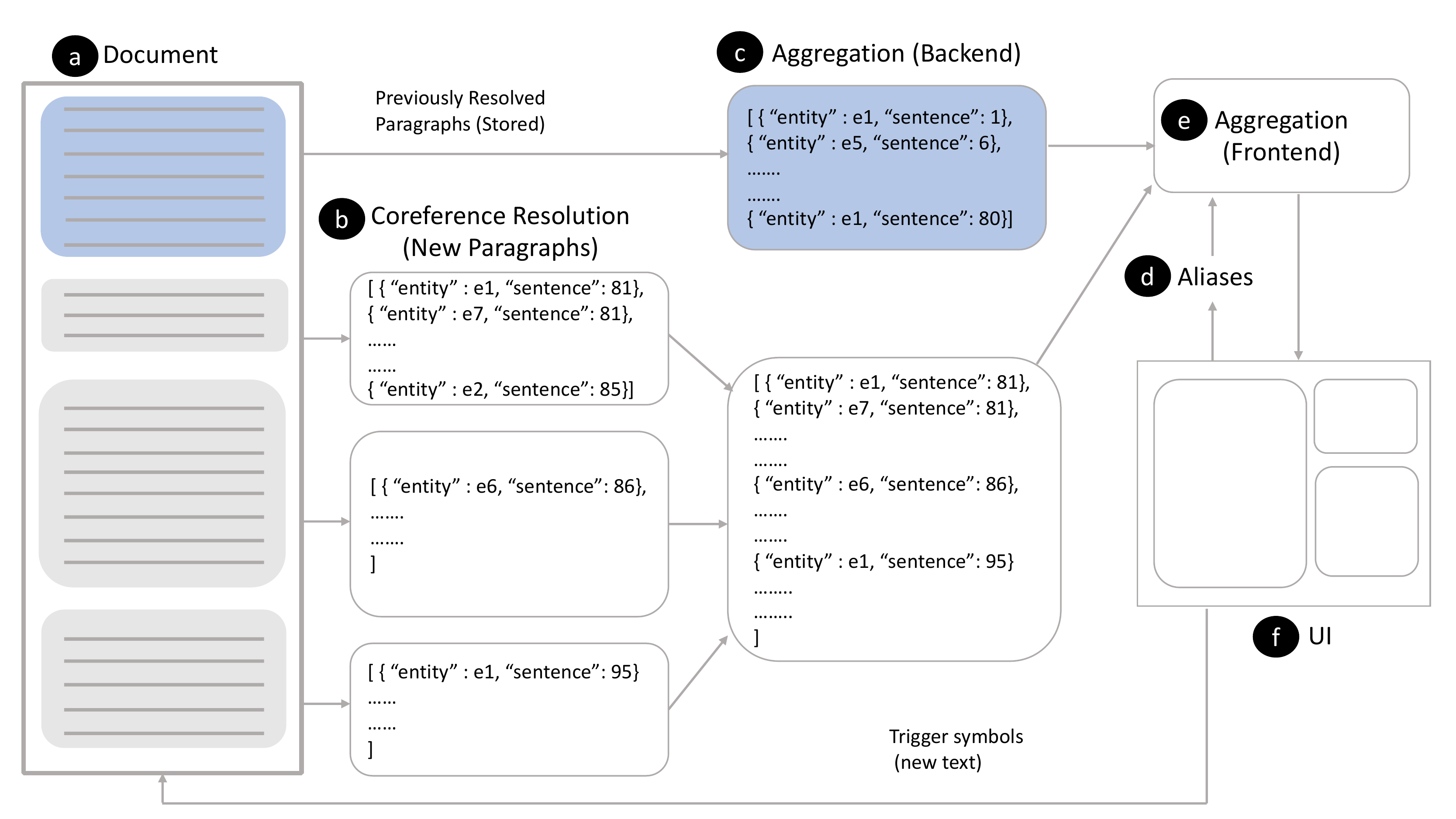}
    \caption{Example workflow of DVP.
    (a) The system detects three new paragraphs in the text editor.
    The blue box represents the previously written text while the grey boxes represent the newly added paragraphs.
    (b) The system runs co-reference resolution and entity extraction independently on the newly added paragraphs.
    (c) They are then aggregated together with the stored entity information of the previous text.
    (e) A second aggregation on the front end is applied based on the aliases (d) identified by the user.
    (f) Finally the JSON formatted data is fed to the UI. The UI listens for the trigger symbols such as ``.'', ``?'', ``!'' or copy/paste event in the text editor for updating the UI. }
    \label{fig:implementation}
\end{figure*}

To increase the scalability of computation and reduce processing time, we adopted a divide-and-conquer method and run the coreference resolution model paragraph-wise, instead of the full document together.
We noticed that it is unlikely that a character would be referenced with a pronoun without the actual noun of the character in a new paragraph.
We reached out to the participants from the interview studies and confirmed that this is the usual case.
Figure~\ref{fig:implementation} presents an example scenario of our method.
On detecting a new trigger symbol such as ``.'', ``?'', ``!'' or a copy/paste event, the UI sends the document and the current Delta~\cite{delta} object to the server.
The Delta object contains information about how many and what characters (symbols) are inserted, deleted, or retained since the last update.

Upon receiving the document, the server splits the document into paragraphs. The paragraphs are identified by double newlines or ``\textbackslash n\textbackslash n''.
Based on the Delta object, the server then determines which paragraphs are retained (same as the previous update), and which paragraphs have new contents (insert or delete).
In Figure~\ref{fig:implementation}, the server detects three new paragraphs (grey boxes).
The server then runs entity recognition and coreference resolution models on the newly detected paragraphs.
Although our current implementation processes the paragraphs sequentially, they can be processed in a parallel fashion since the paragraphs are mutually independent.
The information extracted from the newly added paragraphs is then aggregated together with the stored information of previously processed paragraphs.
The client-side then performs another aggregation to combine aliases.


\subsection{Implementation Notes}

DVP is a web-based writing tool.
We used Python as the back-end,  JavaScript as the front-end language, and D3~\cite{bostock2011d3} for interactive visualization.
Semantic-UI and Bootstrap were used for styling various visual objects. 
 We used Spacy~\cite{spacy} and NeuralCoref~\cite{neuralcoref} for all the NLP tasks. The tool is available here: \url{https://github.com/tonmoycsedu/DramatVis-Personae}.
 


\section{Evaluation}

We evaluated DVP in two parts.
First, we conducted think-aloud sessions with three writers to understand the potential of DVP as a writing tool and identify potential usability issues.
We then conducted a user study with 11 writers to identify the benefits of DVP in comparison to a simple text editor.
We describe each study in turn below.

\subsection{Think-aloud Sessions}

For this study, we invited 3 writers to test our tool.
Two of them participated in our formative study (W2 and W5 from Table~\ref{tab:participant}).
We refer to them as W1 and W2 in the following section.
The other writer (W3) did not participate in the formative study.
W3's self-reported demographics are male, white, 44 years of age, with more than 20 years of writing experience.

Before the start of the study sessions, we emailed each writer independently asking whether they would like to write a new short story using our tool, or they would like to analyze stories that have already been written by them.
W1 and W2 wished to analyze previously written stories while W3 wished to write a story using our tool.
We asked writers that their stories should have at least three significant (named) characters and any number of supporting characters, at least two different scenes, and some dialogue between the characters.
Additionally, we encouraged W3 to ideate and take notes on their story and characters, but asked not to begin the actual writing process until the session.

One author of this work administrated the sessions.
After consent, the administrator demonstrated the tool with a sample story. We encouraged participants to ask questions at this stage.
Once participants felt comfortable with the tool, we asked them to analyze a story (W1 and W2) or generate a new short story (W3).
While using the interface, participants thought aloud and conversed with the administrator regularly. Each session ended with an exit interview.
The sessions with W1 and W2 took around 1 hour while the session with W3 lasted around 2 hours.
All sessions were conducted using Zoom video conferencing and were audio and video recorded for post-session analysis.
At the end of the interview, each participant received a \$15/hour worth Amazon gift card.
The study administrator and another author of this work independently analyzed the recorded videos, and study notes and then met together to sort the feedback from the writers and our observations into the following five thematic categories:

\subsubsection{Intuitive and Easy-to-use}

All participants found the interface to be easy and intuitive.
Overall, the design, especially the incorporation of visualizations in a writing tool, intrigued participants.
Initially, W2 and W3 were slightly confused about the timeline representation but quickly became accustomed to the timeline once they started interacting with the tiles.
Participants appreciated that the visualization and text editor were connected, and that writers can easily use the timeline to go to any part of a story.
We did not have to provide any explanation for the word zones; participants were already familiar with them.
Finally, while it took each participant a few minutes to validate the characters, merge aliases, and then assign social identities to the characters, all participants were enthusiastic about this process.
W3 suggested a few stylistic changes such as auto-indenting a new paragraph and adding a thesaurus.

\subsubsection{Effect of Assigning Social Identities to Characters}

The mere presence of an interface where writers can assign social identities to characters had an effect on the writers.
W3 said: \textit{``This is an important feature for me as I usually plan very little before writing, and this helped me incorporate planning early into the writing and made me think how the characters should be represented in the story.''}
W2 mentioned that assigning social identities to characters could be a good practice for writers, especially when characters do not share identities with the writer.
This will help them think about the characters from the beginning.

\subsubsection{Usage Patterns}

All participants mentioned that DVP does not obstruct their creative process.
They appreciated that using DVP they can explore biases when they want to and concentrate on writing or reading other times.
For example, W3 used the ``Hide'' option to hide the visualizations while writing and only checked them periodically.
W2 also suggested that they will probably check the visualizations after drafting a scene or section.

Both W1 and W2 extensively used the word zone by comparing words used for characters and social identities.
W2 suggested that it would be useful to quickly find the characters that have significantly different adjectives or verbs.

\subsubsection{Errors in Identifying Character Mentions}

DVP is powered by NLP toolkits and their capabilities to extract character names and identify character mentions throughout the story.
The pipeline may sometimes not recognize characters or their relevant mentions.
We noticed such cases created confusion among writers.
One of the animal characters from W2's story was not identified by our tool.
W2 asked whether our tool can identify unusual names since their sci-fi stories often have names that are not common.
Similarly, W1 noticed the tool missing out a few mentions of a character in the timeline.
While we expect the mitigation of such problems as NLP techniques become increasingly powerful, an alternative would be to allow writers to highlight name entities in the text editor and manually add them for tracking.

\subsubsection{Potentials and Impact}

At exit interviews, participants discussed several use cases and the impacts of DVP.
All participants suggested that DVP would be helpful to any creative writer.
They admired its support for different forms of writing (e.g., screenplays, fiction, non-fiction) which makes it usable for a wide range of creative materials.
W1 teaches novice journalists on how to write critiques, and was interested in using the tool to show examples of correct and biased representation to the students.
Further, W1 thought novice journalists can use the DVP tool as a learning tool and editors can use it to quickly check drafts.

\begin{figure*}
\begin{subfigure}{0.48\linewidth}
\centering
\includegraphics[width = \textwidth]{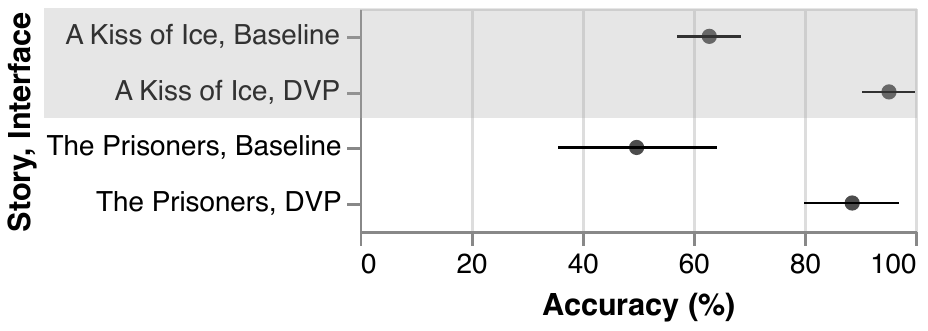}
\caption{}
\label{fig:accuracy_5}
\end{subfigure}%
\hfill
\begin{subfigure}{0.48\linewidth}
\centering
\includegraphics[width = \textwidth]{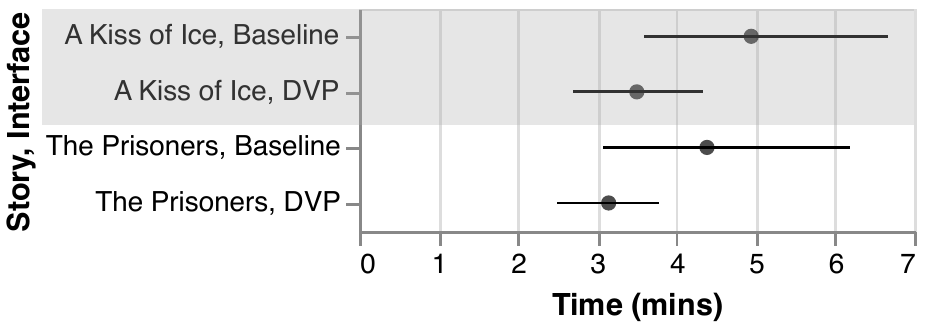}
\label{fig:time}
\caption{}
\end{subfigure}

\begin{subfigure}{0.48\linewidth}
\centering
\includegraphics[width = \textwidth]{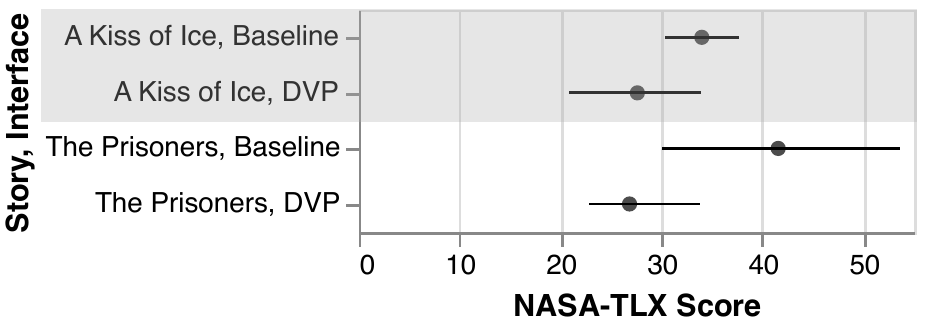}
\label{fig:nasatlx}
\caption{}
\end{subfigure}%
\hfill
\begin{subfigure}{0.48\linewidth}
\centering
\includegraphics[width = \textwidth]{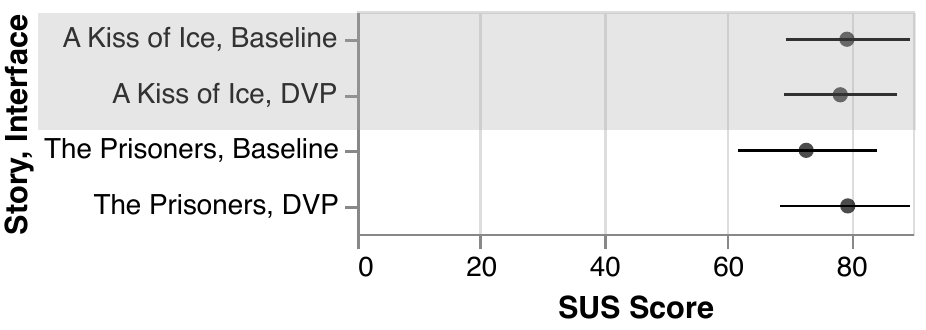}
\label{fig:sus}
\caption{}
\end{subfigure}

\caption{Study Results.
    (a) Participants' accuracy in answering the study questions; (b) Task completion time in minutes; (c) NASA-TLX (perceived workload) score; and (d) SUS score (usability ratings).
    Error bars show 95\% confidence intervals (CIs).}
\label{fig:my_label}
\end{figure*}

\subsection{User Study}


We conducted a user study to measure the effectiveness of DVP in identification of biases. We designed a repeated-measures within-subject study with two interfaces: \textbf{I1. Baseline}, presented a simplified version of DVP, featuring only the text editor (Figure~\ref{fig:teaser_05}a) and demographic information of the characters (Figure~\ref{fig:teaser_05}b); and \textbf{I2. DVP Tool}, presented our interface with all features. 




Before conducting this study, based on the feedback from the think-aloud sessions, we made the following three refinements to our prototype.
First, we added a feature for users to manually highlight a character in the text editor and add it to the Characters and Demographics Panel for tracking.
This ensures that any character missed by the NLP pipeline can be supported in the system.
Second, the study session suggested that while word zone provides an easy way to investigate adjective and verbs, the search space to explore all the combination of entities can be large.
To aid writers in this process, we introduced a word embedding based approach to find potential candidate pairs of entities for investigation.
Let $\vec{g1}$ be the mean vector representing the words from entity $ent1$ and $\vec{g2}$ be the mean vector representing the words from entity $ent2$.
Then, the cosine distance between $\vec{g1}$ and $\vec{g2}$ can indicate how different the words from $ent1$ and $ent2$ are. This method closely matches the \textit{relative norm difference}~\cite{100years, ghai2021wordbias}, previously used to quantify gender bias in word embeddings.
The top 10 pairs of entities sorted by their cosine distance  are fed to the interface as a notification in the word zone control panel.
Finally, we added a few convenience features such as auto indent and word count in the system.

\subsubsection{Participants}

We recruited 11 participants through local mailing lists, university mailing lists, and public posts on Facebook groups and Twitter.  
Our inclusion criteria included prior experience with any forms of creative writing (fiction, non-fiction, short story, screenplays, etc).
We did not require participants to have published articles in their portfolios.
The participants varied in age from 21 to 45 (mean 26.5, s.d.\ 4.32), gender (5 males, 6 females), and race (6 Asian, 4 White, and 1 African/American or Black).
All participants had prior experience with writing creative materials and using a text editor.
None of them participated in any of our prior studies.
Participants received \$15 per hour worth of Amazon gift cards for their participation.

\subsubsection{Data: Short Stories}



We used two newly written short stories for the study. One of the authors of this work who had published fiction in their portfolio wrote one short story (``A Kiss of Ice'') for the study.
The other story (``The Prisoners'') was written by a professional writer, hired from \url{Fiverr.com}.
We compensated the professional writer with \$50.
As requirements, we asked both writers to write a short story that includes at least 4-5 characters and 3-4 different intersectional identities with a significant amount of interaction (having conversations, or participating in a scene together) between them. We also requested them to introduce a few stereotypes in the stories, hoping they will incite meaningful discussion in the exit interviews.

\subsubsection{Study Questions}

DVP is designed to support the identification of two types of biases:
(1) lack of agency for characters;
and (2) stereotypes in describing characters.
Thus, we wanted to test how effective DVP would be in detecting these biases.
To do so, we designed 7 questions for each story that either fall into the first or second bias, or both.
A few examples of questions are
\textit{``Which of the following characters has the least presence in the story?
Which of the following social identities has the least presence in the story?
Which of the following two characters have the least interactions (having conversations, or participating in a scene together) with each other in the story?
How would you categorize how Female characters were portrayed in the story?''}

We provided multiple choices for each question.
The questions were similar for both stories although they were slightly modified to adjust to the nature of the stories. Two authors of this work separately and manually analyzed the stories to get the correct answers and verified them by discussing with each other. The correct answers were also validated by the writers of the stories.
Note that DVP is capable of supporting more demanding tasks such as sorting presence of every character; however, we intentionally did not include these tasks as they might be exhausting and overbearing to perform in a plain text editor (our baseline interface).




\subsubsection{Procedure}

Similar to the think-aloud sessions, we conducted the study sessions via Zoom. The study participants were provided a url where DVP was hosted. They interacted with the system via their own web browser.
A study session began with the participant signing a consent form.
Following this, the participants were introduced to the assigned first condition.
All participants had prior experience in writing and reading in a simple text editor (I1) so they did not require any training for this interface.

For DVP (I2), we provided a brief description of the interface and a demo showing its different features.
The participants then interacted with the system (with a training story), during which they were encouraged to ask questions until they were comfortable.

Each participant was then given a short story for the first interface and asked to read the story. 
On average, it took participants 10 minutes (min = 5 minutes, max = 18 minutes) to read the short story.
After participants finished reading the story, we provided them with the task list (questions) for the first interface and asked them to answer the questions using the interface within 15 minutes. The questions were presented in random order to the participants. We provided them verbal clarification for the questions, if needed.
At the end of the first interface, we administrated the NASA-TLX~\cite{hart2006nasa} questionnaire to measure the participants' perceived workload, and SUS questionnaire to measure user experience and usability of that interface.
The same process was carried out for the second interface.
Each session lasted around 1.5 hours and ended with an exit interview.



\subsubsection{Study Design Rationale}

Our research objective from this study was to determine whether DVP can actually help writers identify biases. 
To evaluate that, we contemplated different study tasks and story types described as follows:

\textbf{Task.} The ideal scenario would perhaps be to ask writers write a long and complicated story using DVP, since that is the primary use case of DVP.
However, given the sensitive nature of bias, it is unlikely that writers would be comfortable analyzing their work critically and answer questions related to bias in the presence of a study administrator.
Additionally, writing a story, even a short one, takes significant planning and effort, and is likely to be a very long and involved process.

An alternative would be to ask writers to use our tool on their own and provide feedback periodically.
While that might lead to insightful feedback, the process is uncertain and it will require a long-term commitment from a group of writers.
We felt that this was impractical at this stage of the research project.
Thus, we decided to avoid asking writers to write new stories in this study.

\textbf{Story Type.} Since we decided to avoid the writing task, we needed  already written stories for the study. We considered using stories written by the writers themselves or freely available stories online.
In both cases, participants may have already read and analyzed the stories.
Thus, using them may not lead us to isolating the effect of DVP.
Additionally, we needed to know the social identities of the characters, which is typically not explicitly provided in published articles. 
Considering these factors, we decided to use newly written stories for the study.

\subsubsection{Results}

To measure performance, we calculated participants' accuracy in answering the questions.
We also measured task completion time, perceived workload (NASA-TLX), and usability (SUS).
Following recent guidelines for statistical analysis in HCI~\cite{dragicevic2016fair}, we intentionally avoided traditional null-hypothesis statistical testing (NHST)
in favor of estimation methods to derive 95\% confidence intervals (CIs) for all measures.
More specifically, we employed non-parametric bootstrapping~\cite{efron1992bootstrap} with $R = 1,000$ iterations.
We also report mean differences as sample effect size and Cohen's $d$ as a standardized measure of effect size~\cite{cohen1988statistical}.


\textbf{Accuracy.}
%
Figure~\ref{fig:accuracy_5} presents participants' accuracy in answering the questions. 
Overall, participants were more accurate when answering the questions using DVP.
For the ``A Kiss of Ice'' story, participants were more accurate on average by 32.49 ($d$=4.27) percentage points using DVP.
Similarly, for ``The Prisoners'' story, participants were more accurate on average by 38.25 ($d$=2.31) percentage points using DVP.
The very large effect sizes suggest strong practical difference.
%
%
We noticed participants performed slightly better for the story ``A Kiss of Ice'' than ``The Prisoners'' using both interfaces.
We anticipated this result as ``The Prisoners'' featured frequent interactions between different characters and social identities which raised the difficulty level for answering the questions.


\textbf{Task Completion Time.} 
%
Figure~\ref{fig:my_label}(b) presents participants' task completion time.
The completion time was measured by calculating number of minutes participants took to answer the study questions, excluding the time to read the stories.
Unsurprisingly, participants spent less time to answer the questions using DVP.
For ``A Kiss of Ice,'' participants spent on average 1.37 ($d$=0.88) minutes less using DVP.
Similarly, for ``The Prisoners,'' participants spent on average 1.24 ($d$=0.72) minutes less using DVP.


\textbf{Perceived Workload (NASA-TLX).}
%
Figure~\ref{fig:my_label}(c) presents NASA-TLX scores, average of the raw ratings provided by the participants on six commonly used workload measures: mental demand, physical demand, temporal demand, effort, performance, and frustration~\cite{hart2006nasa}.
DVP reduced workloads of participants in answering the questions.
For ``A Kiss of Ice,'' on average the perceived workload of the participants were 5.66 ($d$=0.84) points less using DVP. 
Similarly, for ``The Prisoners'' on average the perceived workload of the participants was 14.79 ($d$=1.08) points less using DVP.

\textbf{Usability (SUS Score).}
%
Participants rated both condition similarly in terms of usability.
Figure~\ref{fig:my_label}(d) presents SUS scores calculated from the 10 individual SUS questions.
As observed in Figure~\ref{fig:my_label}(d)
, the extra visual representations of DVP did not warrant any major usability issues.

\textbf{Qualitative Feedback.}
Participants mentioned that DVP adds a new dimension to reading stories (P2, P4, P7-9).
The demographic information of the characters and visualization created a situational awareness among them.
For example, P4 said \textit{``Normally, I do not think about the identities of the characters while reading, beyond what can be inferred from the names, and description of the characters.
The visualization certainly created a new dimension to my reading and I can imagine the characters in new ways that I would not do usually.''}
We noticed several participants used the default search and find feature of browsers when answering the questions using the simple text editor.
In the exit interviews, several participants mentioned that they used the search and find feature as a way to quickly find relevant information for answering the questions.
However, it was not effective in answering the questions as they needed to mentally analyze the search results along with the demographic information which might not be directly encoded as key words in the story.
In contrast, the visualization in DVP supported them in answering the questions by providing summarized information.
Participants also saw value in DVP as a writing tool.
They mentioned that the tool will be most useful for writing long stories with many inter-connected characters and subplots (P1-5, P7, P9).
Participants mentioned several use cases where DVP may be helpful.
For example, P1 said \textit{``I think this tool has a clear application for writing adapted screenplays.
When I write an adapted screenplay, I constantly try to evaluate my writing with the representation in the original book.
The tool can be helpful in such scenarios.'' } 


\section{Discussion, Limitations, and Future Work}

In this section, we discuss potential implications and limitations learned from the development and evaluation of DVP.

\textbf{Long Term Deployment of DVP.}
The think-aloud sessions and user study show promise for DVP as a new medium for writing and critical reading.
Even for short stories that were used in the user study, DVP has made the bias identification process much easier for writers.
Further, all participants were enthusiastic about the tool and thought the tool would be useful for them.
To evaluate DVP to its full extent, our future work will concentrate on a long term deployment of DVP with writers (e.g, writing a full-length story using DVP).

\textbf{Impact on Bias in Creative Writing.}
Bias in creative writing is ubiquitous, as described in Section~\ref{sec:story-bias}.
Despite our success with DVP, we believe it is important to acknowledge that our work will not solve systematic or infrastructural discrimination that exists in the publishing community. 
For example, there is still a lack of writers from marginalized groups in the current writer's community~\cite{gd_equity2021}. 
We consider DVP as a part of the larger movement against biases in creative writing, a probe for reducing biases, and a catalyst for future efforts in this direction.
We hope its adaptability to different forms of writing (novels, children's books, short stories) will attract diverse writers. 



\textbf{Balancing Automation and Artistic Freedom/Agency.}
From the beginning of this project, we were vigilant about the artistic freedom of writers, and how we can balance agency with automation.
We understand writers may want to intentionally write about discrimination and biases against a social group, and we do not want to see the tool preventing them in this process.
That is one of the reasons why we did not use any automation in correcting biases and stereotypes.
However, this leads to a potential limitation of our tool: a writer can still write harmful stereotypes and there is no way to perform corrective measures if the writer is not willing to self-evaluate themselves.


It is also worth noting here that our intention was not to replace critique groups or sensitivity readers.
We believe it is important to have subjective feedback about creative writing, especially from relevant marginalized groups.
Our intention was to help writers during their writing process by offering another set of eyes---albeit electronic ones.

\textbf{Design Implications for Human-centered AI.}
AI systems can inherit harmful biases and stereotypes.
These biases can impact social groups disparately, especially when used as a decision-making platform for critical resources ~\cite{scheuerman2019computers, angwin2016machine, ghai2022cascaded}.
These systems may also lack inclusivity (e.g., lack of supports for non-binary identities).
These limitations motivated several design decisions of our system.

First, based on DG5, we made the \textit{demographics} dropdowns unconstrained so that a writer could add any social identity required for the story.
Second, we provided several functions in the interface for the writers to validate (e.g., merge, delete) the results returned by the NLP pipeline, a safety check against potential biases in coreference resolution~\cite{zhao2018gender}, and dependency parsing~\cite{garimella2019women}.
Additionally, a writer can interactively add an entity for tracking in the case the entity was not recognized by the NLP pipeline.
Thus, our work promotes human-centered AI for creative writing.

\textbf{Limitations of the Underlying AI.}
NLP research has made big strides in recent years but we cannot still expect perfect results for problems such as coreference resolution.
In particular, it is worth noting that different NLP models like coreference resolution, named entity recognition might exhibit biases based on social identities like gender~\cite{zhao2018gender, mehrabi2020man}.
This might result in skewed performance against minority groups~\cite{mehrabi2020man, mishra2020assessing}. 
Moreover, the NLP models used in this work were trained on a set of news articles, weblogs, etc.
Such data might differ from literary texts like novels books, etc. as they might contain longer sentences, more sophisticated language, etc~\cite{ rosiger2018towards}.
Although we did not observe any such issues in our experiments, future work might employ fairness-aware NLP models that are trained on domain-specific datasets like books, novels, etc. for better performance.   

The time to process text increases proportionally with the amount of text.
Higher processing time will add more latency on the front-end which might negatively impact user experience.
In this work, we made an informed decision to choose NLP models which provide good performance while providing fast response time.
Based on the current trend in NLP literature, future NLP models might be more computationally expensive and provide better performance.
To incorporate such models, future work might prefer to conduct text analysis less frequently and/or leverage multiple compute cores.

Finally, the design and development process of this tool has been influenced by English-speaking study participants and NLP models trained over English language corpus.
Hence, our tool can currently only support the English language.
Having said that, we know that social biases based on gender, race, etc, transcend societies and languages~\cite{fr_es, hindi}.
We believe that ideas put forward in this work will help facilitate the development of similar tools for other languages as well.
Future work might support others languages like French, Chinese, etc. by incorporating NLP models trained over different language corpora. \\


In this work, we have presented an in-depth case study on supporting creative writing and combating implicit bias in fiction using interactive technologies, data visualization, and NLP.
In particular, we have reported on an interview study involving 9 creative writers where we asked them about their process, how they navigate harmful stereotypes, and how they think tool support could help in this work. 
Based on these interviews, we design \sysname{} (DVP), a visual analytics tool using NLP to visualize characters, their demographics, and their story arcs in an effort to mitigate implicit bias.
The tool can be used both in an online manner while writing a story, as well as offline during the analysis of an already written story. 


We believe that our work here suggests many interesting future avenues of research. 
For one thing, while creative writing is a notoriously individual and idiosyncratic process, and while the human touch is vital to true art, our moderate success with \sysname{} points to possible ways to augment this human touch to improve even such famously crooked processes.
In particular, we think that our work shows how automatic machine eyes, while certainly less keen and discerning than human ones, can be helpful for certain applications such as mitigating bias if only because they---unlike human eyes---remain unblinking. 
We are not so foolish as to believe in the idea of an ``unbiased algorithm''---all algorithms are created by humans and thus intrinsically carry the biases of their creators---but we do believe in the virtue of training as many different lenses as possible on a creative artifact in the hope of uncovering yet another harmful stereotype or instance of implicit bias.
Thus, we tend to think that our work here is in no way indicative of an end of art, but rather a new beginning.

\let\textcircled=\pgftextcircled
\chapter{Cumulative Effect of Multiple Fairness-Enhancing Interventions}
\label{chap:cascade}

Understanding the cumulative effect of multiple fairness-enhancing interventions at different stages of the machine learning (ML) pipeline is a critical and underexplored facet of the fairness literature. Such knowledge can be valuable to data scientists/ML practitioners in designing fair ML pipelines. This work takes the first step in exploring this area by undertaking an extensive empirical study comprising 60 combinations of interventions, 9 fairness metrics, 2 utility metrics (Accuracy and F1 Score) across 4 benchmark datasets. We quantitatively analyze the experimental data to measure the impact of multiple interventions on fairness, utility and population groups. We found that applying multiple interventions results in better fairness and lower utility than individual interventions on aggregate. However, adding more interventions do not always result in better fairness or worse utility. The likelihood of achieving high performance (F1 Score) along with high fairness increases with a larger number of interventions. On the downside, we found that fairness-enhancing interventions can negatively impact different population groups, especially the privileged group. This study highlights the need for new fairness metrics that account for the impact on different population groups apart from just the disparity between groups. Lastly, we offer a list of combinations of interventions that perform best for different fairness and utility metrics to aid the design of fair ML pipelines.

\begin{figure}
 \centering 
 \includegraphics[width=0.9\columnwidth]{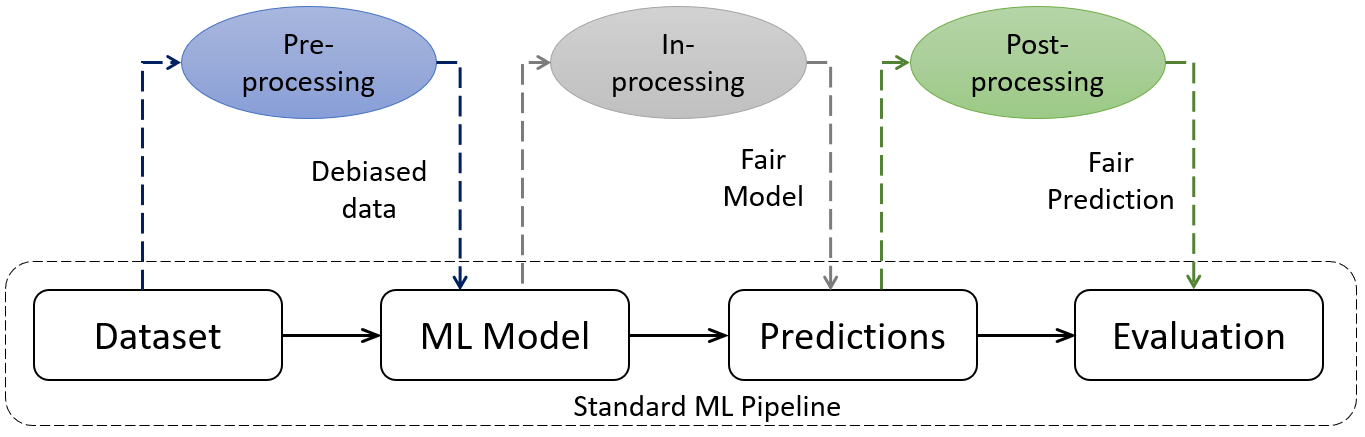}
 \caption{Three different types of fairness enhancing interventions and how they fit into the standard ML pipeline.}  
 \label{fig:types}
\vspace{-1em}
\end{figure}

\section{Introduction}
Algorithmic bias is a complex socio-technical problem whose impact can be felt in all sub-disciplines of machine learning \cite{100word,ghai2020measuring,vision,accent,ekstrand,ghai2021wordbias}. Recent years have seen a huge surge of fairness-enhancing interventions that operate at different stages of the ML pipeline. Some of these interventions are more effective than others at reducing bias as captured by a specific fairness metric. However, the problem is far from being solved 
if that is even possible \cite{friedler2021possibility}. Hence, there is a need for better interventions to reduce bias even further. Moreover, Algorithmic bias can virtually emerge from any single or multiple stage(s) of the machine learning pipeline, right from problem formulation, dataset selection/creation to model formulation, deployment, and so on. \cite{holstein2019improving}. The existing literature focuses on curbing algorithmic bias by intervening at \textit{a} particular stage of the ML pipeline (see \autoref{fig:types}). However, algorithmic bias might still flourish via other stages/components of the ML pipeline. So, our focus should be on ensuring fairness across the ML pipeline instead of a single stage of the pipeline. This issue is also backed by a recent study with ML practitioners that elaborated on the disconnect between academic research and real-world needs \cite{holstein2019improving}. One of the findings was to consider fairness as a system-level property where the focus is on evaluating the impact of ML system as a whole instead of monitoring individual components.
 

An intuitive solution to enhance fairness across the ML pipeline can be to apply multiple fixes (interventions) at different stages of the ML pipeline where bias can emerge from. We will refer to such a series of fairness-enhancing interventions as cascaded interventions. For example, one might choose to debias the dataset, train a fairness-aware classifier over it and then post-process the model's predictions to achieve more fairness. 
This approach is inline with the real world where different laws/policies/guidelines try to alleviate social inequality by intervening at multiple stages of life like education, employment, promotion, etc. Examples include Affirmative action in the US and Caste based reservation in India. 
This begs the question if it were possible to achieve more fairness in the ML world by intervening at multiple different stages of the ML pipeline
and what might be its possible fallouts. 
   


\begin{figure*}
 \centering 
 \includegraphics[width=\columnwidth]{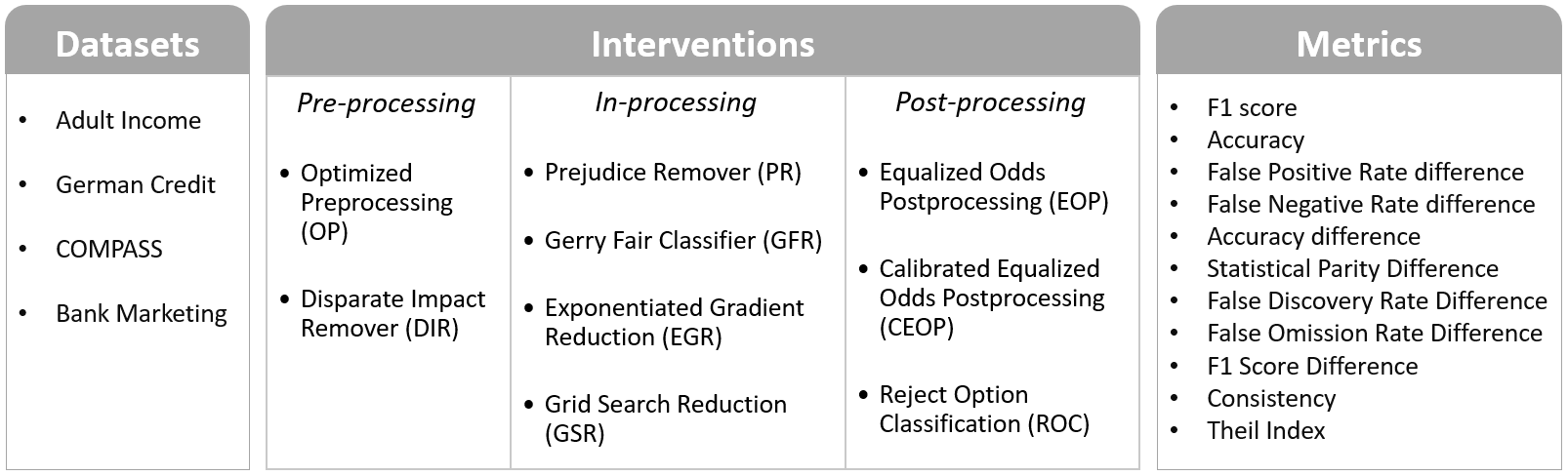}
 \vspace{-5pt}
 \caption{Experimental Setup - Different datasets, interventions and metrics considered for the empirical study}  
 \label{fig:exp_setup}
\end{figure*}

In this work, we undertake an extensive empirical study to understand the impact of individual interventions as well as the cumulative impact of cascaded interventions on utility metrics like accuracy, different fairness metrics and on the privileged/unprivileged groups. Here, we have focused on the binary classification problem over tabular datasets with a single sensitive (protected) attribute. We have considered 9 different interventions where 2 operate at the data stage, 4 operate at the modeling stage and 3 operate at the post-modeling stage. We also consider all possible combinations of these interventions as shown in \autoref{fig:distribution}. To execute multiple interventions in conjunction, we feed the output of one intervention as input to the next stage of the ML pipeline. We simulate multiple 3 stage ML pipelines that are acted upon by different combinations of interventions. 
We measure the impact of all these interventions on 9 fairness metrics and 2 utility metrics over 4 different datasets. Thereafter, we perform quantitative analyses on the results and try to answer the following research questions:

\begin{itemize}
    \item [\textbf{R1.}] Effect of Cascaded Interventions on Fairness metrics \\ Does intervening at multiple stages reduce bias even further? If so, does it always hold true? 
    What is the impact on Group fairness metrics and Individual fairness metrics?
    \item [\textbf{R2.}] Effect of Cascaded Interventions on Utility metrics \\
    How do utility metrics like accuracy and F1 score vary with different number of interventions? Existing literature discusses the presence of Fairness Utility tradeoff for individual interventions. Does it hold true for cascaded interventions?  
    \item [\textbf{R3.}] Impact of Cascaded Interventions on Population Groups \\
    How are the privileged and unprivileged groups impacted by cascaded interventions in terms of F1 score, False negative rate, etc.? Are there any negative impacts on either group?
    
    \item [\textbf{R4.}] How do different cascaded interventions compare on fairness and utility metrics?
\end{itemize}




\section{Background and Literature Review}

\subsection{Fairness Enhancing Interventions}
\label{sec:interventions}
Bias Mitigation techniques can be broadly classified into 3 stages:- Pre-processing, In processing, and Post-processing (Fig. \ref{fig:types}). In the following, we discuss a few interventions that we have considered in this work, in the context of the intervention stage they operate. 

\subsubsection{Pre-processing} Interventions at the Pre-processing stage operate on the raw dataset to generate its debiased version. The debiased dataset can then be fed back into the standard ML pipeline for fairer predictions. Specifically: 


\emph{Optimized Preprocessing (OP)} —  uses convex optimization to transform the underlying dataset such that fairness is enhanced and utility is preserved with limited data distortion \cite{calmon2017optimized}.

\emph{Disparate Impact Remover (DIR)} — edits the feature set of a given dataset such that the predictability of the protected variable is impossible 
while preserving rank ordering within groups \cite{feldman2015certifying}.

\subsubsection{In-processing} Interventions in this stage operate at the data modeling stage to train a fair ML model. Specifically: 

\emph{Gerry Fair Classifier (GFC)} — formulates fairness as a zero-sum game between a Learner (the primal player) and an Auditor (the dual player) to compute an equilibrium for this game \cite{kearns2018preventing}. 

\emph{Prejudice Remover (PR)} — adds a specialized regularization term to the learning objective
such that the classifier becomes independent of the sensitive information
\cite{kamishima2012fairness}.

\emph{Exponential Gradient Reduction (EGR)} — reduces fair classification to a sequence of cost-sensitive classification problems, returning a \textit{randomized classifier} with the lowest empirical error subject to fair classification constraints \cite{agarwal2018reductions}.

\emph{Grid Search Reduction (GSR)} — reduces fair classification to a sequence of cost-sensitive classification problems, returning the \textit{deterministic classifier} with the lowest empirical error subject to fair classification constraints \cite{agarwal2018reductions, agarwal2019fair}.


\subsubsection{Post-processing} Such interventions operate at the model's predictions to yield more fair predictions. Specifically: 

\emph{Calibrated Equalized Odds Postprocessing (CEOP)} — changes classifier results based on calibrated score outputs and an equalized odds goal \cite{pleiss2017fairness}.

\emph{Equalized Odds Postprocessing (EOP)} — solves a linear program to find probabilities whose corresponding labels will optimize the equalized odds goal \cite{hardt2016equality, pleiss2017fairness}.

\emph{Reject Option Classification (ROC)} — reduces discrimination by assigning positive labels to the unprivileged groups and negative labels to the privileged groups for the data points that lie close to the the decision boundary i.e. the points for which the classifier is uncertain about \cite{kamiran2012decision}.

Existing literature has studied the above mentioned interventions in isolation. In this work, we explore if a combination of these interventions can lead to enhanced fairness across the ML pipeline. 

\begin{figure}
    \centering
    \includegraphics[scale=0.55]{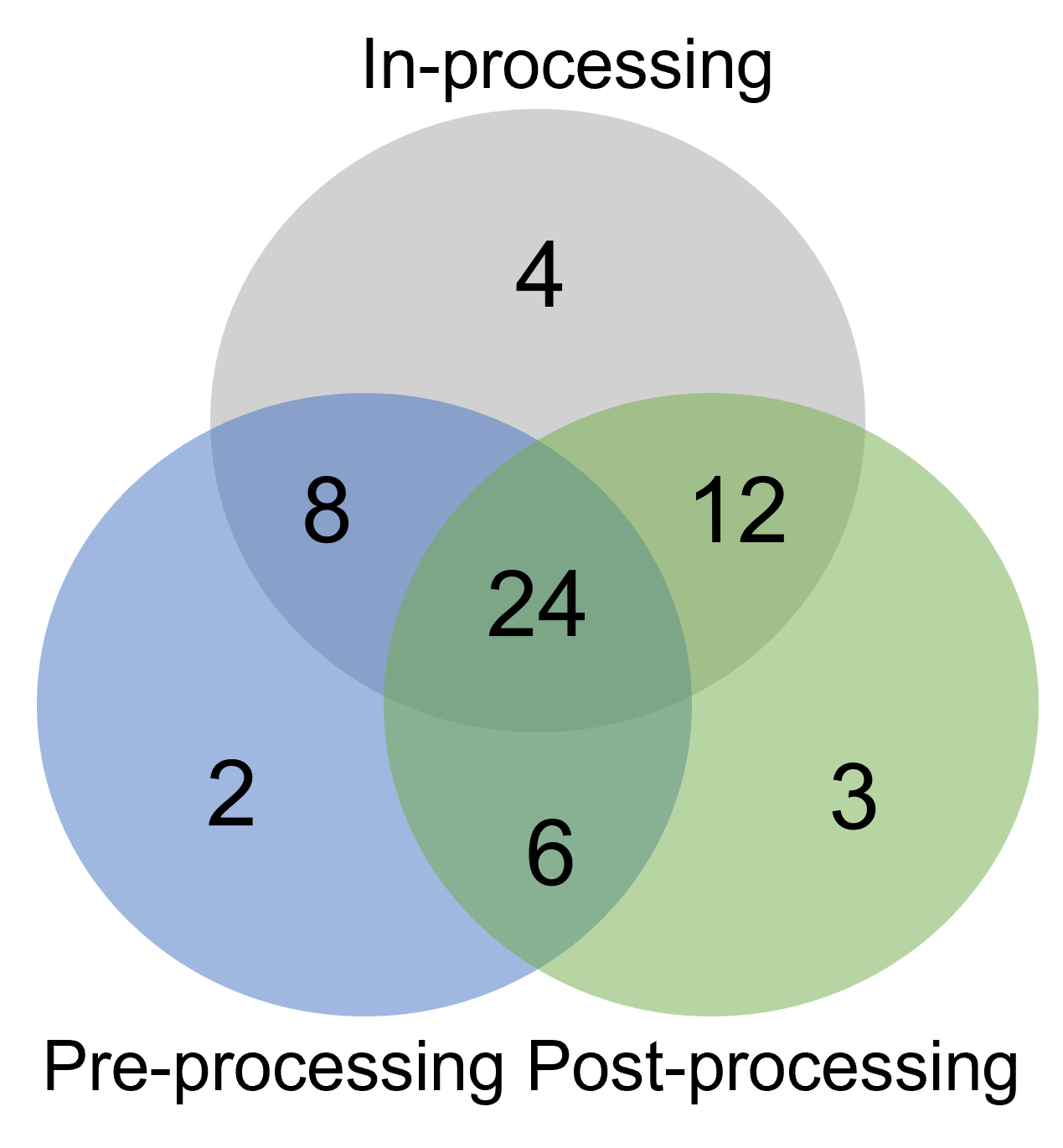}
    \caption{Distribution of fairness enhancing interventions considered in this work. This includes 9 individual interventions and 50 different combinations of interventions.}  
    \label{fig:distribution}
\end{figure}

\subsection{Measuring Fairness}
Quantifying fairness in Algorithms is an active research area. Numerous fairness metrics have been proposed in the literature which mathematically encode different facets of fairness like Group fairness, Individual fairness, Counterfactual fairness, etc. \cite{dwork2012fairness,dwork2018group,aif360, bechavod2017penalizing, kate,kusner2017counterfactual,arvindTalk}. For eg., Group fairness implies that members of one group should receive a similar proportion of positive/negative outcomes as other groups \cite{bechavod2017penalizing, kate}, Individual fairness implies that similar individuals should be treated similarly \cite{dwork2012fairness, dwork2018group}, etc. Another way to classify fairness metrics can be on the level they operate on. For eg., dataset based metrics are solely calculated on the basis of the dataset and are independent of the classifier. On the other hand, classifier based metrics are dependent on the predictions of the classifier like the False negative rate difference. 
In this work, we have opted for a diverse set of fairness metrics to paint a more comprehensive picture. Here, we have not used any dataset based metrics due to their inability to capture the impact of in-processing and post-processing interventions.

The efficacy of different fairness enhancing interventions is typically measured using different fairness metrics. However, these metrics do not present the impact on different population groups. 
For example, the impact on different groups is irrelevant for fairness metrics following the notion of individual fairness. This even holds true for multiple fairness metrics based on the notion of group fairness. Such metrics focus on measuring the disparity between groups without much regard for the impact on specific groups. For example, the fairness metric, False Negative Rate difference, reports the difference in false negativity rate between groups. 
An increase/decrease in this metric does not tell us anything about the specific impact on the privileged or unprivileged groups. In this work, we analyze how interventions impact different fairness metrics and population groups.   



\subsection{Fairness across ML Pipeline}
Research that focuses on fairness in a multi-stage ML system has received some attention and is still in its early stages \cite{martin2021engineering, biswas2021fair, wang2021practical, wu2021fair}. Biswas et al. studied the impact of data preprocessing techniques like standardization, feature selection, etc. on the overall fairness of the ML pipeline \cite{biswas2021fair}. They found certain data transformations like sampling to enhance bias. Hirzel et al. also focus on the data preprocessing stage \cite{martin2021engineering}. They present a novel technique to split datasets into train/test that is geared towards fairness. Wang et al. focused on fairness in the context of multi-component recommender systems \cite{wang2021practical}. They found that overall system’s fairness can be enhanced by improving fairness in individual components. 
Our work focuses on how different combinations of interventions at 3 stages of the ML pipeline can be leveraged to enhance fairness across the ML pipeline.

There is a related line of work that discusses fairness in the context of compound decision making processes \cite{bower2017fair, dwork2020individual, emelianov2019price}. In such data systems, there is a sequence of decisions to be made where each decision can be thought of as a classification problem. 
For eg., in a two stage hiring process, candidates are first filtered for the interview stage and the remaining candidates are again filtered to determine who gets hired. This line of work focuses on fairness over multiple tasks and does not pay much attention towards enhancing fairness of an individual task. 
 Here, a single task (classification problem) can be thought of as a ML pipeline. This is where our work comes in. 
 Our work studies the different combinations of interventions that can together enhance fairness for a single decision making process. 


\section{Experiment Setup}
We have used IBM's AIF 360 \cite{bellamy2019ai} open source toolkit to conduct all experiments for this work. More specifically, we leveraged 4 datasets, 9 fairness enhancing interventions and 11 evaluation metrics from this toolkit as shown in Fig. \ref{fig:exp_setup}. To have a more even comparison, we have used the same ML model i.e., logistic regression (linear model) across the board. Moreover, we only selected those in-processing interventions that are based on or compatible with linear models. 

\begin{figure*}
 \centering 
 \includegraphics[width=0.9\columnwidth]{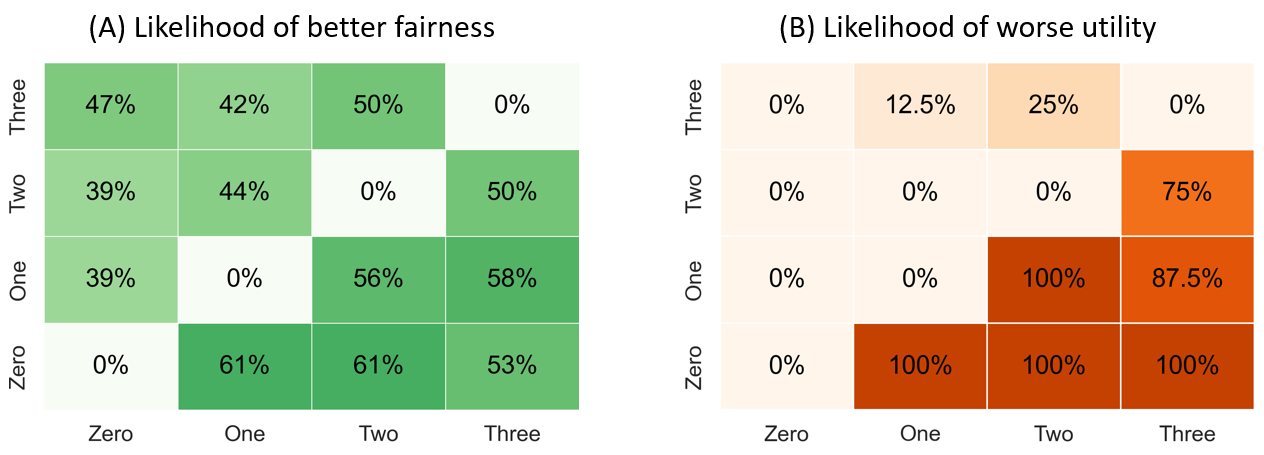}
 \caption{Heatmaps for Fairness and Utility metrics across different numbers of interventions. In Figure (A), a cell (i,j) represents the percentage of cases where j interventions yielded better fairness metrics than i interventions. In Figure (B), a cell (i,j) represents the percentage of cases where j interventions yielded worse utility metrics than i interventions. Here, i represents rows and j represents columns.
 }  
 \label{fig:heatmap}
\end{figure*}

\paragraph{\textbf{Interventions}} Among the 9 interventions, 2 belong to the pre-processing stage, 4 belong to the in-processing stage and 3 belong to the post-processing stage. Apart from these individual interventions, we also execute different combinations of these interventions in groups of 2 and 3. For example, one might choose to intervene at any 2 stages (say a pre-processing intervention followed by a post-processing intervention) or choose to intervene at all 3 stages of the ML pipeline. To form all possible combinations, we cycle through all available options (interventions) for a given ML stage along with a `No Intervention' option and repeat it for all the 3 stages.   
This results in 8 combinations of pre-processing and in-processing interventions, 12 combinations of in-processing and post-processing interventions, 6 combinations of pre-processing and post-processing interventions and 24 combinations of of all 3 types of interventions (see Fig. \ref{fig:distribution}). In totality, we perform 9 individual interventions, 50 different combinations of interventions and a baseline case (No intervention for all stages) for each of the 4 datasets. Here, we have used the default set of hyperparameters for all interventions. In this work, we will refer to the different interventions by their acronyms like PR for Prejudice Remover as defined in \autoref{sec:interventions}. For cascaded interventions, we will concatenate the respective acronyms with a `+' sign. For example, OP + PR means that we performed the Optimized Preprocessing (OP) intervention followed by the Prejudice Remover (PR) intervention. The baseline case is referred as `Logistic Regression'.

\paragraph{\textbf{Evaluation Metrics}} The impact of the different interventions is captured using a diverse set of 11 evaluation metrics. Two of them, namely Accuracy and F1 score, are utility metrics that measure the ability of a ML model to learn the underlying patterns from the training dataset. Here, we have included the F1 score as it can better deal with imbalanced output class distributions. Both of these metrics range between 0 and 1. Higher values mean better performance. The remaining 9 metrics each capture some facet of fairness. Two of the fairness metrics, namely Consistency and Theil index, subscribe to the notion of individual fairness. Higher values for Consistency and lower values for the Theil index mean more fairness. All other fairness metrics subscribe to the notion of group fairness, namely False Positive Rate Difference (FPR Diff), False Negative Rate Difference (FNR Diff), Statistical Parity Difference (SPD), False Discovery Rate Difference (FDR Diff), False Omission Rate Difference (FOR Diff), Accuracy Difference (Accuracy Diff) and F1 Score Difference (F1 Score Diff). All group fairness metrics measure disparity between groups based on some measure such as False Positive Rate (FPR). A lower absolute value for the group fairness metrics means more fairness. The sign of these metrics represents the group that is getting the upper/lower hand. A value of 0 means perfect fairness.

\paragraph{\textbf{Datasets}} Each of the 4 tabular datasets used in this work have been used extensively in the fairness literature. They deal with a binary classification problem and typically contain one or more binary protected attributes such as gender, race, etc. For each of these datasets, we have used the default preprocessing procedure as provided by the AIF360 package (not to be confused with preprocessing interventions). It should be noted that the default preprocessing often involves one hot encoding to deal with categorical variables; this inflates the number of columns compared to the original dataset. We describe the datasets briefly as follows:

\paragraph{Adult Income Dataset} After pre-processing, this dataset consists of 45,222 rows and 99 columns that are derived from the 1994 Census database. Each row represents a person characterized by 
variables like education, gender, race, workclass, etc. These attributes are used to predict if an individual makes more than \$50k a year. Here, we have used gender as the sensitive attribute with males as the privileged group and females as the unprivileged group.  

\paragraph{German Credit Dataset} After pre-processing, this dataset consists of 1,000 rows and 59 columns which was originally prepared by Prof. Hofmann. The task is to predict if an individual has good or bad credit risk based on features like credit amount, credit history, personal status, sex, etc. Here, the sensitive attribute is age. Individuals older than 25 years belong to the privileged group and vice versa. 

\paragraph{COMPAS Recidivism Dataset} After pre-processing, this dataset contains 6,167 rows and 402 columns which pertains to the COMPAS algorithm used for scoring defendants in Broward County, Florida. The task is to predict if an individual will recommit a crime within a two year period based on personal attributes like charge degree, prior count, etc. Here, the sensitive attribute is race with Caucasians as the privileged group and non-Caucasians as the unprivileged group. 

\paragraph{Bank Marketing Dataset} After pre-processing, this dataset consists of 30,488 rows and 53 columns; it pertain to a direct marketing campaign of a Portuguese banking institution. The classification task is to predict if a client will buy a term deposit based on features like type of job, marital status, education, etc. Here, the sensitive attribute is age. Individuals (clients) younger than 25 years belong to the unprivileged group and vice versa. 

\paragraph{\normalfont {For each of these datasets, the positive outcome label refers to the favorable outcome for the recipient. For example, the positive outcome label for the adult income dataset refers to an income greater than \$50k. Similarly, for the COMPAS dataset, positive outcome label refers to \textit{not} recommitting a crime in 2 years. This information will help interpret measures such as false positive rate, false negative rate, etc}}

\begin{figure*}[t]
 \centering 
 \includegraphics[width=0.9\columnwidth]{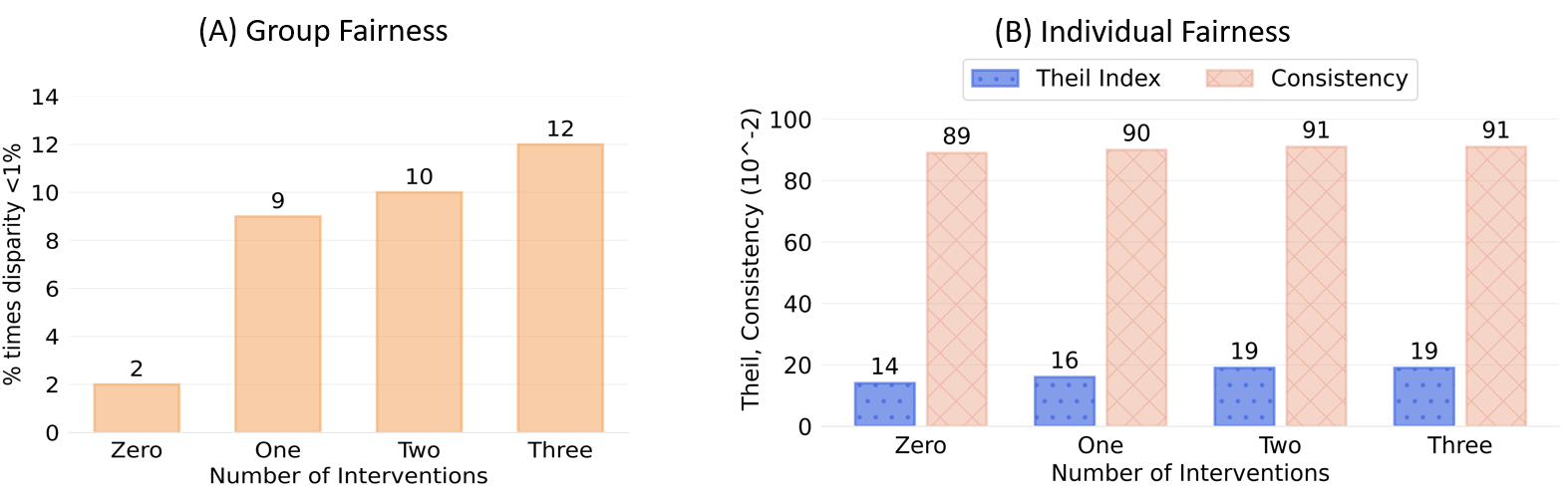}
 \caption{Effect of individual and cascaded interventions on fairness metrics (A) Percentage of times the disparity between the privileged and unprivileged groups was less than 1\% across all group fairness metrics. A higher value means more group fairness. (B) Mean values for Theil index and Consistency across different numbers of interventions. Lower values of the Theil index and higher values for Consistency means more individual fairness.}  
 \label{fig:group_indi_inter}
\end{figure*}

\begin{figure*}
 \centering 
 \includegraphics[width=0.95\columnwidth]{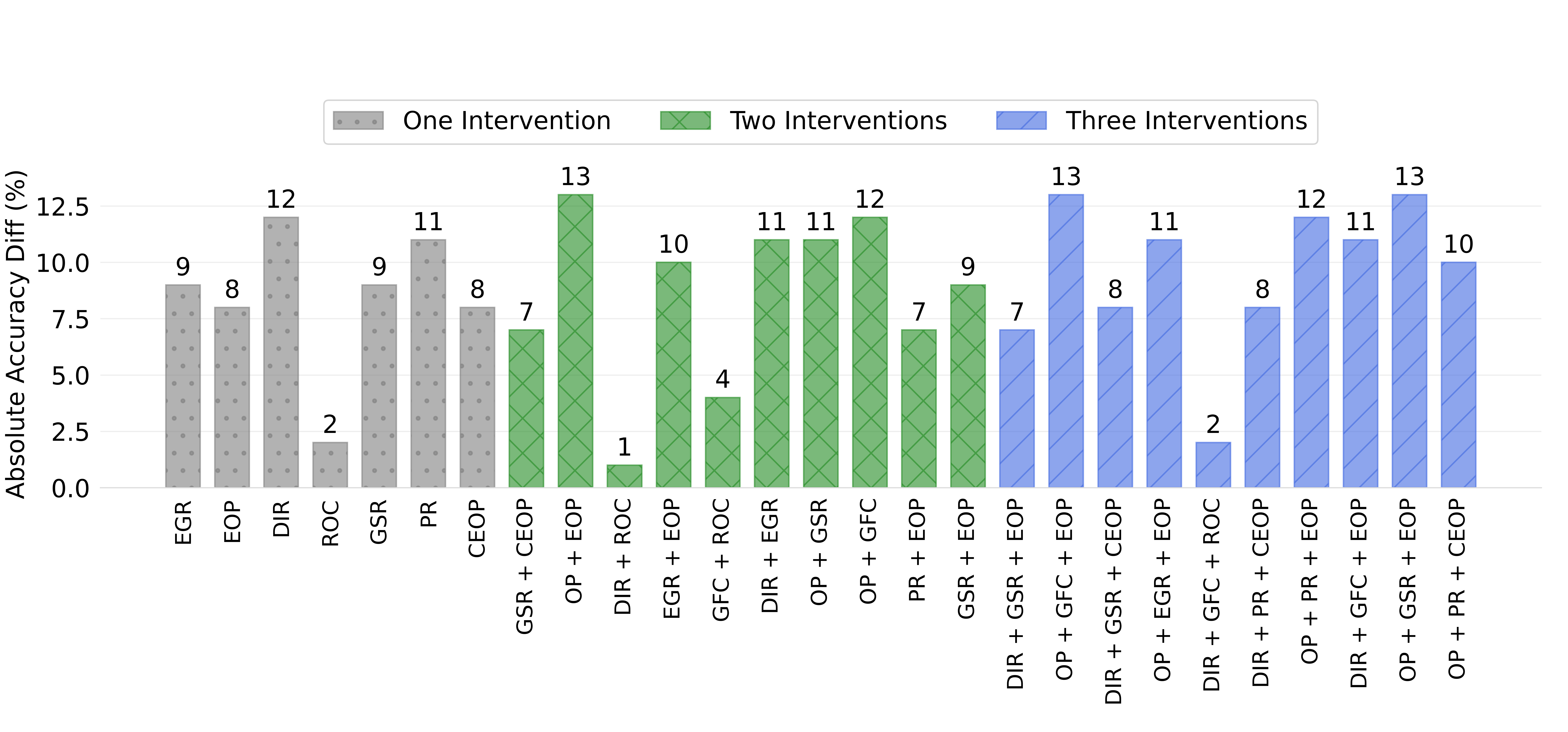}
  \setlength{\abovecaptionskip}{-1pt}
 \caption{Absolute values for the Accuracy difference metric across different interventions for the Adult Income dataset. Here, lower values are desirable. This plot shows multiple cases where values corresponding to more number of interventions are larger than lower number of interventions. Hence, more interventions does not always lead to more fairness.}  
 \label{fig:counter_exp}
\end{figure*}

\paragraph{\textbf{Method}} After default pre-processing, we standardize different features of the dataset so that all non-protected features have the same mean and standard deviation. Thereafter, each dataset is randomly divided into train and test dataset in the ratio 70:30. For the baseline case, we train a logistic regression model on the training dataset and then compute different evaluation metrics using the test data. Next, we execute all individual and cascaded interventions using the train dataset and record their impact on different utility and fairness metrics using the test dataset. Apart from these metrics, we also record statistics like false negative rate (FNR), base rate, etc. for the privileged and unprivileged groups.    
This entire process is repeated 3 times for each dataset with different random splits between train and test dataset to counter sampling bias. Lastly, we compute the mean values of all evaluation metrics across the 3 iterations for each intervention. For each dataset, these results can be represented in tabular format with 60 rows and 11 columns where each row represents an intervention and each column represents an evaluation metric. 



\section{Results}
In this section, we analyze the empirical data from our experiments to understand the possible effect of different cascaded interventions on fairness, utility and population groups. 

\subsection{Effect of Cascaded Interventions on Fairness Metrics}
\label{sec:res_fair}

We first gauge the impact of cascaded interventions on fairness as a whole (across fairness metrics). We start by grouping all interventions into 4 buckets i.e., 0 intervention, 1 intervention, 2 interventions and 3 interventions, respectively. For each bucket, we compute the average score for different fairness metrics and repeat this process for all datasets. It is important to note that different fairness metrics are not directly comparable as they are based on different interpretations of fairness and also vary in terms of their numerical distribution (range, mean, standard deviation, etc.). So, we will compare the mean value of a fairness metric with its counterpart for a different bucket. 
We count the percentage of times one bucket performs better than another across fairness metrics and datasets. This data is visualized using a heatmap of size 4 x 4 in \autoref{fig:heatmap}(A). Each row and column represents a bucket (number of interventions). Here, a cell (i,j) represents the percentage of times j interventions performs better than i interventions. For example, the cell (2,1) is labeled 44\%. It means that a single intervention yielded better fairness scores than two interventions for 44\% of cases. A bucket j will be considered favorable over another bucket i if the value for the cell (i,j) is greater than 50\% and vice versa. 

\begin{figure*}
    \centering
    \begin{minipage}{.33\textwidth}
        \centering
        \includegraphics[width=\textwidth]{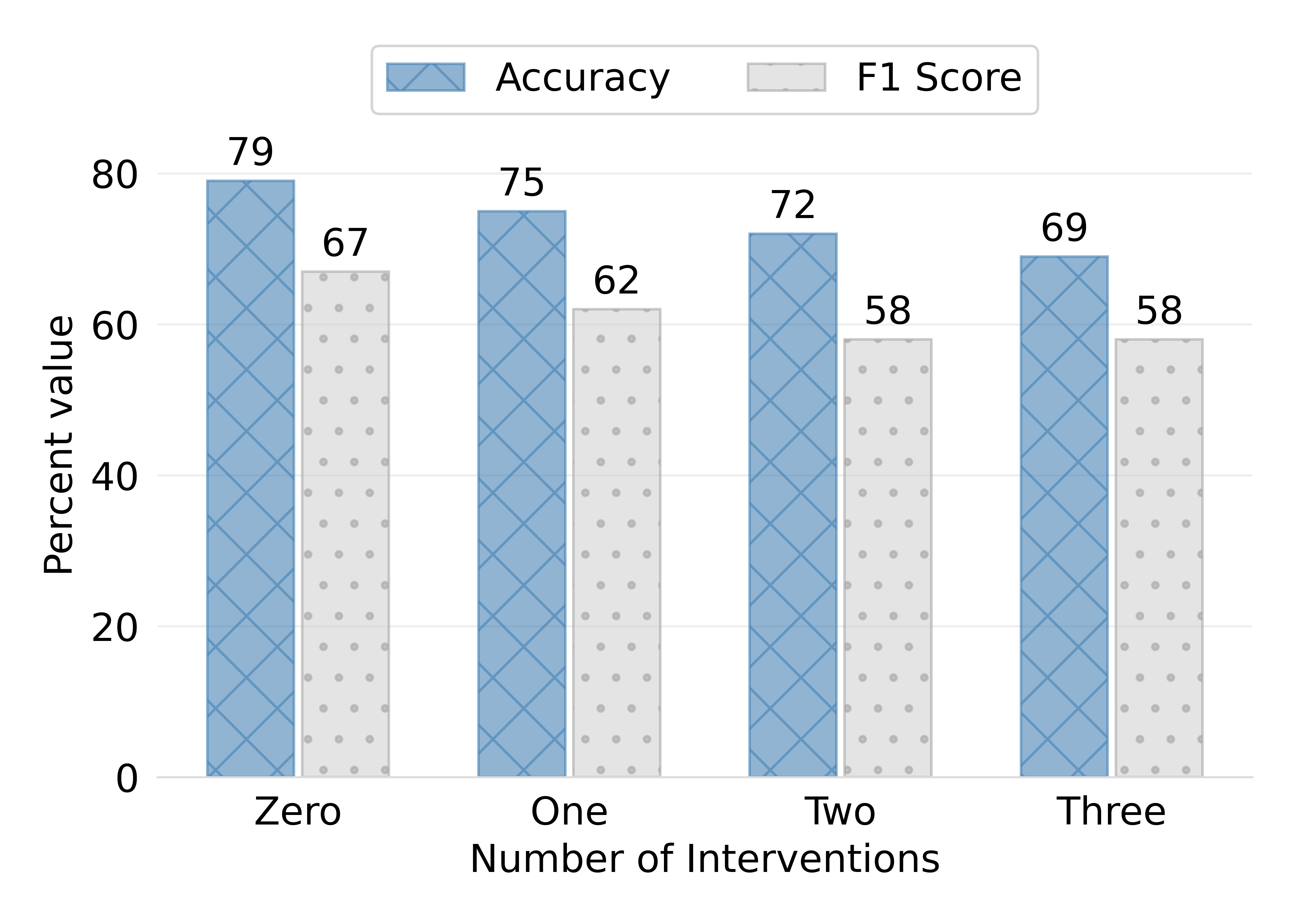}
        \caption{Mean Accuracy and F1 Score for different number of interventions across all datasets. 
        }  
        \label{fig:utility}
    \end{minipage}%
    \hspace{0.01\textwidth}
    \begin{minipage}{0.65\textwidth}
        \centering
        \includegraphics[width=\textwidth]{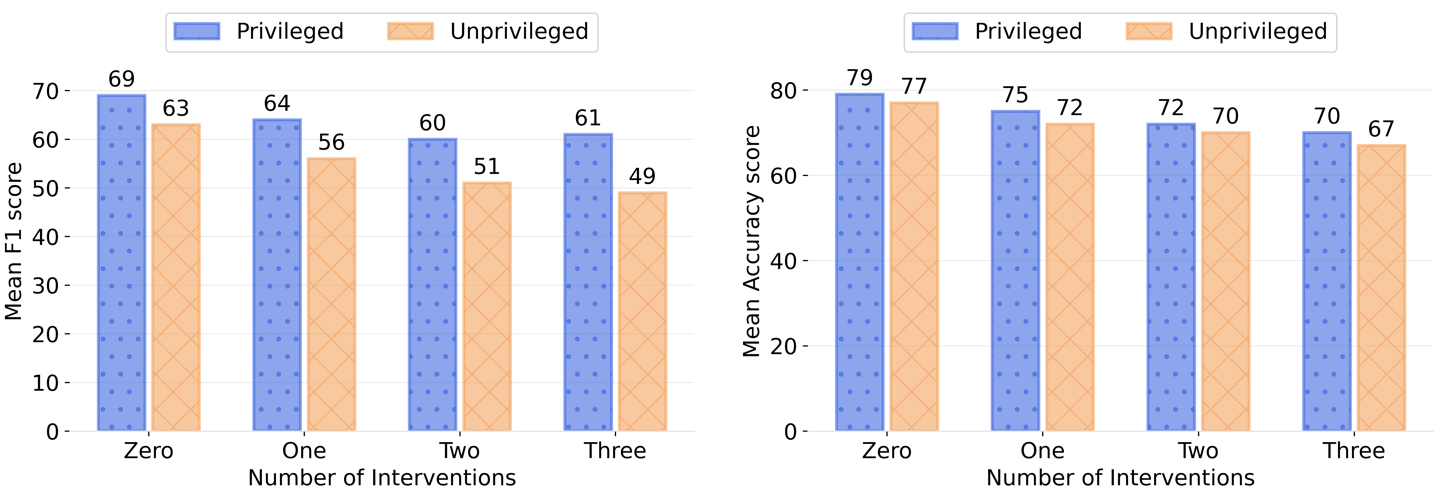}
        \caption{Mean F1 Score (Left) and Accuracy (Right) for different number of interventions. We observe that both metrics decrease for the privileged and the unprivileged groups with more number of interventions.}
        \label{fig:f1_score}
    \end{minipage}
\vspace{-1em}
\end{figure*}

It should be noted that different fairness metrics might be incompatible with each other. So, the net trend (cell values) might appear a bit faded as some fairness metrics might cancel the effect of another. Looking at the row i=0, we find that any number of interventions greater than 0 provide better overall fairness than having no interventions. Looking at the row i=1, we find that the columns j=2 and j=3 have values more than 50\% i.e., two or three interventions yielded better fairness than a single intervention. Moving to the row i=2, we find that the value of the cell (2,3) is 50\%. Perhaps surprisingly, this means that it is equally likely for either buckets to outperform each other. Overall, it appears that fairness improves from 0 to 2 interventions and becomes constant thereafter. However, it is important to note that the heatmap encodes frequency and not the magnitude of difference between fairness metrics. So, it is possible that three interventions reduce bias significantly more (in terms of magnitude and not the count of fairness metrics) than the two interventions case and might still appear to be no better than the two interventions case. 


The heatmap provides an aggregate picture of how fairness metrics vary for different numbers of interventions. Now, let us dig a bit deeper and gauge the impact of cascaded interventions on group fairness and individual fairness. For individual fairness, we plot the mean values for the Theil index and Consistency for different numbers of interventions. For the group fairness metrics, we compute the percentage of times the absolute value of each constituting metric is less than 0.01. As we can see from \autoref{fig:group_indi_inter} (A), the percent of times the group fairness metrics are below a threshold increases steadily with higher numbers of interventions from 2\% to 12\%. In other words, \textit{group fairness improves with more interventions on aggregate.} This observation largely concurs with our findings from \autoref{fig:heatmap}(A). On the other hand, we get mixed signals from the individual fairness metrics. The Consistency metric shows a slight improvement in fairness while the Theil index shows a downfall in fairness with higher number of interventions.

\begin{table*}[]
\caption{Spearman correlation coefficient between F1 score and fairness metrics for different number of interventions (represented as rows). We observe that the correlation coefficient decreases as the number of interventions increase across metrics. }
\resizebox{\columnwidth}{!}{%
\begin{tabular}{rrrrrrrrrr}
\hline
 & FPR Diff & FNR Diff & Accuracy Diff & FOR Diff & FDR Diff & SPD    & F1 Score Diff & Theil Index & Consistency \\
\hline
0                & 0.748    & 0.42     & 0.385         & 0.42     & 0.329    & 0.678  & 0.58          & 0.708       & -0.986      \\
1                & 0.343    & -0.054   & 0.211         & 0.354    & 0.176    & 0.263  & 0.253         & 0.148       & -0.628      \\
2                & 0.228    & -0.11    & 0.15          & 0.24     & 0.148    & 0.109  & 0.115         & -0.144      & -0.481      \\
3                & 0.119    & -0.185   & 0.189         & 0.058    & -0.112   & -0.056 & 0.073         & -0.282      & -0.167 \\ \hline    
\end{tabular}}
\label{table:corr}
\end{table*}

It is important to note that all of these patterns reflect the aggregate trend and may not apply for all cases. For example, \autoref{fig:counter_exp} shows the absolute values for the Accuracy difference metric across different interventions for the Adult Income dataset. In this case, we observe multiple instances where a larger number of interventions did not lead to more fairness. On the other hand, we observed multiple instances where lower number of interventions performed better than higher numbers of interventions. This observation is contrary to the aggregate trend for group fairness that we observed in \autoref{fig:group_indi_inter}(A). So, \textit{it is not always the case that more interventions will result in more fairness. One needs to choose the right combination of interventions to get the best results.} We will discuss which combinations work for different metrics in \autoref{sec:table}.  

\subsection{Effect of Cascaded Interventions on Utility Metrics}

We start off by analyzing how different number of interventions compare against each other on utility metrics as a whole. Following a similar procedure as defined in \autoref{sec:res_fair}, we plot a heatmap for utility metrics instead of fairness metrics (see \autoref{fig:heatmap} (B)). Here, a cell (i,j) represents the percentage of cases where j interventions yielded lower utility than i interventions. As expected, we observe that any non-zero number of interventions results in lower utility than the baseline case (see row i=0). Similarly, we observe that two interventions yields worse utility metrics than one intervention (see cell(1,2)) and three interventions yields worse utility metrics than two interventions (see cell(2,3)). Overall, this reveals a strong downward trend for utility metrics with more number of interventions. Looking at \autoref{fig:heatmap} (A) and (B) in conjunction, we observe that three interventions perform on par with two interventions on fairness. However, 75\% of the times three interventions performed worse on utility metrics than two interventions. This observation hints that one should typically opt for two interventions and go for the third intervention only in specific contexts. 

\begin{figure*}
 \centering 
 \includegraphics[width=\columnwidth]{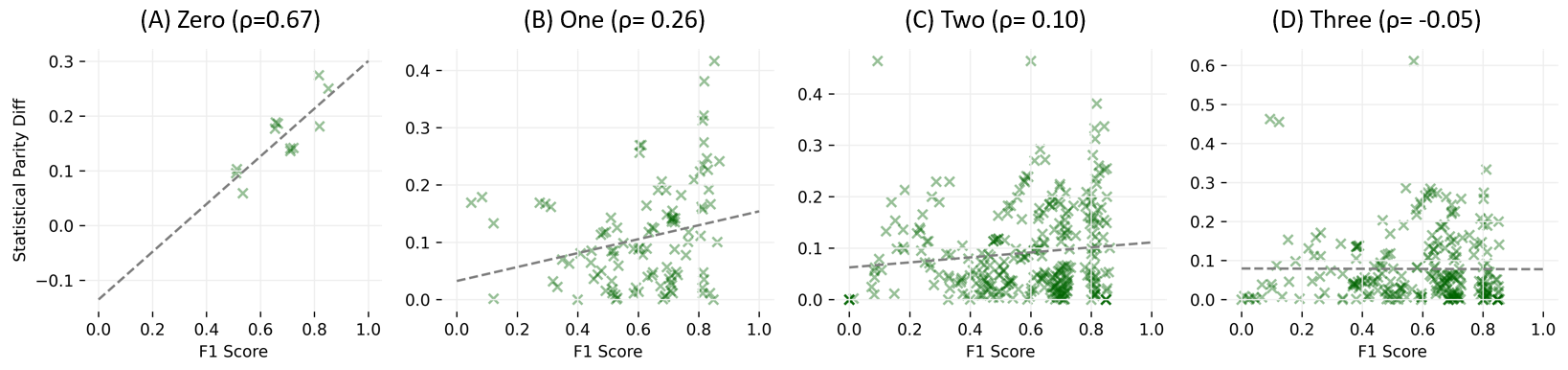}
 \caption{Relation between Statistical Parity difference and F1 score for different numbers of interventions from 0 to 3 (A - D). Each green `x' marker corresponds to a specific intervention executed on one of the 3 random subsets of a given dataset. The grey line represents the regression line that best fits all the points. It visually indicates the strength of the correlation.  }  
 \label{fig:corr}
\end{figure*}

To quantitatively understand the effect on specific utility metrics, we analyzed how Accuracy and F1 score vary across different number of interventions. 
We computed the mean accuracy and F1 score for different number of interventions across datasets.
These mean scores are visualized in \autoref{fig:utility}. In line with our findings in \autoref{fig:heatmap}(B), we observe that both accuracy and F1 score steadily decrease as the number of interventions increase. This downward trend is more pronounced in the beginning than the end. For eg., the mean F1 score drops by 5\% going from no intervention to one intervention and later stabilizes going from two interventions to three interventions. Overall, this trend shows that there is a cost to be paid for adding more interventions. So, \textit{one should not blindly opt for more interventions.} ML practitioners should consider the potential loss in utility metrics while designing fair ML pipelines.

We have looked at the effect of cascaded interventions on utility metrics and fairness metrics in isolation. Now, let us investigate the effect of cascaded interventions on the bivariate relationship between utility metrics and fairness metrics. We start off by grouping all experimental data across datasets by the number of interventions. For each group, we compute the spearman correlation coefficient between different fairness metrics and F1 across as shown in \autoref{table:corr}. We observe that there is a significant positive correlation between fairness metrics and F1 score for the baseline case. For all fairness metrics except for the consistency metric, a higher value means more bias (less fairness). Hence, a positive correlation suggests that F1 score and fairness are negatively linked. In other words, interventions with high F1 score generally result in poor fairness and vice versa. This observation is in line with existing literature which discusses a tradeoff between accuracy and fairness for individual interventions \cite{menon2018cost}. As the number of interventions increases, we observe a steady decline in the correlation coefficient across fairness metrics. Here, the correlation coefficient for consistency moves in the opposite direction as unlike all other fairness metrics higher values means more fairness. \textit{The decrease in correlation suggests that the likelihood of attaining a high F1 score along with high fairness increases with higher numbers of interventions}. As an example, we plot the bivariate relation between Statistical parity difference and F1 score for different numbers of intervention (see \autoref{fig:corr}). We observe that three interventions are able to achieve high F1 score and low bias scores more consistently than one or two interventions. Here, the decrease in correlation coefficient $\rho$ is evident from the decrease in the slope of the regression line as we move towards higher numbers of interventions. If the reduction in F1 score caused by different interventions was in proportion to the corresponding increase in fairness, the correlation coefficient would have remained constant across different number of interventions. Hence, the decrease in correlation coefficient shows the efficacy of cascaded interventions in reducing bias without sacrificing too much on performance (F1 Score). 


\begin{figure*}
    \centering
    \begin{minipage}{.59\textwidth}
        \centering
        \includegraphics[width=\columnwidth]{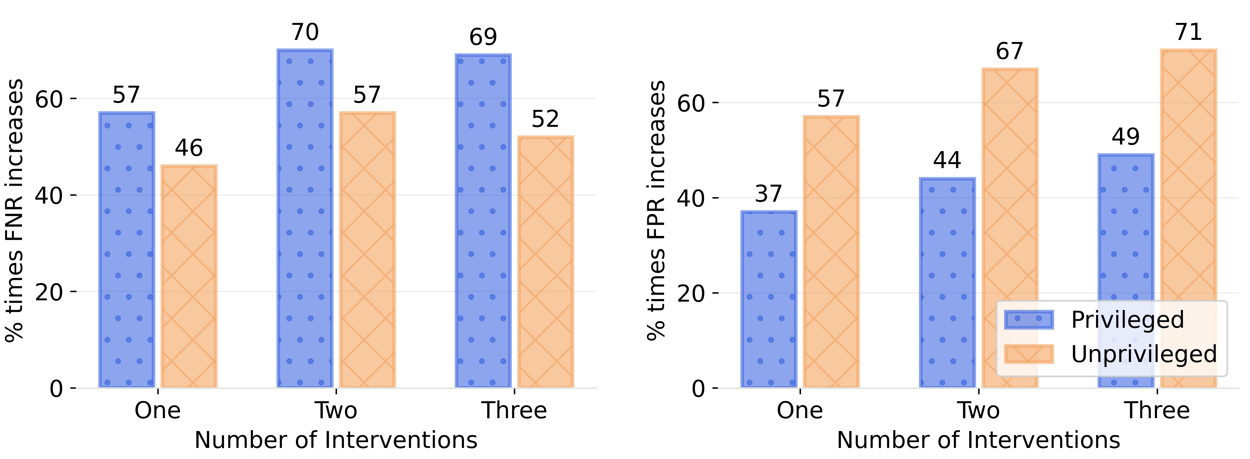}
        \caption{Percentage of times False Negative Rate (Left) and False Positive Rate (Right) increases compared to the baseline (No Intervention) across different number of interventions for all datasets.}  
        \label{fig:group_metrics}
    \end{minipage}%
    \hspace{0.01\textwidth}
    \begin{minipage}{0.36\textwidth}
        \centering
        \includegraphics[width=\columnwidth]{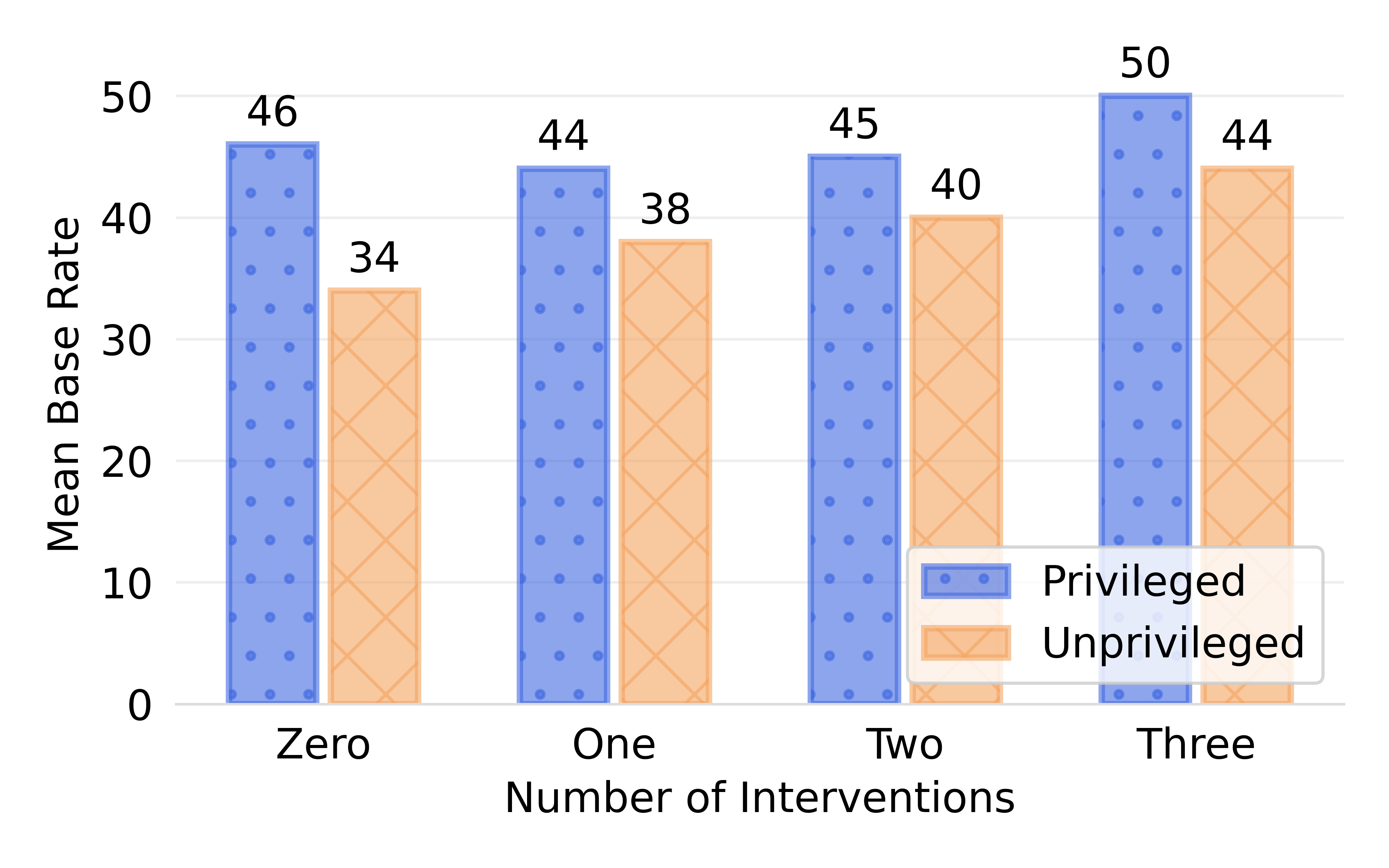}
        \caption{Mean base rate for the privileged and unprivileged groups across different numbers of interventions. 
        }
        \label{fig:base_rate}
    \end{minipage}
\end{figure*}

\subsection{Impact on Population groups}

 In our experimental setup, we  kept a log of different statistics like false positive rate, false negative rate, F1 score, base rate, 
 etc. for the privileged and unprivileged groups across all interventions. We analyzed this data to understand the impact of different interventions on these two groups.
 \autoref{fig:f1_score} shows the aggregate impact of different number of interventions on Accuracy and F1 Score. In line with our earlier finding (see \autoref{fig:utility}), we observe that these utility metrics deteriorate for both groups with more number of interventions. However, the impact on the underprivileged group is more severe than the privileged group for the F1 metric. The F1 score for the privileged group dropped 8 percent points from 69\% to 61\% while it dropped 14 percent points for the underprivileged group from 63\% to 49\%. This disproportionate impact also lead to an increase in disparity between the groups in terms of F1 Score from 6\% to 12\%. 
 In the case of accuracy, the impact on both groups is roughly even and the disparity between groups remains almost constant ($\scriptstyle\mathtt{\sim}$3\%) across different number of interventions.

 The decrease in utility metrics signal an increase in error rates. So, let us look at the impact on the False positive rate (FPR) and False negative rate (FNR).   
 \autoref{fig:group_metrics} shows the percentage of times we observed an increase in FNR and FPR compared to the baseline for different number of interventions. As we can see, there is a large percentage of cases where individual interventions resulted in higher error rates for both the privileged and the unprivileged group than we started out with. As we go for higher numbers of interventions, the percentage of such cases generally increases further. This trend is in agreement with the decreasing trend in utility metrics for more number of interventions. On comparing between groups, we find that interventions are more likely to result in higher FNR for the privileged group than the unprivileged one. It means that individuals from the privileged group are more likely to be misclassified with the unfavorable outcome than the unprivileged group. 
 This trend flips for FPR where the unprivileged group are more likely to have a higher FPR. In other words, individuals from the unprivileged group are more likely to be misclassified with the favorable outcome than the privileged group. Both of these trends persist for different number of interventions and generally deepens with more number of interventions. 
 Looking at these patterns in conjunction with \autoref{fig:f1_score}, it appears that the loss in Accuracy/F1 score can atleast be partially explained by the tendency of the interventions to assign more positive outcomes to the unprivileged group and negative outcomes to the privileged group.  
 
 Next, let us look at the impact on base rate for different groups (see \autoref{fig:base_rate}). 
 Here, base rate is defined as the proportion of positive outcomes for different groups. It is computed over model's prediction for the test data post all relevant interventions. For the no intervention case, we observe a 12\% disparity in favor of the privileged group. With more interventions, the base rate for the unprivileged group steadily increases from 34\% to 44\% (10 \% jump). On the other hand, the base rate for the privileged group decreases a bit for the one and two interventions case and increases by 4\% for the three interventions case. 
 Overall, this leads to a decrease in disparity between groups from 12\% to 5\% for the two intervention case. In a context where equality between base rates is a priority, two interventions seems to be the way to go. It is also important to note the some of the interventions can negatively impact the privileged group. This is evident from the drop in base rate for the privileged group for the one and two interventions case. 
 If we look at base rate over the entire population, we find that the base rate undergoes a modest increment for the one (1\%) and two interventions case (2\%). However, it increases significantly for the three interventions case (7\%). So, ML practitioners should exercise caution while adding the third intervention, especially for contexts where the number of favorable outcomes is fixed such as hiring.  

  \begin{figure*}
 \centering 
 \includegraphics[width=0.9\columnwidth]{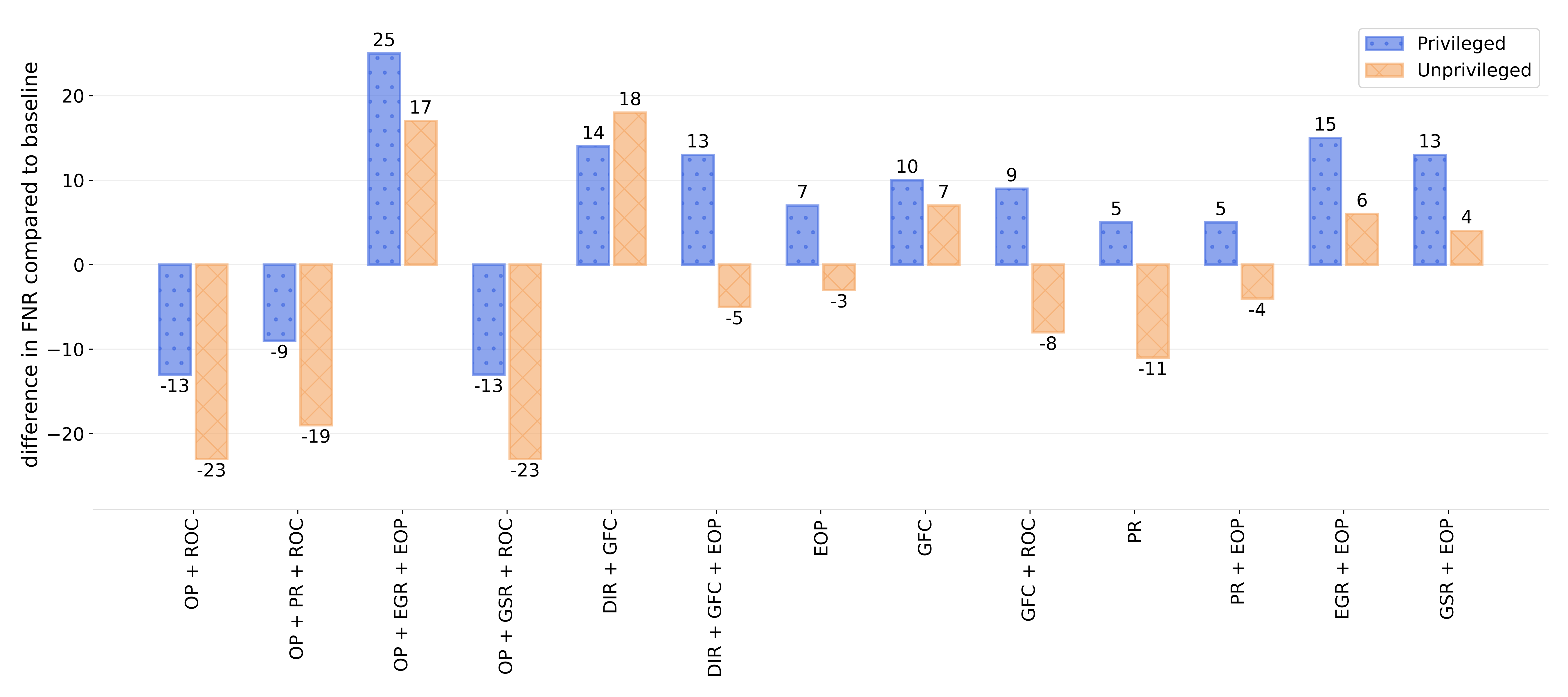}
 \caption{Difference in False Negative Rate compared to the baseline (No Intervention) across different Interventions for the Adult Income dataset. Here, Negative values are desirable and vice versa.}  
 \label{fig:fnr_adult}
\end{figure*}

Fairness metrics provide elegant mathematical representations for different notions of fairness but might obscure the specific effect on different population groups. \autoref{fig:group_indi_inter} shows that group fairness metrics decrease with more interventions on aggregate. However, we do not observe a similar reduction in error rates for individual groups as shown in \autoref{fig:group_metrics}. To investigate this further, let us look at a specific example. \autoref{fig:fnr_adult} shows the difference in false negative rates (FNR) compared to the baseline for the privileged (males) and the unprivileged (females) groups across different interventions. Here, we have only considered interventions that have reduced the magnitude of the FNR difference metric between males and females compared to the baseline for the Adult Income dataset. So, as per the FNR Diff metric, all of these interventions are effective at reducing bias. However, we observe that for many of these interventions FNR has actually increased for one or both groups. In \autoref{fig:fnr_adult}, we observe 10 cases (interventions) where FNR has increased for either or both groups. So, the reduction in the FNR Diff metric is due to the uneven increase in FNR for different population groups.
It can be argued whether it is desirable to reduce disparity between groups by increasing error rates unevenly for different groups. \textit{This finding points to a need for new fairness metrics that account for the specific impact on individual groups apart from just the gap between those groups}.


There can be different ways to interpret the empirical findings based on one's value system. One way to interpret these numbers can be from a pure ML perspective where the 
focus is to train a ML model that best fits the underlying dataset.
Here, the objective of an intervention is to ensure that the model performs equally well for different groups in terms of Accuracy, False positive rate, False negatives rate, etc. 
From this viewpoint, many of the interventions are counterproductive as they increase error rates and decrease the Accuracy/F1 score for different groups (see \autoref{fig:f1_score} and \autoref{fig:group_metrics}). Going from individual interventions to cascaded interventions makes things worse as the error rates for population groups increases further and the Accuracy/F1 score further deteriorates. Ideally, one would reduce disparity by reducing the error rates for different groups but not making it worse for any of them. For example, 3 interventions in \autoref{fig:fnr_adult} reduce disparity (FNR Diff metric) by reducing FNR for both groups. 
Similarly, mitigating discrimination against one group at the cost of another is self-defeating and unethical.
For illustration, we find 5 interventions in \autoref{fig:fnr_adult} where FNR decreases for the unprivileged group and increases for the privileged group compared to the baseline. In other words, some high income individuals from the privileged group were misclassified as low income in an effort to increase fairness. This is not a one off case. Our experiments show an aggregate trend across datasets where privileged groups were disproportionately misclassified with unfavorable outcomes and unprivileged groups were disproportionately misclassified with favorable outcomes (see \autoref{fig:group_metrics}). These observations highlight how interventions can negatively impact the privileged group.
Future interventions should be more considerate towards their possible negative impact on the different population groups.

Another way to interpret these numbers can be from the perspective of social justice where interventions should not only reduce disparity in error rates but also serve as an instrument to right historical wrongs. For example, interventions should enforce equality/equity of outcomes i.e., ML models should assign positive outcomes to different groups at the same rate or even prioritize the unprivileged group irrespective of the patterns in the dataset \cite{mehrabi2020statistical}. In the process of achieving these goals, the loss incurred in terms of accuracy and F1 score is secondary as these metrics are computed over output labels (ground truth) polluted with historical biases. Similarly, the disproportionate increase in FNR for the privileged group is incidental/imperative to serve the larger goal of equal representation. Under this viewpoint, 
cascaded interventions are desirable as they help bridge the gap in base rates.



\subsection{Comparison between different Interventions}
\label{sec:table}
So far, we have largely focused on the aggregate trends which might or might not apply for a given intervention. In this section, we focus on specific interventions and compare how they perform against different utility and fairness metrics. Such knowledge can assist practitioners and researchers in choosing interventions based on their specific context. We begin by computing the mean scores for all evaluation metrics across 4 datasets corresponding to each intervention.
This provided a mean score for each evaluation metric across 60 different interventions. For each evaluation metric, we ranked all interventions from the best performing to the worst performing. For the Accuracy, F1 score and Consistency metric, higher values are desirable so we sorted interventions based on descending order of their corresponding values. For all other metrics, we used ascending order. It is important to note that we used absolute values for all group fairness metrics as we are primarily concerned with the magnitude of bias.

\begin{table*}[]
\caption{Ranking of 10 best and worst performing interventions for different evaluation metrics. Here, we have ranked interventions in ascending order of their corresponding absolute values for all metrics except for Accuracy, F1 Score and Consistency that are sorted in descending order. This ordering schema ensures that desirable interventions are ranked higher for all metrics. Interventions with `+' sign represents combinations of multiple interventions. Here, `Logistic Regression' represents the baseline case i.e., no internvetions for all stages.  }

\resizebox{\columnwidth}{!}{%
\begin{tabular}{llllll}
\hline
Rank & Accuracy            & F1 Score            & Theil Index         & Consistency     & FPR Diff            \\
\hline
1    & Logistic Regression & DIR + EGR + ROC     & DIR + EGR + ROC     & OP + GFC + ROC  & EGR + EOP           \\
2    & DIR                 & DIR                 & DIR                 & OP + GFC + CEOP & OP + GFC + ROC      \\
3    & GSR                 & DIR + GSR           & DIR + EGR           & OP + GFC        & PR + EOP            \\
4    & DIR + EGR + ROC     & DIR + EGR           & DIR + GSR           & OP + GSR        & OP + EGR + EOP      \\
5    & DIR + EGR           & Logistic Regression & DIR + PR + EOP      & OP              & OP + GFC + EOP      \\
6    & DIR + GSR           & OP + CEOP           & DIR + ROC           & OP + EGR + ROC  & OP + GFC            \\
7    & EGR                 & OP + GSR + CEOP     & DIR + CEOP          & OP + PR         & DIR + GFC           \\
8    & DIR + PR + CEOP     & OP                  & Logistic Regression & OP + GSR + CEOP & GFC + EOP           \\
9    & CEOP                & OP + GSR            & DIR + PR + ROC      & OP + CEOP       & DIR + EOP           \\
10   & PR + CEOP           & GSR                 & DIR + GSR + CEOP    & OP + PR + CEOP  & DIR + GSR + EOP     \\
&&&&& \\
51   & OP + GFC + CEOP     & OP + PR             & OP + PR + ROC       & PR + EOP        & Logistic Regression \\
52   & OP + PR + ROC       & OP + GFC + EOP      & OP + PR             & DIR + PR + ROC  & GSR + CEOP          \\
53   & DIR + GFC + EOP     & DIR + GFC + EOP     & OP + ROC            & DIR + EGR + EOP & OP                  \\
54   & DIR + GFC           & GFC + EOP           & OP + GFC            & DIR + EOP       & PR + CEOP           \\
55   & OP + ROC            & OP + GFC            & DIR + GFC           & DIR + GFC + EOP & GFC + CEOP          \\
56   & OP + GFC + EOP      & DIR + GFC           & GFC + CEOP          & EOP             & DIR + GSR + CEOP    \\
57   & OP + GSR + ROC      & GFC + ROC           & OP + GFC + EOP      & OP + GSR + EOP  & DIR + PR + CEOP     \\
58   & DIR + PR + ROC      & DIR + GFC + CEOP    & GFC + ROC           & OP + EOP        & OP + CEOP           \\
59   & DIR + ROC           & GFC + CEOP          & OP + GSR + ROC      & OP + PR + EOP   & DIR + CEOP          \\
60   & OP + PR             & DIR + GFC + ROC     & DIR + GFC + ROC     & OP + GFC + EOP  & CEOP        \\ \hline        \\
\end{tabular}}

\resizebox{\columnwidth}{!}{%
\begin{tabular}{llllll}
\\  \hline
Rank & FNR Diff         & Accuracy Diff   & FOR Diff         & FDR Diff            & SPD \\
\hline  
1    & OP + GFC + EOP   & DIR + GFC + ROC & OP + PR          & Logistic Regression & OP + GFC + EOP          \\
2    & GFC + EOP        & DIR + GFC       & DIR + GFC + CEOP & DIR                 & OP + EGR + ROC          \\
3    & OP + EGR + EOP   & OP + PR + ROC   & CEOP             & GSR + CEOP          & OP + EGR                \\
4    & OP + EOP         & OP + ROC        & PR + CEOP        & DIR + EGR + CEOP    & OP + GSR + EOP          \\
5    & OP + GSR + ROC   & ROC             & DIR + CEOP       & GFC + CEOP          & OP + GFC + ROC          \\
6    & OP + GFC + ROC   & EOP             & PR               & GFC                 & GSR + ROC               \\
7    & OP + PR + EOP    & DIR + GFC + EOP & DIR + PR + CEOP  & DIR + GFC           & DIR + GFC               \\
8    & DIR + GSR + EOP  & OP + GSR + ROC  & OP + GFC + ROC   & DIR + PR            & OP + EOP                \\
9    & GSR + EOP        & GFC + CEOP      & DIR + GSR + CEOP & DIR + EGR           & EGR + ROC               \\
10   & OP + GFC         & PR + ROC        & GFC + CEOP       & PR                  & OP + PR + EOP           \\
&&&&&  \\
51   & GFC + CEOP       & OP + EGR        & GSR + ROC        & OP + GFC + CEOP     & GFC + CEOP              \\
52   & DIR + EGR + CEOP & OP + PR         & GFC + EOP        & DIR + GSR + EOP     & PR                      \\
53   & GSR + CEOP       & OP + PR + CEOP  & OP + ROC         & OP + EGR            & OP + CEOP               \\
54   & DIR + GFC + CEOP & OP + EGR + ROC  & OP + CEOP        & OP + EGR + ROC      & GSR + CEOP              \\
55   & EGR + CEOP       & OP + GSR        & OP + GSR + ROC   & OP + PR             & DIR + PR + CEOP         \\
56   & DIR + GSR + CEOP & OP + GSR + CEOP & OP + PR + ROC    & OP + EOP            & Logistic Regression     \\
57   & DIR + PR + CEOP  & OP + GFC + ROC  & OP + EGR + EOP   & OP + EGR + EOP      & DIR + GSR + CEOP        \\
58   & PR + CEOP        & OP + EGR + CEOP & DIR + GFC + EOP  & OP + PR + EOP       & PR + CEOP               \\
59   & DIR + CEOP       & OP + GFC        & OP + EGR + ROC   & OP + GFC + EOP      & DIR + CEOP              \\
60   & CEOP             & OP + GFC + CEOP & OP + EGR         & OP + GSR + EOP      & CEOP                    \\ 
\hline
\end{tabular}}

\label{table:rq4}
\end{table*}

\autoref{table:rq4} contains the 10 best and worst performing interventions for each evaluation metric. From this table, we can make a few important observations. Logistic regression (by which we mean `No intervention') tops the list for Accuracy. This is in line with existing literature that comments on the accuracy fairness tradeoff \cite{berk2017convex}. As all interventions optimize for some aspect of fairness, they might sacrifice a bit on accuracy. So, one should not use any intervention for achieving the best accuracy.  For the F1 score, we observe that a few interventions rank higher than Logistic Regression, such as DIR + EGR + ROC. However, the difference between them was quite slim (0.3\%) which might be attributed to imbalanced output class distribution. The broader point is that \textit{applying more interventions does not always lead to a significant loss in utility}. In the case of DIR + EGR + ROC, we observe the best performance for the F1 score and the 4th best for accuracy. Among the 10 bottom ranked interventions across all fairness metrics, Logistic Regression occurs only twice. This shows that there are several individual and cascaded interventions that perform worse than the baseline case for at least some fairness metric. Hence, it is important to choose interventions wisely. 
ML practitioners/researchers can leverage resources like \autoref{table:rq4} and prioritize interventions that have worked well for other datasets and hopefully save some time in the process.     

For the fairness metrics, we observe that the best performing intervention is mostly unique for each one of them. In other words, there is no silver bullet for all fairness metrics. 
This observation is in line with the existing literature which proves that no intervention can simultaneously optimize for all fairness metrics \cite{kleinberg2016inherent}. From a practical standpoint, this implies that ML practitioners need to prioritize which metrics are more important to them and then choose interventions accordingly. It is also worth noting that the best performing intervention for any metric is either Logistic Regression (No Intervention) or a combination of two or more interventions. Apart from the top performing interventions, we also observe that the top 10 interventions for all fairness metrics are predominantly cascaded interventions. For example, 9 out of the top 10 interventions for the Consistency metric are cascaded interventions. These observations further motivate the efficacy of cascaded interventions over individual interventions. Among the top 10 interventions across all 10 metrics, OP + GFC + ROC occurs the most number of times (5). Similarly, OP + GFC + EOP occurs the most number of times among the bottom 10 interventions across metrics. It is interesting to see that both of these interventions have much in common (OP and GFC). This shows that certain intervention are more compatible/incompatible with another. Changing an ingredient can drastically impact the outcome. For instance, swapping ROC with EOP resulted in the entire combination (OP + GFC + ROC) to change from being one of the top ranked to one of the worst ranked interventions. 
It should be noted that ranking abstracts the real difference in magnitude. For brevity, we have used ranking in the table. 
We encourage the readers to refer to the source code/experimental data for more details. 

\section{Discussion}
This work explores different research questions in the realm of cascaded debiasing based on comprehensive experiments using multiple benchmark datasets, fairness metrics and interventions. The scope of this work is limited to the binary classification problem for tabular datasets. Future work might conduct similar studies for other data types like text, 
images, etc. and consider other problem types such as regression, clustering, etc. It should be noted that all the insights and analyses presented in this work are based on empirical evidence and so they may or may not generalize to other datasets, interventions or metrics. This study was facilitated by the AIF 360 package that provided easy access to different datasets, interventions and metrics. On the flip side, this package can also be considered as a limiting factor because our choices were limited to the different options it provided. Moreover, we were unable to execute certain interventions like the Optimized Preprocessing intervention for the Bank Marketing dataset due to limited technical support. In the future, researchers might also consider other fairness packages \cite{lee2021landscape} like Fairlearn and possibly include more fairness metrics, datasets, interventions, etc.

On the fairness front, we have considered a respectable set of fairness metrics but there are other popular metrics like counterfactual fairness \cite{kusner2017counterfactual} or more recent metrics like Statistical Equity \cite{mehrabi2020statistical} that can be explored in future studies. Moreover, this work deals with fairness at a group level (say males and females) and at an individual level (through the individual fairness metrics). Future work might study the cumulative effect of fairness enhancing interventions on different subgroups say poor black females, young white males, etc. As far as datasets go, we have conducted experiments on 4 datasets that have been used extensively in the fairness literature to benchmark different fairness enhancing interventions. Recent research efforts have questioned the validity of some these datasets like the COMPAS \cite{bao2021s} and Adult Income dataset \cite{ding2021retiring}. Future work might include more recent datasets \cite{ding2021retiring} and other stages of intervention, such as the data curation stage into the analysis \cite{ghai2020measuring}. Hyperparameters for different interventions can significantly impact the results. In the interest of reducing computational complexity, this work largely uses the default set of hyperparameters. Future work might go for a deeper analysis by optimizing hyperparameters for different interventions \cite{wu2021fair}. 
Lastly, the source code and experimental data has been made publicly available at \href{https://github.com/bhavyaghai/Cascaded-Debiasing}{\textit{github.com/bhavyaghai/Cascaded-Debiasing}} for easy reproducibility and for anyone to analyze the data in their own different ways.
We hope the insights provided by this study will help guide further research and assist ML practitioners design fair ML pipelines.


\let\textcircled=\pgftextcircled
\chapter{Future Work}
\label{chap:conclusion}
In the previous chapters, we saw different research projects that all contribute towards fair and explainable AI. In this chapter, we discuss five promising research ideas that extend or follow up on our previous projects. All of these projects also aim toward fair and explainable AI using a human-centered AI approach. Some of these ideas are already being implemented and we present their progress so far.           

\section{Explainable AI to Expedite Model Training}
We propose a new active learning (AL) framework, \textit{Active Learning++}, which can utilize an annotator's labels as well as its rationale. Annotators can provide their rationale for choosing a label by ranking input features based on their importance for a given query. To incorporate this additional input, we modified the disagreement measure for a bagging-based Query by Committee (QBC) sampling strategy \cite{al_bagging}. Instead of weighing all committee models equally to select the next instance, we assign a higher weight to the committee model with a higher agreement with the annotator's ranking. Specifically, we generated a feature importance-based local explanation for each committee model.
  The similarity score between feature rankings provided by the annotator and the local model explanation is used to assign a weight to each corresponding committee model.
  This approach is applicable to any kind of ML model using model-agnostic techniques to generate local explanation such as LIME \cite{ribeiro2016should}. With a simulation study, we show that our framework significantly outperforms a QBC-based vanilla AL framework.

\subsection{Active Learning++}
Active learning (AL) is a semi-supervised learning technique where the objective is to train a machine learning model using a minimal number of labeled training instances. Pool-based AL achieves this by intelligently selecting/sampling a batch of instances iteratively from a pool of unlabeled instances and getting them labeled by an oracle (human annotator) \cite{settles2009active}. The underlying premise is that some unlabeled instances are more informative than others and help train the ML model faster. This kind of learning technique plays a key role when labeled data is scarce and obtaining new labels is expensive or difficult. Some of the use cases include speech recognition, named entity recognition, text classification, etc. 

Typically, AL algorithms learn solely from the labels provided by the annotator. Our work relates more closely to the class of AL algorithms that queries feature-level input \cite{raghavan2007interactive,zaidan2007using}. The downside to such approaches is that it may be challenging for annotators, who are often not ML experts, to reason about all features of a model and provide robust input. Existing works are limited to text classification problems, where ``keywords'' based features are relatively intuitive to consider. We present a novel approach to elicit rationale in the form of feature ranking 
and incorporate this additional input as a weighing signal in the sampling strategy \cite{al_bagging}. This approach makes it easy for annotators to provide feature-level input, is relatively robust to partial or noisy input, and can be applied to problems beyond text classification.  We call this new AL framework \textit{Active learning++} as it can incorporate both instance labels and feature-level input. Our hypothesis is that richer input should help the model train faster than vanilla AL.
  
\subsection{Our Approach}
\begin{figure}
  \centering
  \includegraphics[width=\textwidth]{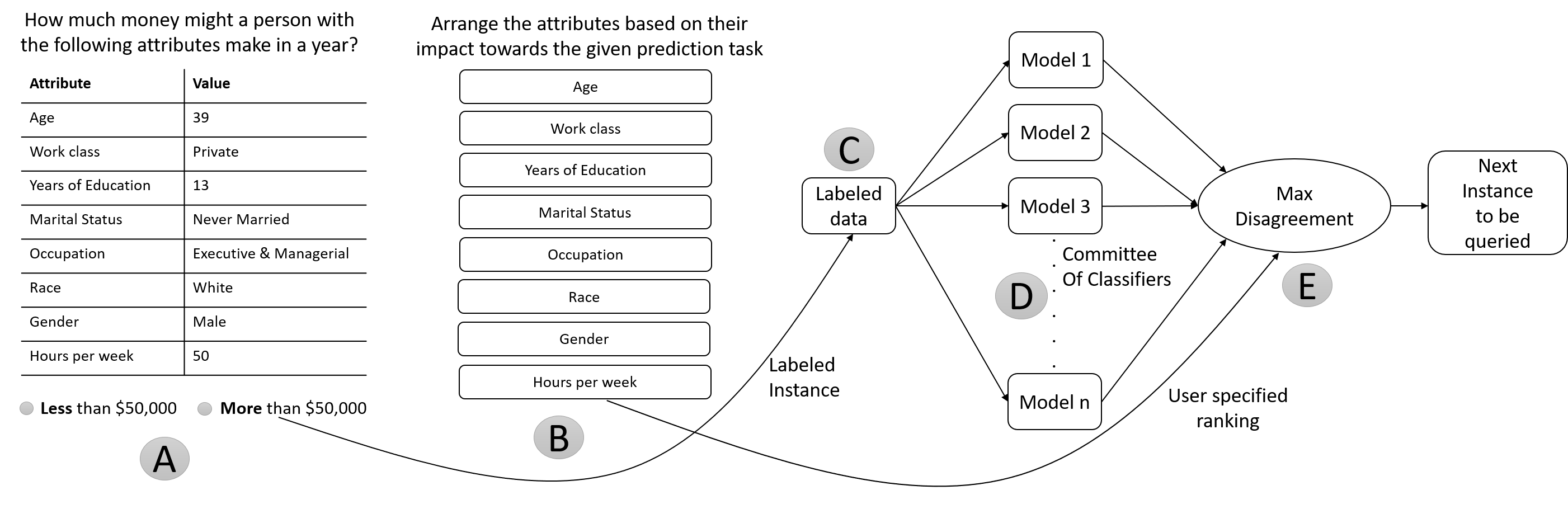}
  \caption{Workflow of our \textit{Active Learning++} framework. (A) represents a query and seeks its corresponding label. (B) elicits an annotator's rationale by arranging input features in decreasing order of importance. (C) represents the labeled dataset which is updated after each query. (D) represents the committee of classifiers used for the QBC sampling strategy. (E) represents the disagreement measure based on the model's prediction, local explanation and the annotator's rationale (as provided in (B)).  }
  \label{fig:teaser_AL}
\end{figure}
We conducted an exploratory user study with an AL system \cite{ghai2020explainable} and found that annotators naturally want to ``teach'' with rationale beyond instance labels. One common type of rationale for justifying their choice of label is by attributing to the important contributing feature(s) of an instance. For example, one may cite job category and/or education as their rationale when asked to judge a person's income level. Hence, we elicit an annotator's rationale by asking them to rank input features based on their contribution in determining the label they give to the instance (see \autoref{fig:teaser_AL}(B)).

To incorporate the rationale, we used a bagging-based QBC sampling strategy \cite{al_bagging} and modified its disagreement measure. QBC uses a set of models (committee) to select the unlabeled instance where the predictions of individual committee members differs the most. 
There are multiple ways to quantify this disagreement, like voting, consensus entropy, etc.\cite{settles2009active}. In this work, we modified the \textit{Max Disagreement} measure to capture the disagreement such that models whose rationale is in higher agreement with the annotator's are given higher weight. 
To generate each model's rationale, we use an Explainable AI method called local feature importance, which characterizes the importance of each feature by its weight for the given prediction task. We then obtain a feature ranking for each model by sorting features based on their importance. Next, we quantify the agreement between a model's and the annotator's rationale by computing the similarity between the two feature rankings (rationales) using the Kendall's tau metric. Lastly, we factor in the similarity score for each classifier in the \textit{Max Disagreement} measure to fetch the next instance to be queried. Our approach is model agnostic as we can use any ML model to train the QBC committee of classifiers and then generate local explanations using model-agnostic techniques like LIME, SHAP, etc. This approach can be applied to any type of data with meaningful input features like text, tabular data, images, etc. 

\subsection{Initial Experiments}
We simulated a pool-based AL pipeline with batch size 1 using the UCI Adult Census Income dataset \cite{adultUCI}. 
The prediction task was to classify the income of an individual given different attributes like Age, Education, Occupation, etc. into less/more than \$50,000 (see \autoref{fig:teaser_AL} (A)). After data pre-processing, we divided the dataset into test and training at a 50:50 ratio. We randomly selected 5 instances from the training data for each output label which acted as our initial labeled dataset. From this pool of labeled instances, we trained 10 logistic regression models by randomly selecting 6 instances with replacement. These models acted as the committee members for the QBC sampling strategy. Furthermore, we used a logistic regression model which acted as the active learner.   

We compared our \textit{AL++} framework with a QBC-based vanilla AL framework.  We made sure that all parameters such as the number of classifiers in the committee, initial labeled instances, etc. were identical for QBC and AL++. The only difference between the two is the disagreement measure (see \autoref{fig:teaser_AL} (E)) where AL++ assign different weights to different classifiers. 

To simulate the oracle, one popular way is to simply use the ground-truth label. However,  AL++ would also require the ranking of input features. Hence, we trained a logistic regression model over the whole training dataset which can provide quality predictions (labels) along with their local explanation. Since it's a linear model, we simply multiplied the normalized feature values by the model coefficients to generate the local explanations. This model will act as the oracle for all sampling strategies. On the test dataset, this model had 84.22\% accuracy and a 0.65 F1 score. 

\subsection{Preliminary Results}
\begin{figure}[tb]
 \centering 
 \includegraphics[width=0.7\columnwidth]{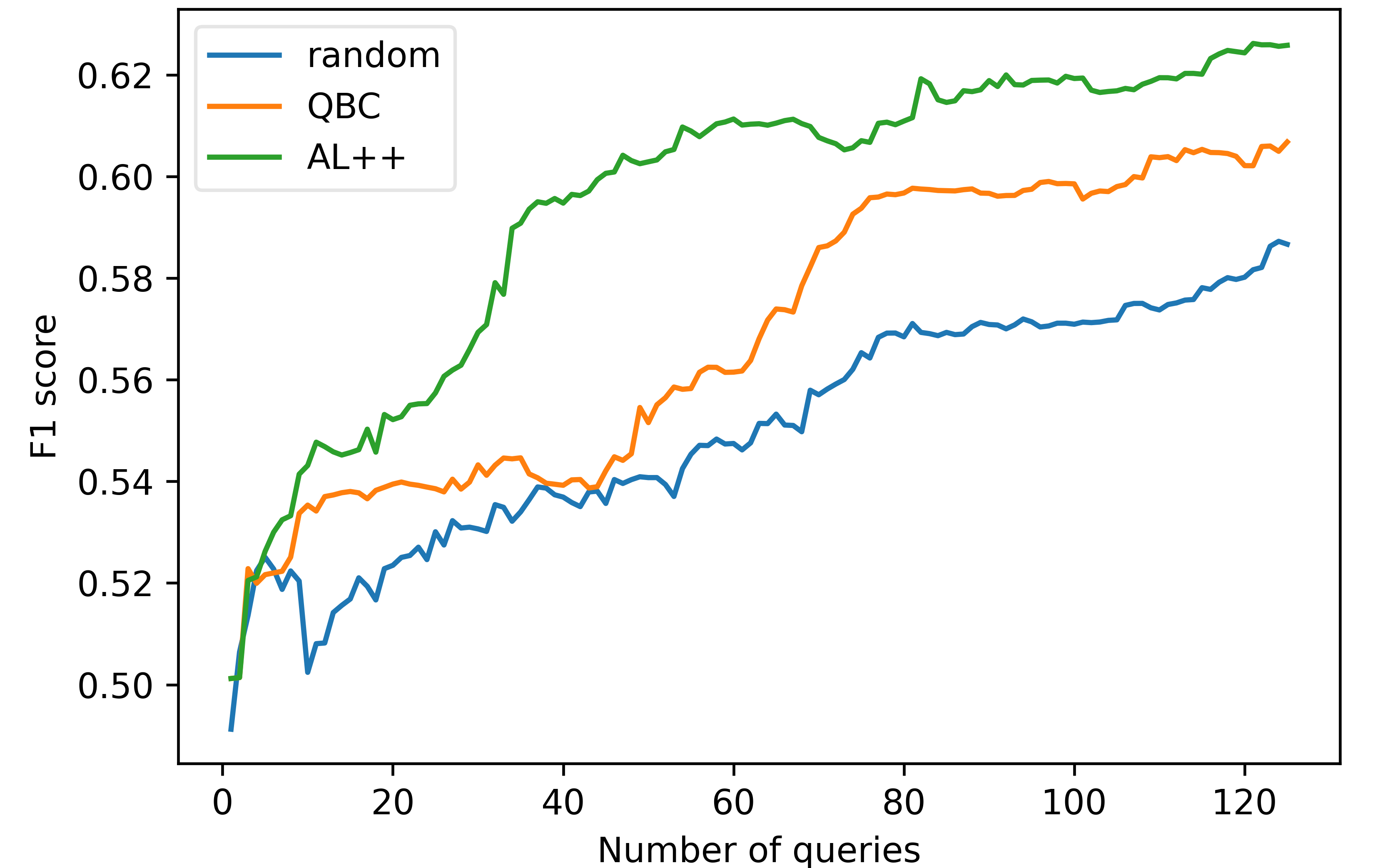}
 \caption{Comparing our proposed AL++ framework with with vanilla AL using Adult Income dataset. 
 }
 \label{fig:result}
\end{figure}
\autoref{fig:result} shows the F1 score for the random, QBC and AL++ strategies, respectively across 125 queries averaged over 10 experiments with different initial labeled datasets. We used the F1 metric over accuracy because the output class has an imbalanced distribution (ratio 25:75). From \autoref{fig:result}, we can clearly observe that AL++ maintains a constant lead over the other two strategies throughout. The mean F1 score across the 125 queries for AL++ (0.594) is much higher than for QBC (0.568). We found this difference to be statistically significant (paired t-test, p<0.01). To reach a F1 score of 0.54, AL++ required 11.8 queries, much lower than QBC requiring 25.9 queries on average. 
It confirms our hypothesis that additional input in the form of feature ranking can help train a model faster. 

In this work, we presented a new AL framework, \textit{Active Learning++}, which can utilize annotator's rationale in the form of feature importance based ranking. 
Our simulation on the Adult Income dataset shows a significant performance jump over QBC based vanilla AL framework. Next, we plan to test our approach on different datasets and different domains such as text classification.
We also plan to conduct a user study to test this framework in practice. It will be interesting to see if our approach can help build fairer ML model given a fair oracle. Finally, we wish to simulate other interaction scenarios where the oracle provides rationale for only the top-k features.

\section{Towards curation of fairer datasets by measuring biases of crowd workers}
Social biases based on gender, race, etc. have been shown to pollute machine learning (ML) pipeline predominantly via biased training datasets. Crowdsourcing, a popular cost-effective measure to gather labeled training datasets, is not immune to the inherent social biases of crowd workers. To ensure such social biases aren't passed onto the curated datasets, it's important to know how biased each crowd worker is. In this work, we propose a new method based on counterfactual fairness to quantify the degree of inherent social bias in each crowd worker. This extra information can be leveraged together with individual worker responses to curate a less biased dataset.    

Algorithmic bias occurs when an algorithm exhibits preference or prejudice against certain sections of society based on their identity. It has been termed as the imminent AI danger faced by our society. Its adverse impact has been seen in domains like healthcare, law, education, etc. \cite{o2016weapons}. 
A major source of bias in the ML pipeline arises from the training dataset. Crowdsourcing is widely used to curate training datasets for ML models . Crowdsourced datasets such as MS-COCO, imSitu, etc. have  been found to contain significant social biases \cite{zhao2017men}. One of the major factors contributing to such biases are the labels provided by the crowd workers. If there was a way to identify biased labelers, we could counter their biases by weighing them down or discarding their labels completely.  \\ 

In this work, we propose a novel technique to estimate social biases of each crowd worker. Our approach is based on the idea of counterfactual fairness (CF) which has been previously used to evaluate ML models for fairness. As per CF, a ML model is considered fair if its prediction for an individual and its counterfactual is the same. Here, counterfactual represents the same individual in a hypothetical world with its sensitive attribute i.e. demographic group changed. We have tried to adopt this approach to evaluate crowd workers instead of ML models. Given a labeling task that consists of a sensitive attribute like gender, race, etc., a crowd worker will be considered fair if she provides the same label for a query and its counterfactual. For instance, a crowd worker is tasked to label a text statement for toxicity. Assuming we want to gauge gender bias, we will generate the counterfactual query by flipping gender-sensitive word(s). A simple example can be "Women are such hypocrites". Its counterfactual will be "Men are such hypocrites". A crowd worker will be considered gender neutral if she provides the same label for the query and its counterfactual.     

The key advantage of this approach is that it integrates seamlessly with the task at hand. Counterfactual queries are added to the existing task in the exact same format as any other query. This leaves the crowd worker unaware that he is being judged for some social bias. So, our hypothesis is that our technique should serve as a better alternative to self-reported surveys because it is more resistant to social desirability bias. 
Each crowd worker is evaluated purely based on his own responses to a query and its counterfactual. So, our approach does not require any ground truth unlike gold questions (a small set of queries whose true labels are known) technique. This characteristic makes our approach more cost effective to implement for a given new task. It also makes it an elegant solution for subjective tasks like rating a movie where there is no ground truth. Lastly, this property makes our approach immune to the biases of the domain experts who provide labels for gold questions.\\

Crowdsourced labeling tasks can be roughly classified as objective tasks like image classification and subjective tasks like rating a tweet. Quantifying worker bias for objective tasks is relatively easier with the presence of objective ground truth. Different conventional fairness metrics like false positive rate difference, true positive rate difference, etc. can be used by comparing worker's responses against ground truth for gold questions. However for subjective tasks, ground truth is often unavailable or costly to obtain, so such fairness metrics can't be leveraged, even though social biases are more detrimental to these tasks. Hence, in this work we will focus on subjective tasks.

\subsection{Related Work}
Our work draws its motivation from the domain of Algorithmic Fairness and Crowdsourcing.

\textbf{Crowdsourcing}
Different quality control techniques like attention checks, gold questions, reputation, argumentation etc. are proposed in the literature to identify and control for spammers, noisy annotators, adversarial annotators, etc. In this work, we are interested in identifying and controlling for biased worker. 
We define biased worker as a human annotator who has strong preference or prejudice against a demographic group which are reflected in his/her labels.
One of the most common ways to gauge social biases like gender bias of crowd workers is via self reported surveys . The downside to such surveys is that they suffer from social desirability bias \cite{sdb}. They are usually distinct from the labeling task at hand. Hence, they make the crowd workers conscious that they are being evaluated. 
  Another popular way to measure inherent social bias is Implicit Association test (IAT) . However, recent studies have questioned the effectiveness of this test \cite{iat_counter}.  So, there is a need for a new better alternative to measure social biases.
Furthermore, there seems to be a disconnect between the crowdsourcing and the algorithmic fairness literature. It will be interesting to see conventional fairness metrics being used to measure crowd worker bias and how it compares with the score returned by self-reported surveys.

\textbf{Algorithmic Fairness}
The existing literature that focuses on mitigating bias at different stages of the ML pipeline can be broadly classified into 3 stages i.e. pre-processing, in-processing and post-processing . In the pre-processing stage, a given dataset is modified such that the social biases with respect to the sensitive attribute are reduced/removed. In the in-processing stage, novel ML algorithms are devised that return fair predictions even when trained on biased data. Lastly, in the post-processing stage, the predictions of the ML model are modified to make it more fair. Another important stage that needs more attention is the data curation stage. This stage falls before the pre-processing stage. Hence, if the biases are firmly dealt with during data curation, then we won't need to deal with it later on. Multiple factors like skewed representation of a demographic group, biased label distribution from human annotators, etc. can pollute the data curation process. Our work focuses on the data curation stage and aims to improve label quality.   


\subsection{Counterfactual Queries}
\begin{figure}[ht]
    \centering
    \includegraphics[scale=0.55]{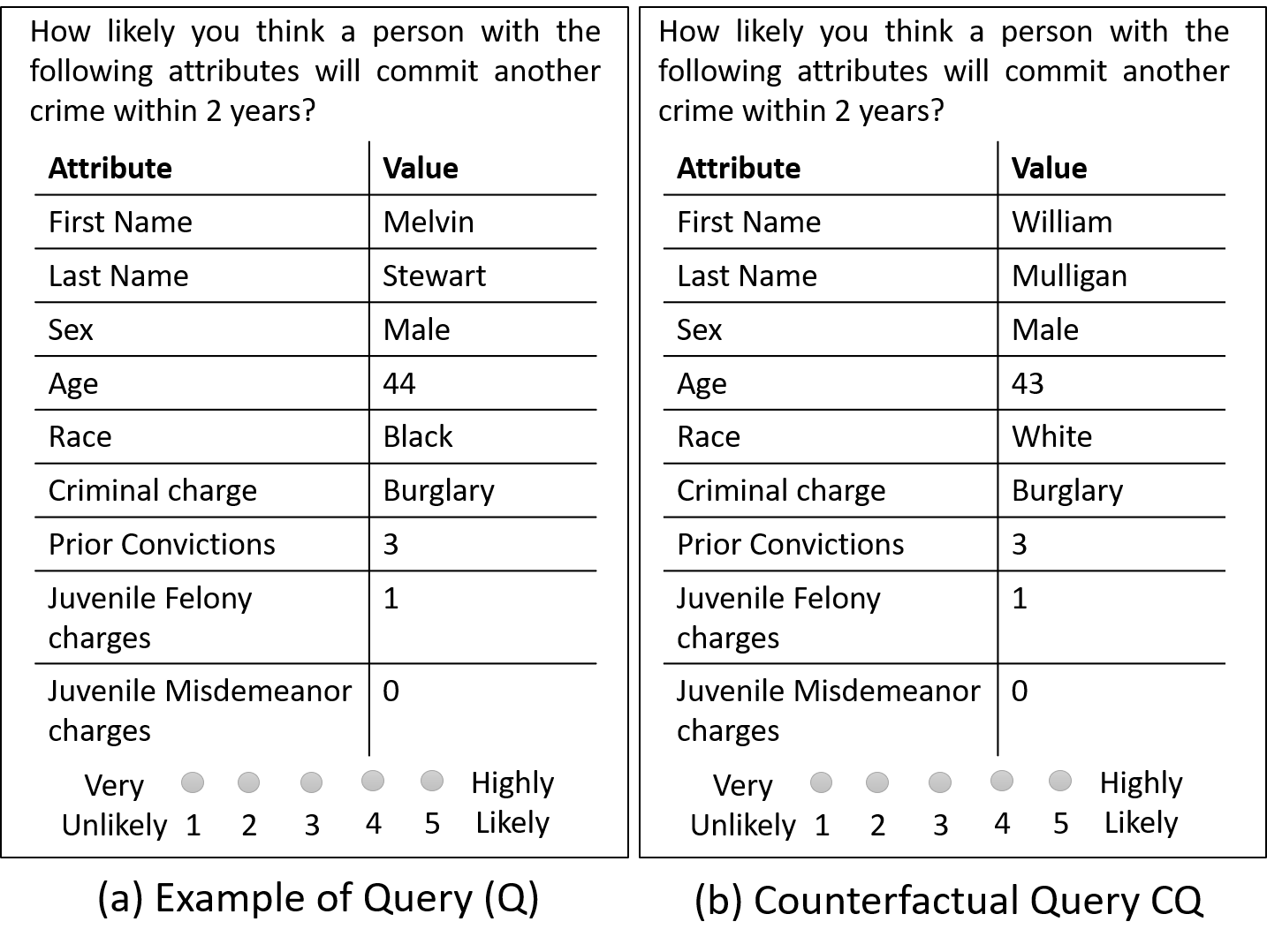}
    \caption{Example of a Query (left) and its Counterfactual (right) for the recidivism prediction task }~\label{fig:query}
\end{figure}

Let's consider a toy problem where crowd workers are asked to predict recidivism. As shown in fig.\ref{fig:query} (a), a crowd worker is presented with a set of features representing a convict. The task is to predict the likelihood of re-offending within 2 years on a 1-5 scale. Let's say each crowd worker is asked to label x such queries. We will choose a subset of size n out of x queries and generate counterfactuals for only those queries Q.  

In this work, we are considering a narrow class of counterfactuals that can be generated by changing the sensitive attribute and keeping other features constant \cite{garg2019counterfactual}. 
Here, we are dealing with racial bias so we will generate counterfactual query CQ by changing the race attribute (see fig.\ref{fig:query} (b)). 
The mean absolute difference in labels provided for all n pairs of Q and CQ will represent the bias score for a crowd worker as shown eq.\ref{eq:worker_bias}. 
\begin{equation}
    \label{eq:worker_bias}
    Worker Bias = \frac{1}{n} \sum_{i=1}^{n}|Label(Q_i) - Label(CQ_i)|
\end{equation}
A zero score characterizes perfect unbiased behavior and higher values symbolize more biased behavior. This extra information will be used in conjunction with the crowd worker responses to yield fairer labels and hence fairer datasets. A simple way to incorporate this information might be to filter out biased workers whose bias score is beyond a threshold. More sophisticated aggregation algorithms can be adopted as well which better utilize the subtle differences in bias scores. We will evaluate our method by comparing the datasets obtained using our approach and self reported surveys on different fairness metrics.  \\
The key challenge in adopting counterfactual fairness to judge crowd workers relative to ML models is that a crowdworker might relate a previously seen query with its counterfactual. If the crowdworker realizes that she has labeled a very similar query before, she might become conscious that she's being judged. This will defeat our purpose of countering social desirability bias. 
To circumvent this issue, we can play with the ordering of the queries such that a query and its counterfactual are placed far from each other. This will ensure that the memory trace created by the original query gets faded by the time the counterfactual query is encountered.   
Furthermore, we can perturb the counterfactual query by adding small noise to certain numeric features. In the above example, we added noise to the 'Age' feature. Lastly, we can add/modify dummy features like 'First Name', 'Last Name', etc. which are irrelevant to the prediction task. Their role is to carve a slightly different identity for the counterfactual query relative to the original query (see fig. \ref{fig:query}). If the query deals with textual data, we can use different paraphrasing techniques like splitting/combining sentences, substituting synonyms, etc.

\section{Tackling Social Biases encoded in Word Embeddings}
Similar to other ML models, word embeddings are not immune to social biases inherent in the training dataset. In Chapter 4, we presented our work on auditing word embedding models such as word2vec for different kinds of social biases based on race, gender, etc. We found that word embedding models can be plagued with biases against groups like females, blacks, etc. as well as subgroups such as black females. Given that word embedding models serve as the foundational unit for many NLP applications, it is important to prevent its inherent undesirable social biases from polluting downstream applications like as sentiment analysis, machine translation, etc. To achieve this, one needs to generate a fair word embedding or debias an existing word embedding before using it for any downstream application.  

Existing literature tries to tackle this problem at different stages of the ML pipeline like making changes to the training corpus to make it more fair, tuning the loss function by adding fairness constraints \cite{zhao2018gender} or making changes to the trained word embedding model \cite{bolukbasi2016man}. The common thread among these approaches is that they are all fully automated approaches. Such approaches might be fast and relatively inexpensive but they might not perform well on human-centric measures like trust, accountability, etc. Even on the fairness front, a fully automated approach can not \textit{precisely} determine if a strong association between words is socially desirable/acceptable. For example, it is desirable to have a strong association between the word embeddings for `she' and `mother' but not for `she' and `teacher'. Moreover, in an effort to curb bias, fully automated approaches tend to go a bit overboard and lose some relevant/desirable semantic knowledge encoded in word embeddings. On the other hand, a human can use their domain knowledge to precisely make the distinction between socially desirable and undesirable/biased associations. Hence, we decided to pursue this problem from a human-centric AI perspective.  

\begin{figure}[h]
 \centering 
 \includegraphics[width=1\columnwidth]{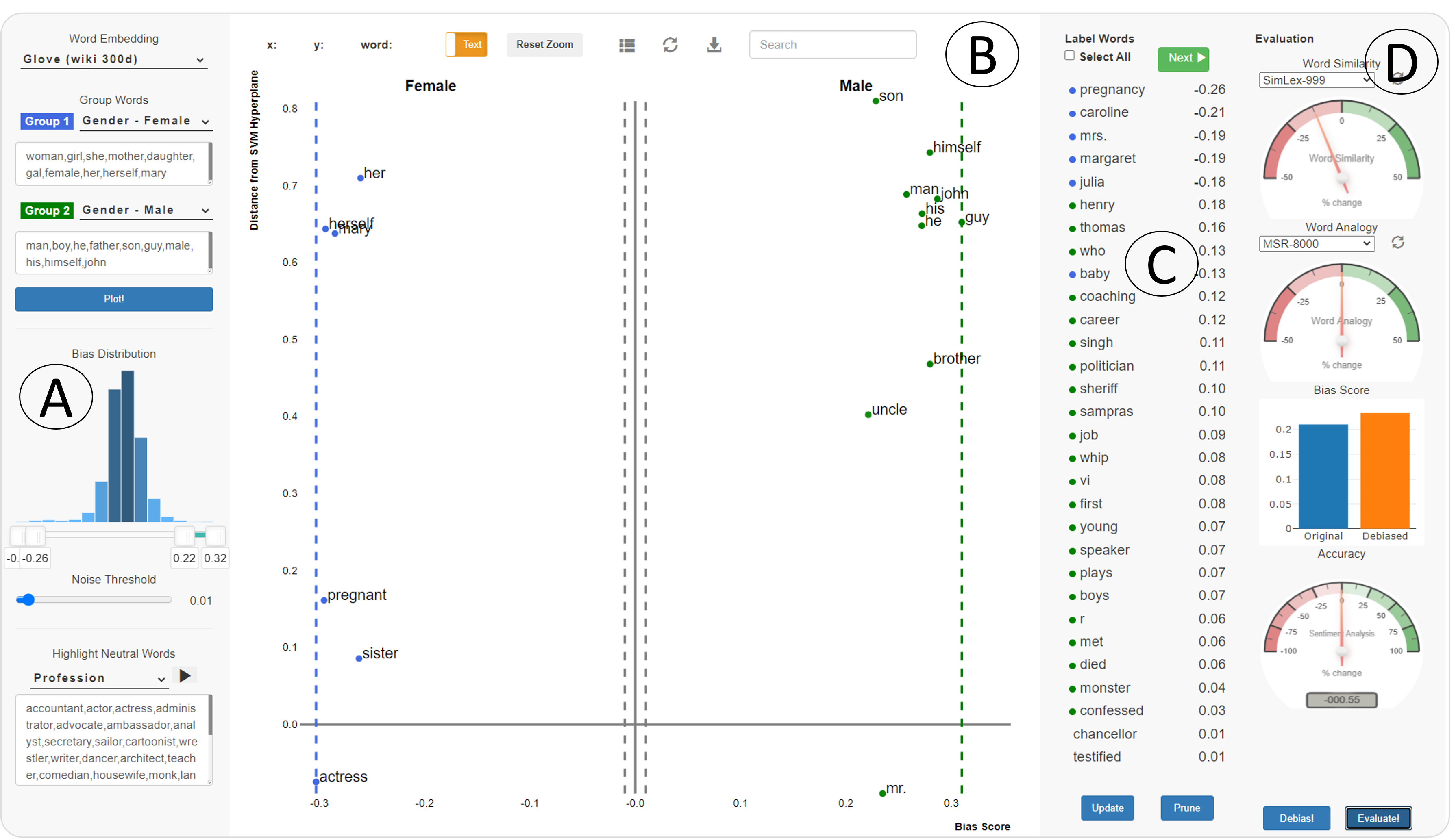}
  \setlength{\belowcaptionskip}{-8pt}
\setlength{\abovecaptionskip}{-4pt}
 \caption{Visual Interface of our \textsc{WorDebias} tool (A) The Generator panel: helps select word embedding, group words, and filter words to be displayed/highlighted (B) The Main view: visualizes different words as a scatterplot where x-axis represents bias score and y-axis represents distance from the SVM hyperplane (C) The Active Learning panel: allows user to infuse domain knowledge by selecting undesirable associations (D) The Evaluation panel: visualizes different metrics pertaining to bias and utility. }
 \label{fig:webvis}
\end{figure}


\subsection{Our Proposed Approach} 
  We propose \textsc{WorDebias}, an interactive visual tool for identifying different kinds of biases like race, age, etc. across different word embedding models followed by real-time mitigation (see fig \ref{fig:webvis}). 
  Given a pre-trained static word embedding as input, \textsc{WorDebias} computes bias scores for all words in the word embedding corresponding to different social categorizations (bias types) like gender, religion, etc. using the relative norm difference metric \cite{100word}, and returns its debiased version.   
  Here, each categorization (bias type) e.g. race consists of two subgroups, say Blacks and Whites. Each subgroup is mathematically defined as the mean of word embeddings for the words that represent that subgroup. 
 As shown in \ref{fig:webvis}(B), different words are visualized as a scatter plot where the x-axis encodes the bias scores. Here, words that lie at the extreme ends represents strong association with either subgroups. A word whose x-coordinate is 0 reflects it’s completely neutral. 
 The user is tasked with identifying if the strong association for different words with either subgroups is legit or biased. For example, the word `actress' has a legit association with female subgroup but the word `nurse' doesn't. Similar to \textsc{WordBias}, \textsc{WorDebias} also presents a set of ideally neutral words to expedite the auditing process. This includes words that represents different professions, emotions, etc. 
 
 If the user finds that the association of a word (e.g. `terrorist') with a subgroup (e.g. `Islam') is socially undesirable, they can choose to unlearn/debias this association. This is achieved by modifying the word's embedding such that bias is mitigated while preserving its semantic meaning as far as possible. In other words, the word's representation in high dimensional space is changed such that it is equidistant from either subgroups. The user is free to manually explore the word embedding space and flag all undesirable associations. Given that word embedding model might have millions of unique words, it will be virtually impossible for a user to go through all words and check if their association is socially acceptable. To address this challenge, we adopt a semi-supervised learning approach, namely active learning. \textsc{WorDebias} starts off with training a SVM model to segregate between legit and biased associations using a few seed words. Instead of randomly labeling different words, the user can provide labels for the words that the SVM model is most uncertain about (also known as uncertainty sampling). Here, we have used the distance of a word from the SVM hyperplane to quantify uncertainty. Smaller the distance, higher the uncertainty. As shown in \ref{fig:webvis} (C), our tool presents the user with the set of words that the model is most uncertain about. The distance of a word's embedding from the SVM hyperplane is encoded on the y-axis of the scatterplot (see \autoref{fig:webvis} (B)). Words on either end of the y-axis represent whether their association with the subgroup is legit/erroneous. For example, in the case of gender bias, words like father, mother, king, nun will be on one extreme while words like boss, emotional, painting will be on the other extreme. As the user provides more labels for different words (legit/biased), the SVM model keeps refining/retraining and becomes more accurate. After each refinement, the user can visually observe how the decision boundary has changed. The user can also choose to debias the word embedding. Based on the current SVM model, all biased word associations be neutralized by making minimal changes to their embedding. The impact of this operation on different utility and bias metrics can be observed in the evaluation panel (see \autoref{fig:webvis} (D)). The user can keep interacting with the tool until they are satisfied with the results, and then download the debiased word embedding. 
 
 \textsc{WorDebias} can be instrumental as an auditing tool for ML researchers who want to employ word embeddings for some downstream task. Its user-friendly interface and intuitive design can facilitate interdisciplinary research among researchers from law, sociology, computer science, etc. Lastly, it can be used for educational purposes and spreading awareness to a non-technical audience.
 

 
\subsection{Evaluation}
After debiasing, the debiased word embedding is evaluated for bias and utility. For evaluating the debiased word embedding, we need to make sure that the semantic meaning encoded in the embedding is retained while mitigating bias. To evaluate semantic structure, there are two broad approaches, i.e. intrinsic and extrinsic evaluation. For intrinsic evaluation, we have used word similarity (WS) and word analogy (WA) tasks. 
From a visual analytics perspective, it's attractive because these metrics are computationally inexpensive and fast.
As intrinsic evaluation does not guarantee better performance on downstream applications so we have used sentiment analysis for extrinsic evaluation. We chose sentiment analysis over other tasks because it is fairly popular and has been shown to exhibit bias towards race and gender. Sentiment analysis is a broad topic in itself. We do not seek to build a sophisticated state-of-the-art model. Our aim is to measure the impact of our debiasing approach on the utility (accuracy) and bias for this downstream application. We plan to train two sentiment analysis models based on original word embedding and its debiased version. If the test accuracy of models trained on debiased embedding is similar to that of original embedding, we can conclude that the debiased embedding preserves semantic meaning. Furthermore, we plan to use the EEC corpus (Equity Evaluation Corpus) to measure the racial and gender biases for the word embeddings trained over english language corpus \cite{200sentiment}.
 
 So far, we have successfully tested \textsc{WorDebias} for popular word embedding models like word2vec and Glove. We found that bias score has decreased significantly after debiasing while utility is virtually the same. Currently, we are experimenting with other embedding models like FastText and other languages like French and Hindi. Also, it will be interesting to evaluate our tool on human-centric measures like trust, accountability, etc. 

\subsection{Limitations}
As shown by Gonen and Goldberg \cite{lipstick}, existing ways to debias word embeddings by removing the bias component from word embeddings doesn't solve the bias problem completely. Bias is reduced but not completely eliminated. 
In the current setup, our approach might also not solve the problem completely but it is a step in the right direction. Moreover, our approach is quite flexible so it can incorporate more effective post-modeling debiasing techniques of the future. Also, our approach only works for static word embeddings. Future work might design visual tools that can also deal with contextualized word embeddings. Lastly, our tool can deal with one kind of bias at a time say gender. Future work might develop tools that can mitigate multiple biases and intersectional biases simultaneously.

\section{AI-Powered Writing Assistant for Gender-Inclusive Writing}
Sexism is a deep rooted social problem whose manifestation in language can be very obvious or can be very subtle. In Chapter 5, we presented our writing tool (\textsc{DramatVis Personae}) that leverages NLP techniques to assist professional writers author books, novels, etc. in a more inclusive manner. However, social biases in text are not limited to professional writers and stories/books. As a natural extension to this work, we intend to build a writing tool for gender-inclusive writing that caters to the general public for their everyday needs like writing emails, tweet, facebook posts, slack, etc. 

\subsection{Writing Tools}
Writing tools have come a long way from typewriters to MS Word. There are a variety of AI-driven writing tools for different tasks and audiences. For example, Fluent \cite{ghai2021fluent} is a writing tool for people who stutter. 
\textit{Grammarly} \cite{karyuatry2018grammarly} is a popular AI writing assistant that provides real-time suggestions for fixing grammatical errors, improving word choice, refining tone, etc. \textit{RECAST} is an AI writing assistant aimed at toxicity reduction. \textit{Textio} \cite{textio} is another AI-powered writing tool that helps with hiring content. It suggests word-level changes for writing inclusive and effective job descriptions. In the context of inclusive writing, there has been some progress in the recent times. For example, Textio helps write more gender-inclusive job postings. This work relates more closely with MS Word which also provides some features to assist gender-inclusive language. Currently, MS word is limited in tackling gendered words such as 'mankind'. For this work, we plan to go a step further by checking for gendered pronouns and sexist phrases. Moreover, MS Word is a closed source proprietary software so researchers/practitioners are restricted in their ability to assess and build upon on it. On the other end, we intend to share all details for our tool including the source code for easy reproducibility.

\subsection{Our Proposed Approach}
We propose \textit{FairText}, an AI-augmented writing assistant for discovering sexism in text, make the author conscious of their gender bias, and nudge them to produce more gender-inclusive text by recommending suitable alternatives. It can be loosely understood as a spell-checker for gender bias. FairText employs different NLP techniques like named entity recognition, coreference resolution, part of speech tagging, and language models like RoBERTa \cite{liu2019roberta} to identify biased instances in text. The utility of such a tool will be threefold: (1) it will help the authors realize their own subconscious biases that might help mitigate them; (2) it will prevent the propagation of implicit social biases from the authors to the readers; and (3) it will prevent ML algorithms from picking up such biases when trained over the resulting fair text.

\begin{figure}[t]
 \centering 
 \includegraphics[width=1\columnwidth]{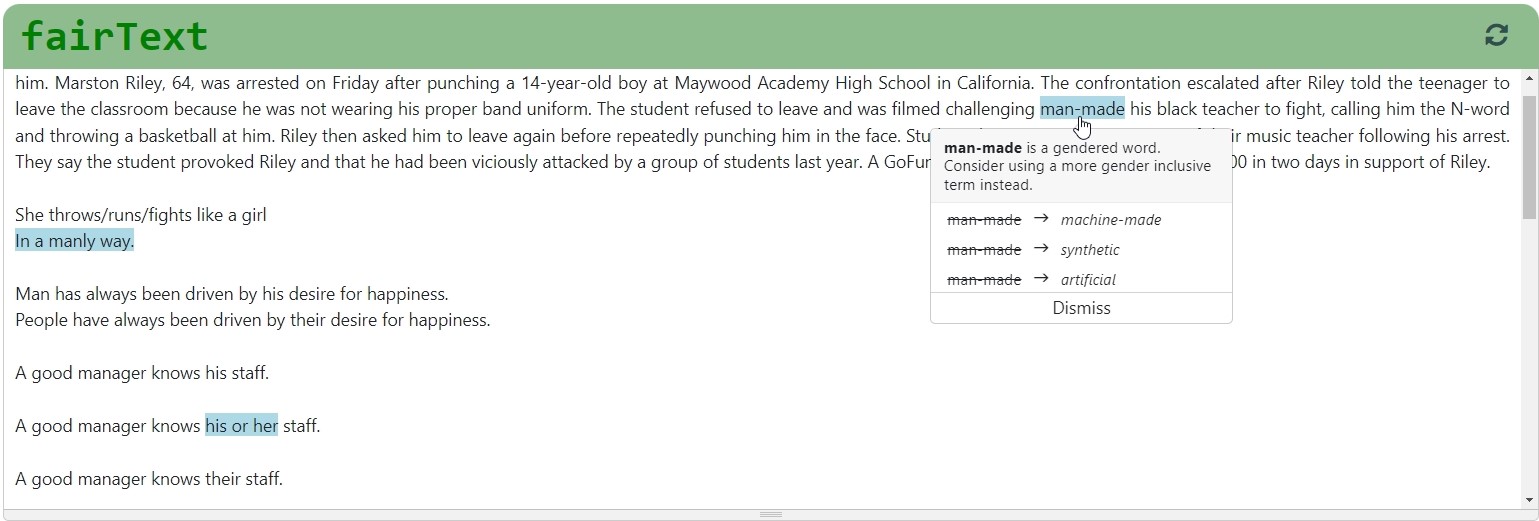}
  \setlength{\belowcaptionskip}{-8pt}
\setlength{\abovecaptionskip}{-4pt}
 \caption{Visual interface of FairText. Text highlighted in blue represents gendered language as detected by our tool. In this
picture, the author hovers over a highlighted word which open a popup that provides a set of alternatives.}
 \label{fig:fairtext}
\end{figure}

FairText is implemented as a web application using python based web framework \textit{Flask}. The visual interface (see Fig.\ref{fig:fairtext}) is built over javascript-based open source library \textit{Summernote}. This makes FairText easily accessible using a web browser across different platforms without needing any third-party software. The default visual interface is designed to look like a generic rich text editor. Words/phrases highlighted in blue represent the biased instances. To display the set of alternatives, we have used a popup mechanism populated with alternatives that appear right below the highlighted word/phrase on hovering. Such a design choice is implemented to mimic that of spell checkers and other popular tools like Grammarly \cite{karyuatry2018grammarly} which most people might have interacted with at some time. This might help users to quickly latch on to our interface without needing additional training 

\subsection{Detect Gender bias}
As per the United Nations, \textit{"Gender-inclusive language means speaking and writing in a way that does not discriminate against a particular sex, social gender or gender identity, and does not perpetuate gender stereotypes."}. On its website\footnote{\url{https://www.un.org/en/gender-inclusive-language/guidelines.shtml}}, UN provides a set of guidelines for gender-inclusive language. We draw some of the best practices/strategies from these guideline and use NLP techniques to incorporate them into our tool. More specifically, our tool identifies 3 kinds of gender biases described as follows:

\paragraph{\textbf{Gendered Nouns:}} There are certain gendered words that should be avoided irrespective of context like `Mankind'. We have compiled a list of such words from different online sources along with their suitable alternatives. For example, `mankind' might be replaced with `humankind'. Our rule-based engine uses regex expressions to search for any word in the compiled list and highlights them. On hovering over such a word, a popup appears that suggests alternative words. 

\paragraph{\textbf{Gender biased Expression:}} We use a dataset that has gender-biased expressions and neutral expressions \cite{grosz2020automatic}. Thereafter, we train a binary classifier based on the RoBERTa embeddings for different sentences. We use this classifier for each sentence in the interface and highlight the sentence if it is biased, i.e., its predicted probability score is greater than a threshold.   

\paragraph{\textbf{Gender Neutral Pronouns:}} We use co-reference resolution to identify the entities which are referred by a specific pronoun like his, her, etc. If the entity is a common noun such as `Professor', we highlight such pronouns and present alternatives such as `they', `them', etc. 

\subsection{Discussion}
There can be wide applications for tools like FairText. There is a need to reduce bias in different spheres like news articles, script writing, job postings, etc. Our proposed tool can be integrated as a plugin into MS-Word, Google Docs, Slack, Mobile keyboard apps, Overleaf, etc. This way it can help journalists, researchers and common public. In the current setup, our work focuses on gender inclusion for the English language. Future work might focus on inclusion based on culture, race, ethnicity, nationality, disability, etc., and support languages other than English and multilingual text. Moreover, it will be interesting to venture into preferred pronouns. Given that there are numerous personal preferred pronouns like Ze, Zir, etc., it can be difficult for an individual to remember one's pronouns. However, a digital tool can easily store such preferred pronouns and prevent the author from misgendering a person (also part of UN Best practices/strategies 1.1).

It is important to note that one can adopt a fully automated approach to detect and replace biased words/phrases. This would mean that the text editor would automatically edit the text as it sees fit. However, this is not a good idea on two accounts. Firstly, current NLP techniques are not advanced enough to identify all biased instances precisely and find optimal replacements. Moreover, the NLP models and techniques used for identifying bias might themselves be plagued with gender bias. Secondly, we want the author to realize their implicit biases as it can have a significant positive impact on mitigating their inner biases. Performing automatic edits might negatively impact this process. Hence, it is important to follow a human-centric AI approach to achieve our objectives. 



At present, we have a developed a basic prototype of the tool. We are working on polishing different components of the tool that deal with identifying bias and recommending alternatives. Next, we intend to conduct a between-subject user study to evaluate the usability and efficacy of our tool. Moreover, we plan to run a big corpus of text through our tool and generate its fairer version assuming all recommendations by the tool were accepted. Thereafter, we can train a word embedding on the original text and its fairer version to evaluate the downstream impact on ML algorithms. 

\section{Towards Inclusive Group Fairness}
Existing literature presents fairness-enhancing interventions that are aimed at optimizing some fairness metric. These fairness metrics are the mathematical manifestations of different notions of fairness like group fairness, individual fairness, counterfactual fairness, etc. Among the different notions of fairness, group fairness has received considerable attention from the research community. Group fairness advocates that different population groups should be treated equally, i.e., each group should receive similar proportion of positive/negative outcomes. Some group fairness metrics include accuracy difference, error rate difference, etc. In this work, we are just focusing on group fairness.

\begin{figure*}
\centering
\includegraphics[width=0.9\columnwidth]{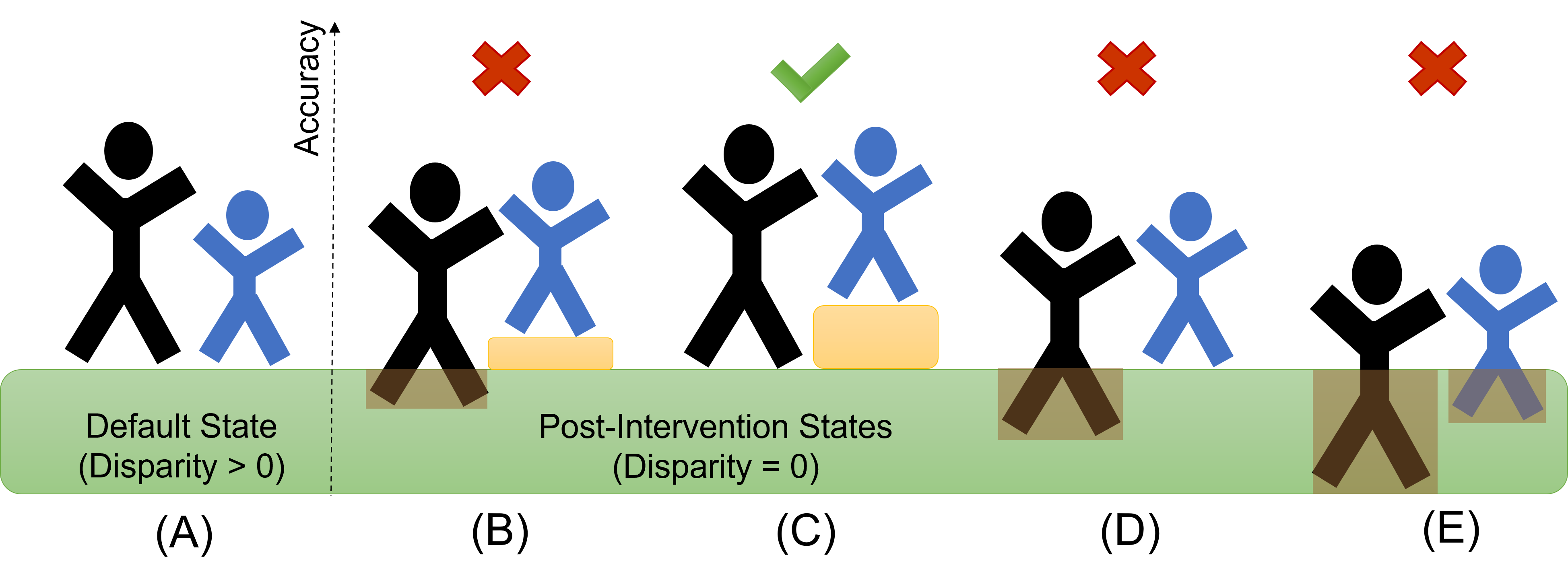}
\caption{Different ways to achieve parity in terms of accuracy between two population groups (shown in Black and Blue). Here, (A) shows the default state where disparity$>$0. (B)-(E) represents the state post different fairness-enhancing interventions. Here, (C) is desirable relative to (B, D and E) because it achieves parity while making minimal changes and without hurting any group. }
\label{fig:intro_gf}
\end{figure*}

Existing interventions aimed at enhancing group fairness solely try to reduce disparity between groups without any consideration of how that is achieved. This can lead to worse outcomes/performance for the privileged and/or the unprivileged group(s). \autoref{fig:intro_gf} shows different ways by which disparity can be mitigated. It should be noted that many of those bias mitigation approaches can be undesirable for either/both population groups. We believe the right way to mitigate disparity is by uplifting the unprivileged group while causing no or minimal harm to the privileged group. Existing techniques do not make a distinction among these different approaches. In our previous work (see Chapter 6) based on real world datasets, we found how different individual and combination of interventions can potentially harm a population group, especially the privileged group \cite{ghai2022cascaded}. Achieving group fairness at the cost of a specific group can be deemed unethical. It can alienate people from that group and can potentially stifle research in this space. So, there is a need to develop a fairness-enhancing intervention that ensures that each population group is not worse off than what they started out with.

\subsection{Our Approach}
We propose a novel model-based intervention that enhances group fairness while ensuring that no population group is harmed in the process.
A popular model-based intervention is to add a specific regularization term to the loss function that aims to enhance group fairness. Similarly, we can try to come up with a regularization term that aims to ensure that each group is not worse off than what they started with. Here, we need to ensure that the added term is differentiable. The advantage of this approach is that it can be easily integrated into existing ML models. The downside is that it frames a multi-objective optimization problem into a single objective one.   
In the following, we present 3 different loss functions: a traditional loss function used extensively in the ML literature (Eq. 7.2), a fairness-aware loss function (Eq. 7.3), and a novel loss function for more inclusive group fairness (Eq. 7.4).
\begin{equation}
    loss_{CE} = BCE\_loss(y, \hat{y})
\end{equation}
\begin{equation}
    loss_{group} = loss_{CE} + \alpha*abs(M^P-M^U)
\label{eq:baseline}
\end{equation}
Here, $loss_{CE}$ represents binary cross entropy loss, $\alpha$ is a hyperparameter, and M is a user-chosen metric which can be F1 score or base rate. $M^P$ and $M^U$ represents the metric M for the privileged group and the unprivileged group.   

\begin{equation}
    loss_{new} = loss_{group} + \beta*(scal(M_{pre}^P - M^P) + scal(M_{pre}^U - M^U)) 
\label{eq:new_loss}
\end{equation}

\begin{equation}
scal(x) = \left\{
        \begin{array}{ll}
            0 & \quad x \leq 0 \\
            e^{px} & \quad x > 0
        \end{array}
    \right.
\label{eq:scal_fn}
\end{equation}

Here, $p$ and $\beta$ are hyperparameters. The second term in Eq. 7.4 is the novel regularization term added for more inclusive group fairness. $M_{pre}$ represents the metric M for a given population group before fairness-enhancing intervention. This term adds exponential penalty when the metric M is lower than what is was before intervention. Larger values of $p$ will more strongly enforce the objective of not hurting any population group. For n population groups, the above equation can be easily generalized as:-
\begin{equation}
    loss_{new} = loss_{group} + \beta \sum_{i=1}^{n} scal(M^i - M_{pre}^i) 
\label{eq:new_loss_n}
\end{equation}
Our hypothesis is that $loss_{new}$ will ensure group fairness while causing minimal harm to any population group in the process. 

The primary contributions of this proposed work are as follows:-
\begin{itemize}
    \item Highlight the limitations of how group fairness is formulated and pursued in the fairness literature. 
    \item Propose a new mechanism to achieve group fairness such that no group is worse off than prior to the fairness-enhancing intervention.
    \item Evaluate our approach by performing experiments on 4 popular tabular datasets.
\end{itemize}

At present, we have undertaken the first two points. Currently, we are conducting experiments on the Adult Income, COMPAS, Bank Marketing, and German credit dataset to evaluate the efficacy of our approach.  


\subsection{Discussion}
This work introduces a novel mechanism for ensuring more inclusive group fairness for tabular datasets. Future work might extend this work for other data types say text, images, etc. and problems say text classification, recommender system, etc. Our proposed regularization term is a function of the user-chosen metric M. At present, we can have M as base rate and F1 score. Future work might come up with innovative ways where M can also be other metrics like false positive rate, false negative rate, etc.      
Fairness is contextual~\cite{wachter2021fairness} and so one should not blindly opt for a given fairness-enhancing intervention, including the one proposed in this work. For example, if one tries to use reduce disparity in base rate using our proposed technique, the total number of positive outcomes will increase as our technique does not redistribute positive outcomes. This might not be desirable/practical for some applications. On the other hand, it might be desirable to reduce disparity in accuracy between groups by uplifting a group.

\chapter{Conclusion}

Explainability and fairness in AI systems have received considerable attention in the past few years and have become important research areas of their own. Both these research areas are interdisciplinary.  
In this work, we pursued these two research areas from the perspective of data visualization and human-computer interaction. We advanced the state of the art by developing novel visual interactive tools and conducted empirical studies 
to identify/mitigate social biases in tabular datasets, text, word embeddings, and classification systems. On the XAI side, we studied how AI explanations can aid the data curation process and expedite ML model training. In this chapter, we briefly present some of our learnings and recommendations.

Owing to our human-centered AI approach, most projects presented in this dissertation have a significant human component to them. Involving a human in the loop to deal with fairness/explainability in AI systems presents different challenges like faithfulness, interpretability, scalability, etc. In our context, faithfulness means how accurately the visualization/visual tool presents the underlying state of the system. It is difficult to address all these challenges simultaneously as they might not be compatible with each other. For example, a visual tool might sacrifice interpretability if it aims to visualize all the nuances and effects of a complex ML model accurately (faithfulness). Similarly, one might be able to achieve interpretability and faithfulness at the cost of scalability. The optimal trade-off among these factors might vary with the given problem at hand, the expertise of the users involved, etc. 

In the case of XAL (Chapter 2), we were able to achieve interpretability and faithfulness in ML model explanations as we were dealing with a simple linear model. Had we used a more complex ML model, we would have sacrificed faithfulness 
with an explanation technique like LIME. For the D-BIAS tool (Chapter 3), we used causal models which are easy to interpret and can deal with larger datasets. However, causal models might falter on faithfulness if their underlying set of assumptions are not met. Fairness is a sensitive issue and one should be careful in assessing/reporting such things. So, we also included various fairness metrics in the D-BIAS tool to present a more comprehensive and accurate picture. Similarly, for the WordBias tool (Chapter 4), we chose a visual design such that the bias scores are visualized without any distortion (ensured faithfulness). The key message here is to be conscious of the possible trade-offs and taking an informed decision. In this work, we gave special attention to faithfulness and would recommend the same. In case a compromise has to be made on faithfulness, one should provide other features/mechanisms by which the end user can assess/estimate the true picture. Under no circumstances should the user be misled to false conclusions. 

Another important challenge is the bias inherent in each individual. Here, we are referring to the intrinsic value system/beliefs of the human users who operate the visual tool and the researchers developing them. We can not stress enough the importance of choosing the users carefully and then training them well. A prime example is the D-BIAS tool where we showed how the system can be rendered ineffective or even be misused to exacerbate existing biases if it is operated by an irresponsible/malicious user. We recommend choosing an individual who is well versed in social and cultural sensitivities and is trusted by a majority of the stakeholders.  
The researchers also have their own biases and value systems. It is important that such biases are not baked into the tools aimed at tackling social biases. In this work, we have tried our best to distance ourselves from propagating any value judgments. For example, a researcher might have a preference for a specific fairness metric. However, we strove to include different perspectives and provide maximum flexibility to end user(s). This is showcased in Chapters 3 and 6 where we have included multiple fairness metrics pertaining to individual fairness and group fairness.

\textit{It is important to note that the different projects presented in this dissertation are research prototypes. One should exercise caution before using them in their current state for any real-world application}. We advise the readers to go through the design goals and the discussion section for each project to understand what each intends to achieve along with their limitations. Each real-world use case might have its own unique requirements that might not have been considered. We have made the source code for most of the projects publicly available. We recommend accessing and tweaking the source code as required.

In conclusion, we hope this dissertation will attract the attention of fellow researchers to this interesting and under-explored research area. We sincerely hope that our work will help inform further research and might lead to some real-world impact in the long run. We highly recommend the readers go through Chapter 7 where we have listed several venues for further research. 
\bibliographystyle{siam}
\renewcommand{\baselinestretch}{1}
\normalsize

\clearpage
\newpage
\phantomsection%
\addcontentsline{toc}{chapter}{\numberline{}{Bibliography}}%
\bibliography{references}
\clearpage
\newpage

\appendix
\section{Appendix A}
This section presents the supplementary material relevant to Chapter 04.  
\subsection{Preprocessing word embedding} 
Before loading the word embedding onto WordBias, we did some prepossessing similar to what is followed in the literature \cite{bolukbasi2016man}. We only considered words with all lower case alphabets and whose length is upto 20 characters long. We then sorted the resulting words by their frequency in the training corpus and picked the most frequent 50,000 words. We made sure to include group words like names, etc. if they don't make it in the final list.

\subsection{Design Rationale}
The problem of visualizing biases against intersectional groups boils down to visualizing a large multivariate dataset where each word corresponds to a row and each column corresponds to a bias type. A straightforward solution for visualizing such high-dimensional data is to use standard \textit{dimensionality reduction} techniques like MDS, TSNE, biplot, etc. and then use popular visualization techniques like scatter plot. However, Algorithmic bias is a sensitive domain; we must make sure that we \textit{accurately} depict the biases of each word (\textbf{G1}). Hence, \textit{dimensionality reduction} and related techniques like the Data Context Map\cite{dataContextMap} are not an option because they almost always involve some information loss. Using such techniques might inflate/deflate real bias scores which might mislead the user. 

Next, we enumerated other possible ways to visualize multivariate dataset, like scatterplot matrix, radar chart, etc. and then started filtering these options based on the design challenges G1-G4. The scatter plot is a popular choice which is also used in Google's Embedding projector \cite{smilkov2016embedding}, but it is limited to three dimensions. A couple of more dimensions can be added by encoding radius and color of each dot yielding a plot that can visualize 5 dimensions; but such a plot will be virtually indecipherable. The scatterplot matrix can also be an option but it is more geared to visualizing binary relationships than the feature value of each point. Moreover, it becomes more space inefficient as the number of dimensions grow. Another alternative can be the biplot but it can be difficult to read and involves information loss. The radar plot provides for a succinct representation to visualize multivariate data but it can only handle a few points before polygons overlap and it becomes unreadable (defeating G4). 
We ended up with the parallel coordinate (PC) plot \cite{inselberg1990parallel} based on our design goals \textbf{G1-G4}. PC can visualize a significant number of points with multiple dimensions without any information loss (G1, G4). It also facilitates bias exploration and adding new bias types. To support plotting large numbers of points, we chose canvas over SVG and also used progressive rendering \cite{progressive_rendering} (\textbf{G4}).

\subsection{Feature Scaling}
\begin{figure}[h!]
  \begin{subfigure}[b]{0.3\textwidth}
    \includegraphics[width=\textwidth]{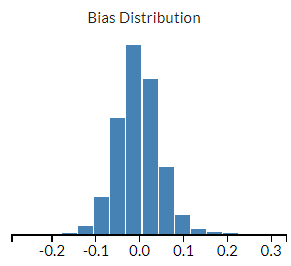}
    \caption{Raw bias scores}
  \end{subfigure}
  \hfill
  \begin{subfigure}[b]{0.3\textwidth}
    \includegraphics[width=\textwidth]{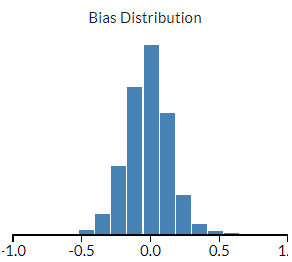}
    \caption{Min-Max Normalization}
  \end{subfigure}
  \hfill
  \begin{subfigure}[b]{0.3\textwidth}
    \includegraphics[width=\textwidth]{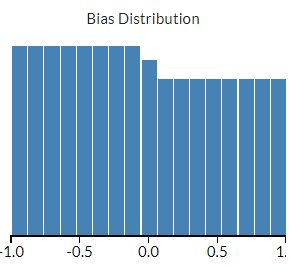}
    \caption{Percentile Ranking}
  \end{subfigure}
  \caption{Distribution of bias scores across 50k words in the Word2Vec Embedding.}
  \label{fig:feat_scaling}
\end{figure}

\begin{figure}[h!]
  \begin{subfigure}[b]{0.45\textwidth}
    \includegraphics[width=\textwidth]{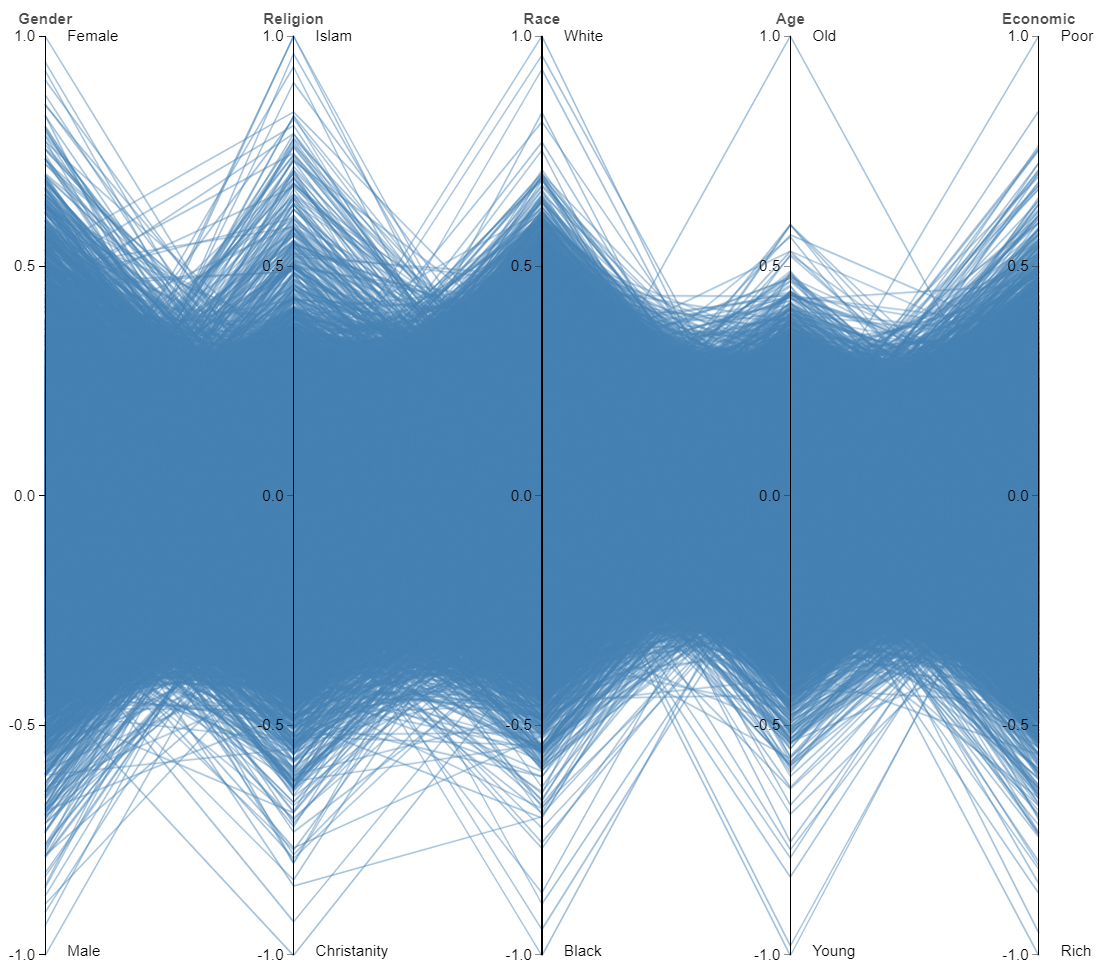}
    \caption{Min-Max Normalization}
  \end{subfigure}
  \hfill
  \begin{subfigure}[b]{0.45\textwidth}
    \includegraphics[width=\textwidth]{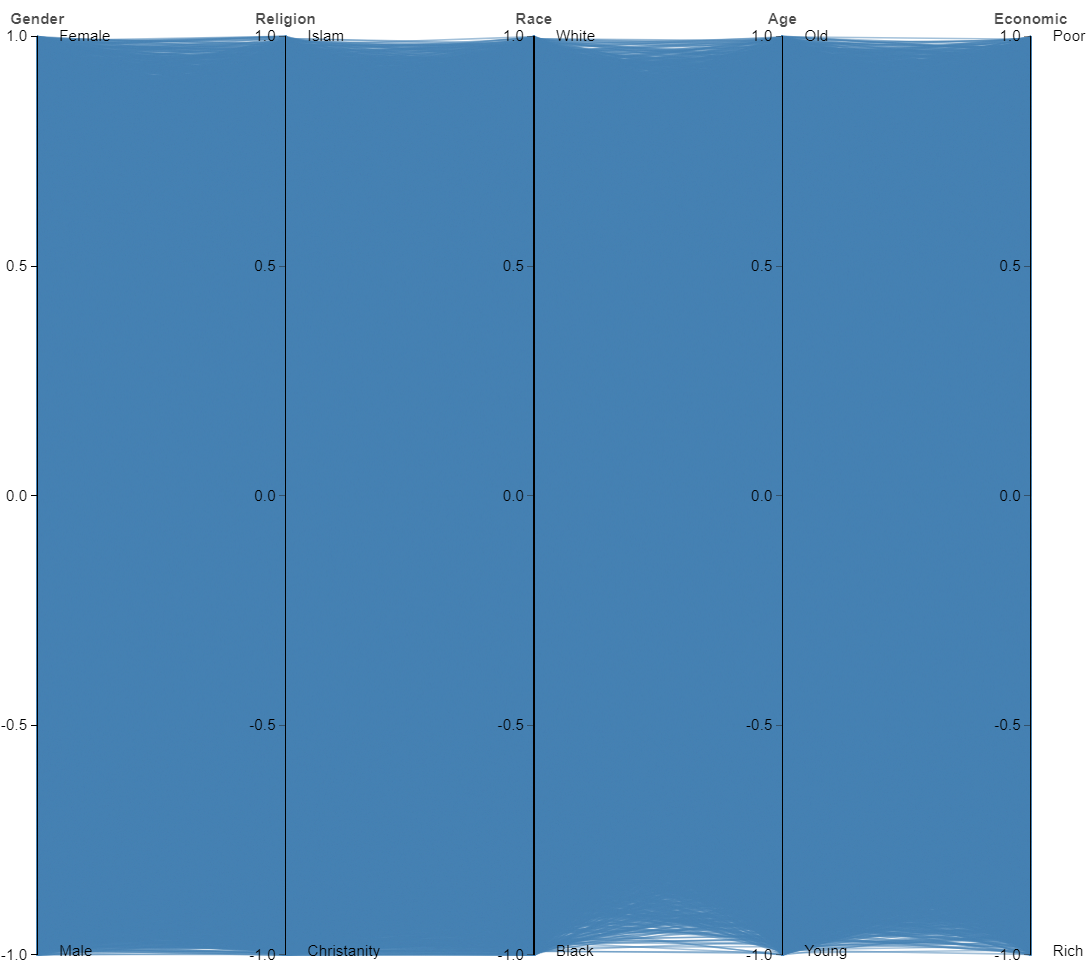}
    \caption{Percentile Ranking}
  \end{subfigure}
  \hfill
  \caption{Parallel coordinate plot for 50k words in the Word2Vec Embedding.}
  \label{fig:pc_feat_scaling}
\end{figure}

WordBias allows the user to choose between raw bias scores and two feature scaling methods namely, Min-Max Normalization and Percentile Ranking. Raw bias scores provides the most accurate representation but it can be a bit difficult to interpret. The other two feature scaling options makes the bias scores more comparable across bias types. \autoref{fig:feat_scaling} shows the distribution of mean bias scores for all 3 options. As we can see, the distribution of bias scores appear similar for (a) and (b) but there ranges on x-axis vary. This is because Min-Max normalization simply stretches the raw bias scores over the range [-1,1]. This signifies that a large majority of words have small bias scores and only a few words on either ends have high bias scores. The distribution for Percentile ranking (\autoref{fig:feat_scaling} (c)) is quite different and interesting. It has the same range on x-axis [-1,1] as Min-Max normalization but the distribution of words across bias scores is much more uniform. We can observe the the bar length is different for bias scores greater than and less than 0. This is because we applied percentile ranking in a piece-wise fashion depending on the sign of the bias scores. \autoref{fig:pc_feat_scaling} further elucidates the difference in distribution of bias scores for Min-Max normalization and Percentile ranking.    

\subsection{Group Words}
\label{group_words}
They play a critical role in computation of bias scores~\cite{group_words}. Hence, if the user chooses to add a new bias type (axis), they should choose the words carefully to get an accurate picture. In our case, we have used group words which have been proposed in existing literature. So far, there is no objective way to choose group words. However, our tool can assist in selecting the most relevant group words from a set of group words. Let's say the user wants to add a new axis for 'political orientation' and they have multiple set of group words to choose from. In such a case, the user can add multiple axes corresponding to each set of group words. Thereafter, the user can explore and compare bias scores for different words across these axes. Group words corresponding to the axis which best aligns with the user's domain knowledge can be chosen.

By default, WordBias shows 5 kinds of biases namely Gender, Religion, Age, Race and Economic. Following are the list of words used to compute bias scores for each of those categories. These words are derived from existing literature \cite{kozlowski2019geometry, 100years, dev2019attenuating}. If any of these words aren't contained in the word embedding, they are ignored.    \\ 

\textbf{Male} (Gender) \cite{100years} \\
he, son, his, him, father, man, boy, himself, male, brother, sons, fathers, men, boys, males, brothers, uncle, uncles, nephew, nephews \\

\textbf{Female} (Gender) \cite{100years} \\ 
she, daughter, hers, her, mother, woman, girl, herself, female, sister, daughters, mothers, women, girls, sisters, aunt, aunts, niece, nieces \\

\textbf{Young} (Age) \cite{dev2019attenuating} \\ 
Taylor, Jamie, Daniel, Aubrey, Alison, Miranda, Jacob, Arthur, Aaron, Ethan \\

\textbf{Old} (Age) \cite{dev2019attenuating} \\
Ruth, William, Horace, Mary, Susie, Amy, John, Henry, Edward, Elizabeth  \\

\textbf{Islam} (Religion) \cite{100years} \\
allah, ramadan, turban, emir, salaam, sunni, koran, imam, sultan, prophet, veil, ayatollah, shiite, mosque, islam, sheik, muslim, muhammad \\

\textbf{Christainity} (Religion) \cite{100years} \\ 
baptism, messiah, catholicism, resurrection, christianity, salvation, protestant, gospel, trinity, jesus, christ, christian, cross, catholic, church \\

\textbf{Black} (Race) \cite{kozlowski2019geometry} \\
black, blacks, Black, Blacks, African, african, Afro  \\


\textbf{White} (Race) \cite{kozlowski2019geometry} \\
white, whites, White, Whites, Caucasian, caucasian, European, european, Anglo \\


\textbf{Rich} (economic) \cite{kozlowski2019geometry} \\
rich, richer, richest, affluence, advantaged, wealthy, costly, exorbitant, expensive, exquisite, extravagant, flush, invaluable, lavish, luxuriant, luxurious, luxury, moneyed, opulent, plush, precious, priceless, privileged, prosperous, classy \\

\textbf{Poor} (economic) \cite{kozlowski2019geometry} \\
poor, poorer, poorest, poverty, destitude, needy, impoverished, economical, inexpensive, ruined, cheap, penurious, underprivileged, penniless, valueless, penury, indigence, bankrupt, beggarly, moneyless, insolvent

\subsection{Neutral Words}
To quickly audit a given embedding for different biases, WordBias provides a set of words which should ideally be neutral for most kinds of biases like gender, race, etc. Following is the list of such neutral words based on different categories which are derived from existing literature \cite{100years, narayan2017semantics}. \\

\textbf{Profession} \\ 
teacher, author, mechanic, broker, baker, surveyor, laborer, surgeon, gardener, painter, dentist, janitor, athlete, manager, conductor, carpenter, housekeeper, secretary, economist, geologist, clerk, doctor, judge, physician, lawyer, artist, instructor, dancer, photographer, inspector, musician, soldier, librarian, professor, psychologist, nurse, sailor, accountant, architect, chemist, administrator, physicist, scientist, farmer \\

\textbf{Physical Appearance} \\ 
alluring, voluptuous, blushing, homely, plump, sensual, gorgeous, slim, bald, athletic, fashionable, stout, ugly, muscular, slender, feeble, handsome, healthy, attractive, fat, weak, thin, pretty, beautiful, strong \\

\textbf{Extremism} \\ 
terror, terrorism, violence, attack, death, military, war, radical, injuries, bomb, target,conflict, dangerous, kill, murder, strike, dead, violence, fight, death, force, stronghold, wreckage, aggression,slaughter, execute, overthrow, casualties, massacre, retaliation, proliferation, militia, hostility, debris, acid,execution, militant, rocket, guerrilla, sacrifice, enemy, soldier, terrorist, missile, hostile, revolution, resistance, shoot \\

\textbf{Personality Traits} \\
adventurous, helpful, affable, humble, capable, imaginative, charming, impartial, confident, independent, conscientious, keen, cultured, meticulous, dependable, observant, discreet, optimistic, persistent, encouraging, precise, exuberant, reliable, fair, trusting, fearless, valiant, gregarious, arrogant, rude, sarcastic, cowardly, dishonest, sneaky, stingy, impulsive, sullen, lazy, surly, malicious, obnoxious, unfriendly, picky, unruly, pompous, vulgar \\

\subsection{Few Examples}
In the following, we list a few words along with their associated subgroups as per WordBias. Here, we have chosen percentile ranking and considered an association significant if its corresponding bias score is $>= 0.5$. 

\newpage
\textbf{(i) nazi} : Male - Christianity - White - Poor
\begin{figure}[H] 
  \centering 
  \includegraphics[width=0.70\columnwidth]{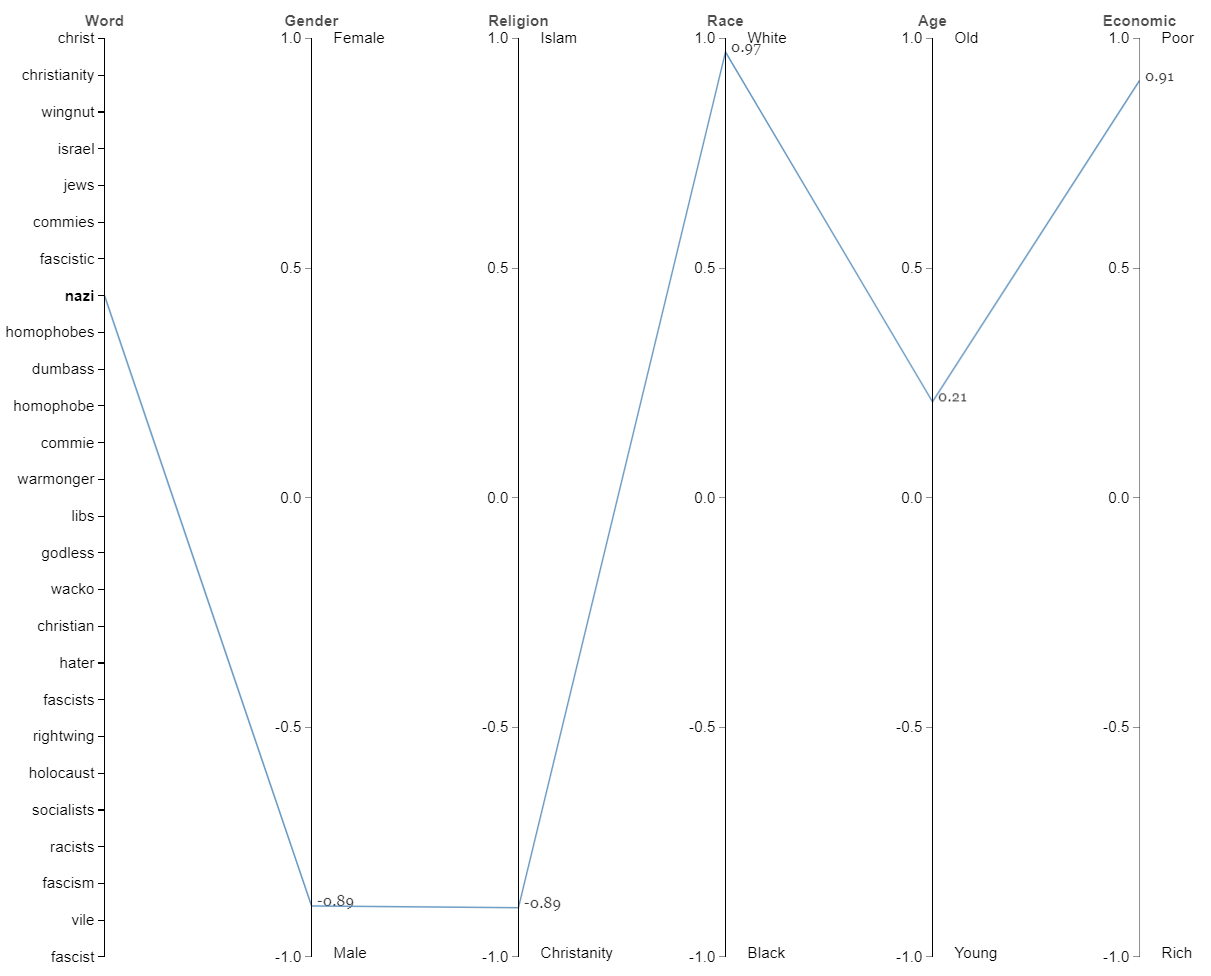}
  \caption{Biases associated with the word `nazi'}
\end{figure}

\textbf{(ii) beautiful} : Female - Christianity - Old - Rich
\begin{figure}[H] 
  \centering 
  \includegraphics[width=0.70\columnwidth]{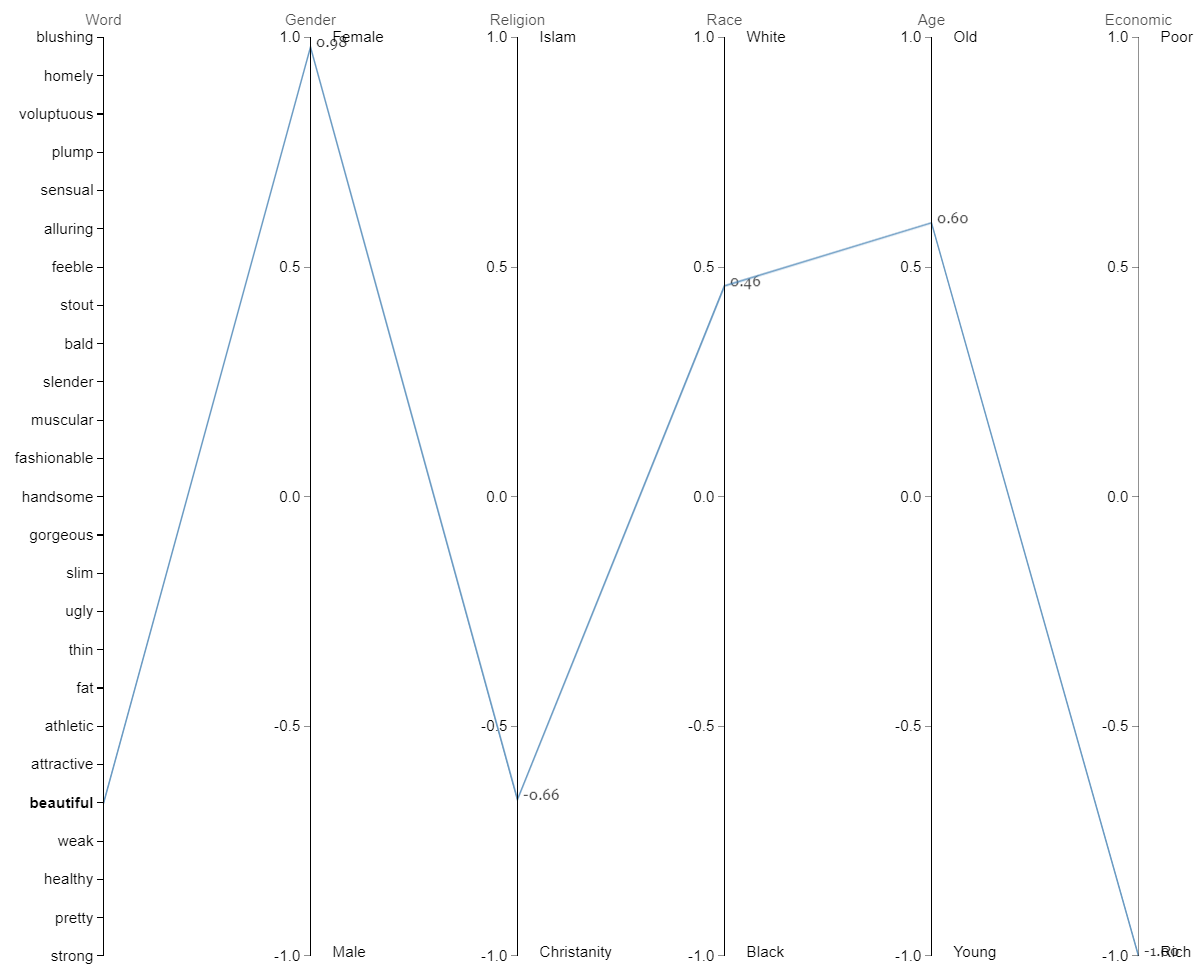}
  \caption{Biases associated with the word `beautiful'}
\end{figure}

\newpage
\textbf{(iii) pretty} : Christianity - White - Young
\begin{figure}[H] 
  \centering 
  \includegraphics[width=0.70\columnwidth]{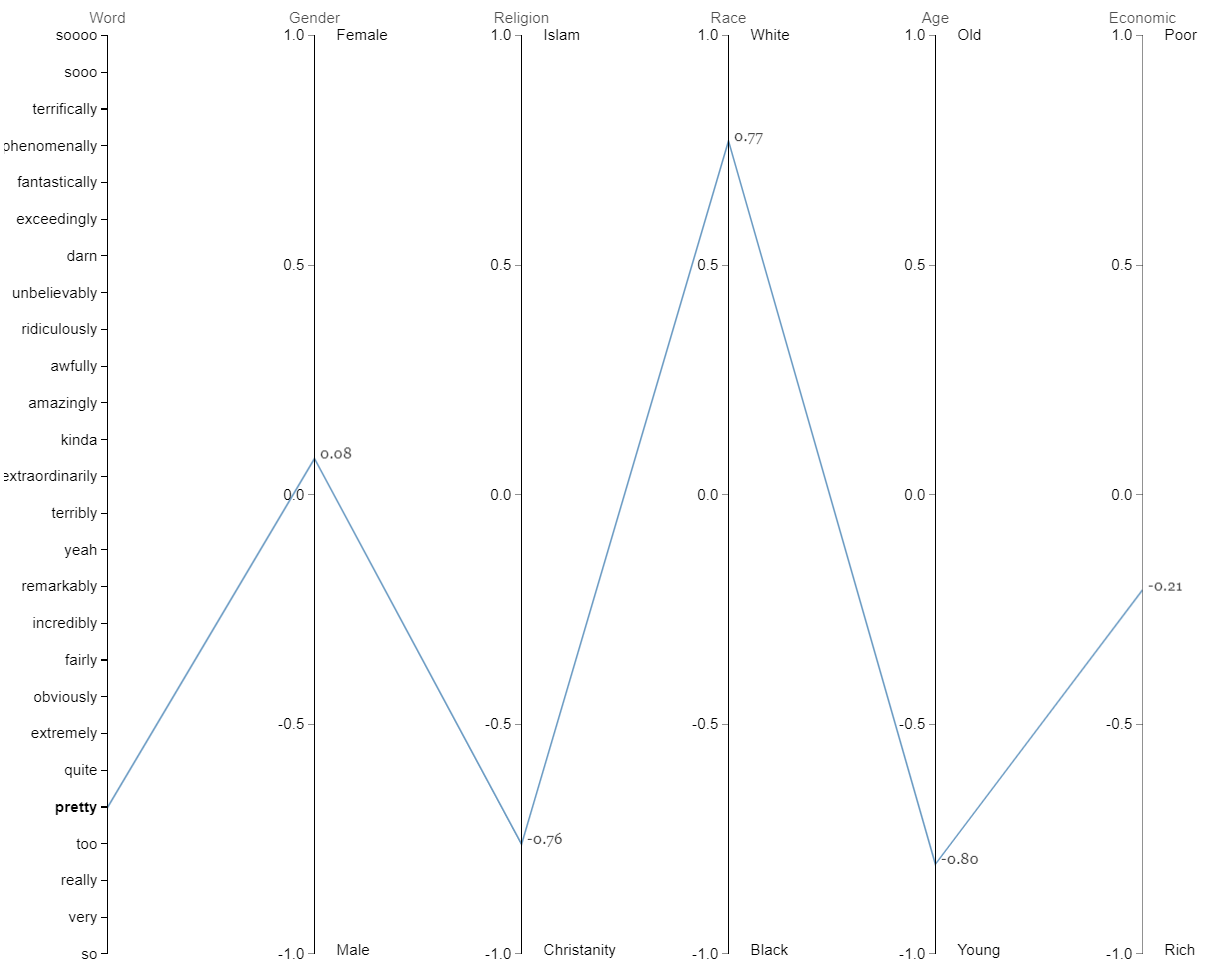}
  \caption{Biases associated with the word `pretty'}
\end{figure}

\textbf{(iv) homicides} : Female - Black - Poor
\begin{figure}[H] 
  \centering 
  \includegraphics[width=0.70\columnwidth]{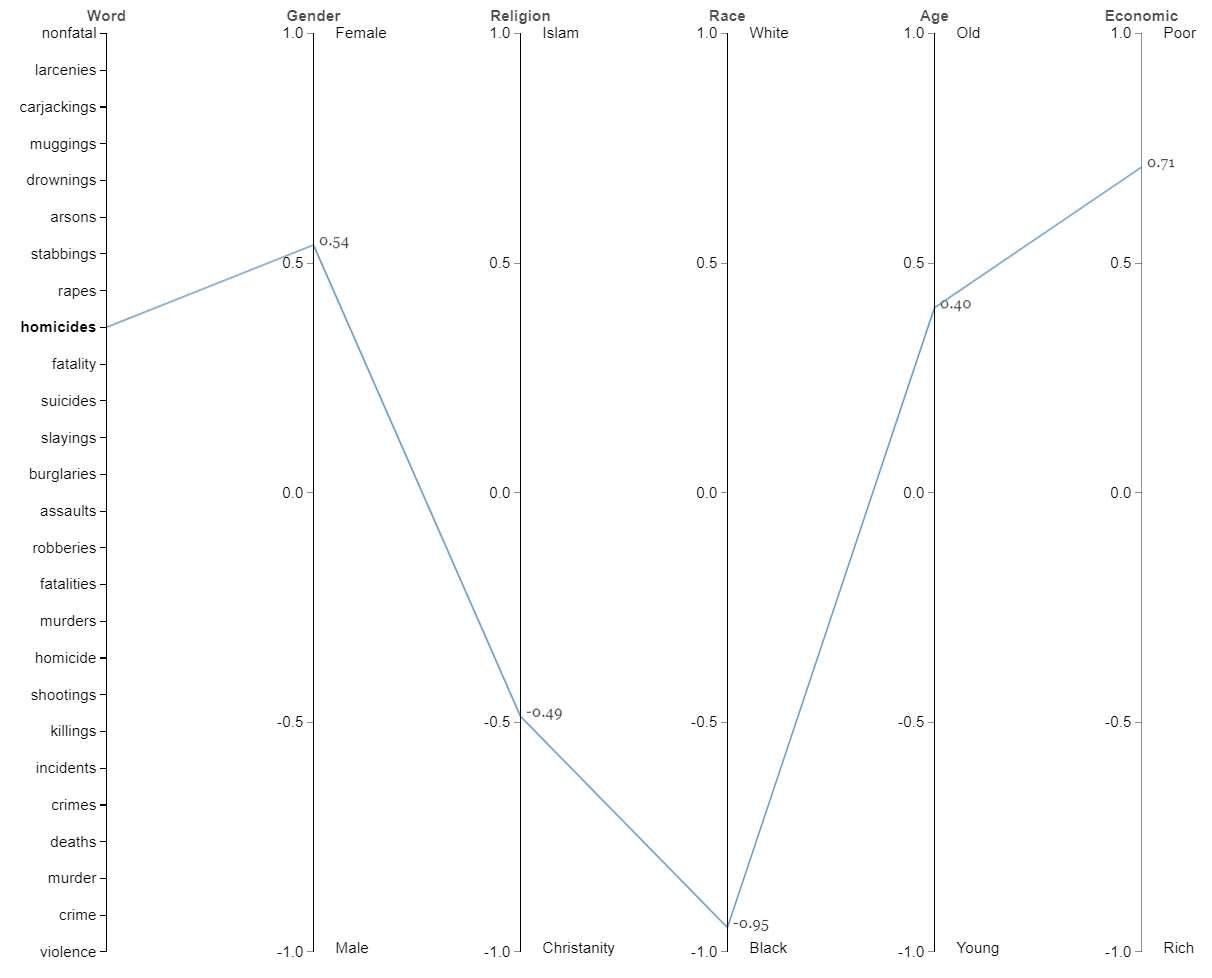}
  \caption{Biases associated with the word `homicides'}
\end{figure}

\newpage
\textbf{(v) picky} : Female - White - Young - Rich
\begin{figure}[H] 
  \centering 
  \includegraphics[width=0.70\columnwidth]{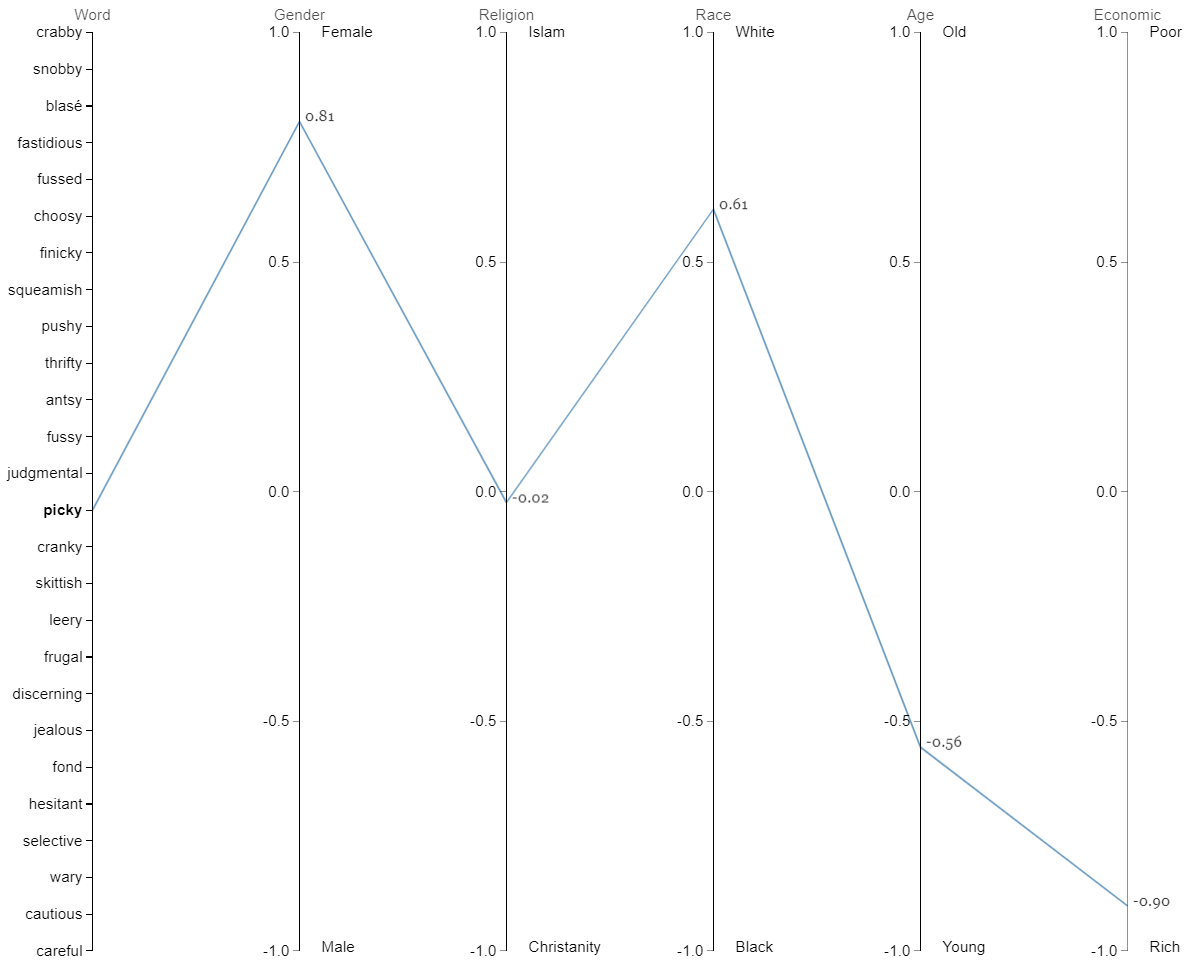}
  \caption{Biases associated with the word `picky'}
\end{figure}

\textbf{(vi) terror} : Male - Islam - Young
\begin{figure}[H] 
  \centering 
  \includegraphics[width=0.70\columnwidth]{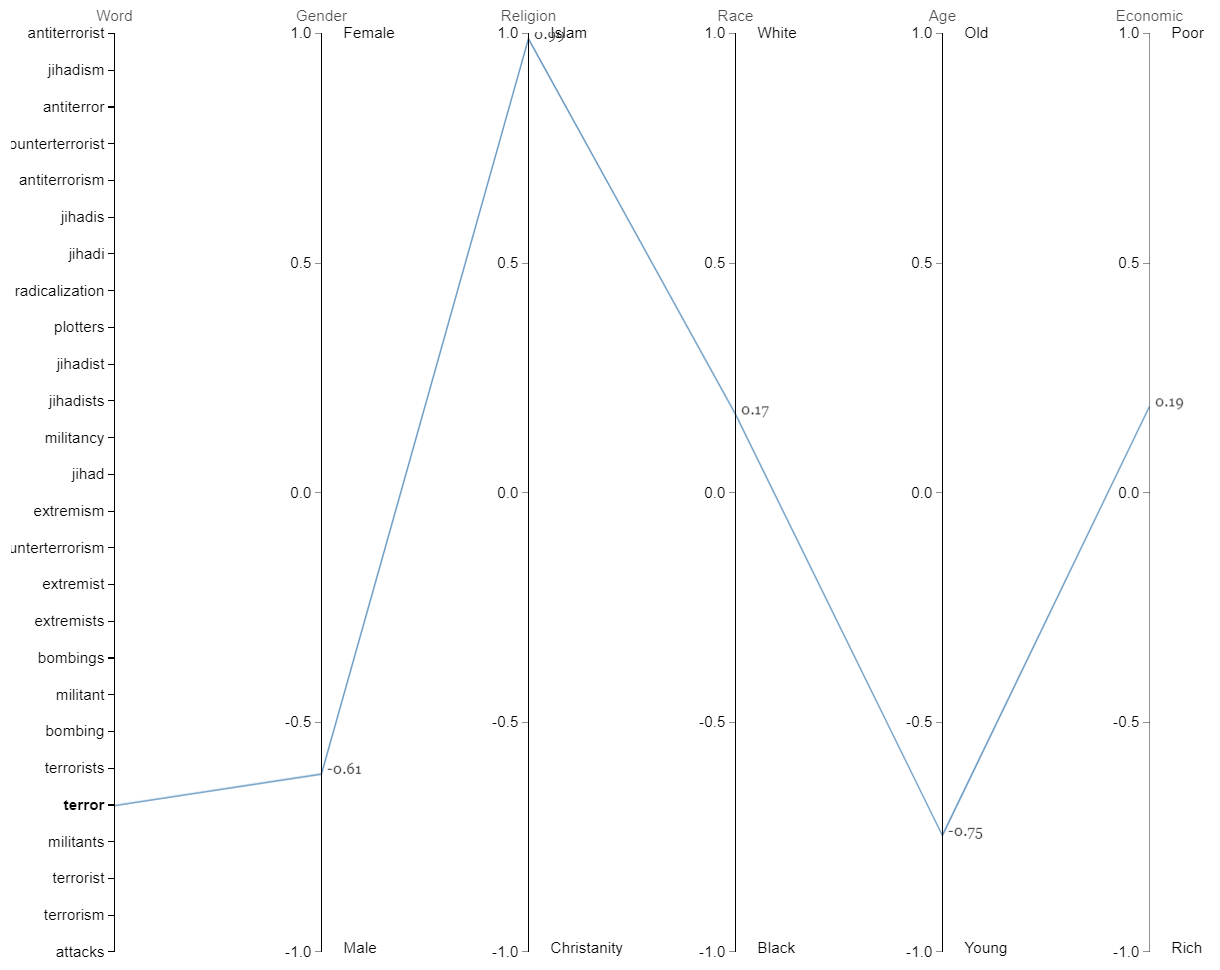}
  \caption{Biases associated with the word `terror'}
\end{figure}

\newpage
\textbf{(vii) prostitute} : Female - Poor
\begin{figure}[H] 
  \centering 
  \includegraphics[width=0.70\columnwidth]{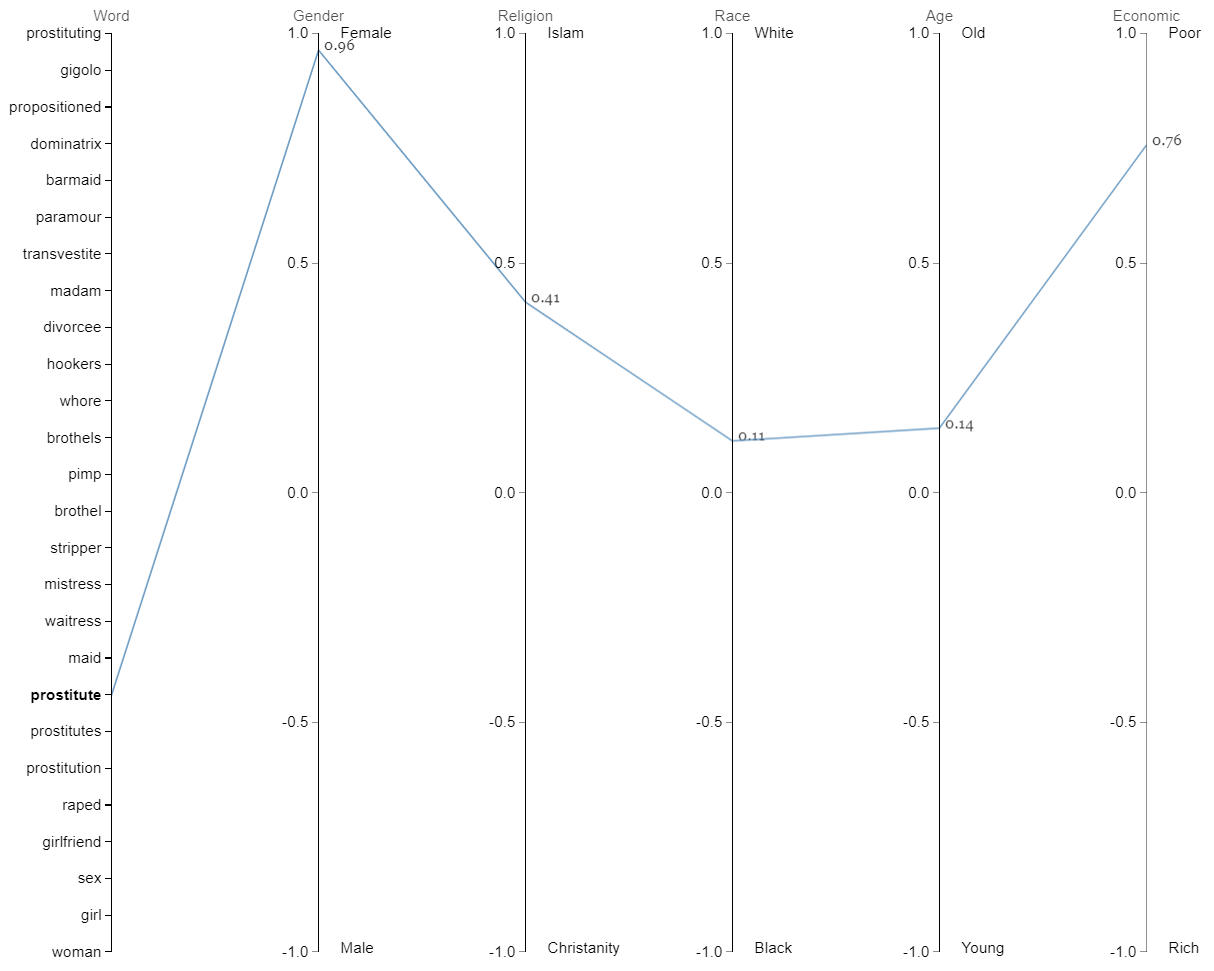}
  \caption{Biases associated with the word `prostitute'}
\end{figure}

\textbf{(viii) clever} : Male - Christianity - Young - Rich
\begin{figure}[H] 
  \centering 
  \includegraphics[width=0.70\columnwidth]{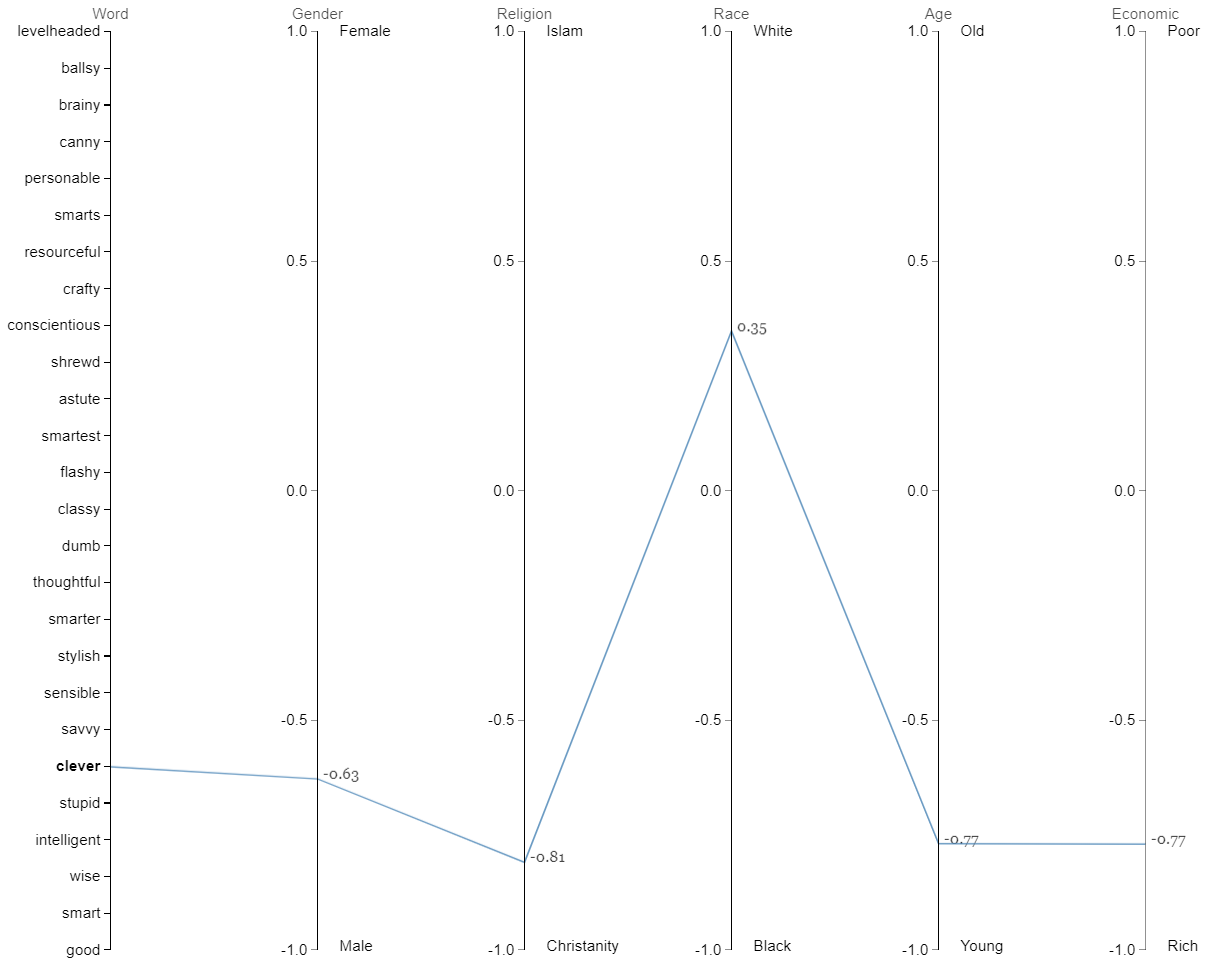}
  \caption{Biases associated with the word `clever'}
\end{figure}

\newpage
\textbf{(ix) dictator} : Male - Islam - Black - Old - Poor
\begin{figure}[H] 
  \centering 
  \includegraphics[width=0.70\columnwidth]{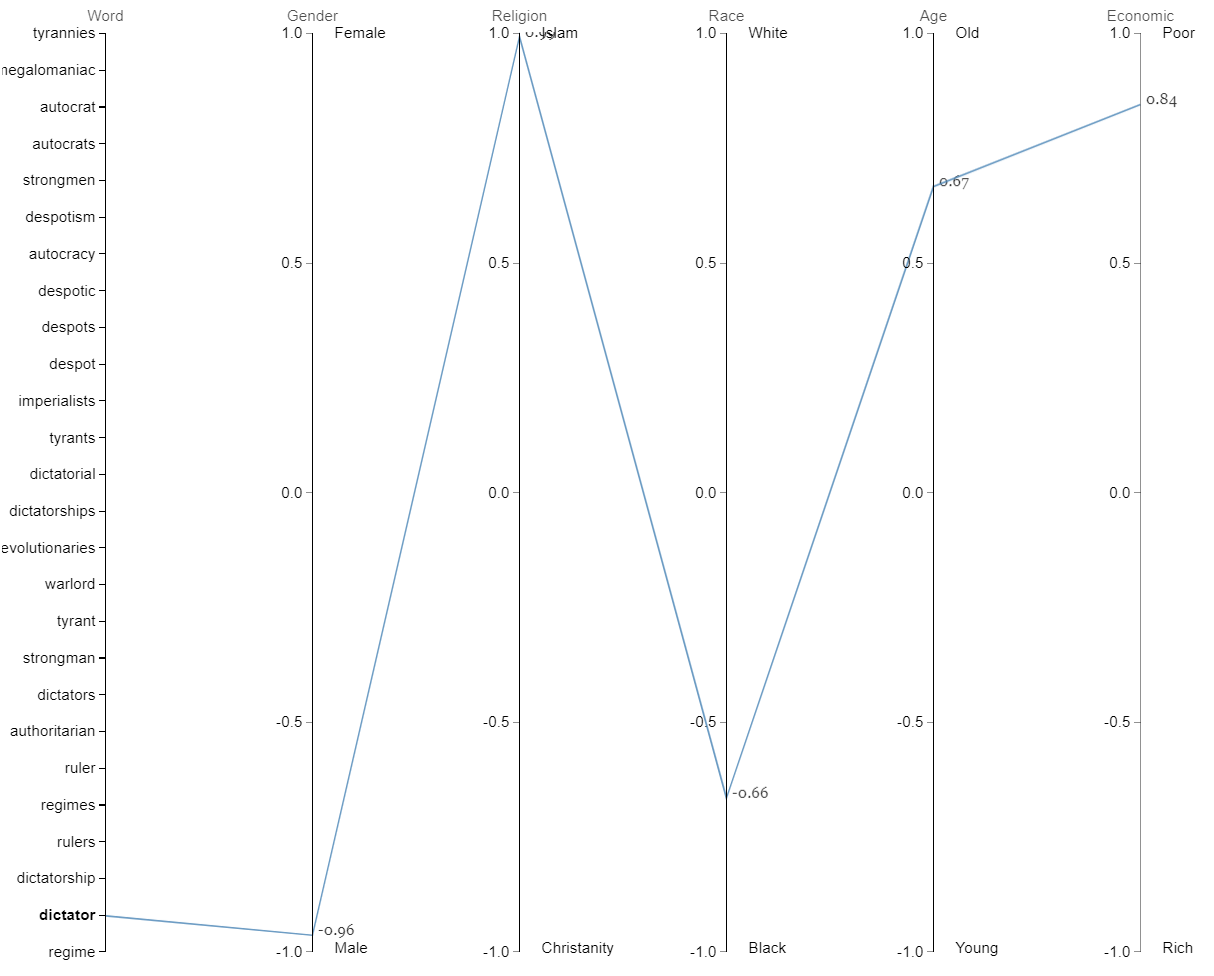}
  \caption{Biases associated with the word `dictator'}
\end{figure}

\textbf{(x) janitor} : Male - Old - Poor
\begin{figure}[H] 
  \centering 
  \includegraphics[width=0.70\columnwidth]{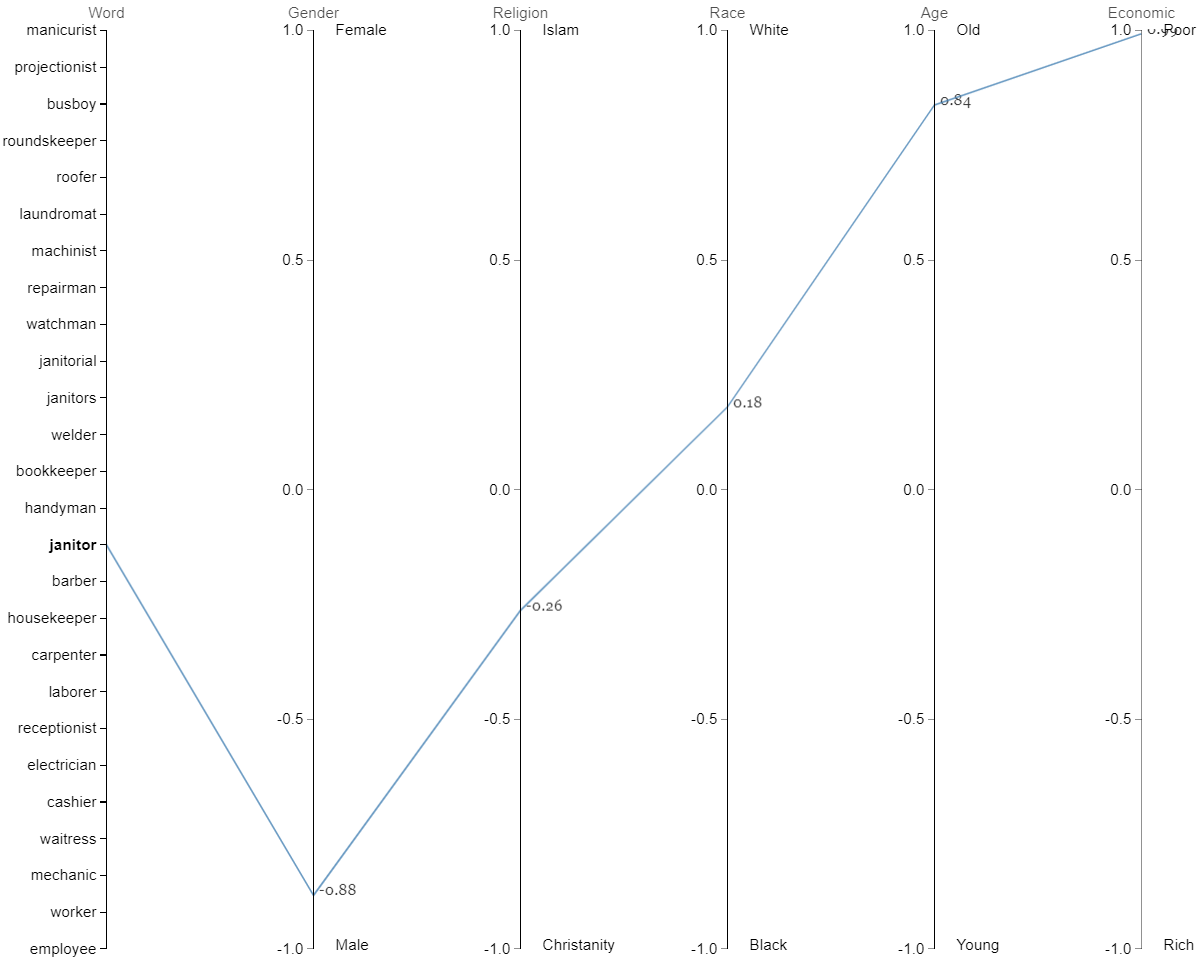}
  \caption{Biases associated with the word `janitor'}
\end{figure}

\newpage
\textbf{(xi) militia} : Male - Islam - Black - Poor
\begin{figure}[H] 
  \centering 
  \includegraphics[width=0.70\columnwidth]{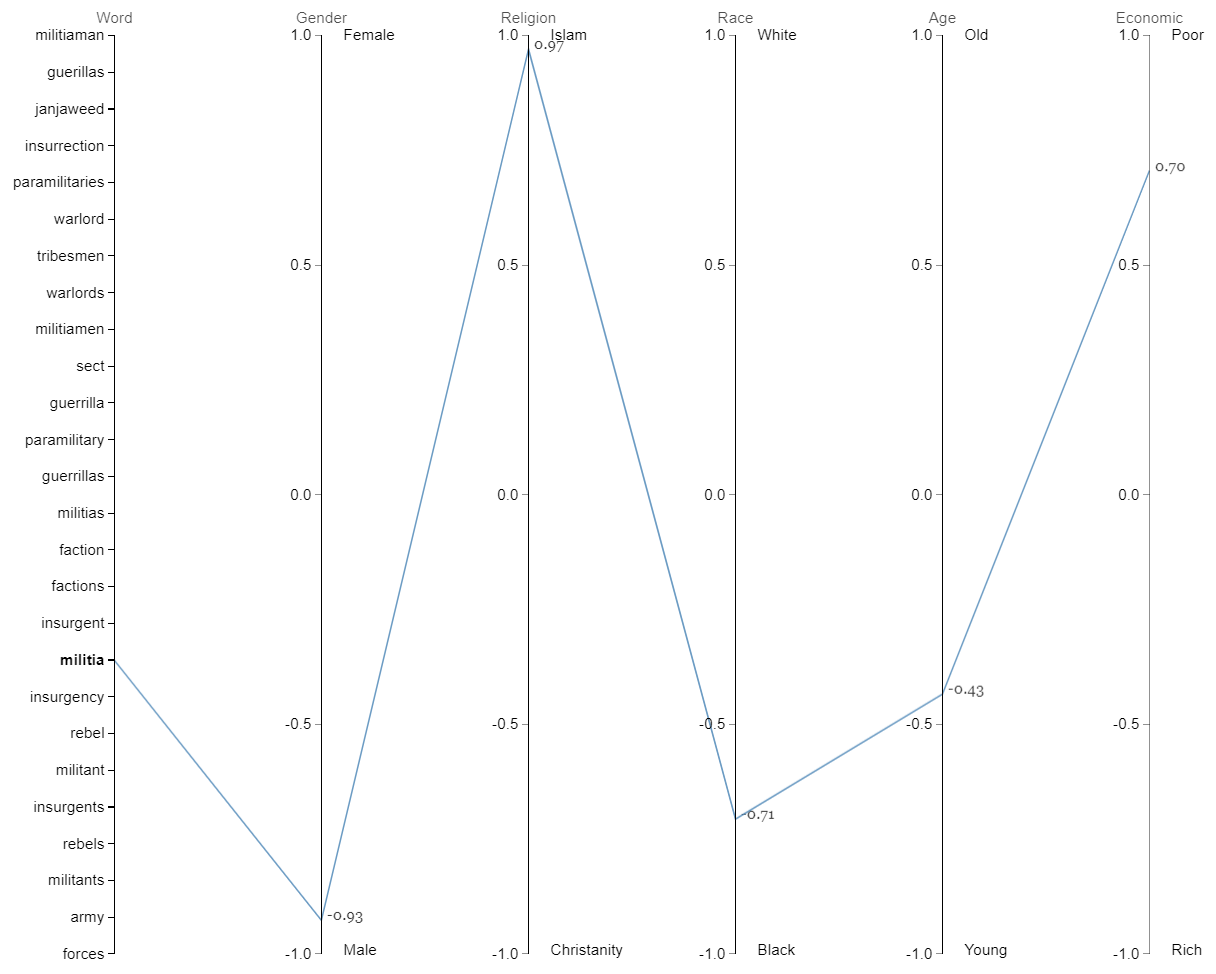}
  \caption{Biases associated with the word `militia'}
\end{figure}

\end{document}